\documentstyle[12pt]{article}
\newcommand{\ket}[1]{| #1 \rangle}
\newcommand{\vev}{\langle\bar{\lambda}\lambda\rangle}
\newcommand{\beq}{\begin{equation}}
\newcommand{\eeq}{\end{equation}}
\newcommand{\bea}{\begin{eqnarray}}
\newcommand{\eea}{\end{eqnarray}}
\newcommand{\bq}{\begin{equation}}
\newcommand{\eq}[1]{\label{#1}\end{equation}}
\newcommand{\ber}{\begin{eqnarray}}
\newcommand{\eer}[1]{\label{#1}\end{eqnarray}}

\newcommand{\be}{\begin{equation}}
\newcommand{\ee}{\end{equation}}

\newcommand{\ba}{\begin{eqnarray}}
\newcommand{\ea}{\end{eqnarray}}

\def\fr#1#2{{#1\over #2}}
\def\du#1{\tilde #1}
\def\mat#1#2#3#4{{\left(\matrix{#1&#2\cr#3&#4\cr}\right)}}

\def\m{\mu}
\def\n{\nu}

\def\s{\sigma}
\def\f{\phi}
\def\F{\Phi}
\def\G{\Gamma}

\def\p{\psi}
\def\th{\theta}
\def\td{\du\th}
\def\o{\omega}
\def\a{\alpha}

\def\d{\delta}
\def\e{\epsilon}
\def\l{\lambda}

\def\S{\Sigma}
\def\k{{\bf k}}         

\def\pa{\partial}
\def\pab{\bar\pa}

\def\ie{{\it i.e.},\ }

\def\cf{{\it cf.}\ }
\def\ind{{1\over 2\pi}\int\!\! d^2\! z\,}

\def\AB{{\bar A}}
\def\JB{{\bar J}}
\def\q{\, , \qquad}
\def\sq{\, , \quad}
\def\rR{\mbox{\scriptsize\bf R}}
\def\zZ{\mbox{\scriptsize\bf Z}}
\def\R{{\bf R}}
\def\C{{\bf C}}
\def\Z{{\bf Z}}
\def\[{\Bigl[}
\def\]{\Bigr]}
\def\gl{\G^1_{ai}}
\def\gz{\G^2_{ia}}
\def\gab{\G_{ab}}
\def\ra{\rightarrow}

\def\t{{\tanh}^2x}

\def\sh{{\sinh}^2x}
\def\ch{{\cosh}^2x}
\def\d{\partial}
\def\db{\bar{\partial}}
\def\G{\Gamma}
\def\Jb{\bar{J}}
\def\a{\alpha}
\def\tha{\theta_1}
\def\thb{\theta_2}
\def\tht{\tilde{\theta}}

\def\l{\lambda}
\def\p{\phi}
\def\e{\epsilon}
\catcode`@=11           
\dimendef\dimen@=0
\def\ialign{\everycr{}\tabskip\z@skip\halign} 
\def\m@th{\mathsurround=\z@}
\def\openup{\afterassignment\@penup\dimen@=}
\def\@penup{\advance\lineskip\dimen@
  \advance\baselineskip\dimen@ \advance\lineskiplimit\dimen@}
\def\eqalign#1{\null\,\vcenter{\openup\jot\m@th
  \ialign{\strut\hfil$\displaystyle{##}$&$\displaystyle{{}##}$\hfil
\crcr#1\crcr}}\,}
\catcode`@=12           

\def\,{\kern .16667em}  

\begin{document}
\addtolength{\baselineskip}{.7mm}
\thispagestyle{empty}
\begin{flushright}
{\sc RI}-1-94\\ {\sc NYU-TH}.94/01/01\\ hep-th/9401139\\ January 1994
\end{flushright}
\vspace{1cm}

\begin{center}
{\Large \bf Target Space Duality in String Theory}
\end{center}
\vskip 1cm

\centerline{\large Amit Giveon\footnote{Email: giveon@vms.huji.ac.il}}
\vspace{0.5cm}

\centerline{\it Racah Institute of Physics, The Hebrew University}
\centerline{\it Jerusalem, 91904, ISRAEL}
\vspace{0.5cm}

\centerline{\large Massimo Porrati\footnote{
On leave of absence from INFN, sez.
di Pisa, Pisa, Italy. Email: porrati@mafalda.nyu.edu}}
\vspace{0.5cm}

\centerline{\it Department of Physics, New York University}
\centerline{\it New York, NY 10003, USA}
\vspace{0.5cm}

\centerline{\large Eliezer Rabinovici\footnote{Email: eliezer@vms.huji.ac.il}}
\vspace{0.5cm}

\centerline{\it Racah Institute of Physics, The Hebrew University}
\centerline{\it Jerusalem, 91904, ISRAEL}

\vspace{1.5cm}

\centerline{\bf Abstract}
\vspace{.2cm}

\noindent
\parbox{15cm}{Target space duality and discrete symmetries in string theory
are reviewed in different settings.}
\vfill
\eject


%
\tableofcontents
\listoftables
\newpage

\begin{section}{Introduction, Summary, and Omissions}
\setcounter{equation}{0}

String theory (see for example \cite{GSW})
assumes that the elementary particles are one
dimensional extended objects rather than point like ones.  String
theory also comes equipped with a scale associated, nowadays, with
the Planck scale ($10^{-33}$ cm).  The standard model describing the
color and electro-weak interactions is based on the point particle
notion and is successful at the Fermi  scale of about 100 Gev
($10^{-16}$ cm) which is $10^{-17}$
smaller than the Planck scale.

The correspondence principle requires thus that string theory when
applied to these low energies resembles a point particle
picture. In fact string theory has an interpretation in terms of
point-like field theory whose spectrum consists of an infinity of
particles, all except a finite number of which have a mass of the order of the
Planck scale. Integrating out the massive modes leads to an effective
theory of the light  particles.

There exists another class of theories
whose spectrum consists as well of an infinite tower of
particles; this is the Kaluza-Klein type.  In such a class of
theories \cite{Kaluza,Klein,KKbook}, gravity is essentially the sole basic
interaction, and
space-time is assumed to have, in addition to four  macroscopic
dimensions, extra microscopic dimensions, characterized by some small
distance scale. The standard model is assumed to be a low-energy
effective action of the light particles resulting from the purely
gravitational higher dimensional system.

String theory can also be
viewed in many cases as representing a space-time with extra dimensions.
Nevertheless, the effective low-energy theory, emerging from string
theory, turns out to be different from that resulting from a field theory
with an infinite number of particles; it possesses many more
symmetries.

String theory shows also differences when physics is probed at a scale
much smaller than the Planck one. In fact, there are various
hints that in string theory physics at a very small scale cannot be
distinguished from physics at a large scale.  A very striking example of
that feature is that a string cannot tell if it is propagating on a
space-time with one circular dimension of radius $R$ (a dimensionless
number) times the Planck scale or $1/R$ the Planck scale (see figure 1.A).
The discrete symmetry apparent in the example is termed {\em target space
duality}. Moreover, there are indications that string theory possesses
an extremely large symmetry of that nature.
A study of that symmetry is the subject of the review.

The term duality has been used time and again in physics, always for
very noble causes. The duality between particle and wave was the
precursor of quantum theory. In the Ising model for magnets it was
found that a spin system at temperature $T$ has identical properties as
the same  system at an inverse temperature, once an appropriate
dictionary is used\footnote{Notice that this temperature duality, outside
the second-order phase-transition point, relates different {\em
non-critical} systems. These systems are {\em not} scale invariant.
The duality we discuss in this review, instead, relates critical systems.}.
This property and its generalizations have been
found in many more statistical mechanical systems, some of
them (such as lattice gauge theories) are of interest to particle
physicists; it
has led to the notion of a disorder variable. In the presence of
monopoles, the electromagnetic interactions posses a duality allowing
the exchange of the magnetic and the electric properties. This type of
duality is very explicit in certain supersymmetric systems. String
theory evolved out of the study of the dual description of two body
interactions in terms of the direct channel ($s$-channel) scattering and
exchange process (such as $t$-channel). Many features of these various
types of duality appear as facets of the duality addressed in this
review; none is identical to it.

The structure of the review is the following:
Section 2 is of an introductory nature; the basic tools and concepts are
introduced in a simple setting. In the other sections they are applied
in more involved cases.

In section 2 we discuss the notion of a string background that is a
an allowed space-time for string propagation\footnote
{Not all such backgrounds need to have a
geometric description even classically.}.
There are restrictions on those space-times as they need to obey various
constraints.  In the early days, it was thought that as a consequence of
such constraints the bosonic string could only propagate in 26
dimensions while a superstring could only propagate in 10 dimensions.
Nowadays, one is  aware of an infinite number of different solutions of
these constraints, however, their complete
classification is not known. It is also not
known if all such solutions  can be  smoothly related or if they consist
of disconnected islands (see figure 1.B).

In the same section we focus on a particular class of such solutions in
which space-time is flat but has a global structure, i.e., several of the
dimensions are compactified on tori.
In that sub-space there are transformations
relating the various solutions. Some of these
transformations turn out to be symmetries, that is they relate
solutions which seem different in terms of the data which classifies
them, but actually describe the same physics; the $R \to 1/R$
duality is one of them (see figure 1.C).
We discuss the group structure of this symmetry of the toroidal
compactifications in detail.
Moreover, we study relations between this symmetry group
and gauge symmetries. The large gauge group thus uncovered
can be viewed as a broken version of an even larger symmetry,
offering a glimpse at the underlying symmetries of string theory.

An attempt to implement this symmetry on the low-energy effective
theory indeed provides an example of what this symmetry could be
like. This symmetry needs to relate massless and massive modes, as in
string theory it turns out that the massive modes cannot really be
decoupled, and they eventually return. This is also worked out in section 2.

Having treated the case of flat compactifications and having introduced
the key ideas, one ventures into less chartered territory.
In section 3 one discusses duality in models which seem closer to
phenomenology. The string is moving on allowed backgrounds which
consist of a special class of compact dimensions called Calabi-Yau
spaces; such spaces manifest themselves in space-time supersymmetric
theories. In order to study such systems, it is useful to introduce the
notion of spaces of an orbifold nature. These are spaces which, in
particular, contain conical singularities.
A compact dimension consisting a closed
line including its two end points is the simplest example of this nature;
it is the $\Z_2$ orbifold of a circle,
constructed by identifying a point $X$ on the circle with $-X$ (see figure
1.D).

Point particle quantum mechanics senses the singular nature of these
backgrounds, however, string theory smoothens these singular effects, a very
encouraging feature. In section 3, string theory and duality on such
objects are described. Another surprise of string theory is that in some
cases a string cannot distinguish if it is moving on an orbifold or on a
circle (see figure 1.E).

The methods developed for orbifolds are applied in section 3  to
discuss additional aspects of the duality invariant low-energy effective
action, as well as some aspects of duality on Calabi-Yau manifolds.
Moreover, in the same
section one approaches phenomenology closer by studying the
string quantum corrections to the tree approximation, and the impact
of the duality symmetry on them.

In section 4, duality properties of curved space-times are discussed.
For example, a string cannot distinguish on which of
the backgrounds displayed in figure 1.F it moves.
The infinite cigar in figure 1.F represents a Euclidean continuation of a
two dimensional black hole. In
order to identify the manifolds related by duality symmetry, another method
is explained in section 4. While progress can be made even for curved
backgrounds, some degree of symmetry is required from the manifold
in order to apply the method. The role of these extra symmetries is
discussed, and the method is then applied to various examples:
\begin{itemize}

\item
Neutral and charged black objects (black holes, black $p$-branes) and other
singular objects.

\item
Cosmological string backgrounds (expanding and
contracting universes, in the presence of matter).

\end{itemize}

\noindent
This method leads to some surprising consequences:

\begin{itemize}
\item
The interchange of singularities with horizons, and the
removal of singularities in string theory.

\item
A relation between neutral objects and charged objects.

\item
Topology change in string theory.

\end{itemize}

\noindent
Moreover, it is shown in section 4
that a particularly interesting target space
duality, termed {\em axial-vector duality}, is related to gauge
symmetries.

Finally, in section 5, we discuss some interplay between the worldsheet and
the target space. In string theory, one is very careful to distinguish
between the worldsheet and the target space. In sections 2-4 a large discrete
global symmetry emerged in many types of target spaces; this symmetry bears
quite some resemblance to global discrete symmetries on the worldsheet
(called ``modular transformations'').
In section 5 we discuss some mathematical facts
which may suggest that the worldsheet and the target space are not
necessarily disjoint.

This concludes the introduction and summary of the review.

In the following we briefly discuss issues that shall not be reviewed
extensively. Our choice of not treating in depth these topics
is due to the review dealing with aspects of duality either more
elementary or more closely related to the authors past work.

Readers less familiar with the subject
may prefer to go first through the bulk of the review before returning to
the omissions:

\begin{itemize}
\item
Mirror symmetry is a particular target space duality that interchanges two
string backgrounds corresponding, to leading order in $\alpha'$, to compact
K\"ahler manifolds with vanishing first Chern class.
Such manifolds are called ``Calabi-Yau'' (CY).
In compactifying the heterotic or supersymmetric strings to four dimensions
one is typically interested in Calabi-Yau manifolds with complex dimension
$d$ smaller than or equal to three.
When $d=3$ mirror symmetry is particularly interesting for various reasons.
Firstly, it is a non-trivial duality relating a pair of backgrounds with no
continuous isometries (and  vastly different from each other). Secondly,
it relates the
complex-structure deformations of one CY background to the K\"ahler-class
deformations of its mirror partner. Therefore, this symmetry is useful,
for instance, for calculating worldsheet instanton
corrections to Yukawa couplings (simply by doing the calculation for the
large radius limit of the mirror background).

Mirror symmetry deserves its own review. Therefore, we mention it only
briefly when we discuss a sub-class of dualities in CY backgrounds in
section 3. This subject is discussed extensively in the literature, for
example, in \cite{GrP,CdO1}, and in \cite{mirbook}.

For a complex torus, mirror symmetry is identical to what is called
``factorized duality'' in this review. We may therefore extend the notion
of mirror symmetry to the case of curved background with toroidal
isometries. The remarkable property that mirror symmetry relates
backgrounds with different topologies remains true in these cases as well.

Mirror transformation
is just one element in the complete discrete symmetry group
of CY backgrounds. Generalized duality in such moduli spaces
was studied in the context of $N=2$ Landau-Ginzburg (LG) theories.
At the infra-red fixed point of the renormalization group flow, and under
certain condition on the superpotential, $N=2$ LG models are conjectured to
be CFTs equivalent, in some sectors, to CY ground
states \cite{Martinec,VW,GVW,LVW,Wcy}.
By studying the deformation space of the superpotential, one
is able to find a subgroup of the complete set of discrete symmetries
\cite{CMLN1,LLW,gs,CMO}.

The rest of the symmetries can be found by studying
the monodromy group associated with the differential equation (of
Picard-Fuchs type) which gives the periods of the three-form $\Omega$ in
terms of the complex-structure deformation moduli \cite{CdO1,LSW,CF,CDFLL,DF}.
We will briefly mention these techniques whenever discussing related
topics.

\item
If one compactifies time as well, one finds new phenomena and new discrete
(stringy) symmetries, relating different backgrounds. For toroidal
backgrounds, one
finds that there exist compactifications with large extended
symmetries, typically associated with Lorentzian lattice
algebras~\cite{GShapere,Moore}. Furthermore, it was shown that in those cases
duality acts ergodically on the moduli space~\cite{Moore}.

The particular element inverting the time-like compactification radius
$\beta$ is referred to as ``temperature duality.''  Indeed, as it is well
known in field theory, the analytical continuation of a compact time yields
a theory at finite temperature $2\pi T=1/\beta$.
This $\beta$ duality was shown
in ref.~\cite{AO} to be an exact symmetry to all orders in string
perturbation theory. This duality implies that two Hagedorn-like phase
transitions (see, for instance \cite{Sath,Koga}) exist,
one at ``low'' temperature ($\beta_1=\sqrt{2}+1$ for the
heterotic string), and
the other at ``high'' temperature ($\beta_2=\sqrt{2}-1$)~\cite{AW,KRo}.

The intermediate-temperature phase of the heterotic string was studied in
ref.~\cite{AK} by generalizing to non-zero temperature the construction of
a duality-invariant effective action proposed in ref.~\cite{GP,GP1}, and
described in sections 2 and 3.
In~\cite{AK} it was claimed that the phase of the
heterotic string above the Hagedorn temperature corresponds to a
non-critical superstring in 7 + 1 dimensions.

\item
The emphasis of this review is on target space duality for the bosonic and
heterotic strings. Yet, there are some interesting
dualities in other types of strings, which are less understood but well
worth mentioning. These dualities typically relate different string
theories.

The first example is provided by the open-string duality. The open string
does not have winding modes, so the usual form of duality, relating
strings compactified on a circle of radius $R$ with strings on a circle of
radius $1/R$, cannot hold in its usual form. Nevertheless, it was claimed in
the literature that some properties of the $R\ra 1/R$ duality survive in
this case as well if, in addition, one interchanges the open-string
boundary conditions, for example, from Neumann to
Dirichelet-like ones~\cite{DLP,Hor,Green}.

The second example of non-standard duality is provided by duality between
type II strings. In this case, there exists a map between the $R\ra 0$
limit (in one compact-space direction) of the type IIA string and the
$R\ra \infty$ limit of the type IIB~\cite{DLP,DHS}.
Type IIA strings are mapped into type IIB since the duality changes the
GSO projection in the Ramond sector~\cite{GSO}.  Therefore, this duality
bears some resemblance with a mirror transformation.
This map gives an equivalence between the
two string theories, however, the corresponding discrete symmetry cannot be
interpreted as spontaneously broken gauge symmetry, in the way discussed in
section 2.

\item
A formulation of the worldsheet action of the bosonic string, in which
target space duality is manifest, was proposed in ref.~\cite{tsey,TW}. This
string action is based on  two sets of coordinates: the usual ones ($X$)
and  the ``dual'' coordinates ($\tilde{X}$). In this way the $O(d,d,\Z)$
duality is a manifest symmetry of the worldsheet action. The price to be
paid is the loss of manifest 2-$d$ covariance.

A doubling of space-time degrees of freedom
in the low-energy effective action (vielbeins, in this case)
occurs also in ref.~\cite{Siegel2}. There, a manifestly $O(d,d,\R)$ form of
the target-space effective action was obtained, and $O(d,d,\R)$ was
realized linearly, again at the price of
loosing manifest Lorentz invariance (this time in target space).

Similarly, the doubling of the dilaton + axion degrees of freedom was
advocated in~\cite{SSen} for the heterotic string.
This doubling allows for the construction of a
four-dimensional low-energy effective action manifestly invariant under
$SL(2,\R)$ transformations of the $S$ field. The $SL(2,\Z)$ subgroup of
$SL(2,\R)$ has been conjectured to be an exact symmetry of string theory
in~\cite{FILQ1,SENs}. This hypothetical exact symmetry, which
includes in particular a strong-weak coupling duality, was used
in~\cite{FILQ1} as a means of fixing non-perturbatively the VEV of the
$S$ field. Even here Lorentz invariance is non-manifest.

Besides the loss of manifest Lorentz invariance, in two last
formulations, it is not yet known how to extend the construction of these
actions to a completely duality-invariant effective action of the heterotic
string. The missing ingredient is the knowledge of how to include
non-Abelian gauge fields. Only after coupling the
theory to non-Abelian fields, are the continuous symmetries of the previous
formulations reduced to discrete (duality) symmetries. This reduction from
continuous $O(6,22,\R)$ to discrete $O(6,22,\Z)$ symmetries was observed
in the context of effective actions in ref.~\cite{GP,GP1}, and
will be discussed in section 2.

\item
Target space dualities are also present in the  space of topological
backgrounds (see for example~\cite{Wtop,DVVtop}).
For the space of twisted $N=2$ topological conformal field
theories it was found that the discrete symmetries are extended to
continuous ones~\cite{GStop}. Mirror symmetry in the space of such
background was discussed in ref.~\cite{W1top}. The study of target space
duality in topological backgrounds is useful, for example,  in calculating
Yukawa couplings~\cite{Ctop,GS1top,CVtop,W1top}.
Moreover, a  duality resembling a strong-weak
coupling duality  has been observed in the
moduli space of a particular topological background~\cite{EFR1}.

\item
The $R\rightarrow 1/R$ duality is also present when the worldsheet is
discretized as it happens in matrix models~\cite{Smat,GKmat}. This symmetry
turns into the regular circle duality in the continuum limit, where one
recovers a string moving in one compactified space dimension.
The one-loop  partition function is proportional to
$R+1/R$~\cite{BKmat,STan,KSmat}, and therefore,
it is manifestly duality-invariant.

\item
In section 4 we discuss a general procedure for finding pairs of curved
backgrounds related by duality. This procedure applies whenever the
backgrounds possess Abelian symmetries. In
refs.~\cite{Bth,FJ,FTdual,dOQ,Venona,GR2,AABL} this procedure was extended
to the case of backgrounds possessing non-Abelian symmetries. Unlike the
case of Abelian duality, here one does not know yet how to deal properly
with global issues.

\item
Finally, let us notice that target space duality can be related to
the well-known electric-magnetic duality of four-dimensional vector fields.
In the context of field theory this duality was studied in a general
setting in ref.~\cite{GZ}.
More precisely, the target space duality acts on the kinetic term of the
low-energy effective action as a particular discrete subgroup of the
electric-magnetic duality. This phenomenon can be extended to other
dimensions. Target space generalized dualities (as mirror symmetry) act on
the kinetic term of the effective action as dualities among
forms. This phenomenon further justifies the name of duality
for the string target-space one since in several ways it is
indistinguishable, from the low-energy point of view, from the old
form-duality. Duality among forms was studied extensively in the
literature, especially in the context of supersymmetric theories, see for
instance~\cite{CJ,dWN,TvN,dWLVP,S,deRoo,dRW,KLRvN,CCDFFM,LRfd,HKLRfd,
GHRfd,RSSfd,dWVP}.
Quantum aspects of
duality were studied, for instance, in \cite{vNV,DGvNW,BvN,DvN}.

\end{itemize}

\end{section}

\newpage

\begin{section}{Duality and Discrete Symmetries of the Moduli Space of
Toroidal Compactifications}

\begin{subsection}{Introduction: 2-$d$ Conformal Field Theories as String
Backgrounds}
\setcounter{equation}{0}

String theories use as building blocks Conformal Field Theories (CFT) (for
a review, see for instance \cite{GSW,GinLH}).
In order to understand the symmetries of the string theories it
is useful to study several properties of CFTs and translate
them to their string theory counterparts.

Assume one is given a certain Lagrangian, $L$, which is an exact
CFT defined on a certain two-dimensional manifold, $\Sigma$.
The Lagrangian is exactly solvable, in particular, the Virasoro
central charge, $c$, the allowed states and the corresponding
operators and their operator product expansion (OPE) coefficients have
been extracted.
Given these data one wishes to investigate if there exist other
CFTs with the same value of $c$ in the neighborhood of $L$.
A neighborhood is defined and constructed in the following
manner. One considers the most general Lagrangian $L'$
\begin{equation}
L'= L+\sum_i g_i f_i (z, \bar z),
\label{ER1}
\end{equation}
where $f_i(z, \bar z)$ are the operators in the spectrum of the theory $L$
and $g_i$ are appropriate coupling constants.
One searches perturbatively for those couplings, $g_i$, which one
can add to $L$ such that the modified $L'$ is also a CFT.

The spectrum of operators associated with $L$ can be
divided, using the classification of statistical mechanics, into
three groups:

\begin{description}

\item[a)]
Operators whose dimension is larger than 2, which are called
``irrelevant operators.''
Each of them  has the property that if one adds it to the Lagrangian
with an appropriate coupling, $g_i$ (which will have a negative
mass dimension), and if one considers the flow of the initial coupling
$g_i$, in the modified Lagrangian, $L'$, one finds that $g_i$
decreases to zero in the infrared limit (thus justifying its
name).

\item[b)]
Operators whose dimension is smaller than 2; these are termed
``relevant operators.''
Under $L'$, $g_i$ (which now has positive mass dimensions) flows
towards large values in the infrared limit.
If the theory is unitary it will flow to a theory with a smaller
value of $c$ \cite{Z}.

\item[c)]
Operators whose dimension is exactly 2. These are termed
``marginal operators.''
If $L'$ differs from  $L$ just by the addition of these marginal operators,
then $L'$ does
not  break classical scale invariance explicitly, as the $g_i$
are dimensionless in this case.
However, the $g_i$ may change under renormalization; this causes the
marginal operators to actually subdivide into three classes:

\item[c1)]
Marginal operators whose couplings $g_i$ in the modified
Lagrangian turn out to be infrared-free on the worldsheet.
Such operators  effectively
belong to the class (a) of irrelevant operators.
They differ from operators that are irrelevant at the classical level
in that $g_i$ decreases logarithmically and not
power-like towards the infrared.

\item[c2)]
Marginal operators whose couplings $g_i$ under the modified
Lagrangian are asymptotically free and thus increase
logarithmically towards the infrared, causing an instability.
This  turns the operators into relevant ones.

\item[c3)]
Marginal operators whose addition to $L$ does maintain the
dimension of the coupling $g_i$, thus promoting $L$ to a family of CFTs.
These operators are called ``truly marginal,'' and it is them who
form a basis for a neighborhood of conformal field theories.

\end{description}

The space of different conformal theories  in the
neighborhood of the theory $L$ is spanned by the corresponding coupling
constants $g_i$.
The number, $N$, of independent operators $f_i(z, \bar z)$,
$(i=1, \ldots, N)$, which can be added simultaneously to $L$, such
that $L'$ is a CFT, counts the local dimension of the space.
In general this is a perturbative statement. In many cases the
exact form of the truly marginal operator, which started out as
being described by $f_i(z, \bar z)$, would need to be modified
so as to maintain $L'$ as a CFT for a finite value of $g_i$.

The space of {\em all} CFTs connected to $L$ by truly marginal deformations
is called the ``(connected) moduli space'' of $L$.
Locally around $L$ the moduli space reduces to the neighborhood
described above.
The dimension $N$ itself is a local concept as it may change for specific
finite values of $g_i$.

Considering the local neighborhood, ${\cal M}$, it turns out that in
some cases one can span the same space ${\cal M}$ by applying a
certain continuous group ${\cal G}$ to $L$.
Moreover, one finds that there also exists a subgroup ${\cal G}_d$
of ${\cal G}$
which acts like a symmetry of the physical theory.
Upon acting with an element $g$ of ${\cal G}$, a theory $L_1$ defined
at one point in
${\cal M}$ is transformed into another theory $L_2$, corresponding to a
different point in ${\cal M}$ (parameterized by different values of the
couplings $g_i$).  When $g\in {\cal G}_d$, these two  theories are actually
physically equivalent.
This identity holds for all the orbits of ${\cal M}$ under  ${\cal G}_d$.
The groups ${\cal G}$ and ${\cal G}_d$ depend on the particular nature of
${\cal M}$. A general
classification is not yet available; we will discuss several
cases in this report.

Let us now return to the language of string theory. The
couplings $g_i$ correspond to allowed target space backgrounds in which the
string may propagate.
For a bosonic string, the couplings are usually collected into the
background metric $\hat{G}_{ij}(X)$, the
antisymmetric-tensor $\hat{B}_{ij}(X)$,
and the  dilaton $\hat{\Phi}(X)$. These couplings are in
general $X$ dependent, where $X$ is a target space coordinate.
A typical worldsheet action $S$ is of the form
\begin{eqnarray}
S&=& \frac{1}{4\pi\alpha'} \int_0^{2\pi} d\sigma \int d\tau \left[\sqrt{g}
g^{\alpha\beta} \hat{G}_{ij}(X) \partial_{\alpha} X^i \partial_\beta X^j
+ \epsilon^{\alpha\beta} \hat{B}_{ij}(X) \partial_\alpha X^i
\partial_\beta X^j
\right.\nonumber \\
&& -{\alpha'\over 2}\left. \sqrt{g} \hat{\Phi}(X) R^{(2)}\right],
\label{ER2}
\end{eqnarray}
where $g_{\alpha\beta}$ is the worldsheet metric, $g\equiv \det
g_{\alpha\beta}$, $\alpha'$ is proportional to the inverse string tension
and $R^{(2)}$ is the worldsheet scalar curvature.

One may wish to recall that the ultimate purpose of the worldsheet action
is to be used in order to calculate the $S$-matrix for scattering in
some target space.
The scattering quanta represent the fluctuations in some target
theory around the classical values $\hat{G}_{ij}$, $\hat{B}_{ij}$ and
$\hat{\Phi}$
defined by
\begin{eqnarray}
\tilde G_{ij} (X) & = & \hat{G}_{ij}(X) + h_{ij}(X), \nonumber \\
\tilde B_{ij} (X) & = & \hat{B}_{ij}(X) + b_{ij}(X), \nonumber \\
\tilde \Phi(X) & = & \hat{\Phi}(X) + \phi(X).
\label{ER3}
\end{eqnarray}
The fluctuations $h_{ij}, b_{ij}$ and $\phi$ describe the massless graviton,
antisymmetric tensor and dilaton in the target space. The existence of truly
marginal operators on the worldsheet corresponds to the existence of
massless particles in the target space.
Similarly, the target space tachyon
corresponds to a relevant worldsheet operator, and massive
states in the target space correspond to irrelevant operators
from the CFT point of view. In
string theory they are always dressed in such a way as to be
vertex operators whose dimension is $(1, 1)$.
The notation $(d_L, d_R)$ means that the left- (right-) handed
conformal dimension of the operator is  $d_L$ ($d_R$).
The right-handed and left-handed  dimensions are defined to be the scaling
weights under the worldsheet coordinate transformation:
$\bar{z}\rightarrow \bar{\lambda}\bar{z},\;\;\; z\rightarrow \lambda z$.

In  this section
we discuss the groups ${\cal G}$ and ${\cal G}_d$ for the case in
which the geometric picture consists of a string moving on a
background where $d$ dimensions are compactified, and
$\hat{G}$, $\hat{B}$, $\hat{\Phi}$ are $X$ independent.
These are termed ``flat compactifications'' or ``toroidal backgrounds.''
In section 2.2 we first describe the $d=1$ case, namely,
a circle compactification, and we study its one-loop partition function.
In section 2.3 we present the higher-genus circle duality.
The general case for the bosonic and the
heterotic strings is presented in sections 2.4 and 2.5.
The identification of the elements of ${\cal G}_d$ with
target space dualities will allow us
to interpret most of the latter as residual broken gauge symmetries. This is
discussed in section 2.6. Finally, in sections 2.7 and 2.8
we present duality invariant effective field theories
realizing this interpretation.

\end{subsection}

\begin{subsection}{$R\rightarrow \alpha'/R$ Duality in the One-Loop
Partition Function}
\setcounter{equation}{0}
The simplest example of duality transformations \cite{KY,SS},
and duality symmetry, is
provided by a single bosonic string coordinate compactified on a circle of
radius $R$. We set the worldsheet metric in (\ref{ER2}) into an orthonormal
form: $g_{\alpha\beta}(\sigma,\tau)=\eta_{\alpha\beta}$. (For the moment
we have suppressed any dependence of the worldsheet metric on the global
structure of the worldsheet).
The worldsheet action is
\begin{equation}
S={1\over 4\pi\alpha'}\int d\tau d\sigma \partial_{\alpha}X\partial^{\alpha}X.
\label{1}
\end{equation}
The compactification is defined by the period identification
\begin{equation}
X\approx X + 2\pi Rm,
\label{2}
\end{equation}
where $m$ is an arbitrary integer.

Since $X(\sigma,\tau)$ satisfies a free wave equation, it admits a
decomposition
in terms of left- and right-movers
\begin{equation}
X(\sigma,\tau)=X_R(\sigma-\tau)+X_L(\sigma+\tau).
\label{3}
\end{equation}
These modes, upon the continuation $\tau\rightarrow -i\tau $, become,
respectively, the analytic and anti-analytic string coordinates.
They have the mode expansion
\begin{eqnarray}
X_R(\sigma-\tau) &=& x_R - \sqrt{{\alpha'\over 2}}p_R(\sigma-\tau) +
i\sqrt{\alpha'\over 2}\sum_{l\neq
0}{1\over l}\alpha_l e^{+il(\sigma-\tau)},
\nonumber \\
X_L(\sigma+\tau) &=& x_L + \sqrt{{\alpha'\over 2}}p_L(\sigma+\tau) +
i\sqrt{\alpha'\over 2}\sum_{l\neq
0}{1\over l}\tilde{\alpha}_l e^{-il(\sigma+\tau)},
\label{4}
\end{eqnarray}
where $x=x_L+x_R$, and the dimensionless quantities $p_L$, $p_R$ read
\begin{eqnarray}
p_R &=& {1\over \sqrt{2}}({\sqrt{\alpha'}\over R}n -{R\over \sqrt{\alpha'}}m) ,
\nonumber \\
p_L &=& {1\over \sqrt{2}}({\sqrt{\alpha'}\over R}n +{R\over \sqrt{\alpha'}}m).
\label{5}
\end{eqnarray}
The canonically conjugate momentum of $X(\sigma)$ is
\bq
P(\sigma)= {1\over 2\pi\sqrt{2\alpha'}}\left[p_L + p_R +
\sum_{l\neq 0}\alpha_le^{+il(\sigma-\tau)}+
\sum_{l\neq
0}\tilde{\alpha}_le^{-il(\sigma+\tau)}\right].
\eq{6}
The total momentum $P=\int_0^{2\pi}d\sigma P(\sigma)$, canonically
conjugate to $x_L + x_R$, is
\bq
P= {1\over\sqrt{2\alpha'}}(p_L + p_R).
\eq{MM6}
{}From eqs.~(\ref{5}, \ref{6}, \ref{MM6})
one derives the canonical commutation relations
\begin{equation}
[x_L,p_L]=[x_R,p_R]=i\sqrt{{\alpha'\over 2}},\;\;\;
[\alpha_m ,\alpha_n ]=[\tilde{\alpha}_m ,\tilde{\alpha}_n ]=m\delta_{m+n,0}.
\label{7}
\end{equation}
The (normal ordered) Hamiltonian reads
\begin{equation}
H=L_{0L} +L_{0R},
\label{8}
\end{equation}
where
\begin{eqnarray}
L_{0R} &=& {1\over 2}
p_R^2 +\sum_{l=1}^{\infty} \alpha_{-l}\alpha_l , \nonumber \\
L_{0L} &=& {1\over 2}
p_L^2 +\sum_{l=1}^{\infty} \tilde{\alpha}_{-l}\tilde{\alpha}_l .
\label{9}
\end{eqnarray}
Here $L_{0L}$ and $L_{0R}$ are separately conserved, due to conformal
invariance, and are the zero modes of, respectively, the left and right
Virasoro operators~\cite{GSW}.

Notice that $L_{0L}$ and $L_{0R}$ are invariant under the
transformation
\begin{equation}
{R\over \sqrt{\alpha'}}\rightarrow {\sqrt{\alpha'} \over R},
\;\;\; m\leftrightarrow n.
\label{10}
\end{equation}
Under eq.~(\ref{10}) $p_R$ transforms into $-p_R$, whereas $p_L$ is
invariant.
The oscillators $\alpha_n$, $\tilde{\alpha}_n$ also transform in a simple
way under $R/\sqrt{\alpha'}\rightarrow \sqrt{\alpha'}/R$:
\bq
\alpha_n\to -\alpha_n,\;\;\; \tilde{\alpha}_n \to \tilde{\alpha}_n
\eq{Mosc}
(so that (\ref{10}) together with (\ref{Mosc}) implies
$\dot{X}_R\to -\dot{X}_R, \; \dot{X}_L\to \dot{X}_L$).

Equations~(\ref{10}, \ref{Mosc}) provide the simplest example of a
{\em Target Space Duality Symmetry}.
This is a {\em stringy} property that
does not exist in field theory. The difference between a point-particle
field theory and string theory may be understood by realizing that an
extended object -- a string moving on a circle -- can wrap around it.
Its winding number $m$,
i.e. the number of times the string wraps around the circle, is
an integer number. In order to wrap around the circle the string must be
stretched, so its energy increases with the winding number and with the
radius of the circle as $m^2 R^2$.
This contribution to the energy of the string is to be added to  the
field-theoretical one, equal to the square of the center-of-mass momentum,
which is quantized in units of $1/R$, and thus proportional to $n^2/R^2$.
The total energy of the string turns out to be
invariant under the transformations~(\ref{10}, \ref{Mosc}),
which inverts the radius of
the circle, and interchanges winding numbers with momenta.

The one-loop partition function of the compactified bosonic string is
\begin{equation}
Z=\int_{\Gamma} d^2\tau \hat{Z}(\tau,\bar{\tau})
\sum_{p_L,p_R}{\rm Tr}\, e^{i\pi\tau L_{0R}-i\pi\bar{\tau}L_{0L}} .
\label{11}
\end{equation}
Here $\tau$ denotes the modular parameter describing conformally inequivalent
tori, ${\Gamma}$ is the fundamental region of the $\tau$-plane under
the action of the torus mapping class group $SL(2,\Z)$, and
$\hat{Z}(\tau,\bar{\tau})$ is the contribution of all coordinates,
other than our compactified $X$, to the partition function.
This partition function was first presented in \cite{GSB};
for more details see e.g. ref.~\cite{GSW}.
The trace in eq.~(\ref{11}) is taken over the Hilbert space spanned by the
oscillators $\alpha_l$, $\tilde{\alpha}_l$. The sum over $p_L$, $p_R$
extends to all momenta of the form~(\ref{5}). Symmetry of the partition
function under the target space duality~(\ref{10}) follows by noticing that
the integers $m$ and $n$ of eq.~(\ref{5}) are dummy variables
in~(\ref{11}).

So far, we have proved that target space duality is a symmetry of
the one-loop partition function, that is of the (free) string {\em
spectrum}.
If symmetry~(\ref{10}) extends to the higher-genus contribution to the
partition
function as well, i.e. to the interacting string, we may claim that
{\em a small compactification radius ($R/\sqrt{\alpha'}\ll 1$)
is completely equivalent
to a very large one ($\sqrt{\alpha'}/R\ll 1$)}.
This statement, pointing out to the existence
of a minimal length in string theory, will be proven in the next
section.

To make contact with the expression for the worldsheet action in eq.
(\ref{ER2}) it is useful to rescale the coordinate $X$ to $RX$.  The new
coordinate is dimensionless and
has periodicity $2\pi$ ($X\approx X+2\pi m$). Now, the action in
(\ref{1}) reads as in eq. (\ref{ER2})
\begin{equation}
{1\over 4\pi\alpha'}\int_0^{2\pi}d\sigma\int d\tau
\hat{G}_{11}\partial_{\alpha}X\partial^{\alpha}X,
\label{MM7}
\end{equation}
with the {\em dimensionful} metric $\hat{G}_{11}=R^2$.
Furthermore, since duality involves explicitly the parameter $\alpha'$
it is convenient to define a dimensionless metric
$G_{11}=\hat{G}_{11}/\alpha'$. In the same way we redefine $R$ to be the
dimensionless radius, namely
\begin{equation}
{R\over \sqrt{\alpha'}}\to R.
\label{MM8}
\end{equation}
The new metric and radius transform in a simple way under duality:
\begin{equation}
G_{11}\to {1\over G_{11}},\;\;\; R\to{1\over R}.
\label{MM9}
\end{equation}

\end{subsection}

\begin{subsection}{Duality in the Higher Genus Partition Function}
\setcounter{equation}{0}
String theory involves a summation over worldsheets of all genera.
For target space duality to be a symmetry of string theory (at least
in perturbation theory) it should be a symmetry on each genus $g$.
In this section we discuss this issue. We
work with the dimensionless coordinate $X$ whose
periodicity is $2\pi$, and with the dimensionless compactification radius
$R$.

In order to study duality on higher-genus Riemann surfaces, it is convenient
to revert from the Hamiltonian representation of the partition function, used
in the previous section, to its functional integral representation.
The partition function on a genus $g$ Riemann surface $M_g$ reads, upon
analytical continuation $\tau\rightarrow -i\tau$~\cite{GSW},
\begin{equation}
Z_g=\int [dg_{\alpha\beta}dXd\hat{X}]e^{
-{R^2\over 4\pi}
\int_{M_g} d\tau d\sigma\sqrt{g}g^{\alpha\beta}\partial_{\alpha}X
\partial_{\beta}X -\int_{M_g}d\tau d\sigma\hat{L}} .
\label{12}
\end{equation}
The integral is extended to all two dimensional metrics $g_{\alpha\beta}$
compatible with the topology of $M_g$. The string coordinates have been
split into $X$, and ``the rest'' $\hat{X}$, with Lagrangian $\hat{L}$.

The compactification of $X$ is defined by generalizing to a genus $g$ surface
the period identification~(\ref{2})
\begin{equation}
\int_{a_i}dX=2\pi m_i, \;\;\; \int_{b_i}dX=2\pi n_i
\label{13}
\end{equation}
The $a_i$ and $b_i$ are the canonical homology cycles of the surface $M_g$;
they are shown in figure 2.A. The label $i$ takes the values
$1,...,g$~\cite{FaKr}.
We denote with $\omega_i$ the standard basis of closed holomorphic
(1,0)-forms spanning
$H_{(1,0)}(M_g,\Z)$, and obeying~\cite{FaKr,A-GMV}
\begin{equation}
\int_{a_i}\omega_j =\delta_{ij},\;\;\; \int_{b_i}\omega_j =\tau_{ij}=
\tau_{1\, ij} + i \tau_{2\, ij}.
\label{14}
\end{equation}
The period matrix $\tau_{ij}$ is symmetric, of dimension $g\times g$, and it is
a function of the $3g-3$ modular parameters of $M_g$~\cite{FaKr,A-GMV}.
The most general field $X$ satisfying eqs.~(\ref{13}) reads~\cite{A-GMV}
\begin{equation}
X(z,\bar{z})=X(z) + \bar{X}(\bar{z})
             +\stackrel{\, o}{X}(z,\bar{z})  .
\label{15}
\end{equation}
The field $\stackrel{\, o}{X}$ is an arbitrary scalar field
{\em globally defined} on $M_g$, whereas $X(z)$ is a holomorphic solution of
the classical equations of motion
\begin{equation}
X(z)=2\pi \sum_{j=1}^g\int_{z_0}^z m_j\omega_j +n_i\tau_{ij}\omega_j .
\label{16}
\end{equation}
The coefficients $m_i$ and $n_i$ are those of eq.~(\ref{13}) and the integral
is
performed on an arbitrary path joining a fixed reference point $z_0$ with $z$.
Equation~(\ref{16}) is
independent of the path because the forms $\omega_i$ are closed.

By a standard choice of the gauge fixing of the general coordinate and
conformal
invariances, and by choosing the appropriate two-dimensional metric
$g_{\alpha\beta}$~\cite{GSW,A-GMV}
the genus $g$ partition function reads
\begin{equation}
Z_g =\int_{{\cal M}_g} dm D(m)R\sum_{m_i,n_i}
e^{-\pi R^2 ({\bf m} + {\bf n}\tau)^t
     \tau_2^{-1}({\bf m} + {\bf n}\tau) }.
\label{17}
\end{equation}
The integral in this equation is performed over the moduli space
${\cal M}_g$ of
$M_g$, and the sum extends to all integer $m_i$, $n_i$.
We denoted by $D(m)$ all contributions to the partition function which
are independent of $R$ and $m_i$, $n_i$. Boldface characters denote
g-dimensional row vectors. The factor $R$ in front of the sum is due to the
$X$ zero mode.

The sum in $Z_g$ can be rewritten in a more compact form by performing a
Poisson
resummation on ${\bf n}$ (see for example~\cite{Serre})
\begin{equation}
\sum_{m_i,n_i}e^{-\pi R^2 ({\bf m} + {\bf n}\tau)^t
     \tau_2^{-1}({\bf m} + {\bf n}\tau) }=R^{-g}
     \det\tau_2\sum_{({\bf m},{\bf k})\in
     \Z^{2g}} e^{i\pi ({\bf p}_R\tau{\bf p}_R - {\bf p}_L\bar{\tau}{\bf p}_L)}
,
\label{18}
\end{equation}
where
\begin{eqnarray}
{\bf p}_R &=& {1\over \sqrt{2}}({1\over R}{\bf k}- R{\bf m}),  \nonumber \\
{\bf p}_L &=& {1\over \sqrt{2}}({1\over R}{\bf k}+ R{\bf m}).
\label{19}
\end{eqnarray}

By sending $R\rightarrow 1/R$ and ${\bf k}\leftrightarrow {\bf m}$, and
noticing
that, as in the previous section, the integers ${\bf k}$ and ${\bf m}$ are
dummy
variables, we find
\begin{equation}
Z_g\left({1\over R}\right)= R^{2g-2}Z_g(R).
\label{20}
\end{equation}
The complete partition function of the string is
\begin{equation}
Z(\Phi, R)= \sum_{g=0}^{\infty} e^{(1-g)\Phi}Z_g(R),
\label{21}
\end{equation}
and $\Phi$ is the (constant) dilaton (see eq.~(\ref{ER2})).
Equation~(\ref{20}) implies therefore \cite{B1,gv}
\begin{equation}
Z\left({\Phi+2\log R}, {1\over R}\right)=Z(\Phi, R).
\label{22}
\end{equation}
Equation~(\ref{22}) is the generalization
to all genera of the result obtained in
the previous section. In contrast to the one-loop case, here the dilaton
VEV $\Phi$ appears explicitly. The equivalence of two seemingly
different partition functions proven here is rather formal, since $Z(\Phi,
R)$
is ill-defined, due to the tachyon divergence. This difficulty however is
not present in the interesting case of the heterotic string, which is
still
duality invariant, as we shall prove later. The conclusion we reach thanks
to eq.~(\ref{22}) is therefore meaningful, and it refines the statement made at
the end of the previous section, namely: {\em a compactification of small
radius is equivalent to one of large radius, provided we also change the
dilaton VEV according to $\Phi \to \Phi + 2\log R$.}
\end{subsection}

\begin{subsection}{$O(d,d,\Z)$
Duality for $d$-Dimensional Toroidal Compactifications}
\setcounter{equation}{0}

The worldsheet action describing a toroidal background is (for a review,
see \cite{GSW})
\begin{eqnarray}
S&=& \frac{1}{4\pi} \int_0^{2\pi} d\sigma \int d\tau \left[\sqrt{g}
g^{\alpha\beta} G_{ij} \partial_{\alpha} X^i \partial_\beta X^j
+ \epsilon^{\alpha\beta} B_{ij} \partial_\alpha X^i \partial_\beta X^j
\right.\nonumber \\
&& \left.-{1\over 2} \sqrt{g} \Phi R^{(2)}\right].
\label{ER4}
\end{eqnarray}
Here $X^i$ are dimensionless  coordinates whose periodicities
are $2\pi$,\footnote{
The geometrical data of the torus, in the presence of torsion,
is encoded in $G_{ij}$ and $B_{ij}$.} namely
\bq
X^i \approx X^i +2\pi m^i.
\eq{ERper}
The metric and antisymmetric tensor $G_{ij}$, $B_{ij}$,
$i,j=1,...,d$, are the dimensionless ones, defined as in the end of
section 2.2.
The simplest case where $d=1$ was already discussed in sections 2.2, 2.3.
In string theory the total Virasoro central charge should vanish.
For the usual bosonic string the matter accounts
for 26 units of $c$. The action above describes only the $d$ compactified
coordinates, and therefore, its central charge is $c=d$.
In general, the model $S$ (\ref{ER4}) is tensored with some other
CFT, for example, a string moving in a ($26-d$)-dimensional
flat Minkowski space, such that the total value of $c$ is indeed
26. Till further notice we will concentrate only on the internal sector
described by (\ref{ER4}).

The number of independent truly marginal
operators for a generic $d$-dimensional background is $d^2$.
This is an exact result as the model is Gaussian.
The $d^2$ marginal operators are composed of the $d(d+1)/2$ operators
\begin{equation}
\sqrt{g} g^{\alpha\beta}\partial_\alpha X^i \partial_\beta X^j,
\label{ER5}
\end{equation}
and the $d(d-1)/2$ operators
\begin{equation}
\epsilon^{\alpha\beta}\partial_\alpha X^i \partial_\beta X^j,
\label{ER6}
\end{equation}
with the corresponding symmetric couplings $G_{ij}$ and
antisymmetric couplings $B_{ij}$, all in all $d^2$ couplings and  operators.
We disregard for a while the dilaton operator and coupling.

The $d^2$ couplings are the data describing the CFT; they are
organized in the form $(G,B)$. We will occasionally find it
convenient to arrange the data in a matrix $E$ whose symmetric
part is $G$ and whose antisymmetric part is $B$,
\begin{equation}
E=G+B.
\label{ER7}
\end{equation}
We call $E$ the ``background matrix.''

We first solve for the spectrum of the theory (\ref{ER4}).
This is done by performing a canonical quantization on the worldsheet.
Since we are studying theories with no Weyl anomaly (as it happens for
strings at criticality) we can set the worldsheet  metric into a diagonal
orthonormal form
\bq
g_{\alpha\beta}(\sigma,\tau)=\eta_{\alpha\beta}.
\eq{ERgauge}
(For the moment we have suppressed any dependence of the worldsheet
metric on the global structure of the worldsheet).

The canonical momentum $P_i$ associated with $X^i$ is given by
\begin{equation}
2\pi P_i = G_{ij} \dot{X}^j + B_{ij} X^{j}{}' = p_i +
{\rm oscillators},
\label{ER10}
\end{equation}
where $p_i$ is the string center of mass momentum.
Since the coordinates $X^i$  have periodicities $2\pi$,
it follows that the momenta $p_i$ is quantized in integer units
\bq
p_i=n_i.
\eq{ERpn}

The Hamiltonian and momentum constraints take the form:
\begin{eqnarray}
H & = & L_{0L}+ + L_{0R} \nonumber \\
& = & {1\over 4\pi}\int_0^{2\pi} d\sigma \left[(2\pi)^2(P_iG^{ij}P_j) +
X^{i}{}'(G-BG^{-1}B)_{ij} X^{j}{}' +4\pi X^{i}{}'B_{ik} G^{kj}P_j\right]
\nonumber \\
& = & \frac{1}{4\pi}\int_0^{2\pi} d\sigma (P_L^2+P_R^2),\nonumber \\
P_{La} & = & [2\pi P_i + (G-B)_{ij} X^{j}{}']e^{*i}_a,\;\;\;
P_{Ra}=[2\pi P_i - (G+B)_{ij} X^{j}{}']e^{*i}_a, \nonumber \\
\int_0^{2\pi} d\sigma PX' & = & L_{0L} - L_{0R} = 0
\label{ER12}
\end{eqnarray}
Here $e$ and $e^*$ are defined by
\bq
\sum_{a=1}^d e_i^a e_j^a = 2G_{ij},\;\;\;
\sum_{a=1}^d e_i^a e_a^{*j} = \delta_i^j, \;\;\;
\sum_{a=1}^d e_a^{*i} e_a^{*j}={1\over 2} (G^{-1})^{ij}.
\eq{ERee}
The vectors $e_i$ are a basis to the compactification lattice $\Lambda^d$,
such that the target space $d$-torus is $T^d=\R^d/\pi\Lambda^d$; the
vectors $e^{*i}$ are a basis to the dual lattice $\Lambda^{d*}$. The
indices $a,b$ label an orthonormal basis to the target space.

We deduce from eqs. (\ref{ER12}) that $P_L$ and $P_R$, even in the
presence of the backgrounds, still decouple and that they
describe left- and right-movers, respectively. The model is indeed
Gaussian for all flat data $E$;
this is instrumental in obtaining exact results.
Now, the Hamiltonian $H$ consists of a zero mode part (for which we
retain the notation $H$), and the part describing the oscillators.
These oscillators -- of each coordinate -- transform similar to
(\ref{Mosc}) under duality, in a way that does not change the spectrum
(as shown at the end of this subsection). We
thus first discuss only the zero mode part, whose  Hamiltonian reads
\begin{eqnarray}
H & = & L_{0L}+L_{0R}={1\over 2}(p_L^2+p_R^2)=  \nonumber \\
& = &\frac{1}{2}\left[ n_i(G^{-1})^{ij} n_j +
m^i(G-BG^{-1}B)_{ij}m^j+2m^iB_{ik}(G^{-1})^{kj}n_j\right].\nonumber\\
& &\label{ER13}
\end{eqnarray}
The integers $n_i$ are the momentum eigenvalues defined in eq.
(\ref{ERpn}) and $m^i$ are the so
called winding modes defined in eq (\ref{ERper}).
The zero-modes $p_L, p_R$ are given by
\ber
p_R&=& [n^t + m^t(B-G)]e^* , \nonumber \\
p_L&=& [n^t + m^t(B+G)]e^*.
\eer{ERplpr}
Here, for simplicity, we suppressed all vector and matrix indices
for all quantities involved.
Note that by eq. (\ref{ER10}) one obtains that a state with a non-zero
winding number but no explicit time dependence still carries momentum.
This phenomenon was detected first in the context of 4-dimensional gauge
theories \cite{Wdyon};
it was shown that a monopole in the presence of a theta term
obtains electric charge.
This was studied in various four- and two-dimensional systems
\cite{CR,Cardy}.
The $B$ field plays a role very similar  to that of a theta term as we
shall see.

The Hamiltonian can be defined for closed worldsheets with genus 0 or 1.
Given the Hamiltonian and its spectrum, the next task is to identify the
group generating the moduli space of the whole set of Lagrangians
\cite{N,NSW},
and the subgroup ${\cal G}_d$ which maintains the physics of the models
\cite{NSSW,DVV1,GRV,sw,GMR}.
The moduli space for toroidal compactifications is isomorphic to
$O(d,d,\R)/(O(d,\R)\times O(d,\R))$~\cite{N,NSW},
where $O(d,d,\R)$ is the non-compact orthogonal group in $d+d$ dimensions.
A convenient way of representing  the elements $g\in O(d,d,\R)$ is
\begin{equation}
g=\left(
\begin{array}{cc}
a & b\\
c & d \end{array}\right),
\label{ER19}
\end{equation}
where $a,b,c,d$ are $d\times d$ matrices, and such that
$g$ preserves the form $J$
\begin{equation}
J=\left(\begin{array}{cc}
O & I \\
I & O
\end{array}\right),
\label{ER17}
\end{equation}
namely,
\begin{equation}
g^tJg=J \;\;
\Rightarrow \;\;a^t c+c^t a=0,\;\;\;b^t d+d^t b=0,\;\;\;\ a^t d+c^t b=I.
\end{equation}
(A useful consequence of the above condition is that if $g\in O(d,d,\R)$,
then $g^t\in O(d,d,\R)$. This is proven as follows: begin with $g^tJg=J$,
and take the inverse of both sides. Using $J^{-1}=J$, one finds
$g^{-1}J(g^t)^{-1}=J$. Multiplying from the left by $g$ and from the
right by $g^t$ one finds
$$
J=gJg^t,
$$
which shows that $g^t\in O(d,d,\R)$).

The $(d,d)$ Lorentzian momenta $(p_{La},p_{Rb})$ in~(\ref{ERplpr})
form an even
self-dual Lorentzian lattice $\Gamma^{(d,d)}$~\cite{N}, namely,
the Lorentzian  length is even
\begin{equation}
p_L^2-p_R^2=2m^in_i\in 2\Z,
\end{equation}
and the Lorentzian lattice is self-dual.
It is known \cite{Serre} that all even self-dual
$(d,d)$ Lorentzian lattices are related by $O(d,d,\R)$ rotations. Obviously,
any $O(d,d,\R)$ rotation of $\Gamma^{(d,d)}$ gives back
an even self-dual Lorentzian lattice. Moreover, to any $\Gamma^{(d,d)}$
corresponds a particular toroidal background. The
{\em Euclidean lengths} $p_L^2+p_R^2$ in two lattices related
by an $O(d,d,\R)$ rotation  are
generally different, and give rise to different spectra.

As $(p_L,p_R)$ transform as  vectors under $O(d,d,\R)$, the
Hamiltonian~(\ref{ER13}) is invariant under rotations by the maximal
compact subgroup $O(d,\R)\times O(d,\R)$. Therefore,
these rotations do not
change the zero-mode spectrum. Thus, the
solution-generating group ${\cal G}$ is $O(d,d,\R)$, and the
moduli space is locally isomorphic to the coset manifold
$O(d,d,\R)/(O(d,\R)\times O(d,\R))$.

Recall that the momenta $p_L$ and $p_R$ are specified by the background
matrix $E$. Therefore,
the way the solution-generating group ${\cal G}$
acts on $E$ is defined by
its action on the vectors $(p_L,p_R)$.

The spectrum is expressed by $(p_L,p_R)$ or, equivalently, by
the bilinear form defined by the $2d\times 2d$ matrix $M$ \cite{GRV}
\begin{equation}
M(E)=\left(
\begin{array}{cc}
G-BG^{-1}B & BG^{-1} \\
-G^{-1}B   & G^{-1}
\end{array}
\right).
\label{ER18}
\end{equation}

The spectrum of the system as obtained in eq.~(\ref{ER13}) is given by
the scalar
\begin{equation}
H={1\over 2}(p_L^2+p_R^2)={1\over 2}Z^tMZ, \;\;\;  Z=(m_a, n_b).
\label{HZMZ}
\end{equation}
Here $Z$ is the vector of integers counting the winding number and
the number of momentum quanta, corresponding to
$(p_L,p_R)$.
Under the $O(d,d,\R)$ transformation $g$, given in eq.~(\ref{ER19}),
the matrix $M$ transforms as
\begin{equation}
M_g\equiv g M g^t.
\label{ER20}
\end{equation}

To find the corresponding
transformation of the background matrix $E$, it is convenient
to first embed the
$O(d,d,\R)/(O(d,\R)\times O(d,\R))$ moduli elements $E=G+B$ in $O(d,d,\R)$.
This is done by looking at the $O(d,d,\R)$ element
\begin{equation}
g_{E=G+B}=
\left(\begin{array}{cc} e & B(e^t)^{-1}\\ 0 & (e^t)^{-1}\end{array}\right),
\end{equation}
where the vielbein $e$ is defined such that $G=ee^t$. Next, we
define the action of $g\in O(d,d,\R)$
(\ref{ER19}) on a $d\times d$ matrix $F$ by
the fractional linear transformation
\begin{equation}
g(F)\equiv (aF+b)(cF+d)^{-1}.
\label{EMA1}
\end{equation}
It now follows that
\begin{equation}
g_E(I)=E=G+B.
\label{EMA2}
\end{equation}
Here $I$ is the $d$-dimensional identity matrix.
Moreover, one finds
\begin{equation}
M=g_E g_E^t.
\label{EMA3}
\end{equation}
{}From eqs. (\ref{ER20}, \ref{EMA1}, \ref{EMA2}, \ref{EMA3})
it follows that
$$
M'=M(E')\equiv
M_g=gMg^t=gg_E g_E^t g^t = g_{E'} g_{E'}^t \;\; \Rightarrow\;\; g_{E'}=gg_E,
$$
and as
$$
E'=g_{E'}(I)=gg_E(I)=g(E),
$$
one finally finds
that the $O(d,d,\R)$ transformation properties of the data $E$ read
\cite{GRV,sw,GMR,GMR1}:
\begin{equation}
E'\equiv g(E) = (aE+b)(cE+d)^{-1}.
\label{ER22}
\end{equation}
(It is easy to verify that this action is consistent with the group
property: $g(g'(E))=(gg')(E)$.)

Now that we know how ${\cal G}=O(d,d,\R)$
sweeps the full moduli space of toroidal backgrounds, we may try
to detect if in the process there still remains some symmetry
group ${\cal G}_d$.
A necessary condition for a group element to belong to ${\cal G}_d$ is
that it leaves the spectrum of $S$ invariant.
It is allowed to change each state separately but the total
spectrum should remain unchanged. The group element should act
like an automorphism on the space of states.
However, more is required in order to prove that both theories
are equivalent: one ought to show that all
correlation functions in one theory can be mapped into
identical correlation functions in the other theory.
In this section we will discuss only the above mentioned
necessary condition on the spectrum,
and in section 2.6 we will show that there
exists a gauge symmetry relating the different points connected
by the symmetry group ${\cal G}_d$. This will guarantee the equivalence
of the theories.

Returning to the spectrum, there are two types of symmetry
operators one can identify explicitly \cite{GRV,sw}:

\begin{description}

\item[1.] {\bf Integer theta-parameter shift $\Theta_{ij}\in\Theta(\Z)$}:\\

One can add to the antisymmetric matrix $B_{ij}$ an
antisymmetric integral matrix $\Theta_{ij}(\Z)$ (that is a matrix composed of
integer numbers) without changing the physics. In other words,
one can act with the $O(d,d,\R)$ elements of the type
\begin{equation}
g_{\Theta}=
\left(\begin{array}{cc} I & \Theta \\ 0 & I\end{array}\right),
\end{equation}
where $\Theta_{ij}\in \Z$ and $\Theta_{ji}=-\Theta_{ij}$.
Indeed this element not only belongs to $O(d,d,\R)$ but to its discrete
subgroup $O(d,d,\Z)$, made of the elements $g$ with integer entries.

In a way, the fact that $\Theta(\Z)$ are symmetries of the spectrum is
evident from the role of $B$ as a theta parameter: the constant
$B$-term in $S$ is a
total derivative, and therefore,
gives only topological contributions. For an integer
$\Theta$-shift the action is changed by an integer multiple of $2\pi$, and
therefore, does not contribute to the path integral.

\item[2.] {\bf Basis change $A\in GL(d,\Z)$}:\\

One can conjugate $E$ by $E'=A E A^t$, where $A$ describes a basis
change of the compactification lattice $\Lambda$.
This can be done by acting on $E$ as in eq.
(\ref{ER22}) with the $O(d,d,\Z)$ elements
\begin{equation}
g_{A}=
\left(\begin{array}{cc} A & 0 \\ 0 & (A^t)^{-1}\end{array}\right),
\end{equation}
where $A\in GL(d,\Z)$.
(For example, some of these operations result in permuting the
space-time dimensions and thus they do not change the physics.)

Explicitly, the invariance of the spectrum
can be shown by considering the new bilinear form
\begin{eqnarray}
Z^t M' Z &=& n^t (A^{-1})^t G^{-1} A^{-1} n  -  m^t A(G-BG^{-1}B)A^tm
\nonumber \\
& &-  2n^t(A^{-1})^t(G^{-1}B)A^tm .
\label{ER23}
\end{eqnarray}
The original form,
\begin{equation}
Z^tMZ=n^tG^{-1}n-m^t(G-BG^{-1}B)m-2n^tG^{-1}Bm,
\label{ER24}
\end{equation}
is obtained by redefining $m'=A^tm$, $n'=A^{-1}n$, which is an
allowed transformation.

In addition to $\Theta(\Z)$ and  $GL(d,\Z)$ in (1),(2)
there is a less obvious element of
$O(d,d,\Z)$ which is a symmetry of the spectrum \cite{GRV,sw,GMR,GMR1}:

\item[3.] {\bf Factorized duality $D_i$}:\\

\bq
g_{D_i}=
\left(\begin{array}{cc} I-e_i & e_i \\ e_i & I-e_i\end{array}\right).
\eq{ERfd}
Here $e_i$ is zero, except for the $ii$ component which is $1$,
and $I$ is a $d$-dimensional identity matrix.
It can be shown straightforwardly that this transformation leaves the
partition function invariant as well.

Factorized duality is a generalization of the
$R\to 1/R$ circle duality in the $X^i$ direction.
Explicitly, $D_i$ takes  $R_i\to {1\over R_i}$ if the
$d$-dimensional background is a direct product of a $R_i$-radius circle and
a ($d-1$)-dimensional background, leaving the latter unchanged.

\end{description}

It turns out that the elements in (1),(2) and (3) generate the discrete
group $O(d,d,\Z)$ \cite{GRV,sw,GMR,GMR1}. An additional symmetry
is the worldsheet parity $\sigma\to -\sigma$ which acts on the background
by $B\to -B$ \cite{GMR1}. It
is not included in $O(d,d,\Z)$ since it interchanges $p_L$ with $p_R$, and
thus flips the sign of the Lorentzian norm $p_R^2-p_L^2$.
This completes the identification of the symmetry group ${\cal G}_d$.

To discuss the transformation of the oscillators in (\ref{ER10})
under $O(d,d,\Z)$ we have
to define the mode expansion of coordinates and conjugate momenta:
\bea
X^i(\sigma, \tau) &=& x^i + m^i\sigma + \tau G^{ij} (p_j -B_{jk}m^k) +
{i \over \sqrt{2}} \sum_{n\neq 0} {1\over n} [ \alpha^i_n(E) e^{-in(\tau -
\sigma)} + \tilde{\alpha}_n^i(E) e^{-in(\tau + \sigma)}],
\nonumber \\
2\pi P_i(\sigma,\tau) &=& p_i + {1\over \sqrt{2}} \sum_{n\neq 0} [ E_{ij}^t
\alpha_n^j(E) e^{-in(\tau-\sigma)} + E_{ij}
\tilde{\alpha}_n^j(E) e^{-in(\tau+\sigma)} ].
\label{KZ1}
\eea
Let us recall that the periodicity of the coordinates $x^i$ is $2\pi$ and
both the winding number $m$ and the oscillator number $n$ are integers.

The commutation relations of $\alpha^i_n(E)$, $\tilde{\alpha}_n^i(E)$,
$x^i$ and
$p_i$ are found by expanding the equal-time canonical commutation relations
\beq
[X^i(\sigma,0),P_j(\sigma',0)]=i\delta^i_j \delta(\sigma - \sigma').
\label{KZ2}
\eeq
The non-zero ones are
\bea
[x^i,p_j]&=&i\delta_j^i, \nonumber \\
{[}\alpha_n^i(E), \alpha_m^j(E)]&=&
[\tilde{\alpha}_n^i(E),\tilde{\alpha}_m^j(E)]=m G^{ij} \delta_{m+n,0}.
\label{KZ3}
\eea
Here the oscillators and their commutation relations are background-dependent
since we fixed the periodicity of the coordinates to be $2\pi$.

By inserting the mode expansion of $X$ and $P$ into eq. (\ref{ER12}),
and after integration and normal ordering, one
finds that the complete Hamiltonian (zero modes plus the oscillator
contribution) is:
\be
H={1\over 2}Z^tM(E)Z+N+\tilde{N} .
\label{KZH}
\ee
$Z$ and $M$ were defined in eqs. (\ref{ER18}) and (\ref{HZMZ}),
and the number operators are
\be
N=\sum_{n>0} \alpha_{-n}^i(E) G_{ij} \alpha_n^j(E), \;\;\;
\tilde{N}=\sum_{n>0} \tilde{\alpha}_{-n}^i(E) G_{ij} \tilde{\alpha}_n^j(E)
{}.
\label{KZNN}
\ee

The transformation of the background matrix $E$ under duality was given in
eq.~(\ref{ER22}). A pair of useful relations between the original metric $G$
and the dual metric $G'$ is:
\be
(d+cE)^t G' (d+cE) = G , \;\;\;  (d-cE^t)^t G' (d-cE^t) = G .
\label{KZGG}
\ee
The first relation is derived by writing $G'=(E'+(E')^t)/2$, and using the
expression for $E'$ from eq. (\ref{ER22}) to evaluate the left hand side.
The second relation can be derived by first writing the analog of
eq. (\ref{ER22}) for $(E')^t$, and then expressing $G'$ in terms of this
$(E')^t$.
Since duality is a symmetry of the string, it has to
act as a canonical transformation on all the oscillators; that is, it has
to preserve the commutation relations~(\ref{KZ2}). This fact, together with
the mode expansion~(\ref{KZ1}) uniquely fixes the action of duality on the
oscillators $\alpha_n^i$, $\tilde\alpha_n^i$. With the help of the
relations (\ref{KZGG}), one finds that the transformations of the
oscillators read~\cite{KZ}
\beq
\alpha_n(E) \rightarrow (d-cE^t)^{-1}\alpha_n(E'),\;\;\;
\tilde\alpha_n(E) \rightarrow (d+cE)^{-1}\tilde\alpha_n(E').
\label{KZ4}
\eeq
The number operators in the Hamiltonian (\ref{KZNN}) are manifestly
invariant under (\ref{KZ4}), together with $G\to G'$ as given in (\ref{KZGG}),
and therefore, we conclude that the entire spectrum is $O(d,d,\Z)$-invariant.

Finally, under worldsheet parity ($B\to -B$) the left-handed oscillators are
interchanged with the right-handed ones:
\be
{\rm worldsheet \; parity}: \;\; \alpha_n\leftrightarrow\tilde\alpha_n.
\label{KZ5}
\ee
The spectrum is manifestly invariant under this transformation.

Now, that we have specified how both the zero modes and oscillators
transform under duality, one can deduce the transformation of states,
$O|p_L,p_R\rangle$.
(Here $O$ is a polynomial in the oscillators, and $|p_L,p_R\rangle$ is
the state corresponding to the operator
${\rm exp} (ip_LX_L(\bar{z})+ip_RX_R(z))$.)

\begin{subsubsection}{Generators of $O(d,d,\Z)$ and $E\to 1/E$ Duality }

We now discuss a particular element of $O(d,d,\Z)$
which is an analog in $d$-dimensions to the $R\to 1/R$
circle duality. It is given by the {\em inversion} of the background
matrix $E$ \cite{GRV,sw}
\begin{equation}
E=G+B\to E'=G'+B'=E^{-1}.
\label{ER25}
\end{equation}
Here $G'$ and $B'$ are the symmetric and antisymmetric parts of $E'$.
{}From eq. (\ref{ER25}) it follows that
the metric and the antisymmetric tensor transform as
\begin{eqnarray}
G &\to& G' = (G-B G^{-1}B)^{-1}, \;\;\;
B \to B' = (B - G B^{-1}G)^{-1}, \nonumber\\
G^{-1}B&\to& -BG^{-1}.
\label{ER27}
\end{eqnarray}
Here $G'$ ($B'$) maintains the symmetry (antisymmetry) of $G$ ($B$).
The form for $B'$ in this equation assumes that $B$ is invertible (it
is possible to write $B'(G,B)$ in a non-singular form even when $B$ has
zero eigenvalues).

Under the transformation (\ref{ER27}) together with interchanging winding
modes with momenta ($n\leftrightarrow m$) the Hamiltonian in (\ref{ER13})
is manifestly invariant, and therefore, the spectrum as well.
Moreover, from the expression for $H$ in eqs. (\ref{ER12}) it follows that
$E\to 1/E$  interchanges $(2\pi\alpha')P \leftrightarrow X'$.
Another way to argue that the interchange of $P$ with $X'$ is
a symmetry is to notice that there exists a canonical
transformation doing precisely that \cite{GRV}.
It is defined by the generating function of the old and new
coordinates (the latter are denoted by a tilde):
\begin{eqnarray}
F & = &-{1\over 2\pi\alpha'}\int_0^{2\pi} X'(\sigma)\tilde
X(\sigma)d\sigma, \nonumber \\
P(\sigma) & = & \delta F/\delta X(\sigma)={1\over 2\pi\alpha'}
\tilde X'(\sigma), \nonumber \\
\tilde P(\sigma) & = & -\delta F/\delta\tilde X(\sigma)=
{1\over 2\pi\alpha'}X'(\sigma).
\label{ER30}
\end{eqnarray}
Note that the canonical transformation (\ref{ER30}) is equivalent to the
simple change of backgrounds (\ref{ER25}), (\ref{ER27}) when these are
constant.

In the absence of torsion $(B=0)$ one obtains the transformation
\begin{equation}
G\to G^{-1}.
\label{ER31}
\end{equation}
Recall that, by our starting point (\ref{ER2}):
\begin{equation}
G_{ij}=(\alpha')^{-1}\hat{G}_{ij}.
\label{ER32}
\end{equation}
Here we have inserted back $\alpha'$, so
$\hat{G}_{ij}$ is the conventional (dimensionful) metric and
$(\alpha')^{1/2}$ is the string length parameter.
We see that the transformation (\ref{ER31}) corresponds to either
$\alpha'\to (\det G)^2 (\alpha')^{-1}$ at fixed $G$ or, more
conventionally, to $\hat{G}\to\alpha'\hat{G}^{-1}$ at fixed $\alpha'$.
Conventionally, one says that the theory is invariant under a
change in $\hat{G}$ (at fixed $\alpha'$).

Duality can be given \cite{GRV} a somewhat
different meaning, which we illustrate in the
case of $d=1$, i.e.\ one compact dimension (discussed in
section 2.2).
Indeed, one may say that, given a string moving on a circle
of radius $R$, the theory has to decide what is the best value
of the ``quantum'' length $(\alpha')^{1/2}$. Such a choice has an
intrinsic ${\Z}_2$ ambiguity given by eq. (\ref{ER31}).
If the value $\alpha'=R^2$ is (is not) picked up the ``vacuum''
does not (does) break spontaneously $\Z_2$.

It is known that, when the symmetric point is picked up, an
enlarged gauge symmetry also follows corresponding to a level
one $SU(2)_L\times SU(2)_R$ representation of the
affine algebra  (for a review, see for example \cite{GinLH}).
Thus a $\Z_2$ preserving vacuum appears to preserve enlarged
gauge symmetries as well.

The duality operator that inverts the background matrix $E$
corresponds to the $O(d,d,\Z)$ element
\begin{equation}
g_D=\left( \begin{array}{cc} O & I \\
I & O \end{array} \right).
\label{ER33}
\end{equation}
As explained before, the oscillators of $X$
also transform under $O(d,d,\Z)$ and, in particular, under the inversion
duality $g_D$; in that case, their transformation is given by eq.
(\ref{KZ4}) with $d=0$ and  $c=I$. For example, at the
self-dual point $E=I$, one finds $\alpha_n\rightarrow -\alpha_n$,
$\tilde\alpha_n\rightarrow \tilde\alpha_n$.

We now consider some examples.
In the $c=d=1$ case only the duality transformation $g_D$ is available. The
group $O(1,1,\Z)$ consists of two elements\footnote{
Actually, $O(1,1,\Z)$ has four elements, but minus the identity acts
trivially as a fractional linear transformation. In general, whenever we
discuss $O(d,d,\Z)$ we actually mean the projective group $PO(d,d,\Z)$.},
only one of which
acts non-trivially on the compactification radius:
\begin{equation}
\left( \begin{array}{cc} 0 & 1 \\
1 & 0 \end{array} \right).
\label{ER34}
\end{equation}
The fractional linear transformation on $G_{11}$ leads to
\begin{equation}
G'_{11}=\frac{1}{G_{11}}.
\label{ER35}
\end{equation}
In this case, the manifold ${\cal M}$
defined in section 2.1 is shown in figure
1.C. The groups ${\cal G}$ and ${\cal G}_d$ defined there are
$O(1,1,\R)$ and $O(1,1,\Z)$, respectively.

In the $c=d=2$ case, corresponding to two compactified dimensions,
the factorized dualities, $D_1$ and $D_2$ (\ref{ERfd}) are represented by:
\begin{equation}
g_{D_1} = \left( \begin{array}{cccc} 0 & 0 & 1 & 0 \\
                           0 & 1 & 0 & 0 \\
                           1 & 0 & 0 & 0 \\
                           0 & 0 & 0 & 1
\end{array}\right);\quad
g_{D_2} = \left( \begin{array}{cccc} 1 & 0 & 0 & 0 \\
                           0 & 0 & 0 & 1 \\
                           0 & 0 & 1 & 0 \\
                           0 & 1 & 0 & 0
\end{array}\right).
\label{ER36}
\end{equation}
When the torus is a direct product of two circles ($G_{12}=B_{12}=0$)
$D_1$ and $D_2$ correspond to
$R_1\to 1/R_1$ with $R_2$ fixed,
and $R_1$ fixed while $R_2\to 1/R_2$, respectively.

The generator $\Theta_{12}$
which causes the shift of the $B_{12}$ field is represented by:
\begin{equation}
g_{\Theta_{12}} = \left( \begin{array}{cc} I & \Theta \\
                              0 & I \end{array}\right);\quad
\Theta = \left( \begin{array}{cc} 0 & 1\\
                             -1 & 0\end{array}\right)
\label{ER37}
\end{equation}

The group $GL(2,\Z)$ is generated by three symmetries:

\begin{description}

\item[a)]
The permutation symmetry $P_{12}$ giving rise to the element:
\begin{equation}
g_{P_{12}} = \left(\begin{array}{cc}P_{12} & 0 \\
                             0 & P_{12}\end{array}\right),\quad
P_{12} = \left( \begin{array}{cc} 0 & 1 \\
                        1 & 0\end{array}\right).
\label{ER38}
\end{equation}

\item[b)]
The reflections on each space component $i$, $R_i$, given by:
\begin{eqnarray}
g_{R_1} & = & \left( \begin{array}{cc} R_1 & 0\\
                               0 & R_1\end{array}\right),\quad
R_1 = \left( \begin{array}{cc} -1 & 0 \\
                         0 & 1\end{array}\right), \nonumber \\
g_{R_2} & = & \left( \begin{array}{cc} R_2 & 0 \\
                                  0 & R_2 \end{array}\right),\quad
R_2 = \left( \begin{array}{cc} 1 & 0\\
                               0 & -1\end{array}\right).
\label{ER39}
\end{eqnarray}
Obviously, one can get $R_2$ from $R_1$ by a permutation:
$R_2=P_{12}R_1P_{12}$.

\item[c)] The transformation $T_{12}$  of the form:
\begin{equation}
g_{T_{12}} = \left( \begin{array}{cc} T_{12} & 0 \\
                              0 & (T_{12}^t)^{-1}\end{array}\right),\quad
T_{12}= \left( \begin{array}{cc} 1 & 1\\
                        0 & 1\end{array}\right).
\label{ER40}
\end{equation}

\end{description}

The generators $P_{12}$, $T_{12}$, $R_1$ generate
$GL(2,\Z)$. Its subgroup $SL(2,\Z)$ is generated by $T_{12}$ and
the symmetry $S_{12}=R_1 P_{12}$.
The labels $S$ and $T$ are supposed to recall the standard $SL(2,\Z)$
generators.

The previous examples are useful in analyzing the general case \cite{GMR1}.
In the general $c=d$ case
the group $O(d,d,\Z)$ is generated by permutations $P_{ij}$,
reflections $R_i$, the symmetries
$T_{ij}$ and $\Theta_{ij}$,  and the factorized dualities  $D_i$,
$i,j=1,...,d$.
In the $d=2$ case the proof will be obtained in subsection 2.4.2
by referring to well known
properties of the modular group $SL(2,\Z)$; in the general case, this can be
obtained \cite{GMR1}
by straightforward, but rather tedious, algebraic manipulations.

Actually, it is enough to consider a single reflection and a single
factorized duality, say $R_1$ and $D_1$, and a single transformation of the
type $T$, and a single $\Theta$, say $T_{12}$ and $\Theta_{12}$.
The other elements can be obtained by repeated application of the previous
ones, together with permutations, $P_{ij}$, $i,j=1,...,d$.

We recall that we have shown the symmetry to occur for
the partition function on the worldsheet torus. We
will see in section 5 that it occurs for any worldsheet Riemann surface.
The moduli space of allowed backgrounds has, therefore, a
large degree of symmetry.
String theories having a  different description in terms of
the conventional target space metric and torsion are actually
physically identical.
{}From a string theory point of view the target space description has
fundamental domains.
Moreover, these fundamental domains have an orbifold structure
due to the occurrences of fixed points, that is there exist points in
moduli space which are fixed under the symmetry group.
It will turn out that these fixed points play an important role
in the identification of the gauge symmetry inherent in the
duality transformations. We thus turn to study some properties
of these fixed points.

For the $c=1$ compactification moduli space, as mentioned before,
the fixed point in the moduli
space of  circle compactifications coincides with the appearance of
enhanced gauge symmetries $SU(2)_L\times SU(2)_R$.

For the more general case, the duality transformation $E\to E^{-1}$
still has
strictly speaking a single fixed point given by
\begin{equation}
G=I,\;\;\; B=0,
\label{ER43}
\end{equation}
which is the unique solution to the equation $(G+B)^2=I$ when $G$ is
positive definite.
At the single fixed point the string forms a
level-one representation of the affine algebra
$SU(2)^d_L\times SU(2)^d_R$.

Let us consider now the slightly more general example of fixed points of
inversion duality, $E\to E^{-1}$, modulo $SL(d,\Z)$ and $\Theta(\Z)$
transformations:
\bq
E^{-1}=M^t (E+\Theta) M, \;\;\; M\in SL(d,\Z),\;\; \Theta\in \Theta(\Z).
\eq{ERfp}
It turns out that any background with a maximally enhanced symmetry falls
into this category \cite{GRV}. (By ``maximal,''
we mean an enhanced semi-simple
and simply-laced symmetry group of rank $d$ corresponding to a level 1
affine Lie algebra).
In those cases the background is \cite{egrs}
\bq
E_{ij}=C_{ij}, \;\;i>j,\;\;\; E_{ii}={1\over 2} C_{ii},\;\;\; E_{ij}=0,\;\;
i<j,
\eq{ERsyb}
where $C_{ij}$ is the Cartan matrix.
Thus,
\begin{equation}
E,\; E^{-1}\in SL(d,\Z),
\label{ER47}
\end{equation}
and $M=E^{-1}$, $\Theta=E^t-E$ (i.e., $\Theta_{ij}=-C_{ij},\; i>j,\;
\Theta_{ij}=C_{ij},\; i<j,\; \Theta_{ii}=0$) solve eq.~(\ref{ERfp}).
This fact shows that a maximally enhanced symmetry point is a
fixed point under some non-trivial $O(d,d,\Z)$ transformation, and
therefore, an orbifold point in the moduli space.
Fixed points corresponding to non-maximal enhanced symmetries can be found
by using factorized duality instead of the full inversion $E\to E^{-1}$.

We have not shown that the condition $E\in SL(d,\Z)$ either implies an
enhanced symmetry or exhausts all solutions.
In fact, already for $B=0$ one knows examples of models which correspond to
physical fixed points, but do not have an enhanced affine algebra.
Such systems are the 24-dimensional Leech lattice and the
8-dimensional $E_8$ lattice (with the full Cartan matrix as the
background metric).
These systems contain new $(n, 0)$ conserved currents, where
$n>1$.

Furthermore, examples where no higher spin conserved currents exist are also
known, as described in the following subsection, where the special
$c=d=2$ case is studied in detail.

\end{subsubsection}

\begin{subsubsection}{The $d=2$ Example}
The symmetry group, ${\cal G}_d$, of $d=2$ flat compactifications is
generated by $O(2,2,\Z)$ and $B\to -B$.
As $O(2,2,\R)$ decomposes into $SL(2,\R)\times SL(2,\R)$ one may expect
$O(2,2,\Z)$ to factorize into $SL(2,\Z)\times SL(2,\Z)$ \cite{DVV1,sw}.
In fact, the moduli space is isomorphic to
$SL(2,\R)/U(1)\times SL(2,\R)/U(1)$ but, instead, the duality symmetry group
turns out to be
isomorphic to $SL(2,\Z)\times SL(2,\Z)\otimes_S [\Z_2\times \Z_2]$
\cite{DVV1,sw,GMR1}.

To show this, one organizes the four real data,
$G_{11}$, $G_{12}$, $G_{22}$, $B_{12}$, into two complex
coordinates $\rho$ and $\tau$ in the following manner \cite{DVV1}:
\begin{eqnarray}
\tau & \equiv & \tau_1+i\tau_2
=\frac{G_{12}}{G_{22}}+\frac{i\sqrt{G}}{G_{22}} \nonumber \\
\rho & \equiv & \rho_1+i\rho_2 = B_{12}+i\sqrt{G}\quad
\mbox{ where } G=G_{11}G_{22}-G_{12}^2
\label{ER48}
\end{eqnarray}
The inverse relation is:
\begin{equation}
E = \frac{\sqrt{G}}{\tau_2} \left( \begin{array}{cc}
\tau_1^2+\tau_2^2 & \tau_1 \\
\tau_1 & 1\end{array}\right) + b\left( \begin{array}{cc} 0 & 1\\
-1 & 0\end{array}\right),
\label{ER49}
\end{equation}
where $b=B_{12}$.
The target space is a two-dimensional torus and as such it
supports a complex structure $\tau$ and a K\"ahler structure
$\rho$.
All in all, as we are going to explain, it turns out that there is a
minimal set of four ``generators'' for the symmetry group
${\cal G}_d$.
``Generators,'' in this context, means a set of group elements from which
the whole group is generated by forming products.

A (non-minimal) set of generators is found as follows. $SL(2,\Z)$ is
generated by two elements,
so a product of two $SL(2,\Z)$ has four ``generators.'' Two other
generators are the one interchanging the two $SL(2,\Z)$, and the reflection
$X^1\to -X^1$. Moreover, the
worldsheet parity $\sigma\to -\sigma$ ($B\to -B$) provides a seventh generator.
The $SL(2,\Z)^2$ generators are
\ber
S:&&\;\;\; \rho\to\rho,\;\;\;\tau\to-\frac{1}{\tau},\nonumber
\\
T:&&\;\;\; \rho\to\rho,\;\;\; \tau\to\tau+1, \nonumber
\\
S':&&\;\;\; \rho\to -\frac{1}{\rho},\;\;\;\tau\to\tau,\nonumber
\\
T':&&\;\;\; \rho\to\rho+1,\;\;\; \tau\to\tau.
\eer{ERsl2gen}
The first two transformations reflect the fact that the target space is a
2-$d$ torus,
and thus unmodified by modular transformations of its complex structure.

The fourth transformation expresses the periodicity in the
antisymmetric torsion $B_{12}$.
The third transformation is a stringy symmetry that for $B=0$ takes the
volume in target space into its inverse.

Writing $(p_L^2, p_R^2)$ in terms of $\tau$ and $\rho$
as
\begin{eqnarray}
p_L^2 & = & \frac{1}{2\rho_2\tau_2} \left| (n_1-\tau
n_2)-\rho (m_2+\tau m_1)\right|^2 \nonumber \\
p_R^2 & = & \frac{1}{2\rho_2\tau_2} \left| (n_1-\tau n_2)
-\bar\rho(m_2+\tau m_1)\right|^2
\label{ER51}
\end{eqnarray}
exposes three more symmetries. The first one exchanges the complex and
K\"ahler structures
\begin{equation}
D_2:\;\;\; (\tau, \rho)\to(\rho, \tau).
\label{ER52}
\end{equation}
This symmetry is of a stringy nature, it is actually the $O(2,2,\Z)$
element we called {\em factorized duality}~(\ref{ERfd}) (here acting on the
second coordinate). This can be seen by setting $B_{12}=G_{12}=0$. In
this case eq.~(\ref{ER52}) corresponds to $G_{22}\to 1/G_{22}$.
The other symmetry is
\begin{equation}
R:\;\;\;\;\; (\tau, \rho)\to (-\bar\tau, -\bar\rho).
\label{ER53}
\end{equation}
This equation corresponds to the $X^1\to -X^1$ reflection we mentioned
above. Indeed, this symmetry changes the sign of $G_{12}$ and $B_{12}$,
while leaving $G_{11},G_{22}$ invariant.
The last symmetry, which corresponds to {\em worldsheet} parity ($\sigma\to
-\sigma$), interchanges $p_L$ and $p_R$ and corresponds to
$B\to -B$, while all other data are invariant. It is given by
\bq
W:\;\;\;\;\; (\tau,\rho) \to (\tau,-\bar\rho).
\eq{ERwsp}
Since this symmetry interchanges $p_L$ with $p_R$, it changes the sign of
the Lorentzian norm $p_L^2-p_R^2$ and thus cannot belong to $O(2,2,\R)$.

The set of generators we presented is not a minimal one.
One can remove the $S'$, $T'$ and $R$ elements from the set of generators,
as they can be obtained from $S$, $T$, $D_2$ and $W$ by the help of
eqs.~(\ref{ER52}, \ref{ERwsp})
($S'=D_2SD_2$, $T'=D_2TD_2$, $R=D_2WD_2W$).
The four symmetries ($S,T,D_2,W$) provide a minimal set of ``generators''
for the symmetry group.

The target space moduli space has the global structure of a product of
the fundamental domains of two tori, that is $SL(2,\R)/U(1)\times
SL(2,\R)/U(1)$, modded out by the discrete symmetries. In figure 2.B we
show, as an example, the slice $\rho=\tau$ of the fundamental domain.
A similarity between worldsheet and target space symmetries emerges.
We will further discuss this issue later in section 5.

Not only  do target space and worldsheet symmetries show
similarity, there are cases where they could even be
interchanged. Let us consider the one-loop
partition function $Z(\tau, \rho;\sigma)$.
It describes a string moving in the
target space background defined by the data $\tau, \rho$ when
the worldsheet is a torus with a given complex structure $\sigma$.
This partition function is symmetric under the
interchange of the complex structures of the target space and
the worldsheet \cite{DVV1}
\begin{equation}
Z(\tau, \rho;\sigma)=Z(\sigma,\rho;\tau).
\label{ER54}
\end{equation}
Together with the factorized duality,
$\tau\leftrightarrow \rho$, this gives rise to a triality.
This result also generalizes (partly)
to higher dimensional backgrounds and higher
genus worldsheets  \cite{GMR} as we discuss in section 5.

We conclude the example by noting several points in moduli space
which are fixed points under some (non-trivial) duality.

Let us describe at first two points with a maximally
enhanced worldsheet symmetry.
The first corresponds to the
$(SU(2)\times SU(2))_L\times (SU(2)\times SU(2))_R$ symmetric point,
described by $E=I$, or equivalently by
\begin{equation}
(\tau, \rho)=(i, i).
\label{ER56}
\end{equation}
The second corresponds to the $SU(3)_L\times SU(3)_R$ symmetric point,
described by
\bq
E=\left(\begin{array}{cc} 1 & 1 \\
                          0 & 1\end{array}\right),
\eq{ER1101}
or  equivalently by
\begin{equation}
(\tau, \rho)=\left(\frac{1}{2} +\frac{i\sqrt{3}}{2},
\frac{1}{2} +\frac{i\sqrt{3}}{2}\right).
\label{ER55}
\end{equation}
Equations (\ref{ER56}) and (\ref{ER55}) are the two ``orbifold singularities''
in the fundamental domain shown in figure 2.B.

Finally, we present a continuous set of fixed points under the $E\to E^{-1}$
duality, combined with the $B\to -B$ worldsheet parity. These backgrounds
correspond to the  special value $\tau=i$, while $|\rho|=1$,
or equivalently
\bq
E(g,b)=\left(\begin{array}{cc} g & b \\
                          -b & g\end{array}\right),\;\;\; {\rm s.t.}
\;\;\; g^2+b^2=1.
\eq{ERcfp1}
It now follows that
\bq
E(g,b)^{-1}=E(g,-b),
\eq{ERcfp2}
and therefore, it is a fixed point of $WD$ (recall that $D=D_1D_2$).
This continuous set of fixed points of an element of ${\cal G}_d$ includes two
fixed points  of $O(2,2,\Z)$. The first is  $b=0, g=1$, that is the
symmetric background (\ref{ER56}). The second is a fixed point of
$T'D\in O(2,2,\Z)$ at the special value $b=\pm 1/2, g=\sqrt{3}/2$.
This last example shows \cite{sw}
that there are orbifold points in the moduli space
that do not have extra conserved chiral currents.

\end{subsubsection}

\end{subsection}

\begin{subsection}{Duality of the Heterotic-String Spectrum}
\setcounter{equation}{0}

The Heterotic-string action in flat background, including the coupling to a
background gauge field $A_\mu^\alpha$ and a background
antisymmetric tensor $B_{\mu \nu}$, $\mu, \nu=1, \ldots, d,
\;\alpha=1,\ldots, 16 $ is (for a review, see \cite{GSW})
\begin{eqnarray}
S & = & \frac{1}{2\pi} \int d^2 z \left[(G_{\mu\nu}+B_{\mu\nu})\partial
X^\mu \bar\partial X^\nu + A_{\mu\alpha}\partial X^\mu
\bar\partial X^\alpha \right. \nonumber \\
 &&  +  \left. (G_{\alpha\beta}+B_{\alpha\beta})\partial X^\alpha
\bar\partial X^\beta \right] + (\mbox{fermionic terms}),
\label{ER57}
\end{eqnarray}
with the constraint that the  $X_\alpha$ are chiral bosons.
Here we work in the orthonormal gauge (\ref{ERgauge}), and
the complex worldsheet coordinates and derivatives are
\bq
z={1\over \sqrt{2}}(\tau+i\sigma),\;\;\;
\pa={1\over \sqrt{2}}(\pa_{\tau}-i\pa_{\sigma}),
\eq{zst}
and therefore, $d^2 z=d\sigma d\tau$, and
$\pa x \pab x={1\over 2}(\pa_{\tau} x)^2+{1\over 2}(\pa_{\sigma} x)^2$.

The index $\alpha$ denotes an internal index.
The internal coordinates live on the weight lattice of $E_8\times
E_8$ or $Spin(32)/\Z_2$.
The indices $(\mu, \alpha)$ label a ($16+d$)-dimensional
orthonormal basis: $G_{\mu\nu}\equiv\delta_{\mu\nu}$,
$G_{\alpha\beta}\equiv\delta_{\alpha\beta}$.

The moduli space of the heterotic string in $d$-dimensional toroidal
backgrounds is isomorphic, locally, to the symmetric space
$O(d+16,d,\R)/(O(d+16,\R)\times O(d,\R))$ \cite{N,NSW}. Here we will study
the global structure of the moduli space following \cite{GRV}, namely, we will
identify the target space discrete symmetries.

Due to chirality the geometrical interpretation of the heterotic background
fields is not manifest.
Therefore, we will ``embed'' the compactified section of the
heterotic string (not including the worldsheet fermions) in
$16+d$ space-coordinates of a bosonic string.
The Lorentzian lattice of the bosonic string is of the form:
\begin{equation}
\left( \begin{array}{c} p_R \\ p_L\end{array}\right) =
\left( \begin{array}{cc} p_R^i, & p_R^I \\ p_L^j, & p_L^J
\end{array}\right),\quad i,j=1, \ldots, d,\quad I,J=1,\ldots, 16.
\label{ER58}
\end{equation}
The embedding is such that the lattice generated by
\begin{equation}
\left( \begin{array}{cc} p_R^i & 0 \\ p_L^j & p_L^J\end{array}
\right)
\label{ER59}
\end{equation}
is a sub-lattice of the original one, and gives the spectrum of
the heterotic string.
This will allow us to get the spectrum of the  heterotic sting from
that of the bosonic string by truncation.
The sub-spectrum of the dual spectrum is given by a duality
transformation of the original sub-spectrum.
Thus, all one has to find is the $E$-matrix of parameters of the
relevant bosonic coordinates.

Let $\{ E_I^{\alpha} | I=1, \ldots, 16\}$, $\{ e_i^\mu | i=1, \ldots, d\}$
satisfy
\begin{eqnarray}
E_I \cdot E_J = 2G_{IJ}, & &  e_i\cdot E_I=0, \nonumber \\
e_i  \cdot  e_j=2G_{ij}, & &
\label{ER60}
\end{eqnarray}
where
\begin{eqnarray}
G^{\alpha\beta}=2G_{IJ}(E^{I\,*})^\alpha (E^{J\,*})^\beta , \nonumber \\
B^{\mu\nu} = 2B_{IJ}(e^{i\,*})^\mu (e^{j\,*})^\nu , & &
B^{\alpha\beta}=2B_{IJ}(E^{I\,*})^\alpha (E^{J\,*})^\beta , \nonumber \\
A^{\mu\alpha} = 2A_{iJ} (e^{i\,*})^\mu (E^{J\,*})^\alpha , & &
\label{ER61}
\end{eqnarray}
and $E^{I\,*}$ $(e^{i\,*})$ is the basis dual to $E_I$ $(e_i)$.

The $A$ field stands for an antisymmetric tensor and a metric in
the coefficients $E_{iI}$, $E_{Jj}$ of the background matrix.
It is impossible for $E_{IJ}$ to depend on $A$, as the
gauge field is a continuous parameter, while there are only
discrete solutions to $E_{IJ}$ which allow truncation of the
spectrum of the $(16+d, 16+d)$ model to a chiral $(16+d, d)$
model.

The operator $E$ of the embedding bosonic string has the form:
\begin{equation}
E\equiv B+G=\left( \begin{array}{cc} \left(G+B+{1\over 4} A^K A_K\right)_{ij}
& A_{iJ}
\\
0 & (G+B)_{IJ}\end{array}\right)
\label{ER62}
\end{equation}
To prove it, let us pick up a basis $\alpha\equiv(\alpha_i,
\alpha_I)$ of a lattice $\Lambda^{(d+16)}$ with the metric $2G$:
\begin{eqnarray}
\alpha_i & = & \big( e_i, \frac{1}{2} A_i^K E_K\big),\nonumber \\
\alpha_I & = & (0, E_I).
\label{ER63}
\end{eqnarray}
The dual basis $\alpha^*$ is
\begin{eqnarray}
\alpha^{i\,*} & = & (e^{i\,*}, 0),\nonumber \\
\alpha^{I\,*} & = & \big(-\frac{1}{2} A_i^I e^{i\,*}, E^{I\,*}\big).
\label{ER64}
\end{eqnarray}
Thus, the bosonic Lorentzian lattice $(p_L, p_R)$ has the form:
\begin{eqnarray}
p_R & = & [(n)^t+(m)^t(B-G)](\alpha^*)\nonumber \\
& = & n_i\alpha^{i\,*} + n_I\alpha^{I\,*} \nonumber \\
& & +(m_i, m_I)\left(
\begin{array}{cc} \left(B-G-{1\over 4} A^K A_K\right)_{ij} & 0 \\
-A_{jI} & (B-G)_{IJ}\end{array} \right)
\left( \begin{array}{c} \alpha^{j\,*} \\ \alpha^{J\,*}\end{array}\right)
\nonumber \\
& = & \big( \big[ n_i+m_j (B-G)_{ji}-\frac{1}{4} m_jA_j^K A_{Ki}
-\frac{1}{2} (n_J+m_I(B+G)_{IJ})A_i^J\big] e^{i\,*}, \nonumber \\
& & [n_J+m_I(B-G)_{IJ}]E^{J\,*}\big).
\label{ER65}
\end{eqnarray}
Here it turns out that $E_{IJ}=(G+B)_{IJ}$ is given by
\bq
E_{IJ}=C_{IJ}, \;\;I>J,\;\;\; E_{II}={1\over 2}C_{II},
\;\;\; E_{IJ}=0,\;\; I<J,
\eq{ERcatr}
where $C_{IJ}$ is the Cartan matrix of $E_8\times E_8$.
Therefore, the vectors $V_J, W_J$ defined by
\begin{eqnarray}
n_J+m_I(B+G)_{IJ} & \equiv & V_J, \nonumber \\
n_J+m_I(B-G)_{IJ} & \equiv & W_J,
\label{ER66}
\end{eqnarray}
are in the root lattice of $E_8\times E_8$.

The truncation to a sub-lattice describing the heterotic string
spectrum is the set of $(n_I, m_I)$ such that $W_J=0$.
Defining $P$ and $L$ by
\begin{eqnarray}
n_i e^{i\,*} & \equiv & \frac{1}{2} P,\nonumber \\
m_j(B-G)_{ji} e^{i\,*} & \equiv & -L-BL,
\label{ER67}
\end{eqnarray}
we get
\begin{equation}
p_R=\big( \frac{1}{2}P-L-BL-\frac{1}{2}VA-\frac{1}{4}A(AL),
0\big).
\label{ER68}
\end{equation}
In the same way
\begin{equation}
p_L = [(n)^t +(m)^t(B+G)](\alpha^*)=\big(\frac{1}{2} P+L-BL
-\frac{1}{2}AV-\frac{1}{4}A(AL), V +AL\big).
\label{ER69}
\end{equation}
The vectors $(p_L,p_R)$ span precisely the $(d+16,d)$-dimensional even
self-dual Lorentzian lattice of the heterotic string \cite{NSW}.
Therefore, the bosonic truncated spectrum coincides with the
heterotic model.

The $A$ field stands for a metric
\begin{equation}
\left( \begin{array}{cc} 0 & {1\over 2} A_{iJ} \\ {1\over 2} A_{Ij} & 0
\end{array}\right),
\label{ER70}
\end{equation}
and an antisymmetric tensor
\begin{equation}
 \left( \begin{array}{cc} 0 & {1\over 2} A_{iJ}\\
  -{1\over 2} A_{Ij} & 0\end{array}
 \right),
\label{ER71}
\end{equation}
in the cross directions.
It also affects the metric in the compactified space:
$G_{ij}\to G_{ij}+\frac{1}{4}A_i^K A_{Kj}$.

The background-inversion
duality transformation for the bosonic string $E'=E^{-1}$,
induces the transformation on the heterotic string background
fields:
\begin{eqnarray}
A'_{ij} & = & -\big( G+B+\frac{1}{4}A^K A_K\big)^{-1}_{ij}
A_{jI}(G+B)_{IJ}^{-1},\nonumber \\
(G'+B')_{ij} & = & \big(
G+B+\frac{1}{4}A^KA_K\big)^{-1}_{ij}-\frac{1}{4}A'{}^K_i A'_{Kj},
\nonumber \\
(G'+B')_{IJ} & = & (G+B)_{IJ}^{-1}\equiv (G+B)_{IJ}\;\;\;{\rm mod}\;
SL(16,\Z)\;\mbox{ transformation }.
\label{ER72}
\end{eqnarray}

So far we have described a particular element of the discrete target space
symmetries of the heterotic string in flat backgrounds.
Following the discussion for the  bosonic string in section 2.4, one can
show that here target space duality is generalized to a group isomorphic to
$O(d+16,d,\Z)$. This group is the subgroup of $O(d+16,d+16,\Z)$ that
preserves the heterotic structure of $E$ in  (\ref{ER62}) (namely, that
transforms the
lower two blocks 0 and $(G+B)_{IJ}$ into themselves), while
acting on $E$ by fractional linear transformations (\ref{ER22}).

\end{subsection}

\begin{subsection}{Duality as a Gauge Symmetry}
\setcounter{equation}{0}

In this section we relate (most of) the discrete symmetry group
${\cal G}_d$ -- the subgroup $O(d,d,\Z)$ for $d$-tori compactifications --
with a gauge symmetry.

\begin{subsubsection}{Motivation}

We have described in some detail the discrete symmetry group ${\cal G}_d$
which relates physically equivalent theories in the moduli space
of flat compactifications; some of these symmetries are of a
stringy nature, and are rather surprising from a field theory point of
view. So far, it was shown that the worldsheet partition functions
of any two
theories related by a $O(d,d,\Z)$ transformation are identical\footnote{
For $d>1$ we have shown explicitly only duality at genus $g=1$;
the proof for $g>1$ is
along the lines of section 2.3, and will also be
discussed later in section 5.}.

We will
show that the theories themselves
are {\em identical} by demonstrating that two
theories related by $O(d,d,\Z)$
are actually gauge equivalent under some gauge group.
Uncovering such a gauge symmetry would afford one a glimpse
into the structure of the large symmetry which is supposedly
associated with string theory.  A gauge symmetry could ensure
that the symmetry group $O(d,d,\Z)$ indeed persists to all orders in string
perturbation theory.  For example, such symmetries in ${\cal G}_d$ would be
protected from explicit breaking had they actually been  residual
gauge symmetries surviving an incomplete gauge fixing of some
continuous gauge group.

The discrete nature of the group ${\cal G}_d$
suggests that if it is a residual gauge symmetry it is to be
associated with large gauge transformations, namely, gauge
transformations which are not connected (in the space of gauge
transformations) to the identity.  Such a
residual worldsheet gauge symmetry is familiar in string theory; it is the
modular group acting on the worldsheet metric.

In string theory a
large discrete symmetry, the modular group of Riemann
surfaces (isomorphic to $SL(2,\Z)$ on the torus, for example) is present;
it reflects the underlying general coordinate  invariance and Weyl invariance
which are there when string theory is viewed as a theory of two dimensional
gravity.  This is the group  of  residual symmetries in
the conformal gauge when the worldsheet is a torus.  There are
traces of general coordinate and Weyl invariance in the Veneziano
and Virasoro-Shapiro amplitudes.

The similarity of the target space
${\cal G}_d$ symmetries with those of the worldsheet
modular group encourages to view
them as some gauged fixed version of a similar larger symmetry
in the space of allowed string theory backgrounds.

Another hint
in that direction emerges from the effective target space
Lagrangian of the theory \cite{DHS}.  Consider the case of the bosonic
string with one compactified dimension ($d=1$).
The target space picture consists of an effective $SU(2)_L\times SU(2)_R$
scalar potential which has flat directions. The potential is constructed in
terms of scalar fields in the (1,1) representation of the $SU(2)_L\times
SU(2)_R$ symmetry group. A special point among all possible vacua
is the one where all scalar fields obtain, say, a vanishing expectation
value; in that case the $SU(2)_L\times SU(2)_R$ symmetry is retained.
For any non-zero expectation values of the scalar fields, the symmetry is
spontaneously broken down to $U(1)_L\times U(1)_R$.

In the spontaneously broken case, the space of classical
minima separates into gauge equivalent ``shells;'' each shell is a
four dimensional surface characterized by its radius, and all points on
a given shell are gauge equivalent (the surface is four dimensional
as it is parametrized by the transformations associated with
the four broken symmetry generators). In a finite scale invariant
field theory (such as a four dimensional $N=4$ SUSY theory), different
surfaces would also be physically equivalent, as they represent
equivalent theories which differ only by their scale; that scale is
spontaneously generated and corresponds to the
expectation value of a scalar field\footnote{
In regular field theory this
picture would be valid for all surfaces.
However, in string theory things are
more complex. The number of compactified dimensions, $d$, is bigger than 1,
and at a finite distance from the point where the scalar
fields had zero expectation value, the $SU(2)_L\times SU(2)_R$ symmetry
is restored; the reappearance of massless states is special to string
theory and will be dealt with in detail in section 2.8.  We will ignore
this extra structure for a while.}.

Using the $SU(2)_L\times SU(2)_R$ symmetry, any
scalar expectation value in the adjoint representation can
be pointed along a single component, for example, in the direction of the
Cartan sub-algebra; let us denote the value of
that component by $h$.  A discrete gauge transformation, such as
a rotation by an angle $\phi$ around an appropriate $SU(2)$ isospin
axis (for example, if $h$ points in the $J^3$ direction, a rotation around
the $J^1$ axis can be used),
rotates the scalar configuration $h$ into $-h$. The
two configurations are gauge equivalent (by a discrete gauge
transformation). From the target space point of view, this
identification corresponds to the $d=1$ circle duality. Let us repeat that the
two configurations are members of the same gauge equivalent
surface, and therefore, are separated by a finite distance.

The correspondence of the above observation
with target space duality is seen from the worldsheet
perspective. The point where $SU(2)_L\times SU(2)_R$ is unbroken corresponds
to the self-dual $R=1$ point in the moduli space of circle compactifications.
The spontaneous breaking of the target space symmetry (by giving an
expectation value $h$ to a target space scalar field) is achieved
by an explicit breaking of the corresponding worldsheet symmetry
\cite{DHS}. The explicit breaking is done by adding
to the worldsheet Lagrangian $L_0$, describing the self-dual point, a set of
compatible marginal operators $f(z,{\bar z})$ to obtain the modified
conformal Lagrangian, $L(\delta R)$, describing a string moving in one of
the dimensions on a circle of radius $R=1+\delta R$,
\be
L(\delta R)=L_0+\delta R\; f(z,{\bar z}).
\label{ERG1}
\ee
Expanding the duality relation $R'=1/R$ around $R=1$, one obtains the dual
Lagrangian $L'(\delta R)$:
\be
L'(\delta R)=L(-\delta R)+{\cal O}((\delta R)^2)=
L_0-\delta R\; f(z,{\bar z})+{\cal O}((\delta R)^2).
\label{ERG2}
\ee
This shows that the theories $L$ and $L'$, obtained by perturbing with
$h\equiv\delta R$ or with $-h$, are dual to each other.
But we have argued that they are
also gauge equivalent by a transformation of the target space
$SU(2)_L$ (or $SU(2)_R$) gauge symmetry!
Therefore, for $d=1$, duality is related to a target space gauge symmetry.

The above consideration has
another immediate worldsheet counterpart.  The affine Lie
algebra on the worldsheet is enhanced at the self-dual point from
$U(1)_L\times U(1)_R$ to $SU(2)_L\times SU(2)_R$, with the appearance of two
more (1,0) chiral currents,
and two more (0,1) anti-chiral currents (the (1,0) and (0,1)
indicate the scaling dimensions of these worldsheet operators).
As will be explained in more detail soon,
the perturbations $h$ and $-h$ are related by a reflection of the
current (or the anti-chiral current) contained in the marginal operator
$f(z,{\bar z})$; this reflection is an element in the Weyl group
of the enhanced symmetry $SU(2)_L$ (or $SU(2)_R$). Therefore, the two
perturbations are related also by a worldsheet symmetry.

The relation between duality among theories near the self-dual point and Weyl
symmetries \cite{can} is valid all along the line of circle compactification
(this will be shown for the general case).
This concludes the discussion in the case of $d=1$ flat compactifications.

As $d$  increases, so does the complexity of
the group ${\cal G}_d$. In particular, the couplings $B_{ij}$ appear, and the
moduli space has an increasing number of special points.  Several
questions arise: is there a generalization of the Weyl group
construction of duality for general values of $d$?  What is the
relation between the symmetry group ${\cal G}_d$ and transformations
appearing in the Weyl construction? Is there a preferred subgroup
of special points to be utilized in the Weyl construction? What
type of gauge group exists for general values of $d$?  These issues
are explored in the following subsection.

\end{subsubsection}

\begin{subsubsection}{The Weyl Construction}

The search for a basis of truly marginal operators in the moduli space,
in the neighborhood of a CFT  defined by some worldsheet
Lagrangian, $L$, was discussed in section 2.1.
For the purpose of the discussion in this section we will
consider expansions around Lagrangians $L(E_0)$ defined at points with
enhanced symmetry in the moduli space of $d$-dimensional
toroidal compactifications (here $E_0$
denotes a background matrix corresponding to a point with enhanced
symmetry). We do not know at this
stage what is the general classification of such points.  It may well
be that all of them are needed to be considered in order to fully appreciate
the larger gauge symmetry.  We limit our discussion to those
characterized by the existence of extra (1,0) currents and (0,1) currents,
in addition to the Abelian ones.
All those have been shown to be fixed points (under some elements of the
duality group).  It will turn out that
they carry quite some information.

The moduli space of flat
compactifications contains points with an enhanced affine symmetry group $G_L
\times G_R$. We will discuss the discrete set of points with a {\em maximally
enhanced symmetry}. By ``maximally''
we mean that $G_L=G_R=G$, where $G$ is semi-simple
and rank$(G)=d$. (It is possible to generalize the study to any $G_L \times
G_R$). These special points are described by a background matrix, which we
denote by $E_M$.
The symmetric part of $E_M$, $G_M$, is one half the Cartan-matrix of a
simply-laced group (modulo the discrete
symmetry group).  The antisymmetric part, $B_M$,  is the Cartan valued torsion
(\ref{ERsyb}) (modulo the discrete symmetry group).
The point $E_M$ can be described by
a level 1 WZNW model on the group manifold of $G$.  Alternatively,
the theory  might be described  in terms of $d$ free bosons on the constant
background $E_M$, with a Lagrangian
$L(E_M)$. These theories contain the necessary  extra (1,0) currents and (0,1)
currents,  in the adjoint representation of $G_L$ and $G_R$.

Thus we are choosing to expand around special points where the
spectrum of $L$ is equipped with some current algebra.  When the
enhanced symmetry is $G_L\times G_R$, with $G_L=G_R=G$,
the truly marginal operators
can be composed out of the $({\rm dim}\,G)^2$ (1,1) operators resulting
from multiplication of the chiral and anti-chiral currents.
The algebraic structure enables one to state a simple condition,
valid to all orders in perturbation theory, on what are the
allowed truly marginal operators.

Denote the chiral (anti-chiral) (1,0) ((0,1))
currents by $J^a$ $(\bar{J}^b)$,  where $a,b=1,...,{\rm dim}\,G$.
The perturbed theory
\be
L(E_M, \epsilon)=L(E_M)+\Delta L,
\qquad   \Delta L=\epsilon_{ab}J^a \bar{J}^b,
\label{ERG3}
\ee
is a truly marginal deformation of $L(E_M)$  whenever the matrix $\epsilon$
satisfies \cite{CS}
\be
\sum_{n, m} \epsilon_{na}\epsilon_{mb} f^{nmc}=0
\label{ERG4}
\ee
for all $a, b, c$. Here $f^{nmk}$ are the structure constants of the group
$G$. The perturbation $\Delta L$ in eq. (\ref{ERG3}) is equivalent to a
perturbation
\be
\Delta L'=\epsilon_{ab}J'^a \bar{J}'^b
\label{ERG5}
\ee
if $J'$ $(\bar{J}')$ are connected to $J$ $(\bar{J})$ by a continuous
transformation in the group $G_L$ $(G_R)$. Taking into
account the ``gauge'' equivalence, the moduli space in the neighborhood of
$E_M$ is spanned by the (1,1) truly marginal operators in the Cartan
sub-algebra of $G_L\times G_R$,
\be
\Delta L=\epsilon_{ij}H^i\bar{H}^j, \qquad    i,j=1,...,d, \qquad
{\rm rank}\,G=d.
\label{ERG6}
\ee
Their number is $({\rm rank}\,G)^2$; a more involved way to appreciate this
number is outlined below.

The number of truly marginal independent
directions is $({\rm dim}\,G)^2$ (generated by the $J^a\bar{J}^b$ operators,
$a,b=1,...,{\rm dim}\,G$).
But, because of the condition (\ref{ERG4}),
the set of critical points that can be
reached from $L(E_M)$ by conformal deformations spans a lower dimensional
surface in the large space; its dimension is $2({\rm dim}\,G-d)+d^2$.
However,  the dimension of the physical moduli space in the
neighborhood of these points is only $d^2$.
This is due to the fact that different
truly marginal perturbations are equivalent under continuous transformations
in the group $G_L\times G_R$.\footnote{
The counting goes as follows: eliminating the gauge equivalence,
by subtracting the $2({\rm dim}\,G-d)$ broken gauge symmetries from
the $[2({\rm dim}\,G-d)+d^2$]-dimensional space, leaves $d^2$ physical
directions.}
For example,
in the $d=1$ case, the space of critical points is
a 5-dimensional surface -- spanned by deformations of
the type $(\sum_{a=1}^3 \alpha_a J^a)(\sum_{b=1}^3 \beta_b \bar{J}^b)$ (the
overall scale is irrelevant) --
in a 9-dimensional  Euclidian space spanned by $J^a\bar{J}^b$, $a,b=1,...,3$.
The physical moduli space is one-dimensional, and generated, say, by the
$J^3\bar{J}^3$ operator.

Suppose one can relate $\epsilon'$
to $\epsilon$, in the tangent space at the point $E_M$,
by gauge transforming the currents $H$ and $\bar{H}$.
Then  the discrete symmetry relating
$L(E_M, \epsilon)$ to $L(E_M,\epsilon')$ is a spontaneously broken gauge
symmetry of the enhanced symmetry group $G$.
The symmetry is extended to the entire moduli space. The products
of such transformations, originating at different enhanced symmetry
points in the moduli space,
correspond to a gauge transformation of a bigger group.
We will discuss these points in more detail in what follows.

\vskip 0.1in
\noindent
{\bf The symmetry group ${\cal G}_d$}:
\vskip 0.1in

\noindent
Let us recall briefly some properties of the discrete group ${\cal G}_d$ (see
section 2.4). ${\cal G}_d$ is
isomorphic to the semi-direct product of  $O(d,d,\Z)$ and a 2-dimensional
parity symmetry.  In the language of background fields,  $2d$ parity
relates the antisymmetric tensor $B_{ij}$ to $-B_{ij}$. The subgroup
$O(d,d,\Z)$ acts on the background fields as a fractional linear
transformation.
The generators have been described in detail in subsection 2.4.1; here we
only recall that one can generate $O(d,d,\Z)$ by permutations $P_{ij}$,
reflections $R_{i}$, the $SL(2,\Z)$ symmetries $T_{ij}$, the antisymmetric
tensor shifts $\Theta_{ij}$, and the factorized dualities $D_i$,
$i,j=1,...,d$.

\noindent
{\bf Weyl symmetries as a part of ${\cal G}_d$}:
\vskip 0.1in

\noindent
We now turn to showing that the operations which are gauge
transformations implemented by Weyl transformations, which themselves
generate a group that we denote $G_{Weyl}$,
can all be expressed as elements in ${\cal G}_d$.
Therefore, many ${\cal G}_d$ elements will be shown to be gauge
symmetries; we will also discuss the other ${\cal G}_d$ symmetries.
In order to show this we need to prove a few statements:

\vskip 0.1in
\noindent
{\bf Statement 1} \\
\noindent
Any element of $G_{Weyl}$, of a type expressing a pair of chiral
and anti-chiral Weyl reflections,
$(W_L,W_R)$, around an enhanced symmetry point, $E_M$,
corresponds to an element of $O(d,d,\Z)$.

The corresponding $O(d,d,\Z)$ transformation is denoted $g(W_L,W_R,E_M)$.
We will expand on this statement below.

\vskip 0.1in
\noindent
{\bf Statement 2}\\
\noindent
A general element in $G_{Weyl}$ can be expressed as a product of Weyl
reflections around enhanced symmetry points,
all sharing the {\em same} symmetry.

Let us elaborate on this statement.
There are many enhanced symmetry points (for $d>1$ there are
infinitely many points with enhanced symmetries for any group with rank $d$);
nevertheless, the products of Weyl reflections around points
with the same enhanced symmetry group (for instance, $SU(2)^d$)
are sufficient to generate the full group $G_{Weyl}$. The
content of the group itself does not depend on the special enhanced
symmetry points used in its generation.
The same is true if one forms products of
Weyl reflections around points with several enhanced symmetry groups.
This is due to the fact that an element of $G_{Weyl}$ may be
represented by many different products of reflections around different
symmetry points; that is, the realization of $G_{Weyl}$ in terms of
reflections is not unique.

To describe the third statement, we first present some definitions. Let
$\Lambda$  be the lattice defining the torus of compactification,
$T^d \simeq {\R}^d / \pi\Lambda$, of an enhanced symmetry point $E_M$, and let
$A$ be an automorphism of $\Lambda$, namely,
\be
A(G_M)A^t=G_M,
\label{ERG7}
\ee
where $G_M$ is the symmetric part of $E_M$.
Each Weyl reflection is, in particular, an
automorphism of $\Lambda$, called {\em inner}. But
there are automorphisms that are {\em not} Weyl reflections; these are
called {\em outer} automorphisms. To an outer automorphism (for instance, a
permutation of the two $SU(2)$'s in $SU(2)\times SU(2)$) corresponds
an element $g(A,A,E_M)$ which is in the $GL(d,\Z)$ subgroup ($A\in GL(d,\Z)$)
of $O(d,d,\Z)$. The third statement is:

\vskip 0.1in
\noindent
{\bf Statement 3}\\
\noindent
If $g\in O(d,d,\Z)$ then
$g=\prod_{i} g(A_{Li}, A_{Ri}, E_{Mi})$ for some
enhanced symmetry points $E_{Mi}$ .

Here $(A_{Li}, A_{Ri})$ is a pair of automorphisms, inner (Weyl) or outer,
of the enhanced-symmetry root lattice $\Lambda _i$. The content of
statement 3
is that {\em any} element of $O(d,d,\Z)$ can be written as a product of the
elements in $G_{Weyl}$
(corresponding to Weyl reflections around points with enhanced
symmetries) and {\em additional} elements relating current-current
perturbations of enhanced symmetry points; the latter act on
the currents as outer automorphisms (we elaborate on that below).

\vskip 0.1in
\noindent
{\bf Statement 4}

\noindent
4a. The group $G_{Weyl}(d)$ is a subgroup of $O(d,d,\Z)$, generated by the
elements in $O(d,d,\Z)$ that contain an even number of the  generators
$P_{ij}, T_{ij}, \Theta_{ij}$ (see their definition in subsection 2.4.1).

\vskip 0.1in
\noindent
4b. The group $O(d,d,\Z)$ is a subgroup of $G_{Weyl}(d+2)$.

Here $G_{Weyl}(d)$ denotes the group $G_{Weyl}$ for $d$-dimensional
toroidal backgrounds.

The statement 4a tells us that the residual symmetries of the broken
gauge group,
\be
\prod_{i} g(W_{Li}, W_{Ri}, E_{Mi}),
\label{ERG8}
\ee
(namely, the products of elements corresponding to pairs,
$(W_{Li},W_{Ri})$, of chiral and anti-chiral Weyl reflections around the
enhanced symmetry points $E_{Mi}$), generate a special (infinite order)
subgroup of $O(d,d,\Z)$.
Moreover, statement 4b tells us that {\em all} the
elements of $O(d,d,\Z)$ can be interpreted as
spontaneously broken gauge symmetries in the
group generated by products of Weyl reflections around points with enhanced
symmetries in $d+2$ dimensions, $G_{Weyl}(d+2)$.

We are now going to elaborate on each of the above statements, and
prove the various assertions.

The meaning of statement 1 is the following.  In vector
notation, the worldsheet Lagrangian $L(E_M, \epsilon )$ is defined to be
\be
L(E_M, \epsilon)=L(E_M) + H^t\epsilon \bar{H}.
\label{ERG9}
\ee
$L(E_M, \epsilon)$ is equivalent, physically, to the theory
$L(E_M, \epsilon')$, where
$\epsilon'=W_L\epsilon W_R^t$. This is so because $L(E_M)$ is unaffected by
Weyl reflections  acting on the (1,0) currents and (0,1) currents.
The transformations
\be
H'=W_L H ,  \qquad  \bar{H}'=W_R^t \bar{H}
\label{ERG10}
\ee
are equivalent to a change of the background matrix deformation
$\epsilon$ to $\epsilon'$. Now recall that
$\epsilon$ and $\epsilon'$ are equivalent background perturbations of $E_M$
if $W_L$ and $W_R$ are Weyl transformations. In that case the theories
described by the background matrices $E_M+\epsilon$ and
$E_M+W_L\epsilon W_R^t$ are connected by a gauge
transformation in $G_L\times G_R$.

The particular element in $O(d,d,\Z)$,
$g(W_L,W_R,E_M)$, which corresponds to this gauge transformation, is found
in the following manner.
It is sufficient to find the $O(d,d,\Z)$ transformation that coincides
with the symmetry $\epsilon\rightarrow\epsilon '$ infinitesimally close to
$E_M$. This is true because two isometries of a complete space,
which coincide on the tangent space at the point $p$,
and which preserve $p$, coincide on the
entire space. A convenient metric on the moduli space is the Zamolodchikov's
metric \cite{Z} (which is a metric of $O(d,d,\R)/O(d,\R)^2$ viewed as a
symmetric space),
\be
ds^2=g_{ab}\ dx^a dx^b=Tr\left( G^{-1} {\epsilon}^t G^{-1} \epsilon \right).
\label{ERG11}
\ee
Any automorphism satisfying eq. (\ref{ERG7})
is obviously an isometry acting on
$\epsilon$ in the neighborhood of $E_M$. The discrete symmetry corresponding to
the action of $W_L$ and $W_R$ is, therefore, an isometry (as it should be by
definition).

We will look for an
isometry in $O(d,d,\R)$ which coincides with the action of the
inner automorphisms $W_L, W_R$ on the tangent space of $E_M$.
Thus, we look for an element (see section 2.4)
\be
g=\left(\matrix{a&b\cr
                   c&d\cr}\right)\in O(d,d,\R), \qquad
g^tJg=J, \qquad J=\left(\matrix{0&I\cr
                                  I&0\cr}\right),
\label{ERG12}
\ee
such that
\be
g(E_M+\epsilon)\equiv [a(E_M+\epsilon)+b][c(E_M+\epsilon)+d]^{-1}=
E+\epsilon ' +{\cal O}(\epsilon^2).
\label{ERG13}
\ee
Solving for $\epsilon'$ one finds
\be
\epsilon '=(a-E_Mc)\epsilon(cE_M+d)^{-1},
\label{ERG14}
\ee
and we demand that $\epsilon '=W_L\epsilon W_R^t$, thus
\be
a-E_Mc=W_L, \qquad cE_M+d=(W_R^t)^{-1} .
\label{ERG15}
\ee
Finally, after some algebra one finds \cite{can,GMR1}
\be
g(W_L,W_R,E_M)=A(E_M)\left(\matrix{(W_R^t)^{-1}&0\cr
                                0&W_L\cr}\right)A^{-1}(E_M),
\label{ERG16}
\ee
where
\be
A(E)=\left(\matrix{E/2&-E^t G^{-1}\cr
                      I/2&G^{-1}\cr}\right), \qquad
A^{-1}(E)=\left(\matrix{G^{-1}&G^{-1}E^t\cr
                       -I/2&E/2\cr}\right).
\label{ERG17}
\ee
As expected, $g$ is a representation,  $g(W_1 W_2)=g(W_1)g(W_2)$. Using the
properties of $E_M$ one finds that $g$ is not only an $O(d,d,\R)$ rotation,
but an element of its discrete subgroup $O(d,d,\Z)$ as well.
This completes the proof of statement 1.

Statement 3 is easily proved finding the product $\prod g(A_L,A_R,E_M)$
for each generator of $O(d,d,\Z)$. We first present the elements in the
$d=2$ case, and then we generalize to any $d$. For $d=2$,
\ba
g_{D_1}=g(r, 1, I), \qquad g_{R_1}&=&g(r, r, I), \qquad
g_{P_{12}}=g(p, p, I), \nonumber \\
g_{\Theta_{12}}&=&g[p, p, E(SU_3)]g(p, p, I), \nonumber \\
g_{T_{12}}&=&g_{R_1}g_{P_{12}}g_D(g_{\Theta_{12}})^{-1}g[p,1,E(SU_3)]
g_{P_{12}}g_{R_1},
\label{ERG18}
\ea
where
\be
p=\left(\matrix{0&1\cr 1&0\cr}\right), \qquad
r=\left(\matrix{-1&0\cr 0&1\cr}\right), \qquad
E(SU_3)=\left(\matrix{1&1\cr
                        0&1\cr}\right);
\label{ERG19}
\ee
$D=D_1D_2$  is the duality symmetry that takes the background matrix to its
inverse.

Modulo the permutation $P_{12}$, which corresponds to an outer
automorphism of $SU(2)\times SU(2)$, the list in eq. (\ref{ERG18}) is
made of gauge transformations.
The $SU(2)\times SU(2)$ outer automorphism cannot be viewed,
on its own, as a gauge symmetry in a larger group, say $SU(4)$. Neither it is
a Weyl symmetry of $SU(4)$ nor of
any other larger gauge symmetry (we shall discuss
soon what this implies about the identification of the points related by
this discrete symmetry).

The automorphisms of the points in moduli space with an
enhanced symmetry group $SU(2)\times SU(2)$ and $SU(3)$ (even for specific
backgrounds: $I$ and $E(SU_3)$) are sufficient to generate all the elements of
$O(d,d,\Z)$. This is true if we use the points with such symmetries in any pair
of target space coordinates with indices $i,j$. In that way one gets
$D_i, R_i, P_{ij}, T_{ij}$ and $\Theta_{ij}$ using eq. (\ref{ERG18}).
The group $O(d,d,\Z)$  is, therefore, completely generated, and that proves
statement 3.

Actually, one can say more. Weyl reflections at the points with an enhanced
symmetry $SU(2)^d$ are sufficient in order to get all the gauge
transformations.  For any simply laced group, $G$, the following was checked:
Let  $E_M$ be the background matrix at an enhanced symmetry point $G$, defined
as in eq. (\ref{ERsyb}) (namely, $E_M$ is
the upper-triangular matrix given by the
sum of one-half the Cartan matrix plus its antisymmetrization).
The $O(d,d,\Z)$ transformation that corresponds to the right-handed
Weyl reflection of the first simple root ($W_L=1, W_R\not=1$) is $R_1D_1$.
The transformation which
corresponds to the left-handed Weyl reflection ($W_L\not=1, W_R=1$)
at the point $E_M^t$ is $D_1$. These are exactly the $O(d,d,\Z)$
transformations that correspond to reflections in the $SU(2)$ case.
By a conjugation in $O(d,d,\Z)$ one gets all
the gauge transformations which correspond to a $G$-enhanced symmetry point.
We conclude,  therefore,  that the duality relations obtained from different
enhanced affine symmetry groups are the same.

The proof of 4a is now very simple. $D_1$ and $R_1D_1$ are trivially generated
by an even number of generators $\Theta, P$ and $T$. A conjugation in
$O(d,d,\Z)$ and multiplication preserve the parity of $\Theta, P,T$
generators. Therefore, every gauge
transformation is even in $\Theta,P,T$. Also, any element which contains
even number of $\Theta,P,T$'s is in $G_{Weyl}$.
The argument in the previous paragraph also  proves statement 2.

We conclude that the group $O(d,d,\Z)$ is generated by the groups $G_{Weyl}$
and $\Z_2$. The $\Z_2$ stands for the
outer automorphism that permutes two coordinates at the point with an identity
background matrix. Therefore, up to a  change of names,
all the elements of the $O(d,d,\Z)$  discrete symmetries are linked to gauge
transformations of the kind that was  described.

The fact that $O(d,d,\Z)$ transformations which are odd in $\Theta,P,T$'s
are not gauge transformations might be a little surprising. However,
there is a simple way to interpret also the odd
elements as gauge symmetries in string theory.
This is the content of statement 4b, which we are now going to prove.

In string theory there are $26-d$ ($10-d$) space-time
coordinates in addition to the $d$-dimensional internal space.
Let $g_o$ $(g_e)$ denote an
element which is odd (even) in $\Theta,P,T$'s. For $g_o \in O(d,d,\Z)$,
$g_e=g_o Y_{23,24}$ is an element in $O(d,d,\Z)\times O(2,2,\Z)$.
$Y_{23,24}$ is any element in $O(2,2,\Z)$ which is odd in
$\Theta_{23,24},P_{23,24}$ and $T_{23,24}$. The
23-24 plane is spanned by two space coordinates. The action of $g_e$ on the
submoduli,  for which the $23$'rd and $24$'th coordinates are constrained to
have an infinite radius, is reduced effectively to $g_o$. Thus, one may
consider $g_o$ as a broken gauge transformation, in the limit where two
coordinates are decompactified.
Note however, that for $d\geq 21$, the consequence of the
identification through a gauge transformation involves the introduction of
local symmetries mixing internal and 4-dimensional space-time coordinates.

We have completed the discussion of $O(d,d,\Z)$ as a spontaneously broken
gauge symmetry in string theory. However, we are left with the $\Z_2$
class of ${\cal G}_d$ for which we have not found a general gauge
interpretation. These are the symmetries corresponding to worldsheet
parity ($B\to -B$ on the background).
To discuss  properties of this symmetry consider the following.

Recall that a convenient isometric embedding of
the moduli space $O(d,d,\R)/O(d,\R)^2$ in $O(d,d,\R)$ is
the set of phase-space matrices (see section 2.4, eqs.
(\ref{ER18})-(\ref{EMA3}))
\be
M=\left(\matrix{G-BG^{-1}B&BG^{-1}\cr
                  -G^{-1}B&G^{-1}\cr}\right).
\label{ERG20}
\ee
Recall also, that the
spectrum of the theory is given by $Z^t MZ/2$, where $Z$ is a
$2d$-dimensional column vector with
integer components. The inner automorphisms (transformations in $O(d,d,\R)$)
that preserve the spectrum generate the group $O(d,d,\Z)$.
The transformation that takes $B$ to $-B$ is the outer automorphism
\be
M\rightarrow \left(\matrix{I&0\cr
                             0&-I\cr}\right)M\left(\matrix{I&0\cr
                                                           0&-I\cr}\right).
\label{ERG21}
\ee
While the symmetry group $O(d,d,\Z)$  corresponds to linear transformations on
the currents $J$ (and/or the currents $\bar{J}$),
the parity symmetry corresponds to the change $J\leftrightarrow \bar{J}$.
Therefore, unlike $O(d,d,\Z)$
it cannot be interpreted as a spontaneously broken gauge symmetry.

Finally, let us say few words about the heterotic string.
For the heterotic string things are a bit more involved;
only elements in the subgroup of $O(d+16,d,\Z)$ generated by
$g(W_L,W_R=1,E_M)$ are gauge transformations. (On the submoduli of a
zero background gauge field, $A=0$,
they will generate a subgroup made of elements with even number
of generators $R, \Theta, P$ and $T$.
The reason is that in the $SU(2)$ case only $D$ is a
gauge transformation; $R$ and $RD$ act on right-handed currents and as a
result the GSO projection in the supersymmetric sector is changed.
Now, conjugating $I$ and $D$ we get elements which are even in
$R,\Theta,P,T$). However, when $W_R$ is non-trivial, the corresponding
symmetry is not linked to a  gauge transformation.
Yet, we conjecture that all the elements of $O(d+16,d,\Z)$ can still be
interpreted as spontaneously broken gauge symmetries, in analogy with the
bosonic string case.

\end{subsubsection}

\end{subsection}

\begin{subsection}{Effective Field Theory and $O(6,6,\Z)$ Duality}
\setcounter{equation}{0}
Having seen the rich structure of duality symmetry one would
attempt to find an explicit target space realization of it.
Here and in section 2.8 we pursue this attempt by studying
heterotic $N=4$ supersymmetric compactifications to $D=4$.
The rational to study local $N=4$ supersymmetry is that this is the
most constrained heterotic background in four dimensions.
We adopt the following strategy:

\begin{description}

\item[a)] We review the construction of $N=4$ supergravity coupled to
matter. The only freedom in this construction is the choice of the gauge
structure constants.

\item[b)] We find the structure constants of the spin-1 fields relevant to
low energy physics by computing their off-shell string 3-point functions.

\item[c)]  We add to those spin-1 fields other (harmless)
degrees of freedom
in order to close a completely duality-invariant gauge algebra.

\end{description}

$N=4$ is the largest
supersymmetry that can be possessed by the heterotic string compactified to
four dimensions. It is realized when the compact space is a torus.
The most general duality group, generated by a gauge symmetry, for
torus compactifications of the heterotic string to 4-$D$ space-time,
as shown in section 2.5, is $O(6,22,\Z)$. This is
the group of $\Gamma^{(6,22)}$
automorphisms which preserve its Lorentzian metric. In particular, when the
Lorentzian lattice takes the special form
\beq
\Gamma^{(6,22)}=\Gamma^{(6,6)}\oplus \Gamma^{(0,16)},
\label{M1}
\eeq
the Narain compactification reduces to a standard toroidal one, where the
background gauge fields in the compactified space are all zero.
An obvious subgroup of the duality group is, in this case, $O(6,6,\Z)$,
namely, the group of length-preserving automorphisms of $\Gamma^{(6,6)}$.

The group $O(6,6,\Z)$ enjoys an interesting property: it transforms
modes that are massless for any $\Gamma^{(6,6)}$ background into themselves
\footnotemark.
\footnotetext{Modes that become
massless only on special backgrounds (self-dual radii) do mix with massive
ones.}
These modes are called ``marginal'' since they persist and thus correspond
to truly marginal deformations of the CFT.
If we are interested in low energy physics
($E\ll \alpha'^{-1/2}$), the only
relevant interactions are those of the massless modes. All these interactions
can be derived from a Low-Energy Effective Action (LEEA), which describes all
physics well below the string scale.
Since $O(6,6,\Z)$ is a symmetry that (generically) does not
mix marginal low energy degrees of
freedom with ultramassive ones, the LEEA of a Narain compactification,
containing those low-energy degrees of freedom,
must be invariant under $O(6,6,\Z)$. When the
$T^6$ torus of the compactification with lattice~(\ref{M1}) is modded out by
an appropriate discrete symmetry group one finds the so-called orbifold
compactification~\cite{DHVW}. In this case, the $N=4$
supersymmetry of the torus compactification is broken down to $N=2$,
$N=1$ or $N=0$.
Likewise $O(6,6,\Z)$ breaks down to some non-compact subgroup, e.g.
$SL(2,\Z)$. This latter group is
still an invariance of the CFT of the
orbifold compactification, and therefore of the resulting LEEA.

In this section we shall examine the $N=4$ LEEA, and show that, for
compactifications described by lattice~(\ref{M1}), the LEEA is indeed
$O(6,6,\Z)$ invariant.

Since the LEEA describes low energy physics, we should include in it only
terms with at most two derivatives of the low energy fields. Indeed, higher
derivative terms are multiplied by powers of ($\alpha'^{1/2}$), and thus
negligible at $E\ll \alpha'^{-1/2}$. Furthermore, the four-dimensional
LEEA is $N=4$
supersymmetric, since it comes from the dimensional reduction of an $N=1$
theory in ten dimensions~\cite{CM,BdRdWvN,CMan}.
The gauge group, on a generic (non self-dual) background,
is either $E_8\times E_8\times U(1)^6$ or $Spin(32)/\Z_2\times
U(1)^6$.\footnote{
The extra $U(1)^6$, coming from the heterotic right-movers, corresponds to
the graviphotons in the $N=4$ gravitational supermultiplet.}
The ten-dimensional  gauge group
$E_8\times E_8$ or $Spin(32)/\Z_2$ is untouched by $O(6,6,\Z)$.
In four dimensions and at special values of the background
(corresponding to self-dual points w.r.t some element of $O(6,6,\Z)$)
the $U(1)^6$ factor in the gauge group can be promoted to a non-Abelian
group.
For instance, at the self-dual point of a circle compactification, $U(1)$ is
promoted to $SU(2)$ as was discussed in section 2.6.
Since the $SU(2)$ states have nonzero winding number on $T^6$, they cannot be
described by a field-theoretical Kaluza-Klein mechanism.
Generically, though, the LEEA containing modes with zero winding or
$T^6$ momenta will offer an adequate description of low energy physics.

The key observation is that an $N=4$ supersymmetric Lagrangian with at most
two derivatives in the fields is completely determined by its gauge
group~\cite{deRoo,dRW,S,W1}.
In other words, the Ward identities of $N=4$ are so strong
as to completely fix the Lagrangian in terms of the gauge-group structure
constants.

The most general $N=4$ supersymmetric Lagrangian coupled to gravity contains
two types of multiplets~\cite{FS}: the gauge matter multiplet and the
gravitational multiplet.
The gauge matter multiplet is
\beq
(A_\mu^A, \psi^{A}_i, Z^A_{ij}),\;\;\; Z^A_{ij}=-Z^A_{ji},\;\;\;
Z^{A\, ij}\equiv (Z^A_{ij})^*={1\over 2}\epsilon^{ijkl}Z^A_{kl}.
\label{M2}
\eeq
The multiplet contains a vector, two Majorana spinors and six real scalars,
all belonging to the adjoint of the gauge group $G$. The indices
$i,j,...=1,2,3,4$ are the so-called extension indices: they label the four
supersymmetry charges, and span the fundamental representation of
$SU(4)$~\cite{FS,WB}.

The gravitational multiplet contains one spin-2 field (the graviton), four
spin-3/2 (gravitini), six spin-1 (graviphotons), four spin-1/2 and one complex
scalar
\beq
(g_{\mu\nu}, \psi_{i\,\mu}, A_{ij\,\mu},\psi_i,\phi),\;\;\;
A_{ij\,\mu}=-A_{ji\,\mu},\;\;\; A^{ij}_{\mu}\equiv (A_{ij\,\mu})^*={1\over 2}
\epsilon^{ijkl}A_{kl\,\mu}.
\label{M4}
\eeq

The most general Lagrangian containing these multiplets can be found using
superconformal techniques~\cite{deRoo,dRW,W1}.
Without entering into details, let us recall some useful notations.
The physical (Poincar\'e) multiplets~(\ref{M2}, \ref{M4})
can be re-expressed in terms
of the following superconformal multiplets:\\
\noindent
a) Superconformal matter
\beq
(A^S_{\mu}, \psi^S_i, Z^S_{ij}),\;\;\; S=1,...,n+6 .
\label{M5}
\eeq
\noindent
b) Conformal-supergravity multiplet
\beq
(g_{\mu\nu},\psi_{i\,\mu},\psi_i,\phi).
\label{M8}
\eeq
Here $n$ is the dimension of the gauge group. Multiplets~(\ref{M5}, \ref{M8})
contain
more degrees of freedom than~(\ref{M2}, \ref{M4}). In order to eliminate
these degrees of freedom one introduces the constraints~\cite{W1}
\beq
\eta_{ST}Z^S_{ij}Z^{T\,kl}=-\delta_i^{[k}\delta_j^{l]},\;\;\;
\eta_{ST}Z^S_{ij}\psi_k^T=0.
\label{M6}
\eeq
Here $\eta_{ST}=diag(-I_6,+I_{22})$ is the Lorentz metric of signature
$(6,22)$. The physical scalar and spin-1/2 fields have index $S=7,...,n+6$,
the first six scalars and spin-1/2 are non-propagating auxiliary
fields~\cite{deRoo,dRW,S,W1}.

The most general $N=4$ Lagrangian can be written in terms of these
superconformal fields~\cite{deRoo,dRW}.
We write down here the bosonic Lagrangian only.
Fermionic terms, obviously, are uniquely fixed by supersymmetry\footnotemark.
\footnotetext{This Lagrangian could be further
generalized \cite{deRoo,dRW} at the price of reducing the kinetic-term
duality group~\cite{GZ}. This goes exactly in the opposite direction of our
program of finding the {\em maximally} duality-invariant LEEA.}
\bea
e^{-1}L &=& \eta_{RS}\left\{ -{1\over 4} {\psi\over \phi}
F_{\mu\nu}^{+R}F^{\mu\nu\, +S} -{1\over 4} D_{\mu}Z_{ij}^RD^{\mu}Z^{ij\,S}
\right. \nonumber \\
& & \left.
+{1\over 2}{\phi^*\over \phi} K_{\mu\nu}^{+R}K^{\mu\nu\, +S}\right\} -{1\over
4} R(\omega) +4\Re D_\mu\phi D^\mu\psi^* \nonumber \\
& & +{1\over 8} E_{ij}E^{ij}
+\eta_{RS}\left\{ {1\over 4}|\phi|^2W_i^{j\,R}W_j^{i\,S} +{1\over 3}
E^{ij}X_{ij}^{RS}\phi^*\right\} \label{M9}
\eea
In eq.~(\ref{M9}) $\Re\phi\psi^*=1$, $e\equiv \det e^a_\mu $,
and we have used the following
notation~\cite{deRoo,dRW}
\bea
F^{+\, R}_{\mu\nu} &=& {1\over 2} F_{\mu\nu}^R +{1\over
4}e^{-1}\epsilon_{\mu\nu\rho\sigma}F^{\rho\sigma\, R},
\;\;\; D_{\mu}Z_{ij}^P=\partial_\mu Z_{ij}^P +
f^P_{\;\;QR}A^Q_{\mu}Z_{ij}^R
\nonumber \\
X^{RS}_{ij} &=& Z_{ik}^T Z^{kl\, R} Z^U_{lj}f^S_{\;\; TU}, \;\;\;
W_i^{j\, R} = Z_{ik}^S Z^{kj\, T} f_{\;\; ST}^R
\nonumber \\
E_{ij} &=& -{4\over 3} \phi^*\eta_{RS} X^{RS}_{ij}, \;\;\;
K_{\mu\nu}^{+\, R} =
{1\over \phi^*} Z^{ij\, R}F_{\mu\nu}^{+\, S} Z_{ij}^T\eta_{ST}.
\label{M10}
\eea
As usual, $f^P_{\;\; QR}$ denote the gauge-group structure constants. The
fields $Z_{ij}^S$ include 36 non-propagating scalars. Notice that the
indices $ij$ parametrize the antisymmetric representation of $SU(4)$ (which is
isomorphic to the vectorial of $SO(6)$). The $n+6$ scalars $Z_{ij}^S$, due
to the constraints~(\ref{M6}) parametrize the coset manifold
\mbox{$O(6,n)/(O(6)\times O(n))$}.
The indices $S$, $T$, etc. are both in the adjoint of the gauge
group, and in the fundamental of $O(6,n)$. Therefore the allowed gauge
groups $G$ of $N=4$ supergravity are those satisfying $Adj \, G\subset Vect\,
SO(6,n)$. Notice that in eq.~(\ref{M9}) the $n$ matter gauge fields and the
six graviphotons group together naturally in $F^S_{\mu\nu}$, $S=1,...,n+6$.
This
grouping is due to the fact that the ``parent'' conformal-supersymmetry spin-1
fields all belong to superconformal matter, and not to the
multiplet of superconformal gravity~(\ref{M8}).

It must be noticed that the gauge group can be non-compact if
$Adj\, G$ does {\em not} belong to $SO(6)\times SO(n)$, that is, if the
gauge group acts
nontrivially on both physical {\em and} compensating multiplets~(\ref{M5}).
A generic toroidal compactification on $\Gamma^{(6,6)}\oplus\Gamma^{(0,16)}$,
on the other hand, yields the compact gauge group ${\cal G} = U(1)^6\times G$,
with $G$ either $E_8\times E_8$ or $Spin(32)/\Z_2$. In this case
\beq
f^P_{\;\; QR}=0,\;\;\; P,Q,R=1,...,12.
\label{M10'}
\eeq
Lagrangian~(\ref{M9}), when all $f^P_{\;\; QR}=0$,
is invariant under the following non-compact transformation
\beq
Z_{ij}^S\rightarrow \Lambda^S_T Z^T_{ij},\;\; F_{\mu\nu}^S\rightarrow
\Lambda^S_T F^T_{\mu\nu},\;\; \Lambda^P_S\Lambda^Q_T \eta_{PQ}=\eta_{ST}.
\label{M11}
\eeq
This $O(6,n)$ symmetry is explicitly broken by the structure-constant
dependent terms. When ${\cal G}=U(1)^6\times G$ the residual non-compact
symmetry is $O(6,6)$.

This result means that the LEEA involving massless modes possesses indeed the
$O(6,6,\Z)$ symmetry we found in the worldsheet analysis of sections 2.4, 2.5.
In fact, the
residual symmetry of the LEEA eq.~(\ref{M9}) is even larger than expected: it
is a continuous group $O(6,6,\R)$
instead of a discrete one\footnote{
The dilaton $\Re S\equiv \Re(\psi/\phi)=|\phi|^{-2}$
in (\ref{M9}) is the duality invariant one
(see subsection 4.2.4 for more details).}.
When higher derivative terms
(${\cal O}(\alpha')$) are included in the LEEA, the invariance group is
expected to break down to $O(6,6,\Z)$. This latter group, being an invariance
of the underlying worldsheet
conformal theory, is an exact symmetry of the LEEA, to all
orders in $\alpha'$.

We have just found how easy it is to write down a LEEA invariant under
the $O(6,6,\Z)$ subgroup of the full duality group $O(6,22,\Z)$.
This happens because $O(6,6,\Z)$ transforms the marginal modes into themselves.
On the other hand, constructing a LEEA invariant under $O(6,22,\Z)$ is a more
complex endeavor. Indeed, within $O(6,22,\Z)$ there exist elements
transforming  states that are massless for any background of the form
$\Gamma^{(6,6)}\oplus\Gamma^{(0,16)}$ into ultra-massive ones.
This fact is due to the non-Abelian character of the gauge group.
Explicitly, $E_8\times E_8$ states outside the Cartan subalgebra are mapped
into massive modes under generic $O(6,22,\Z)$ transformations.
Moreover, the typical
orbit of a string state under $O(6,22,\Z)$ contains an infinite number of
points. Still, since string backgrounds related by duality yield the same
physics, we do expect an $O(6,22,\Z)$-invariant LEEA to exist. This LEEA may
contain an infinite number of fields, and even ghosts, provided their masses
are always of the order of the compactification scale $R$, or larger. The
essential requirement we must impose on the $O(6,22,\Z)$-invariant LEEA is to
reproduce the correct physics at energies $E\ll R^{-1}, \alpha'^{-1} R$.
The next section is devoted to the construction of such LEEA.
\end{subsection}

\begin{subsection}{A Completely Duality-Invariant Low-Energy Effective
Action for the $N=4$ Heterotic String}
\setcounter{equation}{0}

In the moduli space of toroidal compactifications there is an
infinite number of special points where massive
modes become massless. These points correspond to systems with an enhanced
affine symmetry. Massless fields become massive at the neighborhood of such
points by  the Higgs mechanism. The generalized duality group mixes massless
modes with massive modes that might become massless at some  point of the
moduli space. Therefore, the LEEA should contain {\em all}
the fields which correspond to string states that become massless at some point
of the moduli space. The number of such fields is infinite, and on a given
background all except a finite number of them are very massive.
In this section we present the
relevant structure constants of the
infinite dimensional ``Duality Invariant String Gauge algebra'' (DISG).
We justify
the name given to this group by showing that the Higgs mechanism in the LEEA
coincides with the stringy Higgs mechanism. Namely, the masses of the LEEA
scalar
fields coincide with the string spectrum at the particular background. We
also show the invariance of the LEEA under the action of the generalized
duality group $O(6,22,\Z)$, and the structure of the DISG algebra.
Our analysis follows closely ref.~\cite{GP1}

Since
the form of a gauged $D=4$, $N=4$ supergravity (with at most two space-time
derivatives) coupled to matter multiplets is determined by the knowledge of the
gauge group~\cite{deRoo,dRW,W1,S},
one has only the freedom to choose the right
number of matter multiplets  and the correct gauge group in order to figure out
completely the LEEA of the string. Let us start from the solution and present
the matter multiplets and the structure constants of the  DISG algebra which
are
relevant for $D=4$ low energy physics.

The  scalar fields which are relevant for $D=4$ low energy physics
$Z_a^S$, $a=1,...,6$~\footnotemark,
\footnotetext{We used here the isomorphism between the vector representation
of $SO(6)$, indexed by $a$, and the antisymmetric of $SU(4)$,
indexed by $ij$.} $S=1,....,\infty$ will be labeled as
follows:
\begin{eqnarray}
Z^b_a,\;Z^I_a,\; Z^p_a,\; Z^{-p}_a\equiv \bar{Z}^p_a,\;\;\;
\nonumber \\
p\in \Gamma^{(6,22)},\;\;
p\eta p = 2,\;\; a,b=1,..,6,\;\; I=7,..,28,
\label{A8}
\end{eqnarray}
where $\Gamma^{(6,22)}$ is an even self-dual Lorentzian lattice of signature
$(6,22)$. The ``momentum'' $p=(p_L, p_R)$ has $6$ left-handed  components
$p_L^a$, $a=1,...,6$, and $22$ right-handed components $p_R^I$, $I=7,...,28$.
This is the Narain lattice of the string toroidal
compactification~\cite{N,NSW}. The scalar product is Lorentzian,
$p\eta q  = -p_L^aq_L^a + p_R^Iq_R^I$,
where $L(R)$ denote left-handed (right-handed)
momenta.  The indices $(b,I)$ refer to the Cartan Sub-Algebra (CSA),
while the $p$ indices are of the Lorentzian length two generalized roots (and
therefore their number is infinite).
These fields live on the infinite dimensional coset space
\begin{equation}
{O(6,\infty,\R) \over O(6,\R)\times O(\infty,\R)}.
\label{A9}
\end{equation}

The structure constants of the gauge group which are relevant for $D=4$ low
energy physics are
\begin{eqnarray}
& & f^r_{pq}=\varepsilon(p,q)\delta^r_{p+q},\;\;\; f^I_{p,-p}=-ip^I, \;\;\;
f^q_{Ip}=-ip_I\delta^q_p,\;\;\; f^b_{p,-p}=-ip^b,
\nonumber\\
& & f^q_{bp}=-ip_b\delta^q_p,\;\;\;\; p\eta p=q\eta q=r\eta r=2.
\label{A10}
\end{eqnarray}
The two cocycle $\varepsilon(p,q)$ satisfies the identities
\begin{eqnarray}
\varepsilon(p,q)\varepsilon(p+q,r)&=&\varepsilon(p,q+r)\varepsilon(q,r),
    \nonumber \\
\varepsilon(p,q)&=&(-1)^{p\eta q}\varepsilon(q,p),\;\;\; \varepsilon(p,-p)=1,
\label{A11}
\end{eqnarray}
and
\begin{equation}
\varepsilon(p,q)=\varepsilon(q,-p-q).
\label{A12}
\end{equation}
When $p$ and $q$ are the roots of a finite dimensional gauge group $g$,
the $\varepsilon(p,q)$
reduce to the structure constants of $g$ in the Cartan-Weyl
basis~\cite{H,BHN,FK}.
As demonstrated in~\cite{FK} $\varepsilon(p,q)$ can be extended to all
vectors in $\Gamma^{(6,22)}$. An elegant explicit construction of
$\varepsilon(p,q)$ has been given in~\cite{GO}.

The set of structure constants
in~(\ref{A10}) is not sufficient to close a Lie algebra. The Lie algebra
that contains all fields presented in~(\ref{A10}) and preserves the $D=4$
low energy physics is the DISG algebra. We will elaborate  on the DISG algebra
later on. Meanwhile, we will consider only the sub-set of
generators discussed above.

The metric $\eta$ is easily extended from the CSA
to the generators which are relevant for the $D=4$ low energy theory by
defining
$\eta_{p,-p}=\eta_{-p,p}=1$, and setting all other components to zero.
By lowering with this metric all indices, one finds that the structure
constants $f_{QST}$ are completely antisymmetric, due
to~(\ref{A12}).

Using  the infinite dimensional DISG as the gauge group of the low energy
$D=4$, $N=4$
supergravity theory means to use the structure constants~(\ref{A10}) in the
action~(\ref{M9}). The covariant derivatives and the scalar potential
are obtained upon substituting the structure
constants~(\ref{A10}) into equations~(\ref{M9}) and~(\ref{M10}).
Let us show, at first,  that the Higgs
mechanism of the LEEA coincides with the stringy Higgs mechanism.

The simplest zero cosmological constant minima of the action~(\ref{M9})
lie in the CSA. The value of the cosmological constant is found by writing
down the scalar potential
\bq
V={1\over 4} |\phi|^2
Z^{QU}Z^{SV}(\eta^{TW} + {2\over 3}Z^{TW})f_{QST}f_{UVW},\;\;\;
Z^{PQ}\equiv Z^P_aZ^Q_a.
\eq{Pot}
Let us expand the scalar fields around a VEV in the CSA,
\begin{equation}
Z_a^S = C_a^b \delta_b^S + C_a^I \delta_I^S + \zeta_a^S,
\label{A13}
\end{equation}
where $C_a^b, C_a^I$, $a,b=1,..,6, \; I=7,..,28$, are constants and the
constraint~(\ref{M6}) is satisfied.
A rotation of $Z_a^S$ to
\begin{equation}
(Z')_a^{S'} = Z^{S}_a M_{\;\;S}^{S'},
\label{A14}
\end{equation}
where $M\in O(6,\infty,\R)$, preserves the scalar quadratic
constraint~(\ref{M6}) (which now reads $\eta_{ST} Z_a^S Z_b^T=-\delta_{ab}$).
Changing the VEV of the scalar fields to new ones in the CSA can be done by
a rotation  in $O(6,22,\R)\subset O(6,\infty,\R)$.

Under the orthogonal transformation~(\ref{A14}),
the scalar potential $V$  transforms into
\begin{eqnarray}
V' &=&  \frac{1}{4} |\phi|^2 (Z')^{QU} (Z')^{SV}
(\eta^{TW} + \frac{2}{3} (Z')^{TW}) f_{QST} f_{UVW} \nonumber \\
&=& \frac{1}{4} Z^{QU} Z^{SV}
(\eta^{TW} + \frac{2}{3} Z^{TW}) f'_{QST} f'_{UVW},
\label{A15}
\end{eqnarray}
where
\begin{equation}
f'_{QST} = (M^{-1})^{\;\; T'}_T (M^{-1})^{\;\; Q'}_Q f_{Q'S'T'} M^{S'}_{\;\;
S}.
\label{A16}
\end{equation}
Thus changing the VEV of the Higgs fields in the CSA is equivalent to a
transformation of the structure constants~(\ref{A10}), given by an
$O(6,22,\R)$
rotation of $\Gamma^{(6,22)}$ (i.e. a rotation of the ``momenta'' labels $p$).
This transformation is an isomorphism of the gauge algebra. Finding the mass
spectrum of the theory on an arbitrary VEV is now very simple.

Using the field redefinition  described above, one can always choose
a zero cosmological constant minimum to lie at
\begin{eqnarray}
\phi=1,\;\; Z^b_a=\delta_a^b,\;\;\;\; a,b=1,..,6,
\nonumber \\
Z^I_a=0,\;I=7,...,28,\;\;\; Z^p_a=0\;\;\forall p\in\Gamma^{(6,22)}.
\label{A17}
\end{eqnarray}
Around this point the mass spectrum of the effective action
is exactly that predicted by string theory.
This is easily checked by looking at the mass term of the gauge boson
$A_{\mu}^p$.  Using eqs.~(\ref{M10}, \ref{A10}, \ref{A13}) one  finds
\begin{eqnarray}
\frac{1}{2}\eta_{pq} D_{\mu} Z_a^p D^{\mu} Z^{aq} &=&
\frac{1}{2} \partial_{\mu} \zeta_a^p\partial^{\mu} \zeta^{a\; -p} +
i\partial^{\mu} \zeta_a^p f_{pq}^a A_{\mu}^q -
\frac{1}{2} A_{\mu}^q f_{qa}^p A^{\mu r} f_{rp}^a + ... \nonumber \\
&=&  \frac{1}{2} \partial_{\mu} \zeta_a^p \partial^{\mu} \zeta^{a\; -p} +
p^a \partial^{\mu} \zeta_a^p  A_{\mu}^{-p} +
\frac{1}{2} p_a p^a A_{\mu}^p A^{\mu\; -p} + ... \nonumber \\
& &
\label{A18}
\end{eqnarray}
In~(\ref{A18}) we chose the string tension to be $\alpha'=1$, and `...' stands
for higher-order terms.

The gauge-vector kinetic term is $F_{\mu\nu}F^{\mu\nu}/4$,
thus the mass of the gauge boson $A_{\mu}^p$ is
\begin{equation}
m_p^2 = p_a p^a, \;\;\; a=1,...,6.
\label{A19}
\end{equation}
This is exactly the mass of a string vector boson with internal momentum
$p=(p_L,p_R)$, s.t. the components of $p_L$ are $p_a$~\footnote{The string
vertex corresponding to the field $A_{\mu}^p$ is $\psi_{\mu} e^{i(p_{La} x_L^a
+ p_{RI} x_R^I)}$. The mass and left-right level matching condition are given
by:  $m^2 = m_L^2 = m_R^2$, where
$\frac{1}{2} m_L^2 = \frac{1}{2} + \frac{1}{2} p_L^2 - \frac{1}{2} =
\frac{1}{2} p_L^2$ (the first $1/2$ is the
conformal dimension of the worldsheet fermion and the $-1/2$ is the normal
ordering constant in the NS sector) and
$\frac{1}{2} m_R^2 = \frac{1}{2} p_R^2-1$
(the $-1$ is the normal ordering constant of the bosonic right-handed
sector).}.
The field $p^a \zeta^p_a$ in equation~(\ref{A18}) is the Goldstone boson. The
appearance of the correct masses and the Goldstone boson can be rechecked
by expanding the scalar potential~(\ref{Pot}) up to quadratic terms in the
fluctuations around the point~(\ref{A17}). One finds:
\begin{equation}
V=(p^a p_a)\zeta^{bp}\zeta_b^{-p} - p^a p^b\zeta_a^p\zeta_b^{-p} + o(\zeta^3).
\label{A20}
\end{equation}
Equation~(\ref{A20}) gives five scalars with mass $p^a p_a$ and one massless
scalar for each nonzero $p$. The massless scalars are, naturally, the Goldstone
bosons of the broken generators of the gauge group.

It is important to realize that, since the infinite dimensional gauge group
acts nontrivially on the compensators $Z_b^a$, and since the latter can never
vanish, because of constraint~(\ref{M6}), the gauge symmetry is always broken
to
some finite dimensional rank 22 group $g$ times the $U(1)^6$ (originated by
the left-handed sector of the string). The unbroken gauge group is generically
$U(1)^{22}$, however, it is extended to bigger rank $22$ groups at special
VEVs~\footnote{In string theory this corresponds to points in the moduli space
with an enhanced affine symmetry.}. In the neighborhood of such VEV the
$N=4$
supergravity is reduced to a gauged $D=4$, $N=4$ supergravity with a finite
dimensional gauge group $g$.

The condition for a transformation $M$ in~(\ref{A14}) to be a residual symmetry
of the LEEA is
\begin{equation}
f'_{QST} = f_{QST},
\label{A21}
\end{equation}
where $f'_{QST}$ is defined in~(\ref{A16}), i.e.  $M$ is an automorphism of the
gauge group. By construction, the full duality group $O(6,22,\Z)$ is a symmetry
of the LEEA. This is because {\em
$O(6,22,\Z)$ is the Weyl group of the gauge
algebra}~\footnote{For gauged $N=4$ supergravity with a finite
gauge group $g$ the automorphism group of $g$
is a subgroup of $O(6,22,\Z)$ which leaves the enhanced
symmetry point in the moduli space fixed~\cite{GMR1}, as discussed from the
worldsheet point of view in section 2.6.}.

We pass now to the determination of the gauge group of the duality invariant
low-energy effective action. As mentioned before,
duality implies that massive states are to be
included as well. For example, we must include those states which become
massless on some background, together with all states related to them by
duality. It should be stressed, on the other hand, that the LEEA, by
definition,
gives the correct string dynamics only for those processes involving light
particles, with energies far below the string (or Planck) scale. For this
reason
the complete spectrum of the LEEA needs not (and indeed does not) coincide
with the string spectrum. Only states related by duality to massless ones
must coincide.

The strategy for determining the LEEA reads as follows. We determine first the
string-theoretical three-point functions of those spin-one fields $A^S_{\mu}$
that become massless at
some point of the moduli space. We then take the low-momentum limit
of these three-point functions by defining an appropriate off-shell
continuation. From these three-point functions we extract a set of structure
constants $f_{STU}$. Finally, we find a Lie algebra whose structure constants
coincide with $f_{STU}$ when evaluated on the original states $A^S_{\mu}$.

The spin-one states which become massless at some point of the moduli space are
of two kinds.
The first one is made of vectors which are massless on all backgrounds
\begin{eqnarray}
\varepsilon_{\mu}V^{a\mu}({\bf p})&=&\int d\theta d^2 z
\varepsilon_{\mu}D_{\theta}X^a \bar{\partial} X^{\mu}
\exp(ip_{\nu}X^{\nu})(\theta,z,\bar{z}),\;\;\; a=1,..,6, \nonumber \\
\varepsilon_{\mu}V^{I\mu}({\bf p})&=&\int d\theta d^2 z
\varepsilon_{\mu}D_{\theta}X^{\mu} \bar{\partial} X^I
\exp(ip_{\nu}X^{\nu})(\theta,z,\bar{z}),\;\;\; I=7,..,28. \nonumber \\
& &
\label{A1'}
\end{eqnarray}
Here $\varepsilon_{\mu}$ is a polarization vector obeying
$\varepsilon_{\mu}p^{\mu}=0$;
the vector {\bf p} is the space-time momentum with components $p_{\mu}$;
$X$ is
a heterotic superfield, i.e. its left-handed part is 2-$d$ supersymmetric,
$X(z,\bar{z},\theta)=X_L(\bar{z})+X_R(z)+\theta\psi(z)$,
and $D_{\theta}$ is a
spinorial derivative, $D_{\theta}=\partial_{\theta}+\theta\partial$.

The second one includes all spin-one fields which, at particular points of the
moduli space, extend the Abelian $U(1)^{28}$ symmetry generated by~(\ref{A1'})
to larger non-Abelian symmetries. The vertices corresponding to these fields
are
\begin{eqnarray}
\varepsilon_{\mu}V^{p\mu}({\bf p})&=&\int d\theta d^2 z C(p)
\varepsilon_{\mu}D_{\theta}X^{\mu}
\exp(ip_aX^a_L +ip_IX^I_R+ip_{\nu}X^{\nu})(\theta,z,\bar{z}), \nonumber \\
 & & -p_a^2+p_I^2=2.
\label{A2'}
\end{eqnarray}
In eq.~(\ref{A2'}) $C(p)$ is an operator defined for every internal
lattice vector $p$ and obeying~\cite{FK,HC}
\begin{equation}
C(p)C(q)=\varepsilon(p,q)C(p+q).
\label{A2a}
\end{equation}
The two cocycle $\varepsilon(p,q)$ satisfies the compatibility
conditions~(\ref{A11}) and~(\ref{A12}).
Its presence is necessary in order to reproduce the
correct commutators for the unbroken Lie algebra on the vectors $p_a=0$,
$p_I^2=2$~\cite{FK}.

The spin one states~(\ref{A1'}) and~(\ref{A2'}) give rise to (broken) gauge
symmetries, at least in that they give rise to the correct (on-shell) Ward
identities~\cite{GSW,Ven,MVen,Mah}
\begin{equation}
p_{\mu}V^{a\mu}\approx 0 ,\;\;\; p_{\mu}V^{I\mu}\approx 0,\;\;\;
p_{\mu}V^{p\mu} + p_aV^{p a}\approx 0.
\label{A2b}
\end{equation}
Here $\approx $ denotes an on-shell equality, and $V^{pa}$ is the scalar-field
vertex
\begin{eqnarray}
V^{pa}({\bf p})&=&\int d\theta d^2 z C(p)
D_{\theta}X^{a}
\exp(ip_aX^a_L +ip_IX^I_R+ip_{\nu}X^{\nu})(\theta,z,\bar{z}), \nonumber \\
 & & -p_a^2+p_I^2=2.
\label{A2c}
\end{eqnarray}
{}From eq.~(\ref{A2b}) one can recover an off-shell gauge invariance of the
effective action, as shown in~\cite{P}.

Besides~(\ref{A1'}) and~(\ref{A2'}), other fields may become (almost) massless
in
the appropriate regions of the moduli space. They are characterized by lattice
vectors of zero (Lorentzian) norm $-p_a^2 + p_I^2 =0$. These states become
exactly massless at the boundary of the moduli space, that is (for example)
when some of the
compactification radii $R\rightarrow \infty $. Since this limit gives rise to a
higher dimensional theory, we shall not include those fields. Our aim is only
to
build a four-dimensional effective action. This action makes sense only when
the
dimensions of the compactified space are $O(M_{Pl}^{-1})$, where $M_{Pl}$ is
the Planck mass.

Let us evaluate the relevant three-point functions involving the vector
fields~(\ref{A1'}) and~(\ref{A2'}). We define the off-shell three-point
function
according to ref.~\cite{LeCP,LeC} as
\begin{eqnarray}
V_{ijk} &=& \lim_{\epsilon\rightarrow 0} \langle 0 | h^2\circ V_i(\epsilon)
h\circ V_j(\epsilon)V_k(\epsilon)|0\rangle \nonumber \\
&=& \lim_{\epsilon\rightarrow 0}[h'(h(\epsilon))h'(\epsilon)]^{d_i}
h'(\epsilon)^{d_j}\langle 0 | V_i(h^2(\epsilon))V_j(h(\epsilon))V_k(\epsilon)
|0\rangle .
\label{A3'}
\end{eqnarray}
In eq.~(\ref{A3'}) $h$ denotes the $SL(2,\C)$ transformation sending $\infty
\rightarrow 0 $, $0\rightarrow 1 $, $1\rightarrow \infty $, i.e.
\bq
h(z,\bar{z})=(h,\bar{h})=\left( {1\over 1-z},{1\over 1-\bar{z}} \right) ,
\eq{A4'}
and $h'=|\partial h|^2$.
The suffix $d_i$ is the conformal weight of the vertex $V_i$, which also
includes reparametrization (and super-reparametrization) ghosts. Obviously
$d_i=0$ for on-shell vertices.

In eqs.~(\ref{A1'}) and~(\ref{A2'}) we represented all vector vertices in the
zero-ghost picture~\cite{FMS}. Actually it is convenient to use several
different pictures simultaneously. In the $-1$ picture, the vertices in eq.
{}~(\ref{A2'}) read, e.g.
\bq
\varepsilon_{\mu}V_{(-1)}^{\mu p}(z, \bar{z}) =
\varepsilon_{\mu}C(p)\psi^{\mu}\exp(-\phi)
\exp(ip_aX^a_L +ip_IX^I_R+ip_{\nu}X^{\nu})c\bar{c}(z,\bar{z}) .
\eq{A5'}
Here $c$ and $\bar{c}$ are the reparametrization ghosts,
and $\phi(z,\bar{z})$ is
the bosonized supersymmetry ghost~\cite{FMS}. $\psi^{\mu}(z)$ is the fermionic
superpartner of $\partial X^{\mu}(z)$.

Let us evaluate
\bq
V_{PQR} =\lim_{\epsilon \rightarrow 0}\langle 0 |
h^2\circ \varepsilon V^p_{(0)}({\bf p})(\epsilon)
h\circ \varepsilon V^q_{(-1)}({\bf q})(\epsilon)
\varepsilon V^r_{(-1)}({\bf r})(\epsilon)|0\rangle .
\eq{A6'}
We need the following correlation functions
\ber
\langle \exp(-\phi)(\frac{1}{1-\epsilon})\exp(-\phi)(\epsilon) \rangle &=&
1 + O(\epsilon), \nonumber \\
\langle (p\cdot\psi) \psi^{\mu}(1/\epsilon)
\psi^{\rho}(1)\psi^{\sigma}(0)\rangle &=&
\epsilon^2 (g^{\mu\rho}p^{\sigma} - g^{\mu\sigma}p^{\rho}) \nonumber \\
+ O(\epsilon^3), & & \nonumber \\
\langle \exp (iPX)(1/\epsilon)\exp(iQX)(1)\exp(iRX)(0)\rangle &=&
\epsilon^{P^2}[1 + O(\epsilon)], \nonumber \\
\langle \partial X^{\mu}\exp(iPX)(1/\epsilon)\psi^{\rho}\exp(iQX)(1)
\psi^{\sigma}\exp(iRX)(0)\rangle &=& i[-p^{\mu}\epsilon^{P^2+1} + \nonumber \\
q^{\mu}\epsilon^{P^2+2}]g^{\rho\sigma} +O(\epsilon^{P^2+3}). & &
\eer{A7'}
Here $P+Q+R=0$, where $P=(p_a,p_I,p_{\nu})$ is the total momentum vector, and
$g^{\mu\nu}$ is the space-time metric. The $\langle\rangle$ in the first line
of~(\ref{A7'}) includes the insertion of the charge at infinity $\exp 2\phi_0$,
where $\phi_0$ is the zero mode of $\phi$.
Substituting eqs.~(\ref{A7'}) in definition~(\ref{A6'}) we find,
taking into account the transversality of the polarization vectors
$\varepsilon$,
\bq
V_{PQR}=i\varepsilon_{\mu p}\varepsilon_{\rho q}\varepsilon_{\sigma
r}(g^{\mu\rho}p^{\sigma}+g^{\mu\sigma}r^{\rho}+g^{\rho\sigma}q^{\mu})
\varepsilon(p,q) .
\eq{A8'}
In~(\ref{A8'}) the two cocycle $\varepsilon(p,q)$ is a function of only
the internal momenta, as $C(p)$ in eqs.~(\ref{A2'}, \ref{A2a}) depends only on
the
internal lattice vectors.

The other three-point functions we need are
\bq
V_{aQR}=\lim_{\epsilon \rightarrow 0}\langle 0 |
h^2\circ V^a_{(0)}({\bf p})(\epsilon)
h\circ \varepsilon V^q_{(-1)}({\bf q})(\epsilon)
\varepsilon V^r_{(-1)}({\bf r})(\epsilon)|0\rangle ,
\eq{A9'}
and
\bq
V_{IQR}=\lim_{\epsilon \rightarrow 0}\langle 0 |
h^2\circ V^I_{(0)}({\bf p})(\epsilon)
h\circ \varepsilon V^q_{(-1)}({\bf q})(\epsilon)
\varepsilon V^r_{(-1)}({\bf r})(\epsilon)|0\rangle .
\eq{A10'}
By calculations similar to those leading to eq.~(\ref{A8'}), and since
\ber
\langle\bar{\partial}X^{\mu}\partial X^a
\exp(iPX)(1/\epsilon)\psi^{\rho}\exp(iQX)(1)\psi^{\sigma}\exp(iRX)(0)\rangle
&=&
\nonumber \\
-g^{\rho\sigma}q^aq^{\mu} \epsilon^{P^2+2} + O(\epsilon^{P^2+3}), & &
\eer{A11'}
we find
\ber
V_{aQR} &=& -\varepsilon_{\mu p}\varepsilon_{\rho q}\varepsilon_{\sigma
r}g^{\rho\sigma}q^{\mu}q^a ,\nonumber \\
V_{IQR} &=& -\varepsilon_{\mu p}\varepsilon_{\rho q}\varepsilon_{\sigma
r}(g^{\mu\rho}p^{\sigma}+g^{\mu\sigma}r^{\rho}+g^{\rho\sigma}q^{\mu})q^I,
\nonumber \\
P+Q+R &=& 0 ,\;\;\; P=(0,{\bf p}) .
\eer{A12'}

Now, we must compare eqs.~(\ref{A8'}) and~(\ref{A12'}) with the three-point
functions one gets from an $N=4$ supergravity Lagrangian discussed
in section 2.7. By eliminating auxiliary fields and expanding the $N=4$
Lagrangian around the
background~(\ref{A17}) (see for example~\cite{S}), we find the kinetic term
\bq
{1\over 4}\sum_{a}F_{\mu\nu}^aF^{a\, \mu\nu} +
{1\over 4}\sum_{K,K'}F_{\mu\nu}^KF^{K'\, \mu\nu}\eta_{KK'}.
\eq{A13'}
Notice that the graviphotons $A^a_{\mu}$ appear with positive signature. The
index $K=(I,p)$ labels all spin-one gauge fields belonging to vector
multiplets.
The term, trilinear in the vector fields, that we get from~(\ref{A13'}) is
\ber
\partial^{\mu}A^{\nu a}f^a_{KL}A^K_{\mu}A^L_{\nu}+
\partial^{\mu}A^{\nu K}f^{K'}_{aL}A^a_{\mu}A^L_{\nu}\eta_{KK'}+ & & \nonumber
\\
\partial^{\mu}A^{\nu K}f^{K'}_{La}A^L_{\mu}A^a_{\nu}\eta_{KK'}+
\partial^{\mu}A^{\nu K}f^{K'}_{LM}A^L_{\mu}A^M_{\nu}\eta_{KK'} . & &
\eer{A14'}
Recalling that the indices $a$, $K$ are lowered with the matrix $\eta_{ST}$
which obeys $\eta_{ab}=-\delta_{ab}$, $a=1,..,6$, and that $f_{STU}$ must be
completely antisymmetric~\cite{deRoo,dRW,S,GP},
we cast~(\ref{A14'}) in the following form
\bq
-\partial^{\mu}A^{\nu K}f_{aKL}A^a_{\mu}A^L_{\nu}+
\partial^{\mu}A^{\nu K}f_{KLM}A^L_{\mu}A^M_{\nu}.
\eq{A15'}
By substituting to the $A_{\mu}^S$ of eq.~(\ref{A15'}) the appropriate
transverse
plane waves, we find the three-point functions
\ber
V_{aKL} &=& -i\varepsilon_{\mu p}\varepsilon_{\rho q}\varepsilon_{\sigma r}
g^{\rho\sigma}q^{\mu}f_{aKL}, \nonumber \\
V_{KLM} &=& i\varepsilon_{\mu p}\varepsilon_{\rho q}\varepsilon_{\sigma r}
(g^{\rho\sigma}q^{\mu}+g^{\mu\sigma}r^{\rho}+g^{\mu\rho}p^{\sigma})f_{KLM} .
\eer{A16'}
Comparing eq.~(\ref{A16'}) with eqs.~(\ref{A8'}) and~(\ref{A12'}) we get
\bq
f_{aqr}=-iq^a , \;\;\; f_{Iqr}=-iq^I,\;\;\; f_{pqr}=\varepsilon(p,q),
\eq{A17'}
which are the structure constants in eq.~(\ref{A10}).

Next, we ought to find a Lie algebra whose structure constants include those
given in~(\ref{A17'}). This algebra can be defined as
follows~\cite{GO,GP,GP1}.
Let us consider twenty-eight free bosonic fields with correlation functions
\bq
\langle X^A(z)X^B(w)\rangle = -\eta^{AB}\log(z-w),\;\;\; A,B=1,..,28 .
\eq{A18'}
Here $\eta^{AB}={\rm diag}(-1^6,1^{22})$. Notice that these fields are {\em
chiral}.

Let us further compactify the bosonic fields on a $(6,22)$ Lorentzian even
self-dual lattice $\Gamma^{(6,22)}$. Let us denote, as before, with $X^a(z)$
the
six negative-metric bosons, and with $X^I(z)$ the other twenty two positive
metric bosons. For each lattice vector of Lorentzian norm $p^Ap^B\eta_{AB}=2$,
we associate a {\em chiral} vertex operator
\bq
V^p=\oint {dz \over 2\pi i} C(p) \exp(ip_AX^A)(z),\;\;\; p^Ap^B\eta_{AB}=2 .
\eq{A19'}
The vertices~(\ref{A19'}), together with
\bq
V^A=\oint {dz \over 2\pi i} \partial X^A(z),\;\;\; A=1,..,28 ,
\eq{A20'}
give rise to a Lie algebra {\bf g}, whose elements are~(\ref{A19'})
and~(\ref{A20'}) together with all possible multiple commutators of the $V^p$.
This algebra is a particular example of (indefinite-signature) lattice
algebra~\cite{GO}. In some cases lattice algebras reduce to hyperbolic, affine
or finite Kac-Moody algebras~\cite{GO,K}. The vertices $V^A$ span a maximal
commuting subalgebra~\cite{GO}, which plays the role of the Cartan subalgebra.
The rank of {\bf g} is thus finite (and equal to 28, in our example).

Two facts ought to be noticed now.
The first one is that, as shown in the following, there exist a {\bf g}
invariant non-degenerate two form $\eta$, with at least six negative
eigenvalues, which reduces to $\eta_{AB}$ on the Cartan subalgebra. This
implies
the identity
\bq
f^V_{ST}\eta_{VU} + f^V_{SU}\eta_{TV}=0.
\eq{A21'}
Here $S,T,U,V$ label the generators of {\bf g}.
Identity~(\ref{A21'}) is indispensable for the construction of an $N=4$
supersymmetric Lagrangian~\cite{deRoo,dRW,S,W1}.

The second fact we should emphasize is that in {\bf g} there are many states
actually not present in the string spectrum. This is not a serious problem for
our construction, anyway. We are in fact interested in constructing only a
duality-invariant {\em low-energy} effective action. All new states introduced
by the definition of {\bf g} have masses $O(M_{Pl})$ for all values of the
lattice momenta $p^A$ corresponding to four-dimensional compactifications of
the
heterotic string. The only region in which some of the additional states may
become light corresponds to the decompactification limit mentioned earlier in
this section. In this limit some states with momenta obeying
$p^Ap^B\eta_{AB}=0$ may become almost massless. On the other hand, a 4-$D$
low-energy effective action cannot be expected to correctly describe a higher
dimensional theory. The Kaluza-Klein spectrum is indeed truncated down to the
lightest states from the very beginning. A phenomenon similar to the one
encountered here has already been noticed in the past, in connection with the
study of string effective actions. In~\cite{D,CFGPP} e.g., it was shown that by
adding higher derivative terms to the effective Lagrangian of both the
bosonic~\cite{D} and heterotic~\cite{D,CFGPP} strings, additional states of
mass
$O(M_{Pl})$, as well as ultramassive ghosts, may propagate on a generic
background. This is the price paid for truncating the higher spin modes of
the string.

We pass now to a more detailed study of the DISG algebra {\bf g}.
The simplest new states that one finds by taking the commutator of two
vertices~(\ref{A19'}) are
\ber
[V^p,V^q]&=&\varepsilon(p,q)
\oint {dz\over 2\pi i} C(p+q) p_B\partial X^B \exp i(p_A+q_A)X^A(z),
\nonumber \\
& & (p_A+q_A)(p^A+q^A)=0 \;\;\;\; A,B=1,..,28.
\eer{A22'}
In other terms, by taking the commutator of two vertices~(\ref{A19'}),
corresponding to momenta obeying $p\eta q=-2$, one finds ``photon'' states
of the form
\bq
\oint {dz\over 2\pi i} C(p) \xi_A\partial X^A \exp (ip_AX^A)
(z),\;\;\; p_Ap^A=0,\;\;\; p_A\xi^A=0 .
\eq{A23'}
The mass of the states one gets from formula~(\ref{A19}), is only small
in the decompactification limit, in which, as
previously stressed, the whole effective action approach looses meaning. The
low
energy dynamics of strings is therefore not affected by the presence of these
new ``photon'' states.

The states with $p\eta p=-2$, whose mass square is always larger than
$M_{Pl}^2$, read
\bq
V^p(\xi_{AB},\xi_A)=\oint {dz\over 2\pi i} C(p)
(\xi_{AB}\partial X^A\partial X^B+\xi_A\partial^2 X^A)\exp (ip_AX^A)(z),
\eq{A24'}
where the coefficients $\xi_{AB}$, $\xi_A$ obey
\bq
\eta^{AB}\xi_{AB}=0,\;\;\; \xi_{AB}p^B=\xi_{BA}p^B=\xi_A,\;\;\; \xi_Ap^A=0.
\eq{A25'}
Notice that some of the states~(\ref{A23'}) and~(\ref{A24'}) have negative
norm. Their presence is to be expected, since we are dealing with a model
possessing six time-like coordinates. In this case, the Virasoro constraints
(see eq.~(\ref{A27'}) below) cannot, by themselves alone, eliminate all
ghosts.
The same remarks given previously apply to these ghost states:
their mass is always $O(M_{Pl})$.

All elements of {\bf g} are line integrals of primary fields of conformal
weight
one of the Virasoro algebra generated by
\bq
L(z)=-\frac{1}{2}\eta_{AB}\partial X^A \partial X^B (z).
\eq{A26'}
This statement is easily proven, since it holds true for all states given
in~(\ref{A19'}) and~(\ref{A20'}), and the Jacobi identity together with
$[L_n,A]=[L_n,B]=0$ imply
\bq
[L_n,[A,B]]=[[L_n,A],B]-[[L_n,B],A]=0,\;\;\; n\geq 0.
\eq{A27'}

To define an invariant tensor $\eta^{ST}$ on all elements of {\bf g} is equally
easy. Indeed, by writing the elements of {\bf g} as
\bq
V^S=\oint{dz \over 2\pi i} \phi^S(z),
\eq{A28'}
where $\phi^S(z)$ is a primary field of conformal dimension one, and denoting
with $|0\rangle$ the $SL(2,\C)$ invariant vacuum of~(\ref{A26'}), we may
define $\eta^{ST}$ as
\bq
\eta^{ST}=\lim_{\epsilon\rightarrow 0}{1\over \epsilon^2}\langle
0|\phi^S(-1/\epsilon)\phi^T(\epsilon)|0\rangle .
\eq{A29'}
This definition is nothing but the BPZ scalar product~\cite{BPZ}. Hermiticity
of
$\eta^{ST}$ is therefore immediate~\cite{BPZ}.
The invariance of~(\ref{A29'}) is proven by using representation~(\ref{A28'})
for
the elements of {\bf g}, and by use of the Cauchy theorem
\bq
f^V_{ST}\eta_{VU}+f^V_{SU}\eta_{TV}=
\lim_{\epsilon\rightarrow 0}\langle 0|\oint_{C}{dz\over 2\pi i}\phi_S(z)
{1\over \epsilon^2}\phi_T(-1/\epsilon)\phi_U(\epsilon)|0\rangle=0,
\eq{A30'}
where $C$ is a contour encircling both points $\epsilon$ and $-1/\epsilon$.

As we recalled earlier, the invariant metric $\eta_{ST}$ has more than six
negative eigenvalues. This simply signals the limits of validity of the LEEA
approach. Gauged $N=4$ supergravity can be constructed for any invariant tensor
obeying eq.~(\ref{A21'}) and of signature $(6+n,m)$~\cite{deRoo,dRW,S,W1}.
In this case the $N=4$ matter scalars live on the coset manifold
\bq
{O(6+n,m,\R)\over O(6,\R)\times O(n,m,\R)},
\eq{A31'}
and as expected there are scalar ghosts propagating\footnote{Together with
their supersymmetric partners.}.

The fundamental property of the DISG algebra {\bf g} is that its
automorphism group contains
the duality group $O(6,22,\Z)$, as mentioned before.
Since all elements of {\bf g} can be
written as commutators of elements~(\ref{A19'}) and~(\ref{A20'}), we need only
to
show that $O(6,22,\Z)$ acts as an algebra isomorphism on these elements.
Let us define the action of $O(6,22,\Z)$ on {\bf g} by
\bq
X^A(z)\rightarrow \Lambda^A_BX^B(z), \;\;\; p^A\rightarrow
(\Lambda^{-1})^A_Bp^B,\;\;\; \Lambda^A_B\in O(6,22,\Z).
\eq{A31a}
By inspection of eqs.~(\ref{A19'}) and~(\ref{A20'}), $O(6,22,\Z)$ is an
isomorphism if
\bq
C(\Lambda^{-1} p)=\hat{\Lambda}^{-1}C(p)\hat{\Lambda}.
\eq{A31b}
The linear operator $\hat{\Lambda}$ acts on the same Hilbert space on which
$C(p)$ is defined.
Property~(\ref{A31b}) is demonstrated in the appendix of ref.~\cite{GP1},
where an
explicit construction of $\hat{\Lambda}$, $\varepsilon(p,q)$, and $C(p)$ is
given.

The DISG contains infinitely many affine Lie algebras. To see this, let us
select a set of vectors $p$ of $\Gamma^{(6,22)}$ of the form
\bq
p=(0,0,0,0,0,0,\alpha),\;\;\; \alpha^2=2.
\eq{A31c}
Let us choose the $\alpha$ so that they generate a finite dimensional Lie
algebra, say $SO(44)\times U(1)^6$. Let us further select a vector
$p_0\in\Gamma^{(6,22)}$ of zero norm $p_0\eta p_0=0$, and orthogonal to all
$\alpha$. Define now the generators
\bq
E^{\alpha}_n\equiv V^{\alpha +np_0},\;\;\; H^A_n=\oint{dz\over 2\pi i} C(np_0)
\partial X^A\exp(inp_{0A}X^A)(z).
\eq{A31d}
The commutators of these generators read
\ber
[E^{\alpha}_n,E^{\beta}_m]&=&\left\{
\begin{array}{cl}
\varepsilon(\alpha,\beta)E^{\alpha+\beta}_{n+m} &  \alpha\cdot\beta=-1 \\
\alpha_AH^A_{m+n} + m\delta_{m,-n}p_0\eta H_0 &  \alpha=-\beta \\
0 & {\rm otherwise}
\end{array} \right.  \nonumber \\
{[}H^A_n,E^{\alpha}_m{]}=\alpha^AE^{\alpha}_{n+m}, & &
{[}H^A_n,H^B_m{]}=m\delta^{AB}\delta_{m,-n}p_0\eta H_0 .
\eer{A31e}
Here we made use of the equation
\bq
p_{0A} H^A_m=\oint{dz \over 2\pi i}p_{0A}\partial X^A\exp(imp_{0A}X^A)(z)=0,
\;\;\;{\rm if}\; m\neq 0 .
\eq{A31f}
This identity holds because the integrand can be written as a total derivative.
It is immediate to recognize in eq.~(\ref{A31e}) the commutation relations of
an affine Lie algebra of level $p_0\eta H$~\cite{K,M,BH}.

To conclude this section, let us notice that there is a way of avoiding ghosts
in an effective action which includes massive fields. One can indeed pick up
twenty-eight linearly independent positive lattice vectors $p_i$, $i=1,..,28$,
of Lorentzian length two and such that $p_i\eta p_j \geq 0$
and form with them a (hyperbolic)
Kac-Moody algebra. The procedure is standard~\cite{K}. To
any length-two vector $\pm p_i$, one associates a generator $V_{\pm p_i}$.
One then considers the algebra ${\cal G}$ freely generated by the
$V_{\pm p_i}$, the vertices $V^A$, $A=1,..,28$, and the commutation relations
\ber
{[V_{p_i},V_{-p_j}]} &=&\left\{
\begin{array}{cl}
\varepsilon(p_i,-p_j)V_{p_i-p_j} & p_i\eta p_j =1 \\
p_{iA}V^A &  i=j \\      0 & {\rm otherwise}
\end{array} \right.
\nonumber  \\
{[V_{p_i},V_{p_j}]}=0,\;\;\; & &
{[V^A,V_{\pm p_i}]}=\pm p^A_iV_{\pm p_i}, \nonumber \\
{[V^A,V^B]}=0, & & \;\;\; A,B=1,..,28 .
\eer{A32'}
By dividing ${\cal G}$ by the (unique) maximal ideal commuting with all $V^A$
one defines a hyperbolic Kac-Moody algebra $\hat{\cal G}$~\cite{K,M}.

Among the known properties of $\hat{\cal G}$ there is the following: there
exists an invariant Hermitian form $\eta_{ST}$, reducing to $\eta_{AB}$ on the
Cartan subalgebra, and positive definite outside it~\cite{K}. In our case this
means that there exists an $\eta$ with only six negative eigenvalues. Thus, the
algebra $\hat{\cal G}$ gives rise to a ghost-free $N=4$
supergravity~\cite{deRoo,dRW,W1}.
The problem is that this theory is invariant only
under a subgroup of $O(6,22,\Z)$~\footnote{
This fact can be easily proven by noticing
that any root of $\hat{\cal G}$ is either positive or negative~\cite{K}, and
thus  $\hat{\cal G}$ cannot contain all length two vectors of
$\Gamma^{(6,22)}$.}.
We think that complete duality invariance is more important than the presence
of
non-physical ultramassive ghosts, which, anyway, become relevant only at
energies so high as to render the LEEA approach inapplicable. For this reason
we
suggest  that the relevant LEEA gauge algebra is the algebra of the DISG, i.e.
{\bf g}.

Finally, we comment about few relations with Closed-String Field Theory
(CSFT).

In ref.~\cite{KZ} it was shown that $O(d,d,\Z)$ is a symmetry of CSFT of the
bosonic string expanded around flat backgrounds, as is suggested by the
discussion of section 2.6. Moreover, in
ref.~\cite{giveon2} it was shown that a truncation of the string field
gives rise to an effective action
compatible to cubic order with the duality invariant LEEA
for any choice of a cyclic string vertex.
{}From CSFT one may learn how to add the $p\eta p=0$ modes to the effective
action.

As we mentioned before, to construct the complete invariance algebra of the
string, we must add higher spin states to the algebra. To bypass this
difficulty it is easier to study first the gauge symmetries of the
$N=2$ string, as the higher spin modes are never physical
in that case. This was done in ref.~\cite{GShapere}.
A suggestion for the complete invariance algebra of the bosonic string
was put forward along the lines of this section in ref.~\cite{Moore}.

We conclude here our study of the $D=4$, $N=4$ duality-invariant effective
Lagrangian of the heterotic string. In the next section we shall
extend the analysis
to more general compactifications. In particular, we shall study string theory
on orbifolds and Calabi-Yau manifolds from the effective-action perspective
and, briefly,  from the (microscopic) worldsheet viewpoint.
\end{subsection}
\end{section}

\newpage

\begin{section}{Duality and Discrete Symmetries of the Moduli Space of
Orbifold and Calabi-Yau Backgrounds}

The previous section dealt with the case of toroidal compactifications of
the heterotic string. In
order to make contact with phenomenology, on the other hand, strings must
be compactified on more complicated background as, for instance,
orbifolds~\cite{DHVW}
or Calabi-Yau spaces~\cite{CHSW}.
In these cases one may also define duality between seemingly different
backgrounds.

Before discussing more complicated cases let us mention the simplest
example of non-flat compactification and its duality symmetry, namely the
``orbicircle'' example.

The orbicircle is defined by identifying the points $x$ and $-x$ on the
circle $x\approx x + 2\pi Rm$, $m\in \Z$ (see figure 1.D).
This identification gives rise to
a closed segment whose two boundary points correspond to the fixed points
under $x \rightarrow -x$ : $x=0$ and $x=\pi R$.
Even in this case there is a duality transformation relating a length
$\pi R$ orbicircle with another one of length $\pi/R$. This can be shown at
the level of the spectrum \cite{egrs}
of the corresponding CFT along the lines of section 2.2.

The  orbicircle compactifications span a sub-space of $c=1$ backgrounds.
The moduli space of $c=1$ unitary CFTs is shown in figure 3.A
\cite{BRSct,DVVc1,Gc1,Kc1}. It consists of target spaces which are circles
of radius $R_c$, target spaces which are orbicircles of radius $R_o$, and
three special orbifolds \cite{Gc1}. These three points correspond to CFTs
without marginal deformations. The local
dimension of the moduli space is zero at those points.
Along the $R_c$- and $R_o$-line the dimension is one.
There is a special point where the two lines meet. At this point, a string
cannot distinguish if it moves on the
circle or on the orbicircle. This point is the self-dual point of the
orbifold, with a symmetry enhanced from $\Z_2\times \Z_2$ to $U(1)_L\times
U(1)_R$.

We return now to compactifications to 4-$D$, in which case no complete
classification is presently available.
A few particular examples of orbifold compactifications are studied in
some detail in section 3.1. The
subject has been studied in more general or more complex situations in
several papers as, for instance, in
refs.~\cite{LMN1,NO,LZ,MO1,Zucc,EJL,EJN,EJLM,EK,Spa,Spa1,ES,EJSS,SJLS,BLST}.

In section 3.2 we extend the analysis of sections 2.7 and 2.8 to the
$N=1$ and $N=2$ orbifold case. Namely,
we find 4-$D$ effective actions invariant under
the complete duality group or subgroups thereof.
$N=1$ orbifolds are particular (singular) examples of Calabi-Yau spaces. The
effective actions for those compactifications are conveniently described using
the language of ``special geometry,'' which we introduce in section 3.3.
The study of duality in Calabi-Yau spaces is briefly surveyed in section
3.4. In the same section we also review various results
concerning the structure of the superpotential of $N=1$
compactifications; it turns out that duality significantly modifies the
naively expected low energy physics.

\begin{subsection}{Duality in Orbifold Compactifications}
\setcounter{equation}{0}
Reducing $N=4$ to $N=2,1,0$ supersymmetry in four dimensions can be achieved
in a simple way by means of an orbifold construction~\cite{DHVW}.
A (symmetric) orbifold construction based on $T^6$ (the six-dimensional,
compact, internal manifold generating $\Gamma^{(6,6)}$), involves the following
steps:
\begin{enumerate}
\item
Identify points of $T^6$ conjugated by a group element
$g\in G$. The group $G$ is a symmetry of $T^6$, e.g. a discrete rotation.
In the following, for sake of simplicity, we shall take $G=\Z_N$. This choice
gives rise to an Abelian orbifold~\cite{DHVW}. In general, the action of $\Z_N$
on $T^6$ is not free. In other words, there exist points of $T^6$ which are
invariant under $\Z_N$. For instance, as mentioned above,
the one dimensional torus $T^1$ (namely, a radius $R$ circle) can be modded
out by the $\Z_2$ symmetry $x\rightarrow -x$.
This symmetry has two fixed points: $x=0$ and $x=\pi R$.
The existence of fixed points means that the quotient space $T^6/\Z_N$ is not a
smooth manifold, but rather possesses conical singularities. In spite of this
fact, the resulting string theory is consistent, unlike the
corresponding Kaluza-Klein field theory.

Once $\Z_N$ has been defined on the compactified coordinate, one defines its
action on the gauge indices. In the $\Z_2$ example, for the $E_8\times E_8$
heterotic string, one may decompose one of the $E_8$ factors of the gauge
group into representations of the maximal $E_8$-subgroup, $SU(2)\times E_7$.
Then, one may identify the $\Z_2$ acting on the torus coordinate with the
center of $SU(2)$.
\item
Project the Hilbert space ${\cal H}$ of the $N=4$ heterotic string onto $\Z_N$
invariant states. Explicitly, the projection reads
\bq
P={1\over N} \sum_{m=1}^N g^m.
\eq{M13}
Here $g$ is the generator of $\Z_N$, obeying $g^N=1$.
\item
Unlike the field-theoretical case, this is not the end of the story. Indeed,
one must also take into account those string configurations that close only up
to a $\Z_N$ transformation:
\bq
X^A(\sigma + 2\pi,\tau)=(hX)^A(\sigma,\tau),\;\;\; h\in \Z_N.
\eq{M13'}
These configurations generate new string states. The Hilbert space they
span is called the {\em twisted sector}.
\end{enumerate}
The $N=1$ or $N=2$ LEEA of an orbifold compactification contains two types of
fields: the untwisted ones, which form a subset of the $N=4$ fields, and the
twisted fields. The twisted fields are new, additional fields, not present in
the truncation of the $N=4$ LEEA. Their Lagrangian, therefore, is not fixed
by supersymmetry and field-theoretical arguments alone~\footnotemark.
\footnotetext{A remarkable exception is found in the fermionic construction of
four-dimensional strings~\cite{ABK,KLT}. In that case one can figure out a
$\Z_2$ truncation of a four-dimensional $N=2$ theory which breaks supersymmetry
to $N=1$ without introducing additional massless states. The LEEA for the
massless states of the resulting $N=1$ theory is determined by supersymmetry
and a few other properties~\cite{FGKP1}.}

In this section we find out what subgroup of the $O(6,6,\Z)$ duality group
survives the orbifold truncation, and find the transformation laws of
untwisted {\em and twisted} massless states under such duality. The more
ambitious problem of finding a LEEA, invariant under the $O(6,22,\Z)$ subgroup
surviving the orbifold truncation, is (partially) answered in section 3.2.

The orbifold construction sketched above can be extended to a general
$\Gamma^{(6,22)}$ compactification~\cite{NSV}. Here we deal, for simplicity,
with $\Gamma^{(6,6)}\oplus \Gamma^{(0,16)}$ (symmetric) orbifolds.

Let us define explicitly the action of $\Z_N$ on the string coordinates.
To keep the notation simple we will consider the $\Z_3$ orbifold, giving
rise to $N=1$ supersymmetric theory. The generalization to any $\Z_N$ is
done in \cite{DHVW}.

First of all, by complexifying the six internal torus coordinates we may write
\bq
(gX)^i(z,\bar{z})=e^{2\pi i/3}X^i(z,\bar{z}),\;\;
X^{\bar{\imath}}\equiv (X^i)^*, \;\;\; i=1,2,3,
\eq{M14}
namely, from now on we work with the complex torus $T^6=(T^2)^3$.
Notice that
we choose a $\Z_3$ acting in the same way on all coordinates.
By denoting with $X(\bar{z})^A$, $A=1,...,16$,
the 16 chiral left-moving bosons
taking value in the maximal torus of $E_8\times E_8$ (or $Spin(32)/\Z_2$),
we further define
\bq
(gX)^A(\bar{z})= X^A(\bar{z}) + 2\pi \delta^A.
\eq{M16}
The shift $\delta^A$ obeys $3\sum_{A=1}^{16}\delta^Ap^A\in \Z$
for all $p^A$ in $\Gamma^{(0,16)}$.

Equations~(\ref{M14}, \ref{M16}) induce the following transformation on the
$\Gamma^{(6,6)}$ vector ($k^i_L$, $k^i_R$)
\bq
(gk)^i_L=e^{2\pi i/3} k^i_L, \;\;\; (gk)^i_R=e^{2\pi i/3} k^i_R.
\eq{M15}
The $k^i_{L,R}$ are expressed in terms of the momenta and winding numbers
of $X^i$ by eqs.~(\ref{ER68}, \ref{ER69}).

The gauge algebra is realized in terms of $(1,0)$ conformal vertices, as
explained in section 2.8 for the DISG algebra. Here they take the form
\ber
V_p &=& C(p)\oint {d\bar{z}\over 2\pi i} e^{i\sum_{A}X^A(\bar{z})p^A},\;\;\;
\sum_1^{16} p^Ap^A=2,\;\; p^A \in \Gamma^{(0,16)}, \nonumber \\
V_A &=& \oint {d\bar{z}\over 2\pi i} \bar{\partial}X^A(\bar{z}).
\eer{M16'}
The cocycle $C(p)$, in this case, can be written down explicitly as
follows~\cite{HC}. Pick up a basis $\{e_M | M=1,...,16\}$ of $\Gamma^{(0,16)}$,
and choose the $e_M$ such that $e\cdot e\equiv \sum_{A}e_M^A e_M^A=2$.
Then the cocycle reads
\bq
C(p)\equiv (-)^{p*\hat{p}}, \;\;\;
X*Y\equiv \sum_{M>N}X^MY^N e_M\cdot e_N.
\eq{M17}
Here $\hat{p}$ is the momentum operator of the left-moving chiral bosons.

The $\Z_3$ action on vertices~(\ref{M16'}) reads~\cite{DHVW}
\bq
(gV)_p=e^{2\pi i \sum_{A}\delta^A p^A} U V_p U^\dagger,\;\;\; (gV_A)=V_A,
\eq{M18}
where $U$ is a unitary transformation acting on the cocycle $C(p)$
only. This transformation simply defines an equivalent representation of
the cocycle, as noticed in section 2.8.

Let us examine now the behavior of the orbifold compactification under
duality, beginning by analysing the states in the untwisted sector.

Equations~(\ref{M14}, \ref{M16}) completely specify the projection~(\ref{M13})
on all untwisted states. These states can be written as a product of an
oscillator part and a $\Gamma^{(6,6)}$-lattice part. It is useful to
decompose the lattice states in the following way
\bq
\ket{k}_m= \sum_{n=1}^3 e^{2\pi i mn/3}\ket{e^{2\pi i n/3}k}.
\eq{M19}
This linear superposition of lattice vectors transforms as
\bq
g\ket{k}_m=e^{-2\pi i m/3}\ket{k}_m .
\eq{M20}
If we denote by $O_m$ a polynomial in
the oscillators of the string coordinates, transforming as
\bq
(gO_m)=e^{2\pi i m/3}O_m,
\eq{M21}
the invariant states are simply $O_m\ket{k}_{m'}$, with \mbox{$m-m'=0$ mod
$3$}. Notice that at $k=0$ the only invariant states read $O_0\ket{0}$.

The subgroup of $O(6,6,\Z)$ surviving the orbifold truncation can now be
determined quite simply.

Let us find at first what constraints follow from the lattice-dependent
part of the vertex $O_m\ket{k}_{m'}$. By definition, a symmetry transforms
physical states into physical states, so, the duality transform of
$O_m\ket{k}_m$ must be $\Z_3$ invariant too. This requirement implies in
particular that
\bq
g\ket{\Lambda k}_m = e^{-2\pi i m/3} \ket{\Lambda k}_m,\;\;\; \Lambda \in
O(6,6,\Z).
\eq{M22}
In general
\bq
(\Lambda k)^i=\Lambda^i_jk^j + \Lambda^i_{\bar{\jmath}}k^{\bar{\jmath}},
\eq{M23}
thus, the $\Lambda$ satisfying eq.~(\ref{M22}) obey
\bq
\Lambda^i_{\bar{\jmath}}=\Lambda^{\bar{\imath}}_j=0.
\eq{M24}
This equation is quite natural, it says that the surviving duality group is
included in $O(6,6,\Z)\cap SU(3,3,\R)$~\cite{FLT,FFS},
namely, the duality subgroup preserving the complex
structure of $(T^2)^3$. This group, called $SU(3,3,\Z)$, commutes with the
orbifold projection and, therefore, transforms $O_m$, that is the
oscillator dependent part of a physical vertex, into another operator
$O'_{m}$ which transforms under $\Z_3$ as $O_{m}$. This means that
$SU(3,3,\Z)$  maps massless physical
states into themselves. Invariance under this group puts severe constraints on
the form of the $N=1$ LEEA arising from this orbifold compactification.

We have not yet shown that $SU(3,3,\Z)$ is an invariance of the twisted sector
too. Here we shall give a sketch of this proof, carried on in detail in
ref.~\cite{LMN}\footnotemark.
\footnotetext{Ref.~\cite{LMN} deals
with invariance under an $SL(2,\Z)$ subgroup
of $SU(3,3,\Z)$, but its techniques can be  extended to the general case.}

In the twisted sector, the string coordinates obey the boundary conditions
\ber
X^i(\tau, \sigma +2\pi) &=& e^{2\pi i m/3}X^i(\tau,\sigma),\;\;\;
\psi^i_R(\tau,\sigma +2\pi)= e^{2\pi i m/3}\psi^i_R(\tau,\sigma),
\nonumber \\
X^A_L(\tau + \sigma + 2\pi) &=& X^A_L(\tau + \sigma) +2\pi m\delta^A,
\qquad m=1,2.
\eer{M29}
The boundary conditions of the
Neveu-Schwarz fermions $\psi_R^i$ are dictated by worldsheet supersymmetry.
The boundary conditions in ~(\ref{M29}) entail the following mode expansion
\ber
X^i(\tau,\sigma)&=& f^i_a +{i\over \sqrt{2}}\sum_{n\in \Z+m/3}{1\over
n}\alpha_n^i e^{-i n(\tau -\sigma)}+
{i\over \sqrt{2}}\sum_{n\in \Z-m/3}{1\over
n}\tilde{\alpha}_n^i e^{-i n(\tau +\sigma)}, \nonumber \\
\psi_R^i(\tau,\sigma)&=&\sum_{n\in \Z+m/3+1/2-t/2}\psi_n^i
e^{-i n(\tau -\sigma)}, \nonumber \\
X^A_L(\tau +\sigma)&=& (p^A + m\delta^A)(\sigma +\tau) +{i\over
\sqrt{2}}\sum_{n\in \Z} {1\over n}
\tilde{\alpha}^A_n e^{-i n(\tau +\sigma)},\;\;\;
p^A\in \Gamma^{(0,16)}.
\eer{M30}
In this equation $f^i_a$ denote the $T^6$ fixed points of the $\Z_3$
transformation $g^m\equiv \exp(2\pi i m/3)$.
The index $a$ labels these fixed points and the integer $t$ is equal to
0 in the Neveu-Schwarz sector, and to 1 in the Ramond sector. Notice that the
mode expansion of the heterotic chiral boson $X^A$ is unchanged, but for a
translation by $m\delta^A$ of the lattice vector $p^A$. There is a mode
expansion given by eq.~(\ref{M30}), and therefore, a Hilbert space of states
for each one of the fixed points $f_a$. In other words, each twisted sector
(with twist $g^m$) is made of many identical copies, one for each fixed point.

By denoting with $O_k$, as before, a polynomial in the string oscillators
obeying $(gO)_k =\exp(2\pi i k/3)$, and by $\ket{p^A,a}$ a state at the $a$-th
fixed point with heterotic momentum $p^A$, we may write a
$\Z_3$ invariant twisted state as
\bq
O_k\ket{p^A+m\delta^A,a},\;\;\; k + \delta^A(p^A+m\delta^A)=0 \; {\rm mod}\;
3.
\eq{M31}

Notice that the twisted states have $k^i=k^{\bar{\imath}}=0$, since a non-zero
momentum or winding number in $T^6$ is incompatible with boundary
conditions~(\ref{M29}).

It may seem that the state~(\ref{M31}) be mapped into itself by
a duality transformation, since it does not depend on $k^i$,
$k^{\bar{\imath}}$.
This is not true, however. States, of identical oscillator and momentum
content, belonging to different fixed points, are isomorphic (they have the
same conformal weight etc.) and can mix together. In general, under duality
states~(\ref{M31}) transform as~\cite{LMN}
\bq
O_k\ket{p^A+m\delta^A,a }\rightarrow
e^{i\varphi(G,B,\Lambda)}S_{ab}(\Lambda)O_k\ket{p^A+m\delta^A,b},\;\;\;
\Lambda\in SU(3,3,\Z).
\eq{M32}
The phase $\varphi(G,B,\Lambda)$ depends on the background fields $G$, $B$,
but not on the fixed point, whereas the unitary matrix $S_{ab}(\Lambda)$,
transforming different fixed points into each-other, is independent of the
background.

Equation~(\ref{M32}) is sufficient to conclude
that the spectrum of an orbifold
compactification is invariant under duality. Indeed, eq.~(\ref{M32}) maps
physical states into physical states, and defines a
unitary transformation between equivalent Hilbert spaces.

The explicit form of $S_{ab}$ and $\varphi$ was given in
ref.~\cite{LMN}. There it was also shown that not only the twisted-sector
spectrum, but also
its interactions, namely the three- and four-point
correlation functions, are invariant under duality, if
the twisted vertices are transformed as in eq.~(\ref{M32}).

The previous analysis has dealt with a simple question, namely, what is the
duality group of an orbifold compactification. The next question one is
naturally led to is how this duality acts on fields, in particular on the
massless ones.  To answer this question one must find how twisted and
untwisted vertices depend on the background. Here, for sake of simplicity,
we choose
$T^6$ to be a product of three {\em identical} two dimensional tori $T^2$.
The four $T^2$-moduli (background fields) $B_{ij}$, $G_{ij}$, $i,j=1,2$
reduce to two, once the complex structure has been defined by
\bq
X(z,\bar{z})=X^1(z,\bar{z}) + e^{2\pi i/3}X^2(z,\bar{z}).
\eq{M33}
This identification leaves one free complex modulus, since the metric of a
complex 2-$d$ manifold has only one independent component.  Let us define this
modulus by
\bq
\tau=i\sqrt{\det G} + B_{12}.
\eq{M34}
Then, our general formul\ae~(\ref{ER68}, \ref{ER69})
giving $p_L$ and $p_R$ in terms of
the background fields reads
\ber
p_L &=& {i \over (\sqrt{3} \Im \tau)^{1/2}} [ m_2 + e^{-2\pi i /3} m_1 +
\bar{\tau}(n_1 - e^{-2\pi i/3} n_2)], \nonumber \\
p_R &=& {i \over (\sqrt{3} \Im \tau)^{1/2}} [ m_2 + e^{-2\pi i /3} m_1 +
\tau (n_1 - e^{-2\pi i/3} n_2)].
\eer{M35}
Here $m_i, n_i \in \Z$.

There are three moduli $\tau_i$, one for each $T^2$ factor in $T^6=T^2\otimes
T^2\otimes T^2$. The subgroup of $SU(3,3,\Z)$ most frequently studied in the
literature is $SL(2,\Z)^3$. This subgroup acts on the $\tau_i$ as
follows
\bq
\tau_i \rightarrow {a_i\tau_i +b_i \over c_i\tau_i + d_i} ,\;\;\;
a_i,b_i,c_i,d_i \in \Z,\; a_id_i-b_ic_i=1.
\eq{M36}
We notice that the group acts projectively on the moduli, thus, it should more
properly be called $PSL(2,\Z)^3$.

Equation~(\ref{M36}) induces a $SL(2,\Z)^3$ transformation on the
lattice~(\ref{M35}).
This transformation can be realized by acting on the fields
$X_R^i(z)$, $X_L^i(\bar{z})$ as follows
\bq
X_L^i(\bar{z})
\rightarrow \bar{\lambda} \left( {c_i\tau_i+d_i\over c_i\bar{\tau}_i
+d_i}\right)^{1/2}X_L^i(\bar{z}),\;\;\;
X_R^i(z)
\rightarrow \lambda \left( {c_i\bar{\tau}_i+d_i\over c_i\tau_i+d_i}
\right)^{1/2}
X_R^i(z).
\eq{M37}
Obviously, as dictated by worldsheet supersymmetry,
the right-moving fermions $\psi_R(z)$ transform as the coordinates
\bq
\psi_R^i(z)
\rightarrow \lambda \left( {c_i\bar{\tau}_i+d_i\over c_i\tau_i+d_i}
\right)^{1/2}
\psi_R^i(z).
\eq{M37'}
In these equations, $\lambda$ is a $SL(2,\Z)$-dependent, but background
independent phase~\cite{LMN,IL,ILLT}.

Let us find how $SL(2,\Z)^3$ acts on the vertices of massless untwisted
scalars.
The transformation law on target space fermions has been presented in
ref.~\cite{IL,ILLT},
and follows by imposing target space supersymmetry on the scalars.
The scalar vertices read
\ber
\phi_{(\bar{3},\bar{27})} &=&
\int d^2 z \psi_R^i(z) V_L^{(\bar{3},\bar{27})}(\bar{z}),\;\;
\phi_{(3,27)} = \int d^2 z \psi_R^{\bar{\imath}}(z) V_L^{(3,27)}(\bar{z}),
\nonumber \\
\phi^{i\bar{\jmath}} &=&\int d^2 z \psi_R^i(z) \bar{\partial}
X_L^{\bar{\jmath}}(\bar{z}).
\eer{M38}
Here $V_L^{(a,b)}(\bar{z})$ denotes a $E_8\times E_8$ vertex, given by
eq.~(\ref{M16'}), transforming in the $(a,b)$ representation of $SU(3)\times
E_6$. The breakdown $E_8\rightarrow SU(3)\times E_6$ is achieved by
setting the lattice shift $\delta^A$ equal to $(1/3,1/3,1/3,0,...,0)$.

Equations~(\ref{M37}, \ref{M37'})
give the following duality transformation laws for
the massless, untwisted scalar vertices
\ber
\phi_{(\bar{3},\bar{27})}&\rightarrow &
\prod_i\lambda\left( {c_i\bar{\tau}_i+d_i\over c_i\tau_i+d_i}\right)^{1/2}
\phi_{(\bar{3},\bar{27})},\;\;\;
\phi_{(3,27)} \rightarrow
\prod_i\bar{\lambda}\left( {c_i\tau_i+d_i\over c_i\bar{\tau}_i+d_i}
\right)^{1/2}
\phi_{(3,27)}, \nonumber \\
\phi^{i\bar{\jmath}}&\rightarrow &
\prod_i\left( {c_i\bar{\tau}_i+d_i\over c_i\tau_i+d_i}
\right)\phi^{i\bar{\jmath}}.
\eer{M39}

The transformation laws of the  twisted vertices are more involved, since
states corresponding to different sectors transform into each other.
Nevertheless, one may fix the $\tau_i$-dependent part of the duality
transformations  by a few simple considerations.

At first, one notices that transformations~(\ref{M37}, \ref{M37'}) act as
chiral rotations on the string coordinates.
The corresponding chiral charges are
determined by the mode expansion of $X_{L,R}^i$ and $\psi_R^i$ in the
$m$-twisted sector, given in eq.~(\ref{M30})
\ber
Q_F^i &=&
\sum_{n\in \Z+m/3 +1/2 -t/2} \bar{\psi}^{\bar{\imath}}_{-n}\psi^i_n -m/3 -t/2
\nonumber \\
Q_{B,L}^i &=& \sum_{n \in \Z -m/3} {1\over n}
\tilde{\alpha}_{-n}^{\bar{\imath}}
\tilde{\alpha}_n^i + m/3, \;\;\;
Q_{B,R}^i= \sum_{n \in \Z +m/3} {1\over n} \alpha_{-n}^{\bar{\imath}}
\alpha_n^i - m/3.
\eer{M42}
The shift in the ground-state chiral charge is dictated by spectral
flow~\cite{FQS}.

Let us analyse the $m=1$ twisted sector. (The $m=2$ case can be worked out in
complete analogy with the previous one).

In this sector, the massless scalar states are~\cite{DHVW}
\bq
\ket{p^A+ \delta^A,a}_L \otimes \ket{0,a}_R,
\eq{M43}
and
\bq
\tilde{\alpha}_{-1/3}^{\bar{\imath}}
\ket{p^A +\delta^A, a}_L \otimes \ket{0,a}_R.
\eq{M44}
The ground state of the left-movers $\ket{p^A+\delta^A}$ contains a non-zero
vector in the shifted $E_8\times E_8$ lattice. This is a vector in the weight
lattice of $SU(3)\times E_6\times E_8$. In eq.~(\ref{M43}) this vector spans
the $(1,27,0)$ representation of $SU(3)\times E_6\times E_8$, while in
eq.~(\ref{M44}) it spans the $(3,1,0)$~\cite{DHVW}.

The duality transformations~(\ref{M37}, \ref{M37'}) are given by
$\exp i\sum_{i=1}^3( Q_R^i -Q_L^i)\theta_i$ with $2\theta_i=\log [(c_i\tau_i
+d_i)/(c_i\bar{\tau}_i +d_i)]$. By denoting with $\phi^a_{(1,27)}$ and
$\phi^{ia}_{(3,1)}$ the vertices associated to states~(\ref{M43})
and~(\ref{M44}), respectively, and using the formula for the chiral charges
given in eq~(\ref{M42}), one finds that $SL(2,\Z)^3$ acts on the twisted states
as follows
\ber
\phi^a_{(1,27)} &\rightarrow & \prod_i
\left( {c_i\bar{\tau}_i+d_i\over c_i\tau_i+d_i}\right)^{1/3}
S^1_{ab}\phi^b_{(1,27)},
\nonumber \\
\phi^{ia}_{(3,1)} &\rightarrow &
\left( {c_i\bar{\tau}_i+d_i\over c_i\tau_i+d_i}\right)^{5/6}
\left( {c_j\bar{\tau}_j+d_j\over c_j\tau_j+d_j}\right)^{1/3}
\left( {c_k\bar{\tau}_k+d_k\over c_k\tau_k+d_k}\right)^{1/3}
S^2_{ab}\phi^{ib}_{(3,1)},\;\;\; i\neq j \neq k .
\eer{M45}
As in eq.~(\ref{M32}), vertices corresponding to different fixed points
transform into each other. The matrices $S^1_{(ab)}$, $S^2_{(ab)}$,
describing this mixing, depend on $a_i,b_i,c_i,d_i$,
{\em but are independent of the
modular parameter} $\tau_i$. These matrices cannot be determined by our
simple method, they are given explicitly in ref.~\cite{LMN}.

Equations~(\ref{M39}, \ref{M45}) completely fix the moduli-dependent part of
the $SL(2,\Z)^3$ duality on all massless scalars of the $\Z_3$ orbifold
compactification. These formulas can be easily extended to any $\Z_N$ orbifold.

As we previously stated, the LEEA containing massless states must be exactly
invariant under $SU(3,3,\Z)$, and thus under $SL(2,\Z)^3$.
This property, together with the explicit form of duality on massless fields,
imposes powerful constraints on the form of the LEEA.
Some of the consequences of this constraint will be studied in
section 3.4.
\end{subsection}

\begin{subsection}{Duality Invariant Effective Field Theories of $N=2,1$
Compactifications}
\setcounter{equation}{0}
In this section, following~\cite{GP1},
duality in $N=1,2,$ compactifications will be analyzed.
We study the ``large'' duality interchanging massive
and massless modes. This is the symmetry that places most restrictions on the
LEEA. Unfortunately, unlike the $N=4$ case, for $N=2,1,$ supersymmetry
arguments alone are not strong enough to fix the form of the LEEA completely.
The presence of additional twisted sectors in the orbifold construction, for
instance, makes the knowledge of the $N=4$ LEEA insufficient to determine the
complete low-energy $N=2,1$ actions. The $N=4$ LEEA is, on the other hand, all
what we need to find the $N=2,1$ effective actions for the untwisted states.
The derivation of such LEEA's is the subject of this section.
In section 3.4, we shall examine the restrictions imposed on
orbifold compactifications by the ``small'' duality $SL(2,\Z)$,
transforming massless states into themselves. In this case the analysis can be
extended, and it gives constraints on the twisted sector as well.

To study the LEEA of the untwisted sector of orbifold compactifications, we
need some definitions, which extend those of the previous section.

A ($\Z_N$) orbifold projection is defined by picking
an element ${\cal O}$ of $O(6,\Z)\times O(6,\Z)\subset O(6,22,\Z)$.
For simplicity, we consider
here the case of an ${\cal O}$ acting symmetrically on left and right movers,
even though this restriction is by no means necessary~\cite{NSV}. Moreover, we
choose ${\cal O}$ such that ${\cal O}^N=1$ for some integer $N<\infty$. The
action of ${\cal O}$ on the LEEA fields is defined by specifying how ${\cal O}$
acts on the gauge indices as well as the $N=4$ ``extension'' indices. For the
gauge indices, one defines how ${\cal O}$ acts on the chiral fields $X^A(z)$ of
eq.~(\ref{A18'})
\ber
{\cal O}X^A(z)&=&\Lambda^A_BX^B(z)+2\pi\delta^A,\;\;\;
\Lambda^A_B={\rm diag}(\lambda^a_b,\lambda^a_b,0), \nonumber \\
\lambda &\in & O(6,\Z),\;\;\; A,B=1,..,28,\;\;\; a,b=1,..,6.
\eer{A33'}
The shift vector $\delta^A$ obeys
\bq
\delta^A=0,\;\; A=1,..,12,\;\;\; Np_A\delta^A\in \Z,\;\; \forall p_A\in
\Gamma^{(6,22)}.
\eq{A33a}
Equations~(\ref{A33'}, \ref{A33a}) define ${\cal O}$ on all vertices.
For example, since
$\lambda^a_b$ is an element of the duality group,
\bq
{\cal O} V^p=\oint{dz\over 2\pi
i}C(p)\exp[ip_B(\Lambda^B_AX^A +2\pi\delta^B)](z)=
\exp(2\pi ip\eta\delta)V^{\Lambda^{-1}p},
\eq{A34'}
where $V^p$ is defined in eq.~(\ref{A19'}), and the second equality
in~(\ref{A34'}) is modulo a conjugation of $C(p)$.

One can further define the action of ${\cal O}$ on the $N=4$ indices by
specifying
how it acts on the fundamental representation of $SU(4)$.
This is because $\lambda\in SO(6)$,
$SO(6)\simeq SU(4)$, and because the vectorial representation of $SO(6)$ is
isomorphic to the antisymmetric of $SU(4)$. With an appropriate choice of
basis,
and by labelling with $|j\rangle$ the fundamental representation of $SU(4)$, we
get
\ber
\lambda&=&{\rm diag}(\lambda_1,\lambda_2,\lambda_3,\lambda_4),\;\;\; m_j\in
\Z,\;
\;\; \sum_{j=1}^4 m_j=0,\;{\rm mod}\; N. \nonumber \\
{\cal O}|j\rangle &=& \lambda|j\rangle=\lambda_j|j\rangle=\exp(2\pi i m_j/N)
|j\rangle .
\eer{A35'}
By recalling that the $N=4$ matter multiplet of the LEEA reads
\bq
(A_{\mu}^S,\psi_i^S,Z_{ij}^S), \;\;\;Z^{Sij}\equiv
(Z^S_{ij})^*={1\over 2}\epsilon^{ijkl}Z^S_{kl},
\eq{A36'}
we define
\bq
 {\cal O}(A_{\mu}^S,\psi_i^S,Z_{ij}^S)V^S
 \equiv (A_{\mu}^S,\exp(2\pi im_i/N) \psi_i^S,
\exp[2\pi i(m_i+m_j)/N]Z_{ij}^S).
\eq{A37'}
This transformation is induced by the transformation of the
vertex $V^S$ given in eq.~(\ref{A28'}), namely, $({\cal O}V)^S$.

The fields of the untwisted sector of the $N<4$ theories are those left
invariant by the above transformation. If ${\cal O}\in SU(2)\subset SU(4)$,
the resulting orbifold is $N=2$ space-time supersymmetric. If ${\cal O}\in
SU(3)$,
${\cal O}\in\!\!\!\!\!/SU(2)$, the orbifold is $N=1$ supersymmetric.
We turn now to a study of the LEEA for the untwisted sector
\cite{FKP,FGKP,CLO,FP,CMO1}.

A method for studying the untwisted sector of 4-$D$
superstring effective actions
has been proposed in refs.~\cite{FKP,FGKP,FP}. That method has been used
in~\cite{FP} to study systematically the scalar manifolds of $\Z_N$ orbifolds.
The results of~\cite{FP} can be generalized easily to the duality-invariant
LEEA presented in section 2.8. Let us briefly review the methods of~
\cite{FKP,FGKP,FP}. The fundamental fact used there is that the orbifold
truncation acts on the compensating multiplet $(A_{\mu}^a,\psi_i^a,Z_{ij}^a)$
in a well defined way, given by eq.~(\ref{A37'}). For the
compensating scalars one finds, by using eqs.~(\ref{A20'}), (\ref{A33'})
and~(\ref{A37'}),
\bq
({\cal O}Z)^{ij}_{kl}=\exp[2\pi i (m_k+m_l+m_i+m_j)/N]Z^{ij}_{kl}.
\eq{A38'}
The compensators
surviving the projection generated by ${\cal O}$, i.e. those left invariant by
transformation~(\ref{A38'}), determine by themselves alone the geometry of the
manifold of the (untwisted) scalar fields.

As in ref.~\cite{FP} we find five different geometrical structures for
$N=1$
theories, and two for $N=2$. Let us consider first the $N=1$ cases. The five
different structures correspond to the following choice of the coefficients
$m_i$ in eq.~(\ref{A35'}) (all equalities below are modulo $N$, and $m_4=0$)
\begin{eqnarray*}
{\rm a} & & m_1=m_2=m_3=N/3,\;\;\; N\in 3\Z, \\
{\rm b} & & m_1=m_2\neq m_3,\;\;\; m_1+m_2+m_3=0,\;\;\;m_i\neq N/2,\;\;
N\in 2\Z, \\
{\rm c} & & m_1\neq m_2 \neq m_3,\;\;\; m_1+m_2+m_3=0,\;\;\; m_i\neq N/2,\;\;
N\in 2\Z, \\
{\rm d} & & m_1=m_2=N/4,\;\;\; m_3=N/2,\;\;\; N\in 4\Z, \\
{\rm e} & & m_1\neq m_2,\;\;\; m_3=N/2,\;\;\; m_1+m_2+m_3=0,\;\;N\in 2\Z .
\end{eqnarray*}
The compensating scalars surviving the truncation defined by ${\cal O}$ are
\ber
Z^{4i}_{4j},\;\; i,j=1,2,3 & & {\rm for\; case\; a} \nonumber \\
Z^{4i}_{4j},\;\; i,j=1,2,\;\; Z^{43}_{43} & & {\rm for\; case\; b}
\nonumber \\
Z^{4i}_{4i},\;\; i=1,2,3 & & {\rm for\; case\; c}
\nonumber \\
Z^{4i}_{4j},\;\; i,j=1,2,\;\; Z^{43}_{43},\;\; Z^{12}_{43} & &
{\rm for\; case\; d} \nonumber \\
Z^{4i}_{4i},\;\; i=1,2,\;\; Z^{43}_{43},\;\; Z^{12}_{43} & &
{\rm for\; case\; e}
\eer{A39'}
By substituting the fields of eq.~(\ref{A39'}) into the
constraint~(\ref{M6}), we find that the structure of the scalar manifold is
always $SU(1,1)/U(1)\times {\cal K}$, where ${\cal K}$ is
\ber
& & {SU(3+n,m,\R)\over U(1)\times SU(3,\R)\times SU(n,m,\R)}
\;\;\;\;\;{\rm for\; case\; a} \nonumber \\
& &{SU(2+n,m,\R)\over U(1)\times SU(2,\R)\times SU(n,m,\R)}\times
{SU(2+n',m',\R)\over U(1)\times SU(2,\R)\times SU(n',m',\R)}
\;\;\;\;\;{\rm for\; case\; b} \nonumber \\
& &\left[{SU(1+n,m,\R)\over U(1)\times SU(n,m,\R)}\right]^3
\;\;\;\;\;{\rm for\; case\; c} \nonumber \\
& &{SU(2+n,m,\R)\over U(1)\times SU(2,\R)\times SU(n,m,\R)}\times
{O(2+n',m',\R)\over O(2,\R)\times O(n',m',\R)}
\;\;\;\;\;{\rm for\; case\; d} \nonumber \\
& &\left[{SU(1+n,m,\R)\over U(1)\times SU(n,m,\R)}\right]^2\times
{O(2+n',m',\R)\over O(2,\R)\times O(n',m',\R)}
\;\;\;\;\;{\rm for\; case\; e}.
\eer{A40'}
These expressions are a little formal, since in our case
$n=m=n'=m'=\infty$, but they can be
made rigorous. Actually, some of these manifolds have been
used in string theory also in~\cite{AGGR}.

To determine what matter fields actually survive the truncation defined
by~(\ref{A37'}) we divide the group generators in two sets. The first one is
made of the generators $V^A$, $A=1,..,28$. The corresponding scalar fields are
the compensators, for which ${\cal O}$ has been defined in eq.~(\ref{A38'}),
and the scalars $Z_{ij}^I$, $I=7,..,28$. By an appropriate choice of basis,
since we consider only symmetric $\Z_N$ orbifolds, we may write
\bq
Z_{ij}^I=(\tilde{Z}_{ij}^{ab}, Z^{\alpha}_{ij}),\;\;\; \alpha=13,..,28,\;\;\;
i,j,a,b \in 4\; {\rm of} \; SU(4),
\eq{A41'}
and
\bq
({\cal O}Z)^I_{ij} =(\exp[2\pi i(m_i+m_j+m_a+m_b)/N]Z_{ij}^{ab},
\exp[2\pi i(m_i+m_j)/N]Z_{ij}^{\alpha}).
\eq{A42'}
The fields $\tilde{Z}^{ab}_{ij}$, surviving the ${\cal O}$ truncation have
therefore the same index structure of the auxiliary fields in eq.~(\ref{A39'}).

The second set of scalar fields is made of the fields
$Z^{p\varepsilon}_{ij}$. Here the index $p\varepsilon$ labels a generator
of the DISG, of nonzero lattice momentum $p$ and polarization tensor
$\varepsilon$ associated with the vertex operator
$V^{p\varepsilon}$. With an appropriate choice of $\varepsilon$, we can write
the action of ${\cal O}$ on these vertices as
\bq
({\cal O}V)^{p\varepsilon}=\exp[2\pi i k(\varepsilon,\delta)/N]
V^{\lambda^{-1}p\varepsilon}.
\eq{A43'}
Here $k(\varepsilon,\delta)$ is a function of $\varepsilon$, $\delta^A$
(and $p$) taking values
in $\{0,..,N-1\}$. The invariant fields, surviving the ${\cal O}$
truncation are
\bq   \hat{Z}_{ij}^{p\varepsilon}={1\over N}\sum_{l=0}^{N-1} \exp[2\pi i
l(m_i+m_j+k(\varepsilon,\delta))/N]Z_{ij}^{\lambda^lp\varepsilon}.
\eq{A44'}
Notice that, after truncation, it is in general no longer possible to choose
a basis of generators made of eigenvalues of the internal momenta $p$.

By using equations~(\ref{A42'}) and~(\ref{A43'}), we may find how, in $N=1$
and $N=2$
orbifolds, the DISG algebra {\bf g}
breaks down into representations of some sub-algebra {\bf h}.
The action of~({\ref{A42'}) and~(\ref{A43'}) on massless states for the
$E_8\times
E_8$ heterotic string was given in ref.~\cite{FP}. There it was found that the
massless families and gauge groups of cases a through e were
\begin{itemize}
\item Nine 27 of $E_6$, corresponding to the fields $Z^{(i,27)}_{4j},\;i,j=1,2,
3$ for case a.
\item Nine 27 of $E_6$, corresponding to the fields $Z^{(i,27)}_{4j},\;i,j=1,2$
and $Z_{43}^{(3,27)}$ for case b.
\item Three 27 of $E_6$, corresponding to the fields $Z^{(i,27)}_{4i},\;i=1,2,
3$ for case c.
\item Four 27 of $E_6$, corresponding to the fields $Z^{(i,27)}_{4j},\;i,j=1,2$
plus a 27: $Z_{43}^{(3,27)}$ and a $\overline{27}$:
$Z_{43}^{(\overline{3},\overline{27})}$
for case d.
\item Two 27 of $E_6$, corresponding to the fields $Z^{(i,27)}_{4i},\;i=1,2$
plus a 27: $Z_{43}^{(3,27)}$ and a $\overline{27}$:
$Z_{43}^{(\overline{3},\overline{27})}$
for case e.
\end{itemize}
The notation used above is the same as in our previous analysis:
the $E_8$ gauge indices, between parenthesis,
have been decomposed in $SU(3)$ and $E_6$ gauge indices.

One may recover the $N=1$ superpotential of all
$N=1$ models by recalling that in $N=1$
the gravitino mass $m_{3/2}$ is related to
the superpotential $W$ by~\cite{C&al,CFGVP}
\bq
m_{3/2}=\exp(K/2)W,
\eq{A44a}
where $K$ is the K\"ahler potential of the $N=1$ scalar manifold.
In order to write the superpotential $W$ containing all the fields that are
relevant to low energy physics, i.e. those that may become massless on some
background, we introduce the following notation
\ber
y^p_i&=&\exp(-K/6)\hat{Z}^p_{i4},\;\;\;
y^A_i=\exp(-K/6){1\over N}\sum_{n=0}^{N-1}{\cal O}^nZ_{i4}^A, \nonumber \\
 & & i=1,2,3, \;\; p\eta p=2.
\eer{A44b}
Thus the fields $y$ are, up to a rescaling, those invariant under the
transformations
in~(\ref{A42'}), (\ref{A43'}). Notice that here, as in eq.~(\ref{A44'}),
$p$ denotes an orbit of the action of $\Z_N$ on $\Gamma^{(6,22)}$.

In terms of the fields~(\ref{A44b}) the gravitino mass surviving the $N=1$
truncation reads (cfr.~\cite{FP,GP,GP1})
\bq
m_{3/2}=\exp(K/2)\epsilon^{ijk}[\sum_{p,q}\sum_{l=0}^{N-1}
y_i^py_j^qy_k^{p+\lambda^lq}\varepsilon(p,\lambda^lq) -
\sum_{p,A}ip_Ay_i^Ay_j^py_k^{-p}],
\eq{A44c}
where $\epsilon^{ijk}$  is the totally antisymmetric tensor,
$i,j,k=1,2,3$.
The $y^A$ are constrained fields, obeying the appropriate $N=1$
reduction of eq.~(\ref{M6}) \cite{FP}. In order to find the
superpotential in terms of physical fields one must first solve that
constraint.
Confronting eq.~(\ref{A44c}) with eq.~(\ref{A44a}) one easily gets the
superpotential of a $\Z_N$ $N=1$ truncation. On the massless modes it coincides
with that of ref.~\cite{FP}.

The study of the $N=1$ models is completed by specifying the form of the
gauge coupling function, i.e. the function multiplying the square spin-one
field strength term in the Lagrangian~\cite{C&al,CFGVP}.
This function $f$ turns out
to be model independent and equal to $f=\eta^{\alpha\beta}S$, where
$\eta^{\alpha\beta}$ is
the invariant metric restricted to the gauge fields surviving the $N=1$
truncation, and $S=\sqrt{G}\exp \phi+ia$.
Here $\Re S$ is the universal duality-invariant dilaton
field~\cite{Witt,FGKP,FP}, and $a$ is the axion.

We turn now to the study of truncations to $N=2$ orbifolds.
The two geometrically distinct cases of $N=2$ orbifolds can be studied by the
same
methods used for the $N=1$ orbifolds. In $N=2$ supergravity the manifold of
scalar
fields reads $SU(1,1)/U(1)\times{\cal K}\times {\cal Q}$. The manifold ${\cal
K}$ is K\"{a}hlerian and ${\cal Q}$ is quaternionic~\cite{WB,dWVP,dWLVP}.
The first $N=2$
case corresponds to the choice $m_1=m_2=N/2$, $m_3=m_4=0$, while the second one
corresponds to $m_1\neq m_2$, $m_1+m_2=N$, $m_3=m_4=0$.

By repeating the same analysis done for the $N=1$ truncations one finds the
following manifolds
\ber
{\cal K}&=&{O(2+n,m,\R)\over O(2,\R)\times O(n,m,\R)},\;\;\;{\cal
Q}={O(4+n',m',\R)\over
O(4,\R)\times O(n',m',\R)},\;\;\; m_1=m_2, \nonumber \\
{\cal K}&=&{O(2+n,m,\R)\over O(2,\R)\times O(n,m,\R)},\;\;\;{\cal
Q}={SU(2+n',m',\R)\over
U(1)\times SU(2,\R)\times O(n',m',\R)},\;\;\; m_1\neq m_2,\nonumber \\ & &
n=n'=m=m'=\infty .
\eer{A45'}

At this point
we stop our by no means systematic review of the untwisted sector of
$N=1$ and $N=2$ theories resulting from orbifold compactifications of $D=4$,
$N=4$ strings. We recall that we confined our attention to $\Z_N$
symmetric orbifolds only. Many other
types of orbifolds can be constructed. For example, one may study Abelian
orbifolds generated by noncyclic groups like $\Z_2\times \Z_2$~\cite{ABK,KLT},
or non-Abelian orbifolds~\cite{DHVW}.
By repeating the procedure of this section one
can work out what the $N=1$ and $N=2$ truncations look like in these cases.

By construction, each  truncated model is  invariant under the automorphism
group of the corresponding $N=1,2$ DISG sub-algebra.
The automorphism group contains
the subgroup of $O(6,22,\Z)$ which stabilizes the orbifold twist group,
that is the
generalized duality group acting on the untwisted sector moduli space of the
orbifold  compactifications. These automorphism transformations  are the
residual discrete symmetries of the broken gauge group in
the untwisted-sector effective action.
}
\end{subsection}

\begin{subsection}{Special Geometry and Calabi-Yau Compactifications}
\setcounter{equation}{0}
Orbifolds correspond to special points in the space of all possible heterotic
compactifications. A more general class of string vacua is given by the
Calabi-Yau (CY) compactifications, giving rise to 4-$D$ theories with $N=1$
supersymmetry.

CY spaces arise when one reduces the ten-dimensional heterotic string to its
point-field limit, namely, when one studies the 10-$D$ effective action which
describes physics at a scale $R\gg \alpha'^{1/2}$. This action, which is
uniquely fixed by 10-$D$ $N=1$ supersymmetry, and by the knowledge of the gauge
group, is then compactified to yield a 4-$D$ theory. The resulting 4-$D$ theory
is a good approximation of the true 4-$D$ LEEA, obtained by computing
string amplitudes in 4-$D$,
provided the compactification radii are much larger than $\alpha'^{1/2}$.

{}From the point of view of the underlying worldsheet conformal
theory, these compactifications are (2,2) superconformal theories~\cite{Ge}.
Orbifolds which are (2,2) superconformal correspond to particular
(singular) points in the moduli space of a CY manifold~\cite{CHSW}.
Conversely,
by giving nonzero expectation values to some twisted moduli, (2,2) orbifold
singularities are blown up to yield smooth CY spaces.
In general, the structure of the LEEA of CY compactifications is not exactly
known, not in even the field-theoretical (large-radius) limit.
Nevertheless, the
Lagrangian corresponding to the flat directions (moduli) of the CY
compactification obeys severe constraints. These constraints are not enough to
fix the (``small'') duality group, but they are sufficient to determine how
the duality group acts on the moduli.

We shall follow ref.~\cite{CdO} in our review of properties of CY spaces.
Let us recall that a Calabi-Yau manifold is a compact K\"ahler manifold of
complex dimension three and holonomy $SU(3)$. A generic 3-$D$ K\"ahler manifold
has holonomy $U(3)$, instead. The vanishing of the $U(1)$ holonomy is
equivalent to saying that a CY space admits a Ricci-flat metric~\cite{CY}.
The moduli of a CY space ${\cal M}$
are given by those variations of the metric $g_{mn}$ which preserve
Ricci-flatness
\bq
R_{mn}(g +\delta g)= R_{mn}(g)=0.
\eq{M46}
By writing these variations in complex coordinates $\mu$, $\bar{\mu}$ etc.,
thanks to special properties of K\"ahler manifolds, one finds that
$\delta g_{\mu\bar{\nu}}$ and $\delta g_{\mu\nu}$, $\delta
g_{\bar{\mu}\bar{\nu}}$ {\em separately} obey  eq.~(\ref{M46}). Thus, the
deformations of a given CY space are of two types: mixed (one holomorphic, one
anti-holomorphic index), and pure (all holomorphic or all anti-holomorphic
indices). To mixed variations one associates a real (1,1)-form
\bq
i\delta g_{\mu\bar{\nu}} dx^\mu \wedge dx^{\bar{\nu}},
\eq{M47}
whereas to pure variations one associates a complex (2,1)-form by using the
holomorphic (3,0) form $\Omega$ present in all CY manifolds~\cite{CY,CHSW}:
\bq
\Omega_{\kappa\lambda}^{\bar{\nu}} \delta g_{\bar{\mu}\bar{\nu}}
dx^\kappa \wedge dx^\lambda \wedge dx^{\bar{\mu}}.
\eq{M49}
These variations are holomorphic, thanks to eq.~(\ref{M46}). Thus, the
deformations of a CY manifold are elements of the Dolbeault cohomology groups
$H^{(1,1)}$ and $H^{(2,1)}$ (see ref.~\cite{GSW} for a simple introduction to
the subject).

The mixed variations are deformations of the K\"ahler class, in
other words, they leave the complex structure invariant and change the metric.
In particular, they change the {\em compactification radii}.

The pure variations, instead, modify the complex structure of the CY manifold
(that is the split into holomorphic and anti-holomorphic coordinates).

Among the fields of the 10-$D$ effective action of the heterotic string, there
is a two form $B=b_{mn}dx^m\wedge dx^n$. This form and the K\"ahler
form $J=g_{mn}dx^m\wedge dx^n$ can be written as a single complex form
$B+iJ$. A variation of $B$ gives rise to a massless
mode in 4-$D$ iff $\delta B\in H^{(1,1)}$~\cite{CHSW}.
Thus, the most general metric for $\delta g$ and $\delta B$ writes as follows
\bq
ds^2 ={1\over 2V}\int_{\cal M}g^{1/2}d^6x
g^{\kappa\bar{\mu}}g^{\lambda\bar{\nu}}[\delta g_{\kappa\lambda} \delta
g_{\bar{\mu}\bar{\nu}} +(\delta g_{\kappa\bar{\nu}}\delta g_{\lambda\bar{\mu}}
+ \delta b_{\kappa\bar{\nu}} \delta b_{\bar{\mu}\lambda})].
\eq{M50}
This equation has another important consequence:
it says that the metrics of $(1,1)$
and $(2,1)$ moduli are factorized. So, at least locally, the moduli space of a
CY space is ${\cal M}_{(1,1)}\times {\cal M}_{(2,1)}$. This property
was proven in~\cite{Se} using effective-theory arguments. In~\cite{DKL} it was
shown that the factorization of moduli space is exact to all orders in
$\alpha'$, being a property of any
$(2,2)$ superconformal worldsheet field theory.

The factorization of moduli
space is at the root of the particular kind of target space duality
called ``mirror symmetry''~\cite{GrP,CdO1}.
In the present context, this symmetry transforms a CY manifold into another
one, in which the role of $(1,1)$ and $(2,1)$ forms are interchanged.
It also has another important consequence. Since ${\cal M}_{(2,1)}$ does not
depend on ${\cal M}_{(1,1)}$, it can be evaluated in the regime where all
compactification radii $\alpha'^{-1/2}R\rightarrow \infty$. In this limit, the
field-theoretical result is {\em exact} in $\alpha'$, and not merely a lowest
order result. This statement is very strong: it says that the metric of the
$(2,1)$ moduli does not receive sigma-model corrections.

The metric of $(1,1)$ moduli, however, does receive sigma-model corrections.
In some exceptional case though, by using the mirror symmetry,
and by finding a pair of mirror-conjugated manifolds, one
obtains exact results for ${\cal M}_{(1,1)}$ as well~\cite{CdO1,GrP,ALR}.

We shall not discuss further these developments though, but instead restrict
our attention to the (2,1)-moduli space and figure out its special properties.

The relation between (2,1) forms and deformations of the complex structure is
given by~\cite{CdO}
\bq
\chi_{a\,\kappa\lambda\bar{\mu}}=-{1\over 2}\Omega_{\kappa\lambda}^{\bar{\nu}}
{\partial g_{\bar{\mu}\bar{\nu}}\over \partial z^a},\;\;\;
\chi_a={1\over 2} \chi_{a\,\kappa\lambda\bar{\mu}}dx^\kappa \wedge dx^\lambda
\wedge dx^{\bar{\mu}}.
\eq{M51}
The $z^a$, $a=1,..., b_{21}$ are the parameters for the complex structure,
whose number is equal to the dimension $b_{21}$ of $H^{(2,1)}$.
Equation~(\ref{M51})
can be inverted to give $\delta g_{\mu\bar{\nu}}$ in terms of
$\chi_a$. By simple algebra one finds the definition of the metric
of the (2,1)-moduli, $G_{a\bar{b}}$,
\bq
G_{a\bar{b}}\delta z^a \delta z^{\bar{b}}\equiv
{1\over 2V}\int_{\cal M}g^{1/2}d^6x
g^{\kappa\bar{\mu}}g^{\lambda\bar{\nu}}\delta g_{\kappa\lambda} \delta
g_{\bar{\mu}\bar{\nu}} =
-{2i\over V \| \Omega \|^2} \delta z^a \delta z^{\bar{b}} \int_{\cal M}
\chi_a \wedge \bar{\chi}_{\bar{b}}.
\eq{M52}
The (2,1) forms $\chi_a$ can be expressed in terms of the holomorphic three
form as~\cite{CdO}
\bq
{\partial \Omega \over \partial z^a}=\phi_a \Omega + \chi_a,
\eq{M53}
where the $\phi_a$ may depend on $z^a$, but not on the coordinates of the CY
space.

Equation~(\ref{M53}) implies that the metric of (2,1) forms is K\"ahler, i.e.
\bq
G_{a\bar{b}}= -{\partial \over \partial z^a} {\partial \over \partial
z^{\bar{b}}} \log \left( i \int_{\cal M} \Omega \wedge \bar{\Omega} \right).
\eq{M54}

Indeed, this metric enjoys other properties. To study them,
one picks up a (real)
canonical homology basis $(A^i, B_j)$, $i,j=0,...,b_{21}$ of the
{\em integral} homology group $H_3({\cal M},\Z)$. The dual cohomology
basis $(\alpha_i,\beta^j)$ obeys
\bq
\int_{A^i}\alpha_j=\int_{\cal M} \alpha_j \wedge \beta^i =\delta_j^i,
\;\;\;
\int_{B_i}\beta^j=\int_{\cal M} \beta^j \wedge \alpha_i =-\delta^j_i.
\eq{M55}
Notice that this equation defines a symplectic metric on $H_3({\cal M},\Z)$,
given by the wedge product of two three-forms.
In terms of this cohomology basis the holomorphic three-form $\Omega$ reads
\bq
\Omega = x^i\alpha_i - {\cal G}_j\beta^j.
\eq{M56}
The dimension of the (2,1)-moduli space is $b_{21}$, therefore only $b_{21}$
of the $2b_{21} +2$ coefficients in eq.~(\ref{M56}) are independent. One can
choose $x^i=\lambda z^i$, $i=1,..,b_{21}$, $x^0=\lambda$, and express ${\cal
G}_i$ in function of the $x^i$. The parameter $\lambda$ is an irrelevant scale
factor thus, obviously, ${\cal G}_i(\lambda x)=\lambda {\cal G}_i(x)$.
Equation~(\ref{M53}) implies
\bq
\int_{\cal M} \Omega \wedge {\partial \Omega \over \partial x^i} =0.
\eq{M57}
This equation says that ${\cal G}_i$ is the derivative of a homogeneous
function of degree two: ${\cal G}_i=\partial {\cal G} / \partial x^i$,
and ${\cal G}(\lambda x)=\lambda^2 {\cal G}(x)$.
The K\"ahler potential of the (2,1)-moduli space is therefore expressible in
terms of a holomorphic function:
\bq
\int_{\cal M} \Omega \wedge \bar{\Omega}=\bar{x}^i{\partial {\cal G}\over
\partial x^i} -x^i {\partial {\cal G}\over \partial \bar{x}^i}.
\eq{M58}
This particular form of the K\"ahler potential is the one dictated by $N=2$
supersymmetry in 4-$D$, for the scalars of the vector
multiplet~\cite{dWVP,dWLVP}.

The link between $N=2$ supersymmetry and the geometry of (2,1) moduli is not
an accident. As shown in ref.~\cite{Se}, the moduli of a (2,2) heterotic
compactification, with $N=1$ target space supersymmetry, are the same as for a
type-II superstring compactification, with $N=2$ supersymmetry. The map
between heterotic compactifications and type-II superstring
compactifications puts stronger constraints on the (2,1)-moduli Lagrangian
than $N=1$ supersymmetry alone.
These constraints, which follow from $N=2$ supersymmetry, have been studied
in refs.~\cite{Cec,Stro,FBC}.
Moreover, one can show that even the (1,1)-moduli Lagrangian can be obtained
from a type II compactification. Therefore, even for (1,1) moduli, the form of
the K\"ahler potential is of special type:
\bq
K=-\log i\left( \bar{y}^m {\partial {\cal F}\over \partial y^m}-
y^m {\partial {\cal F}\over \partial \bar{y}^m}\right),\;\;\; m=0,...,
b_{11}.
\eq{M58'}
Here the prepotential ${\cal F}$ is corrected by ${\cal O}(\alpha')$ terms,
and the link between duality and CY geometry is less transparent than for
(2,1) moduli. For this reason, we prefer to return to this latter case, and
examine it more closely.

We spent some time on the details of the construction of metrics for the
(2,1)-moduli space. These details allow us to find a general {\em exact}
result about target space duality in CY compactifications. First of all, let
us notice that an infinitesimal change of the coordinates $z^a$ does not
change the cohomology basis in eq.~(\ref{M55}),
since the homology group is a topological invariant of the CY space, and its
coefficients are integers. On the other hand, under a large diffeomorphism
$z^a\rightarrow z'^a(z)$, the homology basis could undergo a symplectic
transformation
\bq
\left(\begin{array}{c} \alpha \\ \beta \end{array}\right) \rightarrow
\left(\begin{array}{cc} A & B \\ C & D \end{array} \right)
\left(\begin{array}{c} \alpha \\ \beta \end{array}\right), \;\;\;
\left(\begin{array}{cc} A & B \\ C & D \end{array} \right) \in
Sp(b_{21} +1,\Z).
\eq{M59}
The coefficients $A,B,C,D$ are integer valued, since the homology is integral,
and they define a symplectic matrix, due to eq.~(\ref{M55}).
Equation~(\ref{M59}) leads to a transformation rule for the
periods of the three-form $\Omega$ given, in our notations, by
the vector $(\partial_i{\cal G}, x^i)$
\bq
\left(\begin{array}{c} {\partial {\cal G}\over \partial x^i} \\ x^i
\end{array}\right)
\rightarrow
\left(\begin{array}{cc} A & B  \\  C & D \end{array}\right)
\left(\begin{array}{c} {\partial {\cal G}\over \partial x^i} \\ x^i
\end{array}\right).
\eq{M60}
This is a duality: two different backgrounds, related by
transformation~(\ref{M60}), yield the same physics, since they correspond to a
reparametrization of the (2,1)-moduli space. The form of the transformation on
the $z^a$ coordinates is highly non-trivial, nevertheless, eq.~(\ref{M60})
states that it acts as a linear symplectic transformation on
$\partial_i {\cal G}$, $x^i$.

The previous result is totally general, valid for {\em any} CY
compactification, and it is exact in $\alpha'$, since it concerns (2,1)
moduli. Given a CY manifold ${\cal M}$, a similar result would hold for
the (1,1) moduli, if a mirror manifold $\tilde{\cal M}$ could be found.

Clearly, eq.~(\ref{M60}) is not sufficient to fix the duality group ${\Gamma}$
of a CY space: it only says that ${\Gamma}$ is a subgroup of
$Sp(b_{21}+1,\Z)$. Equation~(\ref{M60}) also defines the action of
${\Gamma}$ on the periods $x^i$, $\partial_i {\cal G}$.

Our analysis of the CY moduli space has linked the action of the duality group
to algebraic geometric properties of the CY space. The explicit form of the
duality group ${\Gamma}$ can be found in some cases, again by use of algebraic
geometry. In ref.~\cite{CdO1}, ${\Gamma}$ was found in the simple case of
CY manifolds described by a one complex-parameter family of
quintic hypersurfaces in ${\bf CP}_4$.
Even in that simple case, it turns out that ${\Gamma}$ is not $SL(2,\Z)$,
as previously conjectured in the literature.

A further development, that we shall only mention here,
is given by the relation between ${\Gamma}$ and the monodromy of Picard-Fuchs
equations for the periods of CY spaces~\cite{CdO1,LSW,CF,CDFLL,DF}.
These are the differential
equations giving the periods $x^i$, $\partial_i{\cal G}$ as functions of the
complex-structure moduli $z^a$. These equations are singular at some
point(s). By transporting $z^a$ about that singular point, the solutions of
the differential equations transform into each other. The group describing
this transformation is called the monodromy group $\Gamma_M$. Almost all known
examples of CY manifolds are representable as algebraic hypersurfaces in
(some) ${\bf CP}_n$ by the equation
\bq
W_i(y,z)=0,\;\;\; y\in {\bf CP}_n,\;\;\; z\in {\cal M}_{(2,1)}.
\eq{M60a}
Here $W_i$ are quasi-homogeneous polynomials in $y,z$. Associated with
eq.~(\ref{M60a}) there exists an invariance group $\Gamma_W$, whose elements
are quasi-homogenous transformations of $y,z$ leaving the
equation invariant. In this case, the authors of ref.~\cite{LSW} proposed a
general formula for the duality group, namely
$\Gamma/\Gamma_W=\Gamma_M$. In other words, they conjectured that the
duality group is a semi-direct product of the invariance group of
equation~(\ref{M60a}), and the monodromy group of the periods
$\Gamma=\Gamma_W\otimes_S\Gamma_M$. This conjecture holds true
in the case of a single modulus~\cite{CdO1,LSW,DF}.

The proposed formula for the duality group
was shown to hold true for $K_3$ compactifications, relevant
for $N=2$ heterotic backgrounds, in ref.~\cite{gs}. Moreover, as was
studied in~\cite{gs}, the
invariance group $\Gamma_W$ is the symmetry group of the corresponding
$N=2$ Landau-Ginzburg (LG) model.
This LG model and its corresponding CY manifold are  ``different phases'' of
an underlying $N=2$ superconformal field
theory~\cite{Martinec,VW,GVW,LVW,Wcy}.

Finally,
we stress again that informations about ${\cal M}_{(1,1)}$ cannot be obtained
simply by algebraic geometry. Only in the case when a
mirror to the CY manifold ${\cal M}$ exists, one can extract exact results
about (1,1) moduli.

\end{subsection}
\begin{subsection}{Duality Invariant $N=1$ Actions: Superpotential, Threshold
Effects and Integration of Massive Modes}
\setcounter{equation}{0}
Any LEEA describing a 4-$D$ compactification must be invariant under the
appropriate duality group. In particular, even the ``small'' duality group,
transforming massless modes into themselves, dictates useful restrictions of
the form of the LEEA.

In section 3.1, we have examined the form of duality transformations on
twisted and untwisted fields in some orbifold compactifications. For the
simple case of the $\Z_3$ orbifold we found that all massless vertices
transform in the following way, up to a moduli-independent phase
\bq
\phi_s\rightarrow \left( {c_i\bar{\tau}_i + d_i \over c_i\tau + d_i}
\right)^{-n^s_i/2}\phi_s.
\eq{M61}
Here, $s$ labels the massless scalar
fields and the charges $n^s_i$ are given in
eqs.~(\ref{M39}, \ref{M45}). The fields $\phi_s$
are those linear superposition of the vertices of section 3.1 that
diagonalise the unitary matrices $S^1_{ab}$, $S^2_{ab}$ of eq.~(\ref{M45}).
The fields $A_s$, defined by
\bq
\phi_s=i\prod_{i=1}^3(\tau_i -\bar{\tau}_i)^{n^s_i/2} A_s,
\eq{M62}
transform under $SL(2,\Z)^3$ as modular forms of weight $(n^s_1,n^s_2,n^s_3)$
\bq
A_s\rightarrow \prod_{i=1}^3 (c_i\tau_i +d_i)^{n^s_i}A_s.
\eq{M63}
An $N=1$ Lagrangian is determined, as we recalled in section
3.2, by the K\"ahler potential $K$, giving rise to the kinetic term of the
scalar-multiplet fields, the superpotential $W$, and the field-dependent gauge
coupling constant $f^{AB}$. The superpotential is a holomorphic function of
the scalar fields, and the indices $A$,$B$ range on the generators of the
gauge group. The K\"ahler potential for a LEEA of a (2,2) compactification of
the heterotic string can be expanded, around a given value of the
moduli, in powers of the fields $A_s$. This potential can be evaluated at
tree-level by computing string scattering amplitudes on the sphere~\cite{DKL}.
The expansion to
quadratic terms in the charged fields $A_s$ reads, for
the case of our $\Z_3$ orbifold~\cite{DKL,FLT,IL1,ILLT}
\bq
K=-\sum_{i=1}^3 \log (2\Im\tau_i) + \prod (2\Im \tau_i)^{n_i^s} |A_s|^2.
\eq{M64}
This result coincides with the K\"ahler potential one finds by dimensional
reduction of 10-$D$ supergravity coupled to $E_8\times E_8$ gauge
matter~\cite{FKP}.

Notice that the K\"ahler potential in eq.~(\ref{M64}) transforms as follows
under $SL(2,\Z)^3$ duality
\bq
K \rightarrow K + \Lambda +\bar{\Lambda},\;\;\; \Lambda=\sum_{i=1}^3
(c_i\tau_i + d_i).
\eq{M65}
The kinetic term for the gauge fields is diagonal within each simple factor
$G_a$ of the gauge group $G=\prod_a G_a$.
At tree-level one finds~\cite{Gin}
\bq
f_a^{AB}= k_a \delta^{AB} S.
\eq{M66}
The integer $k_a$ is the level of the affine Lie algebra associated with
$G_a$, and $S$ is the axion+dilaton field.
In the case of the $\Z_3$ orbifold
$G=SU_3\times E_6\times E_8$, and all levels are
equal to one. An $N=1$ Lagrangian depends on the superpotential $W$ and
K\"ahler potential $K$ only through the combination
\bq
{\cal G}= K+ \log |W|^2.
\eq{M67}
Thus, the LEEA of our orbifold compactification is duality invariant if
\bq
W(A_s,\tau_i) \rightarrow
\lambda W(A_s,\tau_i)\prod_{i=1}^3(c_i\tau_i+d_i)^{-1},
\eq{M68}
with $\lambda$ an arbitrary {\em moduli independent} phase.
This property has been verified at tree-level in $\Z_N$ orbifold
compactifications in refs.~\cite{LMN,CMLN,LLW}.
One may also use the transformation
properties of the fields $A_s$, and the form of the K\"ahler potential, to
predict the form of the superpotential~\cite{FLT}. More importantly, since
duality is an exact {\em all-loop} symmetry, radiative corrections should be
duality invariant~\cite{FLST,FLT}.

If one further assumes that duality, besides being
an all-order symmetry, is also preserved non-perturbatively, one may deduce
useful constraints on supersymmetry breaking
mechanisms~\cite{FILQ,CFILQ,FMTV,NO,BG,GTay}.

Let us find how duality constrains the form of the superpotential. We shall
at first give general arguments, based on duality invariance alone.
To simplify the discussion as much as possible, let us keep only one
modulus $\tau$ in the LEEA, and consider the corresponding $SL(2,\Z)$ duality.
In this case, there are two possibilities of implementing
$SL(2,\Z)$~\cite{FLST}.
The first one is that the K\"ahler potential $K$ and the
superpotential $W$ be separately $SL(2,\Z)$-invariant. A possible solution to
this constraint is to modify $K$ (and $W$) so as to recover the tree-level
form only in the large $\Im \tau $ limit. If the tree-level K\"ahler potential
is chosen, in analogy with~(\ref{M64}) to be
\bq
K= -n\log (2\Im\tau),
\eq{M69}
then a possible modular invariant completion is~\cite{GRV,FLST}
\bq
K=-n \log \left[ \sum_{m,n\in \Z} \exp \left( -{\pi \over \Im \tau } |m
+n\tau|\right)\right].
\eq{M70}

The second possibility is that both
$K$ and $W$ transform under duality, and only the combination
$K+\log |W|^2$ is invariant. When $K$ is given by  eq.~(\ref{M69}),
duality invariance demands that $W(\tau)\rightarrow (c\tau + d)^{-n}W(\tau)$.
This means
that the superpotential must transform as a modular form of weight $-n$.
If we want $W$ to be analytic at any finite value of the
modulus $\tau$, inside the fundamental domain of the upper half-plane
$ |\tau|\geq 1$, $-1/2 \leq \Re \tau \leq 1/2$, we find~\cite{FLST}
\bq
W(\tau) = {\rm const}\,\eta(\tau)^{-2n}= e^{n\pi \Im\tau /6}[ 1 +2ne^{-2\pi
n\Im\tau} + 2n(n+1)e^{-4\pi n\Im\tau}+...].
\eq{M71}
This superpotential has the form expected from worldsheet instantons, and
integration of massive modes. In ref.~\cite{GP1} it was suggested that modular
forms as in eq.~(\ref{M71}) may originate from the integration of massive terms
in $N=1$ LEEAs given by eq.~(\ref{A44c}). Indeed, those LEEAs contain an
infinite number of (untwisted) massive fields, say $M_U$. Some of these fields
may appear linearly in the superpotential: $W=M_U L_T L_T$. By $L_T$ we
denote the massless (light) twisted scalars. Upon integration of the massive
fields, the original trilinear superpotential of the LEEA is modified. This
modification must still be invariant under the ``small'' duality group
$SL(2,\Z)$. The resulting superpotential,
which depends only on the massless fields,
must therefore be a modular form of appropriate weight, as in eq.~(\ref{M71}).

Our discussion has been so far quite general. No input on the form of the
LEEA, besides duality and supersymmetry, has been used. In particular, no
physical mechanism has been proposed for the appearance of automorphic
modular forms. Now, we would like to study in some more detail a mechanism
whereby non-trivial modular functions modify tree-level physics.

The most dramatic modification to tree-level results affects the gauge
couplings. The tree-level result~(\ref{M66}), indeed, states that these
couplings depend on the dilaton field $S$ only. In particular one has the
tree-level identification $1/g_a^2=k_a \Re S$ for the gauge coupling constant
$g_a$. At one loop this equation is modified (as it happens in field theory
too). In fact,
one must specify the scale $\mu$ at which the coupling constant is defined,
and take into account the renormalization-group running of $g_a$ as well as
threshold effects due to the integration of massive modes~\cite{Ka}
\bq
{1\over g^2_a(\mu)}=k_a \Re S + {b_a\over 16\pi^2} \log {M_{string}\over
\mu^2} + \Delta_a.
\eq{M75}
Here $b_a$ is the one-loop coefficient of the beta function, equal to
$-3C(G_a)+ h_j\sum_jT(R_j)$. $C(G_a)$ is the quadratic Casimir of the adjoint
representation of the gauge group, $T(R_j)$ is the Casimir of the $R_j$
representation, $h_j=1$ for spin-1/2 fermions, and $h_j=1/4$ for scalars. The
sum runs over all massless fields. The coefficient $\Delta_a$ represents the
threshold effects at the string unification scale $M_{string}$. Its form has
been found explicitly in ref.~\cite{DKL1} for the case of (2,2) orbifolds:
\bq
\Delta_a(\tau)={1\over 16\pi^2} b'_a \log 2\Im \tau |\eta(\tau)|^4,\;\;\;
b'_a=-b_a -2\sum_j h_jT(R_j)(1+n_j).
\eq{M76}
The $n_j$ denote the modular weights of the matter fields.
Equation~(\ref{M76}) is invariant under $SL(2,\Z)$ duality. This happens
because $\Im \tau$ transforms under duality so as to cancel the transformation
of $|\eta(\tau)|^4$ functions.

It is interesting to notice that the
eta-function contribution in $\Delta_a$ is the real part of an holomorphic
function, and comes from the loop integration over massive modes~\cite{DKL1}.
The $\Im \tau$ contribution instead, arising from massless loops,
cannot be written as the real part of an
analytic function. This seems to contradict supersymmetry, which requires the
gauge-coupling function $f^{AB}$ to be holomorphic (chiral). The conundrum is
solved by noticing that a non-holomorphic coupling is compatible with
supersymmetry, if one allows non-local terms in the effective
Lagrangian~\cite{DFKZ}.
Namely, one finds that it is possible to supersymmetrize
the gauge kinetic term
\bq
\int d^4x h(z,\bar{z})F^a_{\mu\nu}F^{a\,\mu\nu},\;\;\; z={\rm scalar\; field},
\eq{M77}
even when
\bq
{\partial \over \partial z} {\partial \over \partial \bar{z}}h(z,\bar{z})\neq
0.
\eq{M78}
The superfield expression for the gauge kinetic term is~\cite{DFKZ}
\bq
{1\over 4}\int d^2\theta d^4x {\cal P} h(z,\bar{z})W^{a\a}W^a_{\a} +
{\rm h.c.} , \qquad
{\cal P}=-{1\over 16}\Box^{-1}\bar{\cal D}\bar{\cal D}{\cal D}{\cal D}.
\eq{M78'}
Here $z$ is a chiral superfield, $W^a_{\a}$ is the supersymmetric gauge
field-strength,
and ${\cal P}$ is a chiral projector. This expression reduces to the standard
one when $h(z,\bar{z})= h(z)$.
A non-local term may arise from the integration of massless modes, but not
of massive ones. We have just seen that the only nonlocal term
in~(\ref{M78}) is $\log \Im \tau$, which indeed arises from massless
loops~\cite{DKL1}.

The threshold correction $\Delta_a(\tau,\bar{\tau})$ given by eq.~(\ref{M76})
gives rise to a significant modification of the gaugino-condensate mechanism
of supersymmetry breaking~\cite{DRSW}.
The main feature of this mechanism is that
in the presence of a hidden sector, and if some gauge
interactions become strong, a non-vanishing condensate of strongly interacting
fields may form. This means that if
$E_8'$ is unbroken, for instance, gauginos may condense \cite{FGN}. This
condensate modifies the effective superpotential below the condensation scale
$\Lambda_c$, and may lead to supersymmetry breaking at zero
cosmological constant.
The gaugino-condensate dependence on the coupling constant is dictated by
renormalization-group invariance
\bq
\vev = \Lambda_c^3\exp\left(24\pi^2b_H{1\over g_H^2}\right).
\eq{M79}
Here the only relevant gauge coupling is the one of the hidden strongly
interacting gauge group (e.g. $E_8'$), called $g_H$.
By substitution of eqs.~(\ref{M75}, \ref{M76}) into formula~(\ref{M79}), one
finds a VEV for $\vev$ which depends on the modulus $\tau$. Since the threshold
correction function $\Delta_a$ is modular invariant, so is $\vev$.
Assuming again that
no fields in a chiral representation of the hidden group $G_H$ is present,
one finds $b_H=-b'_H$.
In this case the gaugino condensate~(\ref{M79}) produces a very
simple modification of the superpotential~\cite{FILQ,CFILQ}
\bq
W(S,\tau)={e^{24\pi^2 S/b_H}\over \eta^6(\tau)}.
\eq{M80}
This superpotential has modular weight $-3$. Thus,
together with the K\"ahler potential $-3\log (2\Im
\tau)$, it gives rise to a $SL(2,\Z)$ invariant LEEA.
Equation~(\ref{M80}) can be generalized so as to take into account a possible
non-perturbative modification of the $S$-dependent part of the superpotential
\bq
W(S,\tau)={\Omega(S)\over \eta(\tau)}.
\eq{M81}
For instance, in ref.~\cite{DRSW} it was assumed $\Omega(S)=c+\exp(24\pi^2
S/b_h)$. Notice that modular invariance does not give any constraint on
$\Omega(S)$.\footnote{In ref.~\cite{FILQ1} $\Omega(S)$ was determined by
conjecturing a new duality invariance, acting on the field $S$. The existence
of this new symmetry has not yet been proven within the present
formulation of string theory. Calculations supporting this conjecture were
performed in ref.~\cite{STW,SENs,SSen}.}
The superpotential~(\ref{M81}), together with the standard K\"ahler potentials
for $\tau$ and $S$, gives rise to the scalar potential~\cite{FILQ}
\bq
V={1\over 16 \Re S \Im \tau |\eta(\tau)|^{12}}
\left[|2\Re S \Omega_S -\Omega|^2 +3|\Omega|^2\left({(\Im \tau)^2\over
\pi^2}|\hat{G}_2|^2 -1\right)\right].
\eq{M82}
Here $\Omega_S=\partial_S\Omega$, and $\hat{G}_2$ is the weight-2
(non-holomorphic) Eisenstein function (see for instance~\cite{Serre}).
The functional dependence
of~(\ref{M82}) on $\tau$ implies that the potential diverges both at
$\Im \tau\rightarrow\infty$ and $\Im \tau\rightarrow 0$. By recalling that
$\Im\tau$ is the compactification radius of the internal manifold (orbifold)
one draws an important conclusion: gaugino condensation {\em and
duality} force string theory to compactify on spaces of size ${\cal
O}(\alpha'^{1/2})$. The potential in eq.~(\ref{M82}) has been generalized to
the case when $G_H$-chiral matter is present in ref.~\cite{LM}.

In ref.~\cite{IL1} it was also noticed that potential~(\ref{M82}) gives rise
to {\em real} soft supersymmetry-breaking terms, owing to the fact that minima
lie at $\Re\tau$=integer. This fact guarantees the smallness of CP-violating
terms in the low-energy effective action.

Duality invariant gaugino
condensation has also been studied in~\cite{FMTV,BG,NO}. In particular, in
ref.~\cite{FMTV}, gaugino condensation was analysed by
introducing an effective Lagrangian including a composite chiral
superfield $U=W^{a\a}W^a_{\a}$, which describes
the degrees of freedom arising from the condensate.
The resulting Lagrangian is the analog of the pion Lagrangian in
QCD, with $U$ playing the role of the composite pion field.
The scalar potential found in ref.~\cite{FMTV} and the resulting
supersymmetry breaking substantially agree with~\cite{FILQ,CFILQ}
and~\cite{BG,NO}. The analysis of non-perturbative supersymmetry breaking
in string theory has further been studied, for instance, in
refs.~\cite{BG,LT,CCM,KLN,DFKZ}.

Duality-invariant functions, as well as functions transforming with a definite
weight under duality (automorphic dual functions) have entered our discussion
of threshold effects and gaugino condensation in several places. For instance,
we saw that threshold effects in the one-loop gauge coupling constant are
represented by eq.~(\ref{M76}) as $\log \| \eta^{-3}\|^2$,
with the norm $\|..\|^2$
given by $e^{K}|..|^2$. Threshold effect, therefore, can be thought as
given by a non-holomorphic norm of the modular form $\eta$.
The construction of this norm, and of generalized modular forms, has been
extended to the case of (2,2) compactifications in
ref.~\cite{FKLZ}. There it was observed that, even though the duality group of
(2,2) compactifications is unknown in general, yet its action on the moduli
is known. Indeed, the moduli of $(2,2)$ compactifications
parametrize a special K\"ahler manifold, due to
the correspondence between (2,2) $N=1$ heterotic compactifications, and type
II, $N=2$ compactifications. By calling $M$ the dimension of the moduli space,
and $y^m$ its homogeneous coordinates, the K\"ahler potential writes as in
the CY case
\bq
K=-\log i\left( \bar{y}^m {\partial {\cal F}\over \partial y^m}-
y^m {\partial {\cal F}\over \partial \bar{y}^m}\right), \;\;\; m=0,...,M.
\eq{M72}
In order to respect the symplectic structure of~(\ref{M72}), duality
transformations must act on $y^m$,$\partial_m {\cal F}$ as $Sp(M+1,\R)$
transformations. By analogy with the case of (2,1) moduli in CY spaces, we may
suppose that the duality group ${\Gamma} \subset Sp(M+1,\Z)$.
Therefore, one may write down a natural ansatz for a ${\Gamma}$-invariant
function
${\cal G}$ as
\bq
{\cal G}= \sum_{m_i, n^i} \log i{|m_iy^i + n^i{\cal F}_i|^2
\over y^i\bar{\cal F}_i - \bar{y}^i{\cal F}_i}, \;\;\; {\cal
F}_i\equiv \partial_i {\cal F}.
\eq{M73}
Here the sum ranges over all integers in an orbit of the group ${\Gamma}$.
Notice that by writing $\exp {\cal G}=\exp(K)|\Delta(y)|^2$, one would get,
formally
\bq
\Delta(y)=\prod_{m_i,n^i}(m_iy^i + n^i{\cal F}_i).
\eq{M74}
Here too, the sum ranges on an orbit of $\Gamma$.
Equation~(\ref{M74}) needs a regularization to be made rigorous. If this
regularization exists (as in the case of $\Z_3$ orbifolds~\cite{FFS}) then
equation~(\ref{M74}) defines an automorphic function of the duality group
${\Gamma}$.
In these formulae,  ${\cal G}$ has the same structure of the
threshold-effect function~(\ref{M76}), whereas $\Delta$ is the analog of the
Dedekind eta function $\eta(\tau)$.
\end{subsection}
\end{section}

\newpage

\begin{section}{Duality in Curved Backgrounds with Abelian Symmetries}

In this section, we discuss the discrete symmetry group acting on the space
of $D$-dimensional curved backgrounds that are independent of $d$
coordinates, and some of its applications. For such backgrounds
systematic knowledge is available.

We begin, in section 4.1, with a target space geometry with a compact
Abelian symmetry. A
duality transformation is then performed \cite{B3,B1,B2}
by gauging this symmetry and
adding a Lagrange multiplier term that constrains the gauge field to be
pure gauge. We show that such a duality transformation acts on the
(curved) background matrix as a {\em factorized duality} (\ref{ERfd}).

In section 4.2,
we show \cite{RV} that this factorized duality is a target space
duality, namely, it relates (different) backgrounds that correspond to the
same CFT. Moreover, we find a duality (sub-)group acting on the space of
$D$-dimensional curved
backgrounds with $d$ commuting, compact Abelian symmetries \cite{GR}.
In section 4.3 we discuss the action of the continuous group, called
${\cal G}$ in section 2, generating a local neighborhood of the moduli
space. In the case  considered here, of backgrounds with $U(1)^d$
symmetries, ${\cal G}=O(d,d,\R)$.
Based on analogies with the flat case, we conjecture, in section 4.4,
how these results extend to the heterotic string.

In sections 4.5-4.7 we present some applications.
An explicit string background \cite{BCR,EFR,MSW,RSSfd}
turns out to have an interpretation of a
2-$d$ black hole \cite{Wbh}.
It is instructive to study stringy features of
propagation in this background. In section 4.5 we discuss the
duality of the two-dimensional black hole \cite{giveon,DVV,bars}.
In section 4.6, we study two examples: duality
between compact black strings and charged black holes in the bosonic
string, and duality between neutral and charged black holes in the
heterotic string \cite{GR,hhs}.
In section 4.7, we discuss cosmological backgrounds with
compact Abelian symmetries \cite{KL,NW,GPa}.

In section 4.8, we interpret duality in curved backgrounds
as a spontaneously broken gauge symmetry of string theory \cite{GKi};
this is in analogy with the discussion of the
flat case (see section 2.6). Finally, in section 4.9,
we discuss duality and topology change in string theory \cite{GKi}.

\begin{subsection}{Factorized Duality in Non-Linear $\s$-Models}
\setcounter{equation}{0}

In this section we follow ref. \cite{RV}. Let us
consider the $\s$-model action
\be
S=\ind (G_{\m\n}(x)+B_{\m\n}(x))\pa x^{\m}\pab x^{\n} ,
\label{Soriginal}
\ee
where $G$ is the metric on some manifold and the antisymmetric background
$B$ is the potential for the
torsion three-form $H={3\over 2}dB$.
The complex worldsheet coordinates and derivatives are
\bq
z\equiv{1\over \sqrt{2}}(\tau+i\sigma),\;\;\;
\pa\equiv{1\over \sqrt{2}}(\pa_{\tau}-i\pa_{\sigma}),
\eq{zzst}
and therefore, $d^2 z=d\sigma d\tau$, and
$\pa x \pab x={1\over 2}((\pa_{\tau} x)^2+(\pa_{\sigma} x)^2)$.

Suppose that the action (\ref{Soriginal}) is invariant
under the isometry
\be
\delta x^{\m}=\e \k^{\m}.
\ee
This happens when the vector $\k^{\m}$ satisfies the Killing equation
${\cal L}_{\k}G_{\m\n}=\k_{\m;\n}+\k_{\n;\m}=0$ and, in addition,
${\cal L}_{\k}H=0$, where
${\cal L}_{\k}$ is the space-time Lie derivative. The latter implies that
locally
\be
{\cal L}_{\k}B=d\o
\label{omega}
\ee
for some one form $\o$. Here $d$ is the space-time exterior derivative.
For $\s$-models that have, in addition, a dilaton term
\be
S_{dilaton}=-{1\over 8\pi}\int d^2z \phi(x) R^{(2)},
\label{dilor}
\ee
the dilaton field $\f$ must satisfy $\k^{\m}\f_{,\mu}=0$.

With these conditions it is possible to choose coordinates
$\{x^{\m}\}=\{x^0,x^a\}$  such that the isometry acts by translation of
$x^0\equiv \th$,  and all the background fields, $G,B$ and $\f$, are
independent of $\th$. (Here one has to use the fact that $B$ is defined
only up to (space-time) gauge transformations, $B\rightarrow B+d\l$,
for some one-form $\l$; the one-form $\o$ defined in (\ref{omega})
transforms under this gauge symmetry as
$\o\rightarrow \o+{\cal L}_{\k}\l$.)
In the coordinates $\{\theta, x^a\}$, the action
(\ref{Soriginal}, \ref{dilor}) takes the form
\ba
S[x^a,\th]=\ind \[ G_{00}(x^c)\pa\th\pab\th +
(G_{0a}(x^c)+B_{0a}(x^c))\pa\th\pab x^a +
(G_{a0}(x^c)+B_{a0}(x^c))\pa x^a \pab\th + \nonumber\\
(G_{ab}(x^c)+B_{ab}(x^c))\pa x^a \pab x^b -
{1\over 4}\int d^2z \phi(x^c) R^{(2)}\].\nonumber\\
\label{SAx}
\ea

To get the dual theory, we gauge the Abelian symmetry by minimal coupling,
$\pa\th\to\pa\th+A$, $\pab\th\to\pab\th+\AB$, and add a Lagrange
multiplier term $\td F$,
where $F=\pa \AB - \pab A$ is an Abelian field strength.
Choosing a gauge $\th=0$, one finds the first-order action $S[x^a,A,\td]$
\cite{siegel,FJ,B1}
\ba
S[x^a,A,\td]=\ind \[ G_{00}A\AB + (G_{0a}+B_{0a})A\pab x^a +
(G_{a0}+B_{a0})\pa x^a \AB + \nonumber\\
(G_{ab}+B_{ab})\pa x^a \pab x^b -{1\over 4}\int d^2z \phi R^{(2)}+\td F \].
\label{SAxd}
\ea
The integration of the Lagrange multiplier field $\td$ gives $F=0$, implying
that on a topologically
trivial worldsheet the gauge fields are pure gauge, $A=\pa\th$,
$\AB=\pab\th$. Thus, one recovers the original model (\ref{Soriginal})
(modulo some important global issues that will be discussed below).
Integrating out the gauge fields $A,\AB$ one finds the {\em
dual} model; this is a {\em new} theory with action $S'$ \cite{B1,B2}
\be
S'=\ind \[(G'_{\m\n}(x^a)+B'_{\m\n}(x^a))\pa y^{\m} \pab y^{\n}
-{1\over 4}\int d^2z \phi'(x^a) R^{(2)}\] ,
\label{S'dual}
\ee
where $\{y^{\m}\}=\{\td,x^a\}$, and the dual background is related to the
original one by the
action of factorized duality in the $x^0$ direction (exactly as in the flat
case in section 2.4 (\ref{ERfd})).
Explicitly:
\ba
G'_{00}&=&{1\over G_{00}}, \qquad G'_{0a}={B_{0a}\over G_{00}}, \qquad
G'_{ab}=G_{ab}-{G_{a0}G_{0b}+B_{a0}B_{0b}\over G_{00}} \nonumber\\
B'_{0a}&=&{G_{0a}\over G_{00}},\qquad
B'_{ab}=B_{ab}-{G_{a0}B_{0b}+B_{a0}G_{0b}\over G_{00}}.
\label{facdua1}
\ea
The dilaton field  receives corrections from
the Jacobian that comes from integrating out the gauge fields. At one loop,
this leads to a shift in the dilaton \cite{B3,B1,B2}:
\be
\f'=\f+\log G_{00}.
\label{facdua2}
\ee
With this shift, if the original theory is conformal invariant, at least
to one-loop order, the dual theory is conformal invariant as well.

The dual theory is independent of the coordinate $\td$, and therefore, it
also has an Abelian isometry. However, in
general, the geometry of the $\s$-model is changed by the duality
transformation.

Are the dual theories equivalent as CFTs? To answer this question in the
affirmative, one ought to show that the two theories are equivalent on
worldsheets of any genera. For this reason one has
to consider global issues of the procedure \cite{RV,GR2,AABL}.
Let us assume that the isometry
corresponds to a {\em compact} $U(1)$ group, so that the coordinate $\th$ is
periodic with periodicity $2\pi$.
To  recover the original model after integrating out
$\td$, namely, to  get the correct periodicity for $\th$, the holonomies
$h=\oint A$ along any non-trivial homology cycle of the worldsheet
should be restricted to integer numbers.
To restrict $h$ one should add $\sum_{n} \delta(h=n)$ to the path integral.
For example, for a worldsheet with the topology of a torus one may write
\be
\sum_n \delta(h=n)=\sum_{n_a,n_b} e^{i(n_a\oint_a A+n_b\oint_b A)} .
\ee
In other words, the term  $n_a\oint_a A+n_b\oint_b A$ is added to the action
$S[x,A,\td]$, and the integration measure of the path integral contains a
sum over $n_a,n_b$. Integrating out $\td$ gives the original model
(\ref{Soriginal}).  Now, to integrate out $A$ (by first doing partial
integration on $\td F$ to get $\pa\td \AB -\pab\td A$) we must interpret
$n_a,n_b$ as the winding modes of $\td$ around the $a$- and $b$-cycles on
the torus. This fixes the periodicity of $\td$ to be $2\pi$. With this
periodicity the action (\ref{S'dual}) indeed is equivalent to the original
model without the gauge field.

\end{subsection}

\begin{subsection}
{$O(d,d,\Z)$ in Curved Backgrounds Independent of $d$ Coordinates}
\setcounter{equation}{0}

We will now show that the
factorized duality (\ref{facdua1}, \ref{facdua2}) in the $\th$ direction is
a symmetry of the CFT. This symmetry can be generalized to
all the elements of $O(d,d,\Z)$ if the curved background possesses
$d$ Abelian symmetries~\cite{GR}.

To show this we first construct, in subsection 4.2.1, general (conformal)
curved backgrounds in $D$ dimensions,
that are independent of $d$ coordinates, as  Abelian quotients
\cite{GKOct,BRSct,GKct,GMMct,ABRct,KPSYct,Sct,KSct}
of a ``parent'' CFT in a $(D+d)$-dimensional background.
Then the proof will be done \cite{RV,GR}
relating different backgrounds in $D$ dimensions, by
gauging in different ways -- {\em axial} or {\em vector} -- the
$d$ Abelian symmetries of the parent $\sigma$-model,
and combining the different gaugings with some manifest
symmetries of the resulting $D$-dimensional quotients.

In subsection 4.2.2, we use the construction of
subsection 4.2.1 to find the discrete symmetries of the space of these
$D$-dimensional backgrounds.
In subsection 4.2.3, we explore the group structure
of these symmetries, and find a group isomorphic to $O(d,d,\Z )$, as well
as a simple expression for its action on the backgrounds \cite{GR},
analogous to the flat case (see section 2.4). In subsection 4.2.4,
we focus on the dilaton and its transformations.

\begin{subsubsection}
{$\s$-Models Independent of $d$ Coordinates as Quotients}

Following \cite{GR}, we start with a  CFT
with $d$ Abelian chiral currents $J^i$ and anti-chiral currents $\JB^i$.
The parent action is
\be
\label{Dd}
\eqalign{
S_{D+d}=&\, S_1+S_a+S[x] \, ,\cr
S_1=&\, \ind \[\pa\th_1^i\pab\th_1^i + \pa\th_2^i\pab\th_2^i+
2\S_{ij}(x)\pa\th_2^i\pab\th_1^j
+\gl (x)\pa x^a\pab\th_1^i + \gz (x)\pa\th_2^i\pab x^a\]\cr
S_a=&\, \ind\[\pa\th_1^i\pab\th_2^i - \pa\th_2^i\pab\th_1^i\]\cr
S[x]=& \,\ind\[\gab (x)
\pa x^a \pab x^ b -\fr14\F(x) R^{(2)} \]\, ,}
\ee
where $i,j = 1 ,\dots ,d$ and $a, b = d+1 ,\dots ,D$,
and $\S_{ij},\gl ,\gz , \gab$ are
components of arbitrary $x$-dependent
matrices, such that, together with the dilaton $\F$, the worldsheet
theory described by the action $S_{D+d}$ is conformal.

The antisymmetric term $S_a$ is (locally)
a total derivative, and therefore, may give only topological contributions,
depending on the periodicity of the coordinates $\th$.
We define
\begin{equation}
\label{th}
\th^i=\th_2^i-\th_1^i \q \du\th^i=\th_1^i+\th_2^i \, ,
\end{equation}
and specify their  periodicity:
\begin{equation}
\label{per}
\th^i\equiv\th^i+2\pi \q \du\th^i\equiv\du\th^i+2\pi \, .
\end{equation}
In these coordinates, $S_a$ becomes
\be
\label{Sa}
S_a = \ind \fr12 \[\pa\du\th^i\pab\th^i -\pa\th^i\pab\du\th^i\]
\, ,
\ee
which takes half-integer values, and therefore, contributes to the
path-integral.

The action $S_{D+d}$ (\ref{Dd}) is invariant under the
$U(1)_L^d\times U(1)_R^d$ affine symmetry generated by chiral currents
$J^i$ and anti-chiral currents $\JB^i$ given by
\be
\label{cur}
\eqalign{
J^i\, =\, &\pa\th_1^i+\S_{ji}\pa\th_2^j +\fr12\G^1_{ai}\pa x^a\cr
\, =\, &\fr12\[-(I-\S )_{ji}\pa\th^j +(I+\S )_{ji}\pa\td^j +\G^1_{ai}\pa
x^a\]\,
    ,
\cr
\JB^i\, =\, &\pab\th_2^i+\S_{ij}\pab\th_1^j +\fr12\G^2_{ia}\pab x^a\cr
\, =\, &\fr12\[(I-\S )_{ij}\pab\th^j +(I+\S )_{ij}\pab\td^j +
\G^2_{ia}\pab x^a\]\, .\cr
}
\ee

Let us now  gauge the $d$ anomaly-free axial combinations of the symmetries
(\ref{cur}) by minimal coupling $\pa\du\th^i \ra \pa\du\th^i+A^i$,
$\pab\du\th^i \ra \pab\du\th^i+\AB^i$.
(Other options are generated by discrete symmetries discussed
later.)
The gauged action is
\begin{equation}
\label{gauge}
S_{gauged}=S_{D+d}+\ind\[A^i\JB^i+\AB^iJ^i+
\fr12A^i\AB^j(I+\S )_{ij}\]\, .
\end{equation}
Integrating out the gauge fields $A^i,\AB^i$ gives:
\be
\label{D}
\eqalign{
S_D\, =\, &\ind \[{\cal E}_{IJ}(x)\pa X^I \pab X^J -\fr14\f (x) R^{(2)}\]\cr
\, =\, &\ind\[E_{ij}(x)\pa\th^i\pab\th^j+F^2_{ia}(x)\pa\th^i\pab x^a+
F^1_{ai}(x)\pa x^a\pab\th^i \cr
 &\qquad\qquad\qquad\qquad
+ F_{ab}(x)\pa x^a \pab x^b -\fr14 \f(x) R^{(2)}\]\, .}
\ee
Notice that $S_D$ is independent of $\du\th$, as expected.
In eq. (\ref{D})
$$
\{ X^I | I=1\dots D\}=\{\th^i,x^a\ | i=1\dots d,a=d+1\dots D\}.
$$
In the first line in (\ref{D}), we have combined the background fields into a
$D\times D$ background matrix ${\cal E}$
\be
\label{calG}
{\cal E}_{IJ} = {\cal G}_{IJ}+{\cal B}_{IJ}
=\left(\matrix{E_{ij}&F^2_{ib}\cr
                F^1_{aj}&F_{ab}\cr}\right)\, .
\ee
Here  $\cal G$ and  $\cal B$ denote the symmetric part and antisymmetric
part of  $\cal E$, respectively. The four block components in (\ref{calG}) are
\be
\label{ij}
E_{ij}=(I-\S )_{ik}(I+\S )^{-1}_{kj} \, ,
\ee
and
\be
\label{F}
F^2_{ia}=(I+\S )^{-1}_{ij}\G^2_{ja}\sq F^1_{ai}=-\G^1_{aj}(I+\S )^{-1}_{ji}
\sq F_{ab}=\gab-\fr12\gl (I+\S )^{-1}_{ij} \G^2_{jb} \, .
\ee

The  $D$-dimensional background ${\cal E}(x)$ is independent of
the $d$ coordinates $\th^i$. It is the most general $\s$-model
with $d$ commuting, compact Abelian symmetries, since for {\em any}
such background the relations
(\ref{ij}, \ref{F}) can be inverted to solve for
$(\S , \G )$ in terms of $(E,F)$. Moreover, if the
original model $S_{D+d}$ (\ref{Dd}) is conformal, then $S_D$ is conformal
as well \cite{RV},\footnote{Although $S_D$ is
conformal only to one loop, higher-order
corrections to the background that give an {\it exact\/} CFT with a
$D$-dimensional background exist \cite{RV,GR,K2};
these corrections come from the integration measure.} with a dilaton field
\begin{equation}
\label{f}
\f=\F+\log \det (I+\S )\, .
\end{equation}
This relation is also invertible. Therefore, a $\s$-model of a
$D$-dimensional background, which is
independent of $d$ coordinates, can be described as a quotient of a
worldsheet theory in a $(D+d)$-dimensional background with $d$ chiral
currents and $d$ anti-chiral currents.

Finally, if the theory in $D$-dimensional background
is conformally invariant, then the parent theory in $(D+d)$-dimensional
background is conformal
as well \cite{RV}, and hence, {\it any\/} CFT with a background that is
independent of $d$ coordinates can be described as a quotient of a CFT with
a $(D+d)$-dimensional background.

\end{subsubsection}

\begin{subsubsection}{Discrete Symmetries}

The construction in the previous subsection  allow us to understand
(some of) the global structure in the moduli space of string vacua in curved
backgrounds that are independent of $d$ coordinates.
Here we will study discrete symmetries that relate different backgrounds
$(E(x),F(x),\f (x))$ (\ref{D}) which correspond to the same CFT.

We first
discuss transformations of $(E(x),F(x),\f (x))$ that follow from
manifest symmetries of the action $S_{D+d}$ (\ref{Dd}).  We then combine
them with transformations that are manifest symmetries of $S_D$ (\ref{D})
to find a discrete symmetry group isomorphic to $O(d,d,\Z )$.

The $S_{D+d}$ theory is invariant under the coordinate transformations
\be
\label{OxO}
\th_1 \ra O_1\th_1 \q \th_2 \ra O_2\th_2 \q
\ee
together with the transformations of the background
\be
\label{OSO}
\S\ra O_2\S O_1^t\q\G^1\ra\G^1O_1^t\q\G^2\ra O_2\G^2\q\G\ra\G\, ,
\ee
where
\be
\label{cond}
O_1, O_2 \in O(d,\Z ) \, , \qquad \fr12 ( O_1 \pm O_2 )_{ij} \in \Z \, .
\ee
Here $O(d,\Z )$ is the group of matrices $O$ with integer
entries satisfying $OO^t=I$.

These symmetries can be found as follows:
The action $S_{D+d}$ is invariant, up to total derivatives,
under $O(d,\R )\times O(d,\R )$ acting on
$(\th_1 ,\th_2)$ as in (\ref{OxO}), together with (\ref{OSO}) for
the backgrounds.  The periodic coordinates $\th , \td$ (\ref{th})
transform as:
\be
\label{mat}
\left(\matrix{\th\cr\td}\right)\ra
\fr12\left(\matrix{O_1+O_2&O_1-O_2\cr O_1-O_2&O_1+O_2\cr}\right)
\left(\matrix{\th\cr\td}\right)
\ee
To preserve the periodicities of $\th ,\du\th$ (\ref{per}), the condition
(\ref{cond}) must be satisfied.  In particular, this implies
$O_1, O_2\in O(d,\Z )$.

A total derivative comes from the transformation of $S_a$ (\ref{Sa}),
and is
\be
S_a[O_1\th_1,O_2\th_2]-S_a[\th_1,\th_2] = \ind \[ M_{ij}(\pa\td^i
\pab\td^j-\pa\th^i
\pab\th^j )+N_{ij}(\pa\th^i\pab\td^j-\pa\td^i\pab\th^j)\] \, ,
\ee
where
\be
M=\fr14(O_1^t-O_2^t)(O_1+O_2) \q N=\fr14(O_1^t-O_2^t)(O_1-O_2) \, .
\ee
The condition (\ref{cond}) implies that $M_{ij},N_{ij}\in \Z$, and hence
the total derivative is an integer, and does not contribute to the path
integral.  This concludes the proof that (\ref{OxO}, \ref{OSO}) are symmetries
of the action $S_{D+d}$.

The transformations of the background (\ref{OSO}) induce transformations of
the background $(E(x),F(x),\f (x))$ (\ref{ij}, \ref{F}, \ref{f})
in $S_D$ (\ref{D}):
\be
\label{tre}
\quad \,\, \, E\quad\ra\quad E'\quad =
\quad (I-O_2\S O_1^t)(I+O_2\S O_1^t)^{-1}
\ee
\be
\label{trf}
\matrix{F^1&\ra &F^1{}'&=& -\G^1(O_1+O_2\S )^{-1}\cr
&&&&\cr
F^2&\ra &F^2{}'&=& (O^t_2+\S O^t_1)^{-1}\G^2\cr
&&&&\cr
F&\ra& F'&=&\G-\fr12\G^1(O^t_2O_1+\S )^{-1}\G^2\cr
&&&&\cr}
\ee
\be
\label{trd}
\f\ra\f' =\F+\log \det(I+O_2\S O^t_1 ) \, .
\ee
These transformations can be rewritten as
\be
\label{Etrans}
E'=\[(O_1+O_2)E+(O_1-O_2)\]\[(O_1-O_2)E+(O_1+O_2)\]^{-1}
\ee
\be
\label{Ftrans}
\eqalign{
F^1{}'\, =\, &2F^1\[(O_1-O_2)E +(O_1+O_2)\]^{-1}\cr &\cr
F^2{}'\, =\, &\fr12 \[(O_1+O_2)-E'(O_1-O_2)\] F^2\cr &\cr
F'\, =\, &F-F^1\[(O_1-O_2)E +(O_1+O_2)\]^{-1}(O_1-O_2)F^2\cr
&\cr}
\ee
\be
\label{dtrans}
\f'=\f+\fr12\log \[\fr{\det{\cal G}}{\det{\cal G}'}\]
=\f+\fr12\log \[\fr{\det G}{\det G'}\]\, ,
\ee
\vskip .15in
\noindent
where $\cal G$ is the background metric as defined in (\ref{calG})
and $G$ is the symmetric part of $E$ (\ref{ij}).  We discuss the dilaton
transformation (\ref{dtrans}) in detail in subsection 4.2.4.

The transformations (\ref{OxO}, \ref{OSO}) induce a non-trivial
action on the currents (\ref{cur}), and hence, $S'_D$ ($S_D$
with a transformed background $(E',F',\f')$) is derived from $S_{D+d}$
(\ref{Dd}) by a different quotient.  For example, some symmetries
simply change the sign of a $J^i$ without changing $\JB^i$; this
corresponds to a {\it vector\/} gauging
(as opposed to an axial gauging) of the $i$'th $U(1)$.
We thus refer to such symmetries as {\em axial-vector duality} \cite{Kir}
(and we show that these are exact symmetries in section 4.8).

We now consider the additional transformations that are manifest
symmetries of the action $S_D$ itself.  The first are integer
``$\Theta$"-parameters that shift $E$:
$$
E_{ij}\ra E_{ij}+\Theta_{ij}\q \Theta_{ij}=-\Theta_{ji}\in\Z
$$
\be
\label{Th}
F^1\ra F^1 \q F^2\ra F^2 \q F\ra F \, .
\ee
We refer to the group generated by these transformations as $\Theta (\Z )$.
These are obviously symmetries, as they shift the action $S_D$ (\ref{D})
by an integer, and therefore, do not change the path integral.

The second type of transformations are given by homogeneous transformations
of $E,F^1,F^2$ under $A\in GL(d,\Z )$:
\be
\label{gl}
E\ra A^t E A \q F^2\ra A^t F^2\q F^1\ra F^1 A\q F\ra F\, .
\ee
These are obviously symmetries of the theory described by $S_D$, as they
generate a change of basis in the space of $\th$'s that preserves their
periodicities.

Neither the $\Theta (\Z )$ nor the $GL(d,\Z )$ transformations affect the
dilaton, since they do not change the integration measure of the path
integral; consequently, they are symmetries of the exact CFTs, and receive
no higher-order corrections.

The group generated by all the symmetries discussed is isomorphic to
$O(d,d ,\Z )$.  A natural embedding of $O(d,d,\Z )$ in $O(D,D,\Z )$
acts on the background $\cal E$ by fractional linear transformations,
analogous to the flat case and as explained in the following subsection.
\end{subsubsection}

\begin{subsubsection}{The Action of $O(d,d,\Z )$}

We begin by establishing our notation following section 2.4 and refs.
\cite{GMR1,GR}. As explained in section 2.4
the group $O(d,d,\R )$ can be represented as ($2d\times 2d$)-dimensional
matrices $g$ preserving the bilinear form $J$:
\be
\label{g}
g=\left(\matrix{a&b\cr c&d\cr}\right) \q
J=\left(\matrix{0&I\cr I&0\cr}\right)\, ,
\ee
where $a,b,c,d,I$ are ($d\times d$)-dimensional matrices, and
\be
\label{abcd}
g^tJg=J\quad \Rightarrow \quad a^tc+c^ta=0\sq
b^td+d^tb=0\sq a^td+c^tb=I\, .
\ee
This has an obvious embedding in $O(D,D,\R )$ as
\be
\hat g=\left(\matrix{\hat a&\hat b\cr \hat c&\hat d\cr}\right)\, ,
\ee
where $\hat a, \hat b, \hat c, \hat d$ are ($D\times D$)-dimensional
matrices of the form
\be
\hat a={\left(\matrix{a&0\cr 0&I}\right)}\sq
\hat b={\left(\matrix{b&0\cr 0&0}\right)}\sq
\hat c={\left(\matrix{c&0\cr 0&0}\right)}\sq
\hat d={\left(\matrix{d&0\cr 0&I}\right)}
\ee
(here $I$ is the $(D-d)\times (D-d)$-dimensional identity matrix).

We define the action of $\hat g$ on $\cal E$ by fractional
linear transformations:
\be
\label{ghat}
\eqalign{
\hat g({\cal E}) =& {\cal E}'=
(\hat a {\cal E} + \hat b)(\hat c {\cal E} +\hat d)^{-1}
\cr &\cr
=&\left(\matrix{E'&(a-E'c)F^2\cr F^1(cE+d)^{-1}&F-F^1(cE+d)^{-1}cF^2\cr}
\right)}\, ,
\ee
where
\be
\label{tE}
E'=(aE+b)(cE+d)^{-1}
\ee
is a fractional linear transformation of $E$ under $O(d,d,\R)$.
If ${\cal E}$ is constant we recover the action of $O(D,D,\R)$ on the
$D$-dimensional toroidal background matrix, as discussed in section 2.4.

Let us recall here that the group $O(d,d,\R)$ is generated by:

\noindent
$GL(d,\R)$:
\be
\label{GLd}
\left(\matrix{a&b\cr c&d\cr}\right)=
\left(\matrix{A^t&0\cr 0&A^{-1}\cr}\right) \quad {\rm s.t.}
\quad A\in GL(d,\R) \, .
\ee
$\Theta(\R)$:
\be
\label{THe}
\left(\matrix{a&b\cr c&d\cr}\right)=
\left(\matrix{I&\Theta\cr 0&I\cr}\right) \quad {\rm s.t.}\quad
\Theta=-\Theta^t \, .
\ee
Factorized duality:
\be
\label{FAd}
\left(\matrix{a&b\cr c&d\cr}\right)=
\left(\matrix{I-e_1&e_1\cr e_1&I-e_1\cr}\right) \quad {\rm s.t.}\quad
e_1=diag(1,0,\dots ,0 )  \, .
\ee
The maximal compact subgroup of $O(d,d,\R)$ is $O(d,\R)\times O(d,\R)$
embedded as
\be
\label{Odd}
\left(\matrix{a&b\cr c&d\cr}\right)=
\fr12\left(\matrix{o_1+o_2&o_1-o_2\cr o_1-o_2&o_1+o_2\cr}\right)
\quad {\rm s.t.}\quad o_1, o_2\in O(d,\R) \, .
\ee
This subgroup includes factorized duality (\ref{FAd}).

We now turn
from definitions to the actual symmetries of the CFT.  These
form an $O(d,d,\Z )$
discrete subgroup of $O(d,d,\R )$ that acts on the background as
above.   The elements of the subgroup $O(d,d,\Z )$ are given by matrices
$g$ of the form (\ref{g}, \ref{abcd})
with {\it integer\/} entries.

Just as in the
continuous case, the discrete group is generated by $GL(d,\Z )$,
$\Theta (\Z )$, and factorized duality; these are given by
(\ref{GLd}, \ref{THe}, \ref{FAd}) with integer entries.
The subgroup $O(d,\Z )\times
O(d,\Z )$ is given by the matrices (\ref{Odd}), again with integer entries.
Clearly, the $O(d,\Z )\times O(d,\Z )$ symmetries
(\ref{Etrans}, \ref{Ftrans}), the $\Theta (\Z )$ symmetries (\ref{Th}), and
the $GL(d,\Z )$ symmetries
(\ref{gl}), that we found in the previous subsection, act on the
background by the $O(d,d,\Z )\subset O(D,D,\Z )$ fractional
linear transformations (\ref{ghat}) with the matrices $a,b,c,d$ given
by (\ref{Odd}), with $o_{1,2}=O_{1,2}$,
(\ref{THe}), with $\Theta\in\Z$, and (\ref{GLd}), with $A\in GL(d,\Z )$,
respectively.

These results are compatible with the discrete symmetries of
the space of flat $D$-dimensional toroidal backgrounds \cite{GRV,sw,GMR},
described in section 2.4.
In that case, the $O(d,d,\Z )$ symmetries, described above, are simply
a subgroup of the full $O(D,D,\Z )$ symmetry group which
acts as in (\ref{ghat}),
for {\it any\/} $\hat g\in O(D,D,\Z )$.  For curved $D$-dimensional
backgrounds, we expect to find some large symmetry group analogous to
$O(D,D,\Z )$; here we have described its $O(d,d,\Z )$ subgroup that
is associated with a $d$-dimensional toroidal isometry of the background.

In the flat case, the fractional linear transformation is an exact map
between equivalent backgrounds; in the curved case, in general, one expects
higher-order corrections to the transformed background
\cite{B1,B2,B3,SP,tseytlin}.
For $\Theta (\Z )$ and $GL(d,\Z )$, the transformations are exact; however,
factorized duality receives corrections from the path-integral measure.
Nevertheless, because the transformation is exact in the ($D+d$)-dimensional
model, we know that non-perturbative correction must exist such that
factorized duality is exact\footnote{$N=4$ supersymmetry can protect
duality transformations, and there are examples where the
one-loop transformations are exact even in curved backgrounds.
Exact duality was also described for superstrings and
$(1,1)$ heterotic backgrounds corresponding to cosets in \cite{GRT}, and in
some examples of bosonic string backgrounds \cite{GKi,Tsey2,KT}.}.

We close this subsection with a general remark. The group $O(d,d,\R)$ has
disconnected components; in particular,
factorized duality has $\det=-1$, and hence is not connected
to the identity.  Therefore, one expects that in general the moduli space is
made of several disconnected components mapped into each other by
$O(d,d,\R)$ transformations.
In the flat case the moduli space is connected and this phenomenon does not
occurs. However, in other cases one may
get various disconnected components of backgrounds
with different topologies
\footnote{Factorized duality
is similar to mirror symmetry \cite{GrP}, which also identifies
two (possibly) disconnected components of
moduli space corresponding to backgrounds
with different topologies. For $N=2$
toroidal (orbifold) backgrounds,
mirror symmetry and factorized duality are identical~\cite{gs}.}.
This issue will be addressed later.
\end{subsubsection}

\begin{subsubsection}{The Dilaton}

To complete the previous discussion of the transformation of the
background under $O(d,d,\Z )$, we consider the transformation
of the dilaton \cite{B1,gv,GMR,AO,tseytlin,GR,ST}.
This is summarized in eq. (\ref{dtrans}), which is derived
by proving that the quantity
\be
\label{them}
\du\f=\f+\fr12\log \det{\cal G}
\ee
is invariant under $O(d,d,\Z )$ transformations~\cite{GR}.  This implies
\be
\label{tdil}
\f'=\f+\fr12\log\Big( \fr{\det{\cal G}}{\det{\cal G'}}\Big)\, .
\ee
The second equality in (\ref{dtrans}) follows from the proof of
(\ref{them}) given below.

We begin with the identity
\be
\left(\matrix{G_{ij}&G_{ib}\cr G_{aj}&G_{ab}}\right)=
\left(\matrix{G_{ik}&0\cr G_{ak}&I_{ac}}\right)
\left(\matrix{I_{kj}&(G^{-1})_{kl}G_{lb}\cr 0&G_{cb}
-G_{cm}(G^{-1})_{ml}G_{lb}}\right)
\ee
which implies
\be
\label{det}
\det({\cal G})=\det(G_{ij})\det(G_{ab}-G_{ak}(G^{-1})_{kl}G_{lb})\, .
\ee
We next prove that the following two quantities are separately
invariant under $O(d,d,\Z )$ transformations:
\be
\label{lem}
{\rm~the~invariant~fiber~dilaton:~~} \hat\f=\f+\fr12\log \det(G_{ij})\, ,
\ee
\be
\label{quot}
{\rm~the~quotient~metric:~~} G_{ab}-G_{ak}(G^{-1})_{kl}G_{lb}\, .
\ee
Geometrically, a $D$-dimensional space whose metric is independent
of $d$ coordinates $\th^i$ can be thought of as a bundle $M$ with fiber
coordinates $\th$.  The metric on the fiber is $G_{ij}$, and hence
we refer to $\hat\f$ as the ``invariant fiber dilaton.''
The induced metric on
the quotient space $M/\{\th^i\}$ is the quotient metric (\ref{quot}).

To prove the invariance of (\ref{lem}), we consider the action of
the generators of $O(d,d,\Z )$ separately.  The $GL(d,\Z )$
and  $\Theta (\Z )$ transformations trivially leave $\f$
and $\det(G_{ij})$ invariant, and hence $\hat\f$ as well.
To show invariance under factorized duality, we write $\hat\f$
explicitly in terms of $\S$:
\be
\label{fres}
\eqalign{
\hat\f\, =\, &\F+\log \det(I+\S )+\fr12\log \det\fr12\[(I-\S )(I+\S )^{-1}
+(I+\S^t)^{-1}(I-\S^t)\]\cr
=\, &\F+\fr12 \log \det(I-\S^t\S ) \, .}
\ee
To get the second equality, we need to split $\log \det(I+\S )$ as
$\fr12\log \det(I+\S ) +\fr12\log \det(I+\S ^t)$.  Under $\S\ra O_2\S O_1^t$
(\ref{OSO}), $(I-\S^t\S )\ra O_1(I-\S^t\S )O_1^t$, and hence $\hat \f$
is invariant.  This completes the proof of the invariance of the
fiber dilaton $\hat\f$.

The proof of the invariance of (\ref{quot}) is entirely parallel; again,
the $GL(d,\Z )$
and  $\Theta (\Z )$ transformations trivially leave (\ref{quot}) unchanged.
To prove invariance under factorized duality, we express
(\ref{quot}) explicitly
in terms of $\S , \G^1,\G^2,\G$ (see (\ref{ij}, \ref{F})):
\be
\label{abmess}
\eqalign{
G_{ab}&-G_{ak}(G^{-1})_{kl}G_{lb}\cr
=&\[\fr12(\G+\G^t)-\fr14\{\G^1(I+\S )^{-1}\G^2+
\G^{2t}(I+\S^t )^{-1}\G^{1t}\}\cr
&\,\, +\fr12\{\G^1(I+\S )^{-1}+\G^{2t}(I+\S^t)^{-1}\}
\{(I-\S )(I+\S )^{-1}+(I+\S^t)^{-1}(I-\S^t)\}^{-1}\cr
&\qquad\qquad\qquad\qquad\qquad\qquad\qquad\quad\times
\{(I+\S )^{-1}\G^2+(I+\S^t)^{-1}\G^{1t}\}\]_{ab}\cr
=& \[\fr12(\G+\G^t)+\fr14\{\G^1(I-\S^t\S )^{-1}(\G^{1t}+\S^t\G^2)
+\G^{2t}(I-\S\S^t)^{-1}(\G^2+\S\G^{1t})\}\]_{ab}\, .
\cr}
\ee
The final expression is manifestly invariant under (\ref{OSO}),
which completes the proof of the invariance of the quotient metric
(\ref{quot}).

The invariance of $\du\f$ in~(\ref{them}) follows from
the invariance of (\ref{lem}, \ref{quot}) together with the
identity (\ref{det}). This is compatible with results
in low-energy effective field theories \cite{GP1,MV}, as discussed in
section 2.8, and string field
theory \cite{sen,KZ}, and physically implies that the
string coupling constant
$g_{string}^{-1}=<e^{\du\f}>=<\sqrt{\det{\cal G}}\, e^\f >$
is invariant under $O(d,d,\Z )$.  The invariance
of (\ref{lem}, \ref{quot}) actually proves that
$\sqrt{\det(G_{ij})}\, e^\f$ and the quotient metric (\ref{quot}) are
separately invariant; of course, this holds only for $O(d,d,\Z )$, and
not for the full $O(D,D,\Z )$ in which the former group is embedded.

\end{subsubsection}

\end{subsection}

\begin{subsection}{The Action of $O(d,d,\R)$}
\setcounter{equation}{0}

In the previous subsection we have discussed the action of the discrete group
$O(d,d,\Z)$. The {\em continuous} group $O(d,d,\R)$ also maps a conformal
background \cite{GMR1,ven2,MV,GMV,MSch},
that is independent of $d$ coordinates, onto (the leading
order of) a conformal background. Namely,
given a conformal background ${\cal E},\phi$ and an element $g$ in $O(d,d,\R)$
(\ref{g}), then the background $g({\cal E},\phi)=({\cal E}',\phi')$ given
by eqs. (\ref{ghat}) and (\ref{tdil}) is the leading order of an exact
conformal background.

This is true because
not only the discrete groups $\Theta(\Z)$ and $GL(d,\Z)$, but also
$\Theta(\R)$ and $GL(d,\R)$, transform
exact backgrounds to exact backgrounds.
Moreover, we have shown that there is a CFT
corresponding to an exact background whose leading order is given by acting
on the original one with factorized duality, as explained in the previous
section. Now, because the full $O(d,d,\R)$ group is
generated by $\Theta(\R)$, $GL(d,\R)$ and factorized duality, we conclude
that the action of $O(d,d,\R)$ on a conformal background with $d$
commuting, compact Abelian
symmetries generates the leading order of an exact background.

The  moduli space of curved string backgrounds with $d$ Abelian
symmetries (generated by $O(d,d,\R)$ rotations) was suggested at first by
studies of both low-energy effective actions \cite{GMR1,Duff,GP,GP1,MV,GMV}
and worldsheet actions on the sphere~\cite{FTdual,CFG,MO}.
These studies suggested that there is a moduli subspace of backgrounds,
independent of $d$ coordinates, isomorphic to $O(d,d,\R )/G$.
Here $G$ is at least the diagonal subgroup $O(d,\R )_{diag}$
of $O(d,\R )\times O(d,\R)$ (the maximal compact subgroup of $O(d,d,\R )$).
(Needless to say, for flat backgrounds
$G=O(d,\R )\times O(d,\R)$~\cite{N,NSW}).
This is consistent with results from string field theory \cite{sen}.
This moduli subspace is only correct when the $d$ coordinates have
the topology of a torus.  Though in general the local structure of
the moduli space is unknown, when the background is independent of $d$
periodic coordinates, we can still identify a discrete symmetry
group that acts on the moduli space, as  described in the previous section.

The method of $O(d,d,\R)$ transformations is useful
to generate a large class of curved string backgrounds. Generating
solutions in this way was discussed extensively in the literature.
For example, in refs.~\cite{S1bh,HSbh,ils,hhs,GR,Sbh,KKbh,KKKbh} $O(d,d,\R)$
rotations were used in order to generate
many kinds of (charged) black objects (black holes, black strings etc.);
in refs.~\cite{MV,GMV,GPa,Vremsin,KMcos,AKcos,KM1cos}
this method was used to generate cosmological solutions.
In sections 4.5, 4.6 we will describe examples of black-object solutions, and
in section 4.7 we will describe
examples of cosmological string backgrounds with Abelian symmetries.
\end{subsection}

\begin{subsection}{The Heterotic String}
\setcounter{equation}{0}

In this section, we discuss the explicit form of the discrete symmetries
of the heterotic string following ref.~\cite{GR}.
This form can be shown to hold to leading order in
$\alpha'$ for $N=(1,1)$
cosets and their deformations following refs.~\cite{GRT,GKi}.
Its validity for the general case was conjectured in ref.~\cite{GR},
by requiring compatibility
with the flat limit \cite{GRV,sw}, described in section 2.5, and with
the bosonic case.
We start with a {\it curved\/} heterotic background,
which we assume is a consistent, conformally
invariant, heterotic string theory, with action:
\be
\label{het}
\eqalign{
S_{heterotic}=
&\ind\[{\cal E}_{IJ}(x)\pa X^I\pab X^J+{\cal A}_{IA}(x)\pa X^I\pab
Y^A \cr
&\qquad\qquad + E_{AB}\pa Y^A\pab Y^B -\fr14\f (x)R^{(2)}
+({\rm fermionic~terms})\]\, ,}
\ee
and with the constraints that the $Y^A$ are
covariantly chiral bosons: $\pa Y^A+{\cal A}^A_I\pa X^I=0$.
As before  $\{X^I\}=\{\th^i,x^a | i=1\dots d\;\;{\rm and}\;\;a=d+1\dots D\}$.
In addition, we have $d_{int}$ internal
chiral bosons $Y^A$: $A=1\dots d_{int}$.  In flat space we have
$d_{int}=16$, but more generally, we may find other
solutions \cite{pol,mny}.  The space-time background is given by
($\cal E$,$\f$), as in the bosonic case (\ref{calG}, \ref{f}),
and, in addition, the gauge field $\cal A$.
We assume that this curved background is independent of the $d$
coordinates $\th^i$.  The constant internal background is
\be
\label{GAB}
E_{AB}=G_{AB}+B_{AB}\, ,
\ee
where $G_{AB}$ is the metric on the internal lattice
(one half the Cartan matrix of the internal symmetry group when
the lattice is the root lattice of a group)
and $B_{AB}$ is its antisymmetrization \cite{egrs}, \ie $E_{AB}$ is
upper triangular.  In the
space-time supersymmetric flat case,
the symmetry group is $E_8\times E_8$.\footnote{The $Spin(32)/\Z_2$
string can be described as the $E_8\times E_8$ string with a particular
gauge field background \cite{ginnsw,N,NSW}.}  In curved space, this group is
in general different \cite{pol,mny}.

Following section 2.5
\cite{GRV}, the expected symmetry group is isomorphic to
$O(d,d+d_{int},\Z )$ embedded in $O(D,D+d_{int},\Z )$ as in section 2.5,
and acting by
fractional linear transformations on the $(D+d_{int})\times (D+d_{int})$
dimensional matrix
\be
\label{xi}
\Xi (x)=\left(\matrix{
{\cal E}_{IJ}+\fr14{\cal A}_{IA}(G^{-1})_{AB}
{\cal A}_{JB}&{\cal A}_{IA}
\cr0&E_{AB}
\cr}\right).
\ee
Here $(G^{-1})_{AB}$ is the inverse of $G_{AB}$ in (\ref{GAB}).  The matrix
$\Xi$ is the embedding of the heterotic background into a bosonic
$(D+d_{int})$-dimensional background, and the group $O(D,D+d_{int})$
is the subgroup
of $O(D+d_{int},D+d_{int})$ that preserves the form (\ref{xi}).  Note that
the space-time metric ${\cal G}_{IJ}$ (the symmetric part of
${\cal E}_{IJ}$ in (\ref{xi})) is the quotient metric of the
$(D+d_{int})$-dimensional space modulo $\{ Y^A\}$ (here, $G_{AB}$
is the fiber metric).  This leads to a
simple expression for the transformation of the dilaton:
\be
\label{hetdil}
\f'=\f+\fr12\log\Big( \fr{\det{\cal G}}{\det{\cal G'}}\Big)\, ;
\ee
note that this is independent of the gauge fields.
\end{subsection}

\begin{subsection}
{Application I: the D=2 Black Hole Duality}
\setcounter{equation}{0}

In this section, we discuss the first nontrivial example of curved-space
duality, namely, a $D=2$
background that is independent of one coordinate ($d=1$).
Such a background can be
derived as a gauged WZNW model $SU(2)/U(1)$ or $SL(2,\R)/U(1)$.
The exact conformal field theory corresponding to an
$SL(2,\R)/U(1)$ gauged WZNW model
\cite{BCR} describes a black hole in two dimensional
space-time \cite{Wbh}. We will describe the duality of the
$D=2$ black hole~\cite{giveon,DVV}
in some detail to examine explicitly the procedure presented in section
4.2 for this simple example.
Moreover, we will discuss some of its interesting
consequences, namely: interchanging the horizon
with the singularity, and taking seriously the space-time region beyond the
singularity. This example also illustrates the possibility to generate
$D$-dimensional curved backgrounds, independent of $d$ coordinates,
from $G/H$ coset CFTs and their duals.

\noindent
{\bf The $G_k$ WZNW model and its gauging}

\noindent
At first, let us describe briefly a general WZNW model on a group manifold
$G$ at level $k$ (we call it the $G_k$ WZNW model), and its gauging.
The action of the WZNW model
is \cite{Wwznw}
\be
S[g]={k\over 4\pi}I[g]+{k\over 6\pi}\Gamma[g],
\label{Sofg}
\ee
\be
I[g]=\int_{\Sigma} d^2\s \sqrt{h} h^{ij}{\rm Tr}(e_ie_j), \qquad
\Gamma[g]=\int_{B,\;\partial B=\Sigma} d^3\s \e^{ijk} {\rm Tr} (e_ie_je_k),
\label{IGam}
\ee
where $\Sigma$ is a worldsheet with coordinates $\s_i$ and metric $h_{ij}$,
and
\be
e_i\equiv g^{-1} \partial_i g,
\label{egg}
\ee
where the field $g$ is a map from the worldsheet to the group manifold $G$;
it can be taken to be a matrix in the fundamental representation of $G$.
The trace in (\ref{IGam}) is an invariant form on the Lie algebra of
$G$, and the Wess-Zumino term, $\Gamma[g]$, produces an antisymmetric
background on the worldsheet $\Sigma$, after integration over the third
coordinate of $B$.
Explicitly, in the matrix representation $T$ for the Lie algebra,
\be
[T_a,T_b]=f_{ab}^c T_c, \qquad {\rm Tr}(T_aT_b)=\eta_{ab}, \qquad
a,b,c=1,...,{\rm dim}\, G,
\ee
and parametrizing $g(X)\in G$ in a set of coordinates $X^A(\s_i)$,
$A=1,...,{\rm dim}\, G$, one finds
\be
S[X]={1\over 4\pi}
\int d^2\s \left[ \sqrt{h} h^{ij} G_{AB}(X)\partial_i X^A \partial_j X^B
+\e^{ij} B_{AB}(X)\partial_i X^A \partial_j X^B \right],
\ee
where $G_{AB}(X),\; B_{AB}(X)$ are given by the left-invariant vielbeins,
$e_A(X)$,
\ba
e_A\equiv g^{-1}\partial_{A} g = e_A^a T_a, & &\qquad
\partial_A e_B^c -\partial_B e_A^c = e_A^a e_B^b f_{ab}^c ,\nonumber\\
G_{AB}&=&k\eta_{ab}e_A^ae_B^b, \nonumber\\
k{\rm Tr}\left(e_A[e_B,e_C]\right)&=&
\partial_A B_{BC}+\partial_B B_{CA}+\partial_C B_{AB}.
\ea

The WZNW model can serve as a starting point to generate other CFTs. From
an algebraic point of view this was done in what is termed the ``coset
model''~\cite{BH,GKOct}.
Given a $G_k$ WZNW model with central charge $c_G$, a subgroup $H_{k'}$
(of appropriate level $k'$ and central charge $c_H$) could be used to
generate a new Virasoro algebra of central charge $c_{G/H}=c_G-c_H$.

A Lagrangian formulation for this method was found by gauging an
anomaly-free subgroup $H\subset G$~\cite{Nahm,BRSct}. The Lagrangian does
not contain a kinetic term for the gauge fields. In the absence of these
terms, the gauge theory is also conformal. Modular invariance of the $G_k$
model ensures modular invariance of the coset. The coset Lagrangian
reproduces the central charge $c_{G/H}$, and the appropriate spectrum.
The gauged WZNW model has as basic fields the matter ones, appearing in the
$G_k$ model, and the gauge fields. One may rewrite the gauged model in terms
of non-linear $\s$-model variables. This is done by integrating out the
gauge fields \cite{BCR}. To one loop the backgrounds need to be
supplemented by a dilaton field \cite{Wbh}.

\begin{subsubsection}{The Euclidean $SL(2,\R)/U(1)$ Black Hole and its Dual}

To write the action for the gauged $SL(2,\R)$ WZNW model we will parametrize
the group manifold by ``Euler angles.'' Following the notation in
\cite{DVV} we introduce real coordinates
$x,\theta_1,\theta_2$, and write $g\in SL(2,\R)$ as
\be
g=e^{i\theta_1\sigma_3} e^{x\sigma_2} e^{i\theta_2\sigma_3} ,
\label{sl2par}
\ee
with $\sigma_a$ the Pauli matrices. The coordinates $\theta_1,\theta_2$
are such that $\th=\th_2-\th_1$ and
$\td=\th_1+\th_2$ have periodicity $2\pi$, and $0\leq x < \infty$.
This parametrization is convenient for describing the CFT of the Euclidean
$SL(2,\R)/U(1)$ coset; for the Lorentzian case one should simply replace
$\sigma_3\rightarrow \sigma_1$ and $\theta_{1,2}\rightarrow t_{1,2}$.

Inserting (\ref{sl2par}) into  (\ref{Sofg}), one finds that
in these target space coordinates, and in complex worldsheet coordinates,
the $SL(2,\R)_k$ WZNW action is
\be
S_{WZNW}[x,\theta_1,\theta_2]={-k\over 2\pi}
\int d^2z(-\pa x \pab x +
\pa \th_1 \pab \th_1 + \pa \th_2 \pab \th_2
+ 2\cosh 2x\; \pa \th_2 \pab \th_1).
\label{sl2wzw}
\ee
This action has the same structure as $S_1+S[x]$ in (\ref{Dd})  with
$\Sigma=\cosh 2x$, $\Gamma=-1$ and $\Gamma^1=\Gamma^2=0$
(except for an overall $k$ factor\footnote{
The generalization of the procedure of section 4.2 to the case when there
is an overall  $k$ factor in front of the action (\ref{Dd}) can be done
straightforwardly; we have presented only the $k=1$ case in section 4.2
because this is already sufficient  to generate {\em all} the
$D$-dimensional backgrounds with $d$ Abelian symmetries.},  and the
antisymmetric term $S_a$ that is understood to be present).

The action (\ref{sl2wzw}) is invariant under the $U(1)_L\times U(1)_R$
affine  symmetry generated by the currents
\be
J=-k(\pa \th_1 + \cosh 2x\; \pa \th_2), \qquad
\bar{J}=-k(\pab \th_2 + \cosh 2x\; \pab \th_1).
\label{sl2J}
\ee
These currents are the same as in (\ref{cur}) up to the overall $k$ factor
discussed above.
Therefore, the gauged WZNW action takes the form (\ref{gauge})
\ber
S_{gauged}&=&S_{WZNW}[x,\th_1,\th_2]\nonumber\\
&&+\frac{k}{2\pi}\int d^2z
[-A(\pab \th_2 + \cosh 2x\;\pab\th_1)-\bar{A}(\pa\th_1 + \cosh 2x\;\pa\th_2)
+{1\over 2}A\bar{A} (1+\cosh 2x)].\nonumber\\
&&
\eer{sl2gauged}
Integrating out the gauge fields and taking $k\rightarrow -k$ gives:
\be
S[x,\th]={1 \over 2\pi}\int d^2z [k(\pa x \pab x + \tanh^2 x \pa \th \pab
\th) - \frac{1}{4} \phi(x)R^{(2)}],
\label{sxth}
\ee
where $\th$ is defined in (\ref{per}), the $\tanh^2 x$ is found by using eq.
(\ref{ij}), and the dilaton is derived by eq. (\ref{f}):
\be
\phi(x)=\phi_0+\log (\cosh^2 x).
\label{sl2dil}
\ee

Up to $1/k$ corrections, the metric in (\ref{sxth}) and
the dilaton field  in (\ref{sl2dil}) satisfy
the one-loop beta-function equation~\cite{EFR,MSW} of ref.~\cite{callan}
\be
R_{ab}=D_aD_b\phi,
\label{cur6}
\ee
where $R_{ab}$ is the Ricci tensor of the target space.

The central charge of the $SL(2,\R)/U(1)$ model is
\be
c=2+\frac{6}{k-2}=2+3\epsilon^2+{\cal O}(\epsilon^4),
\label{cur8}
\ee
where
\be
\epsilon=\sqrt{2/k}.
\label{cur9}
\ee
It is convenient to absorb the overall factor $k$ in the coordinate $x$.
Defining $r=x/\epsilon$, one finds the action
\be
S[r,\theta]=\frac{1}{2\pi}\int d^2\sigma
\sqrt{h} h^{ij}\left( \partial_i r\partial_j
r+\frac{\tanh^2 \epsilon r}{\epsilon^2}
\partial_i\theta\partial_j\theta\right)
-\frac{1}{8\pi}\int d^2\sigma \sqrt{h} \phi(r,\theta) R^{(2)},
\label{cur10}
\ee
where
\be
\phi(r)=\phi_0+\log \cosh^2\epsilon r .
\label{cur11}
\ee
For completeness and clarification we wrote the action in (\ref{cur10}) with
the worldsheet coordinates $\sigma_i$ (\ref{zst})
and worldsheet metric $h_{ij}$ appearing explicitly, as in section 2.1.

The action in (\ref{cur10})
describes a $\s$-model on a background with scalar curvature
\be
{\cal R}(r)=\frac{4\epsilon^2}{\cosh^2\epsilon r},
\label{cur12}
\ee
and a metric described by the line element
\be
ds^2=dr^2+\frac{\tanh^2\epsilon r}{\epsilon^2}d\theta^2.
\label{cur13}
\ee
This background defines a semi-infinite cigar (see figure 1.F) with radius
\be
R_0(r)=\frac{\tanh \;\epsilon r}{\epsilon}.
\label{cur14}
\ee
This target space can be interpreted \cite{Wbh} as a Euclidean black hole
(see for example~\cite{blackhole,bh2}) with a
temperature $1/\beta$, which is
given by the inverse of the radius at infinity, namely
\be
\beta={1\over \epsilon}.
\ee

We now turn to the target space duality transformation of the Euclidean black
hole \cite{giveon,DVV,bars}.
Let us consider a background of the type
\be
G_{ab}=\left(\matrix{1&0\cr
0&R^2(r)\cr}\right).
\label{cur18}
\ee
As it was shown in section 4.2,
the metric in (\ref{cur18})  and its appropriate
dilaton (that solve eq. (\ref{cur6})) transform under target space duality
by
\be
R(r)\rightarrow R'(r)=\frac{1}{R(r)},\;\;\;\;\;\phi\rightarrow
\phi'=\phi+ \log R(r)^2.
\label{cur19}
\ee
This symmetry also survives  when $r$ is a time-like coordinate
\cite{SP}. In the limit
of a constant background, the duality transformation in (\ref{cur19})
coincides with the one in (\ref{10}), describing the simple circle duality
$R\rightarrow 1/R$~\cite{KY,SS}.

In the case of a Euclidean black hole $R(r)$ is given in eq. (\ref{cur14}),
and therefore, the dual action takes the form
\be
S'[r,\theta]=\frac{1}{2\pi}\int d^2\sigma
\sqrt{h} h^{ij}(\partial_i r\partial_j
r+\epsilon^2 \coth^2 \epsilon r\;
\partial_i\theta\partial_j\theta)-\frac{1}{8\pi}\int d^2\sigma \sqrt{h}
\phi'(r,\theta) R^{(2)},
\label{sl2dual}
\ee
where
\be
\phi'= \log \sinh^2\epsilon r + \phi_0 - \log \epsilon^2.
\label{dualdil'}
\ee
The dual target space has the form of an infinite trumpet (see figure
1.F), with an $r$-dependent radius
\be
R'_0(r)=\epsilon \coth \epsilon r.
\ee
When $r$ approaches 0 the radius diverges, while at the limit
$r\rightarrow\infty$
the radius approaches $\epsilon$ asymptotically. The ``temperature''
of the ``dual black hole'' is therefore
\be
{1\over \beta'}=\beta={1\over \epsilon}.
\ee
It is remarkable that while the Euclidean black hole background has a
trivial fundamental group, the dual model has $\pi_1({\rm trumpet})=\Z$.

To clarify the property of the symmetry, let us discuss
the $\epsilon\rightarrow 0$ limit. The line element (\ref{cur13}) in this
limit describes a flat open two dimensional space in polar coordinates
\be
ds^2_{\epsilon\rightarrow 0}=dr^2+r^2d\theta^2.
\label{cur24}
\ee
The dominant states in the Hilbert space for small $\epsilon$ are therefore
``momentum'' states. Those may be
interpreted as particles sent towards the black hole.
The target space of the dual model, although complicated in terms of the
$r,\theta$ coordinates, for which
\be
(ds')^2_{\epsilon\rightarrow 0}=dr^2+r^{-2} d\theta^2,\;\;\;\;\;
\phi=\phi_0+\log r^2,
\label{cur25}
\ee
is similar, as a CFT, to a two dimensional torus which partly shrinks
to a point. The dominant states in the
Hilbert space for small $\epsilon$ are now ``winding modes.''
Such states do not
have an analog in standard general relativity or in Kaluza-Klein theory; they
are a direct consequence of a string wrapping around circles while
propagating in a compact target space. The duality transformation is a quantum
symmetry of string vacua which interchanges momentum modes with winding
modes\footnote{
In ref. \cite{K2} the duality of $D=2$ flat space in polar coordinates
was further studied in the mini-superspace approximation. This is an
attempt to understand duality in case the target space $M$ has no
$\pi_1(M)$.}.

To lowest order in worldsheet perturbation theory,
the graviton-dilaton system can be described by a space-time effective action
\ba
L &=& \int d^2X \sqrt{G}e^{\phi}(
{\cal R}+G^{ab}\partial_a\phi\partial_b\phi+\frac{8}{k-2})\nonumber\\ &=&
\int d^2X Re^{\phi}\left(-\frac{\partial^2R^2}{R^2}+2\left(\frac{\partial
R}{R}\right)
^2+\partial \phi \partial \phi - \frac{8}{k-2}\right).
\label{cur26}
\ea
The additive constant $8/(k-2)$ originates from the familiar $D-26$ of the
bosonic string. The action in (\ref{cur26}) is invariant under duality:
$R\rightarrow R^{-1},\; \phi\rightarrow \phi+\log R^2$. Thus, the
graviton and dilaton field equations derived from this action are duality
invariant as well.
\end{subsubsection}

\begin{subsubsection}
{The Lorentzian $SL(2,\R)/U(1)$ Black Hole and its (Self-)Dual}

We now turn to the study of
the analytic continuation of the black hole to a Lorentz signature.
Let us first discuss the analytic continuation and the Kruskal
extension~\cite{blackhole,bh2} of the original black hole
following~\cite{Wbh,giveon}. Naively, one simply sets
$\theta=it$~\footnote{
Instead of obtaining the Lorentz signature black hole by
a formal analytic continuation from a Euclidean space,
one can obtain it directly
as a conformal field theory by gauging a different $U(1)$ subgroup of
$SL(2,\R)$~\cite{Wbh}.},
whereupon the line element (\ref{cur13}) is changed to
\be
ds^2=dr^2-\frac{\tanh^2 \epsilon r}{\epsilon^2}\;dt^2.
\label{cur30}
\ee
This metric seems to have a singularity at $r=0$, but this must be purely a
coordinate singularity, since the scalar curvature (\ref{cur12})
is regular at $r=0$.
The analytic continuation past the coordinate singularity can be found by
an extension, similar to the one used for the
Schwarzschild solution~\cite{blackhole,bh2}.
Defining the $u,v$ coordinates
\be
v=\sinh \epsilon r\; e^{t},\;\;\;\;\; u=-\sinh \epsilon r\; e^{-t},
\label{cur33}
\ee
one gets
\be
ds^2=-\frac{1}{\epsilon^2}\frac{dudv}{1-uv}.
\label{cur35}
\ee
The dilaton field is given by (\ref{cur11})
\be
\phi=\phi_0+\log (1-uv).
\label{cur36}
\ee
This solution is similar to that of a black hole in Kruskal-Szekeres
coordinates as discussed in~\cite{Wbh,MSW}.

A space-time diagram for the $u,v$ coordinates is shown in figure 4.A.
This diagram has the same causal structure as the corresponding
four-dimensional black hole. In an astrophysical black hole,
which forms from a spherically
symmetric collapsing star, some of the regions are missing, but one still has
parts of regions I,II and V. The
causal structure of the extended space-time is easily seen from the diagram in
figure 4.A since, by construction, the $u=0$ and $v=0$ lines (the horizons)
are null geodesic lines.
Region I in figure 4.A corresponds to the original ``Lorentzian cigar''
($\sinh^2 \epsilon r > 0$ i.e. $uv<0$), and represents the space-time
region outside the past and future horizons. Region II represents the
space-time between the future horizon and the future singularity.
In general relativity a signal sent from region V
(where time direction goes sideways) cannot cross the horizon towards
an observer in region I.
The black hole singularity at $uv=1$, rather than being the
future as  appearing to an observer in regions I,II,
occurs at the end of the spatial world to an observer in region V.
(Region III has exactly the ``time reversed'' properties of region II, and is
referred to as a white hole. Region IV has properties identical to
region I, and region VI is a continuation over the past singularity.
An observer in region I cannot communicate
with any observer in region IV.)

We now come to a striking property of the  stringy black hole: we
will argue that {\em target space duality interchanges the horizon with
the singularity}~\cite{giveon,DVV}. Therefore,
signals can pass smoothly through the singularity $uv=1$.
Such signals should be associated with the winding modes of the string
as was discussed in the  Euclidean case.
To understand this point let us present the duality transformation of the
original Lorentzian background.

The dual Lorentzian black hole is given by the analytic continuation of the
dual Euclidean black hole. The metric background is thus described by the line
element
\be
(ds')^2=dr^2-\epsilon^2 \coth^2 \epsilon r\;dt^2.
\label{cur38}
\ee
We will repeat a  procedure similar to the coordinate transformation
done for the original black hole. With $u,v$ coordinates defined by
\be
v=\cosh \epsilon r \; e^{\epsilon^2 t},\;\;\;\;\;
u=\cosh \epsilon r \; e^{-\epsilon^2 t},
\label{cur41}
\ee
one gets
\be
(ds')^2=-\frac{1}{\epsilon^2}\frac{dudv}{1-uv}.
\label{cur43}
\ee

The line element $(ds')^2$ of the dual black hole is thus identical to the line
element (\ref{cur35}) of
the original black hole. In the $u,v$ coordinates
one cannot
distinguish between the original background and its duality partner. This is
not bad since the two theories must be equivalent as conformal field
theories.
What is therefore new in regarding the dual black hole?
The answer is remarkable:
duality interchanges the asymptotically flat space-time,
corresponding to the Lorentzian cigar,
with the space-time inside the black hole singularity; that is,
regions I and V  in figure 4.A are
interchanged. (Region II is transformed to itself; regions IV and VI are
interchanged and region III is transformed to itself).
As a consequence,
the physics studied by an observer in region I is equivalent to the physics
in region V.
In particular, the dual transform of a signal leaving region I will be a
signal leaving region V. As it was discussed before, this is related to the
quantum symmetry which interchanges momentum modes with winding modes,
a result of target space duality.

So far we have described the Lorentzian $SL(2,\R)/U(1)$ black hole and its
target space duality partner in the graviton and dilaton background.
To conclude the study of this example,
let us now consider the propagation of a small ``tachyon'' disturbance in the
Lorentzian black hole \footnote{The propagation of a tachyon disturbance in
the Euclidean black hole was studied in \cite{EFR}.},
and verify explicitly its invariance under duality.
In the black hole space the tachyon effective action to
lowest order in $\alpha'$ is
\be
L(T)=\int\; dudv[(1-uv)\partial_uT\partial_vT-8\epsilon^2T^2].
\label{cur51}
\ee
The tachyon field equation is therefore
\be
\partial_u[(1-uv)\partial_vT]+
\partial_v[(1-uv)\partial_uT]+16\epsilon^2T=0.
\label{cur52}
\ee
Define
\be
z=1-uv,
\label{cur53}
\ee
then, with the ansatz $T=T(z)T(t)$,
the equation for $T(z)$ is a hypergeometric differential equation
\be
z(1-z)\frac{\partial^2}{\partial z^2}T(z)+(1-2z)\frac{\partial}{\partial z}T(z)
+8\epsilon^2 T(z)=0.
\label{cur54}
\ee
The differential equation is invariant under the duality transformation
$z\rightarrow 1-z$. There are two linearly independent solutions to eq.
(\ref{cur54}) given by
\be
T_1(z)=F(\alpha, 1-\alpha;1;z),\qquad \alpha=-8\epsilon^2 +
{\cal O}(\epsilon^4),
\label{cur55}
\ee
\be
T_2(z)=T_1(1-z).
\label{cur56}
\ee
Here  $F(\alpha,\beta;\gamma;z)$ is the hypergeometric function. The solution
$T_2$ is dual to the solution $T_1$.
The function $T_2(z)$ has a
singularity at $z=0\;\;(uv=1)$, although regular at $z=1$ ($uv=0$),
and as a result, a physical singularity is created
at $uv=1$.
The dual function $T_1(z)$ has a
singularity at $z=1\;\;(uv=0)$, although regular at $z=0$ ($uv=1$).

To summarize, the target space duality transformation of the $SL(2,\R)/U(1)$
space-time leads to strange stringy properties of the black hole.
Although we have
discussed a particular black hole in string theory, these properties shall be
generalized to other black hole solutions (as well as cosmological
solutions, monopoles, dipoles and other topological backgrounds).
There is an extensive discussion on the generalization of the $D=2$ black
hole duality to other backgrounds in the literature (for example, see
\cite{BDDF,tseytlin,tv,Cdu,BSdu,GR,CQRdu,CDdu,Hdu,HHdu,Ndu,H2du,GQdu,KMdu}
\cite{GPa,KKSdu,HWdu,GVdu,KKLdu}.)
In the next section we will describe some more applications of duality.

\end{subsubsection}
\end{subsection}

\begin{subsection}{Applications II: Charged Black Objects in D=2,3}
\setcounter{equation}{0}

In this section we explore  more consequences of the discrete
duality symmetries \cite{GR}.  We first discuss an exact
$D=3$ closed string background that is independent of $d=2$
coordinates \cite{ils}.  We then turn to $D=2$ heterotic backgrounds
\cite{pol,mny}.  In both cases, we find that uncharged black compact
objects (strings or holes) are equivalent to charged $D=2$ black
holes \cite{hhs,GR}.

\begin{subsubsection}{The Closed String Example}

The simplest nontrivial example after the $D=2$ black hole duality,
presented in section 4.5, is
a compact black string obtained by attaching a circle to every point of the
$D=2$ black hole space-time ($SL(2,\R )_k/U(1)\times U(1)$). To leading order,
the action is
\be
\label{bstr}
S_{Black String}=\ind \[k(\pa x \pab x + \t\pa\th^1\pab\th^1)
+\a\pa\th^2\pab\th^2 -\fr14\f (x)R^{(2)}\]\, ,
\ee
where
\be
\f (x)=\f_0 + \log (\ch)\, .
\ee
The first term in $S_{Black String}$ is  a $\s$-model in
the Euclidean black hole metric, and the second
term corresponds to a free scalar field compactified on a circle of
radius $\sqrt\a$.
This $D=3$ background is independent of $d=2$ coordinates $\th^i$, and is
described by the background matrix (\ref{calG})
\be
\label{back}
{\cal E} =\left(\matrix{E&0\cr
                0&F\cr}\right)\, ,
\ee
where
\be
\label{start}
E=\left(\matrix{k\, \t&0\cr 0&\a}\right)\q F=k \, .
\ee

The group of generalized duality transformations $O(2,2,\Z )$ maps this
background into other backgrounds that (in general) have different
space-time interpretations.  {Since}
$F_1=F_2=0$ in (\ref{back}, \cf \ref{calG}),
$O(2,2,\R)$ acts on the background by transforming only $E$ and $\f$ as
given in (\ref{tE}) and (\ref{tdil}). A particularly interesting point on
the orbit of $O(2,2,\Z )$ is reached by acting with the element
\be
g=\mat I\Theta 0I\mat p00p\mat 0II0 \mat I{-\Theta}0I
=\mat{\mat 100{-1}}{\mat 0000}{\mat 0110}{\mat 100{-1}}\, ,
\ee
where
\be
I=\mat 1001 \q \Theta=\mat 01{-1}0 \q p=\mat 0110 \, .
\ee
One finds that $E$ and $\f$ are transformed to
\be
\eqalign{
E'=g(E)&=\fr1{1+\a k\, \t}\mat {k\,\t}{\a k\,\t}{-\a k\, \t}{\a} \cr
&\cr
&=
\fr1{\ch -\l}\mat {k(1-\l )\sh}{\l\sh}{-\l\sh}{\l\ch /k}\cr}
\ee
\be
\label{dualdil}
\f'(x)=\f_0-\log (1-\l ) +\log (\ch -\l ) \, ,
\ee
where
\be
\label{lam}
\l=\fr{k\a}{1+k\a} \, .
\ee
This gives the worldsheet action
\be
\label{charge}
\eqalign{
S_{charged} &= \ind \[k\left(\pa x \pab x + \fr{(1-\l )\sh}{\ch-\l}
\pa\th^1\pab\th^1\right)\cr
&\cr
\qquad\qquad&+
\fr{\l\sh}{\ch-\l}(\pa\th^1\pab\th^2-\pa\th^2\pab\th^1)
+\fr{\l\ch}{k(\ch-\l )}\pa\th^2\pab\th^2 -\fr14\f'(x)R^{(2)}\]\, ,\cr}
\ee
and, after Wick-rotating $\th^1\ra it$, it
corresponds to a charged black hole of the type found in \cite{ils} with
mass
\be
\label{mass}
M=(1-\l )M_0=\sqrt{\fr2k}e^{\f_0}\q M_0=\sqrt{\fr2k}e^{\f'_0}\, ,
\ee
and charge
\be
\label{q}
{\cal Q}=\sqrt{\l (1-\l )}\fr{2M_0}k=\sqrt{\fr\l{1-\l}}\fr{2M}k \, ,
\ee
where $\f'_0=\f_0-\log (1-\l )$ is the constant part
of the dual dilaton (\ref{dualdil}).
The action (\ref{charge}) is related to the $\sigma$-model
action for the coset
$(SL(2,\R )_k\times U(1))/U(1)$ \cite{ils} by rescaling
$\th^2\ra k\th^2$.\footnote{Our $k$ matches \cite{Wbh}, which is $2k$
of \cite{ils}.}

This shows that {\em the charged black hole is equivalent to the compact black
string} as a conformal field theory.  We now consider some particular
limits of this solution.

The limit $\a\ra 0$ in (\ref{bstr}) corresponds to  the direct product of
the $2D$ black hole and a circle with radius $r\to 0$,
and the limit $\a\ra \infty$ is related to the previous
one by the $R\ra 1/R$ circle duality.
These limits correspond to the limits $\l\ra 0,1$ in (\ref{lam}). In
$S_{charged}$ (\ref{charge}), the background at
$\l\ra 0$ is the direct product of
the $2D$ black hole background and the same $r\to 0$ circle;
however, the $\l\ra 1$ limit gives
the action (modulo an integer total derivative term)
\be
S_{\l\ra 1}=\ind \[k\pa x \pab x + \fr1k\coth^2x\pa\th^2\pab\th^2
+(1-\l )\pa\th^1\pab\th^1 -\fr14\f' (x)R^{(2)}\]\, ,
\ee
$$
\f' (x) = \f_0-\log (1-\l ) + \log(\sh ) \, .
$$

\noindent
The background in this action is the direct product of the dual $2D$
black hole background and a $r=0$ limit of a circle.  In both
cases, the degenerate limits are equivalent as CFTs to a
{\it noncompact\/} black string.
\end{subsubsection}

\begin{subsubsection}{The Heterotic String Example}

Following \cite{GR},
we focus on the example of \cite{mny}, which is a $D=2$ heterotic string, with
internal degrees of freedom taking values in a standard 12-dimensional lattice
(the vector weights of $SO(24)$)\footnote{
More precisely, only space-time bosons
have internal quantum numbers in the vector representations of $SO(24)$;
space-time fermions have internal quantum numbers in the spinor representations
of $SO(24)$.}. We find that {\em a family of  charged
black holes (and naked singularities) is dual to a neutral one},
which is the exact CFT given by
the heterotic $D=2$ black hole \cite{mny}.

We start with a heterotic $D=2$ action:
\be
\label{shet}
\eqalign{
S_{Heterotic}=\ind \[ k(\pa x\pab x +\t\pa\th\pab\th) +\pa Y^A\pab Y^A
-\fr14\f (x) R^{(2)}&\cr
 + \rm{(fermionic~terms)}&\]\, ,\cr}
\ee
where $A=1,\dots ,12$, $k=5/2$ (for criticality), and $\f=\f_0+\log (\ch)$.
This action describes
a neutral heterotic $D=2$ black hole.
The conformal field theory (\ref{shet}) corresponds to a background
(\ref{xi})
\be
\label{bhet}
\Xi=\left(\matrix{k&0&0\cr 0&k\,\t&\cr 0&0&I\cr}\right)\, ,
\ee

\noindent
where the internal background $I$ is the $12\times 12$ identity matrix
corresponding to the vector  weights of $SO(24)$.  Only a
$2\times2$ block $E$
of the matrix $\Xi$ is affected by the discrete transformations we discuss
in this example:
\be
\label{toy}
E=\mat{k\,\t}001\, .
\ee

By
transforming $E$ and $\f$ with a group element $g_n\in O(1,2,\Z )
\subset O(1,13,\Z )$ (where $n $ is an arbitrary integer):
\be
\label{gggg}
g_n=\mat0II0\mat I{n\Theta}0I\mat{A_n^t}00{A_n^{-1}}=
\mat{\mat0000}{\mat10{-n}1}{\mat1n01}{\mat{-n^2}n{-n}0}\, ,
\ee
where
\be
I=\mat1001\q \Theta=\mat01{-1}0\q A_n=\mat10n1 \, ,
\ee
one finds
\be
\label{Ehet}
g_n(E)=E_n'=\mat{(n^2+k\,\t )^{-1}}{-2n(n^2+k\,\t )^{-1}}01\, ,
\ee
\be
\label{dhet}
\f' (x)=\f_0+\log (n^2+k)+\log (\ch -\fr k{n^2+k}) \, .
\ee

After rescaling
\be
\th\ra \fr{k+n^2}{\sqrt{k}}t\, ,
\ee
and defining the coordinate $r$
to be a linear function of the dilaton $\f'$ (\ref{dhet}), given by
\be
\label{rho}
r  = {1\over Q}\log (\ch -\fr k{n^2+k}) \, ,
\ee
where $Q$ is a constant determined below,
the background
(\ref{Ehet}) gives rise to the action\footnote{Recall that
${\cal G}_{tt}=\Xi_{tt}'-\fr14 {\cal A}^2$, see eq. (\ref{xi}).}
\be
\label{Lchar}
\eqalign{
S_{charged}=\ind \[ f(r )\pa t \pab t +
f(r )^{-1} \pa r\pab r -A(r)\pa t \pab Y^1 + \pa Y^A\pab Y^A&\cr
-\fr14\f' (r) R^{(2)} + {\rm (fermionic~terms)}&\]\, . }
\ee
Here
\be
f(r)=1-2me^{-Qr}-q^2e^{-2Qr},
\ee
\vskip .05in
$$
A(r)=nQ+2qe^{-Qr} \q \f' (r)=Qr+\f_0+\log (n^2+k)\, ,
$$
\vskip .1in
\noindent
where $Q=2/{\sqrt{k'}}$ is determined by the normalization of
${\cal G}_{rr}$ in (\ref{Lchar}), and
\be
\label{QMq}
 2m=\fr{n^2-k'}{n^2+k'} \q
 q=\fr{n\sqrt{k'}}{n^2+k'}\, .
\ee
Following~\cite{Wbh} we have replaced $k$ with $k'=k-2=1/2$ in (\ref{QMq}).
By Wick-rotating $t\ra it$, along with
$q\ra -iq$ (necessary to maintain hermiticity of the action), the
theory (\ref{Lchar}) with $|n|>1$ describes
a $D=2$ charged black hole with mass and charge \cite{mny}
\be
M=Q(n^2-k')e^{\f_0}\q {\cal Q}=n\sqrt{8}e^{\f_0}\, .
\ee
For $n=-1,0,1$, the theory
(\ref{Lchar}) describes a naked singularity.

We emphasize that these
backgrounds, for all $n$, are different space-time
interpretations of the {same\/} CFT: the CFT given by the neutral
heterotic $D=2$ black hole.
\end{subsubsection}
\end{subsection}

\begin{subsection}{Applications III:
Cosmological  Backgrounds with Abelian Symmetries}
\setcounter{equation}{0}

In this section we discuss a
class of cosmological solutions in string theory (in the presence of
Maxwell fields) that are obtained by $O(d,d,\R)$ transformations of simple
backgrounds with $d$ Abelian symmetries, following~\cite{GPa}.
Similar cosmological solutions are discussed extensively in the literature
(for examples, see \cite{Mcosmo,KL,NW,Vremsin,KMcos,AKcos,KM1cos}).
In some of the examples we will find  a
(closed) expanding universe. In all the cases for which the
universe has a smooth and complete initial value hypersurface,
a naked singularity can form only at the time when the universe collapses.
The discrete symmetry group $O(d,d,\Z)$ identifies different
cosmological solutions with a background corresponding to a
(relatively) simple conformal field theory (CFT), and therefore,
may be useful in understanding the properties of
naked singularities in string theory. The consequences of duality in
cosmological string backgrounds were also discussed, for instance, in
\cite{BVcosmo,GSVYcosmo,tseytlin,tv}

To find realistic cosmological solutions in
string theory, we shall represent a classical
solution by $M\times K$, where $M$ is a $2d$ CFT
with a four dimensional target space-time, and with a
central charge $c=4$, and $K$ is some internal space
represented also by a CFT. Moreover, although the formation of
singularities is one of the interesting questions in such solutions,
we do not want the
singularities to be ``built in'' the initial conditions. Namely,
we want $M$
to have a smooth and complete initial value hypersurface.

We shall first construct new solutions by $O(d,d,\R)$ rotations
with respect to Abelian symmetries of the space-time background $M$.
Let us consider the 4-$D$ line element
\be
d{\bf s}^2=-dt^2+ds^2+g(t)^2 d\th_1^2+f(s)^2 d\th_2^2.
\label{ds2}
\ee
We want this background to correspond to a CFT\footnote{
{}~More precisely, we should start with
$d{\bf s}^2=k(-dt^2+g(t)^2d\th_1^2)+k'(ds^2+f(s)^2d\th_2^2)$,
and choose $k$ and $k'$ such that $c=4$.
To leading order in $\a'$ this condition is $k = k'$, and for
simplicity we take $k=k'=1$.}.
There are four nontrivial possibilities for
$g(t)$ and four possibilities for $f(s)$. To leading order in $\a'$ the
possibilities are\footnote{
{}~All these possibilities correspond to the exact CFT given by a
direct product of two cosets $\frac{G}{H}$, where $G$ is either $SU(2,\R)$
or $SL(2,\R)$ and $H$ is either $U(1)$ or ${\bf R}$ (see section 4.5).}:
$$g(t)=\tan(t), \cot(t), \tanh(t) \;\; {\rm or} \;\;\coth(t),$$
and
$$f(s)=\tan(s), \cot(s), \tanh(s)\;\; {\rm or} \;\;\coth(s).$$
The dilaton is then given by
\be
\phi(s,t)=\phi_0+\log(\bar{f}(s)^2\bar{g}(t)^2),
\label{phits}
\ee
where $\phi_0$ is a constant and
$$\bar{g}(t)=\cos(t),\sin(t),\cosh(t)\;\; {\rm or} \;\; \sinh(t),$$
$$\bar{f}(s)=\cos(s), \sin(s), \cosh(s)\;\;{\rm or} \;\;\sinh(s),$$
respectively.

The background in (\ref{ds2}),(\ref{phits}) is a $D=4$ curved background
that is independent of $d=2$ coordinates. It has one time-like
coordinate $t$,
and three space-like coordinates $s,\th_1,\th_2$.
The background is time dependent, and
therefore, we say that
it describes a cosmological solution to Einstein's equations.
The $2\times 2$ matrix $E_{ij}(s,t)$ (\ref{calG})
in the worldsheet action (\ref{D}) (with $i,j=1,2$ and $a,b=s,t$) is
\be
E(s,t)=\left(\matrix{g(t)^2 &0\cr
                     0&f(s)^2 \cr}\right).
\label{Efg}
\ee

We can now generate new cosmological solutions by acting on ($E,\phi$)
in (\ref{Efg}) and (\ref{phits}) with $O(2,2,\R)$
transformations as described in section 4.2.
Let us discuss a particular one parameter sub-family
of $O(2,2,\R)$ rotations.
By transforming $E$ and $\phi$ with the group element $g_b\in O(2,2,\R)$
(where $b$ is an arbitrary real number):
\be
g_b=\left(\matrix{0&I\cr
                  I&0\cr}\right)\left(\matrix{I&b\Theta\cr
                                              0&I\cr}\right),
\label{gb}
\ee
where
\be
I=\left(\matrix{1&0\cr
                0&1\cr}\right),\qquad
\Theta=\left(\matrix{0&1\cr
                    -1&0\cr}\right),
\label{IT}
\ee
one finds that the new background matrix, $E'_b$, is
\be
g_b(E)=E'_b=(E+b\Theta)^{-1}=
\frac{1}{f(s)^2 g(t)^2+b^2}\left(\matrix{f(s)^2 & -b\cr
                                             b &g(t)^2 \cr}\right).
\label{NW}
\ee
Namely, the new background matrix is given by adding a constant
antisymmetric background $\left(\matrix{0&b\cr
                                       -b&0\cr}\right)$ to $E$, and then
using duality  to invert the background matrix.
{}From eq. (\ref{dtrans}) and the discussion in subsection 4.2.4
one finds that the new dilaton is
\be
\phi^\prime (s,t) = \phi_0 + \log \left[\bar{f}(s)^2\bar{g}(t)^2
\left(f(s)^2g(t)^2 + b^2\right)\right]\ .
\label{NWphi}
\ee

\noindent
For the particular choice
\ba
f(s)=\tan(s)\ &,&\qquad  g(t)=\cot(t)\ ,\nonumber \\
\bar{f}(s)=\cos(s)\ &,& \qquad \bar{g}(t)=\sin(t)\ ,
\label{fgtan}
\ea
the conformal background corresponds to a
closed expanding universe \cite{NW}\footnote{
To get the solution in  \cite{NW} (corresponding to the $\sigma$-model of
the $c=4$ coset
${SL(2,\R)\times SU(2,\R) \over \R\times \R}$)
one should rescale $\theta_1\to b\theta_1$, reintroduce $k$ and $k'$, and take
$k=k'$ very large such that the maximal size of the universe is of
order $k$, and the central
charge of $M$ is $c=4$ (to leading order in $1/k$).}.

Let us consider now the case in which the internal space $K$
also has Abelian symmetries. Then by $O(d,d,\R)$ rotations one can turn
on gauge fields, namely, non-trivial
elements of the background matrix interpolating between the space-time
($M$) indices and the internal space ($K$) indices.

The action  (\ref{D}) can be
generalized adding internal degrees of freedom, {\it i.e.} extra
compactified coordinates $Y^A$, $A=1,...,d_{int}$.
(The coordinates $Y^A$ can be regarded as
either the internal coordinates of a
bosonic string or the extra coordinates of a heterotic string, as in
section 4.4 \cite{GR}). We start from the action
(\ref{het}), and we choose the initial structure of the background as before,
i.e., the block $E_{ij}$ is again given by eq. (\ref{Efg}), $F^1=F^2=0$,
and $F=\left(\matrix{-1 & 0\cr 0 & 1 \cr}\right)$, and we start with a
vanishing background gauge field: ${\cal A}=0$ in (\ref{het}).

It is now possible to make a more
general rotation using the group $O(2,2+d_{int},\R)\subset
O(2+d_{int},2+d_{int},\R)$ acting by the fractional
linear transformations (\ref{tE})
on the $(2+d_{int}) \times (2+d_{int})$ matrix (see eq. (\ref{xi}))
\be
\Xi= \left( \matrix{ E_{ij} & 0 \cr 0 & E_{AB}\cr}\right),
\ee
such that
\be
\Xi^\prime = \left( \matrix{ E_{ij}^\prime +
{1\over 4}{\cal A}_{iA}(G^{-1})^{AB} {\cal A}_{jB} & {\cal A}_{iA} \cr
0 & E_{AB}\cr}\right).
\label{Xip}
\ee
Here $G_{AB}$ is defined as in eq. (\ref{GAB}).
The structure of $\Xi^\prime$ is such that $E'_{ij}=G'_{ij}+B'_{ij}$
gives the correct metric after a dimensional reduction from $4+d_{int}$
dimensions to four space-time dimensions. The structure of the
second line in the matrix $\Xi^\prime$
allows the internal coordinates to be the extra coordinates of a
heterotic string as explained in sections 2.5 and 4.4.

The new background in (\ref{Xip})  corresponds to the
heterotic worldsheet action  given in (\ref{het}) with
non-vanishing background gauge field ${\cal A}$, or to a bosonic action
given by (\ref{het}) (neglecting worldsheet fermions in the bosonic case).
Both the old background and the new metric,
antisymmetric tensor, dilaton and gauge fields are
solutions of the equations of motion derived from
the effective action
\be
S_{eff}=\int d^4 X \sqrt{-G} e^{\phi} \left[
R^{(4)} + (\nabla \phi)^2 -
\frac1{12} H^2 - \frac{1}{16} {\rm Tr} F^2
\right]\ .
\label{Seffprime}
\ee
(The old background has $H_{\mu\nu\rho}=F_{\mu\nu}=0$.)

For simplicity,
we now consider the case in which $d_{int}=1$, {\it i.e.} we
start from a five dimensional background where the fifth coordinate is
compact and by $O(2,3,\R)\subset O(3,3,\R)$
rotations we  introduce a Maxwell field ${\cal A}_i$.
(The discussion can be easily generalized to the internal dimension
$d_{int}$ that is needed for criticality.)

We fix $E_{AB} = 1$ and we rotate $\Xi$ by the $g_{a,b,c}\in O(2,3,\R)\subset
O(3,3,\R)$ matrix
\be
g_{a,b,c} = \left(\matrix{0 & I\cr I & 0 \cr}\right)
\left(\matrix{I & \Theta\cr 0 & I \cr}\right)
\left(\matrix{A & 0\cr 0 & (A^t)^{-1} \cr}\right),
\label{gabc}
\ee
where
\be
\Theta = \left(\matrix{0 & b & c \cr -b & 0 & a \cr -c & -a & 0 }
\right),  \qquad\quad
A = \left(\matrix{1 & 0 & c \cr 0 & 1 & a \cr 0 & 0 & 1 }\right)\ ,
\ee
$I$ is the $3\times 3$ identity matrix and $a,b,c$ are arbitrary
real numbers.

After the rotation, the non-zero components of the metric $G^\prime_{IJ}$,
antisymmetric tensor $B^\prime_{IJ}$, gauge field  ${\cal A}_I$ and
dilaton $\phi^\prime$, are:
\ba
&& G^\prime_{tt} = -1 \qquad\qquad\qquad\qquad\qquad \ \
G^\prime_{ss} = 1 \nonumber\\
&& G^\prime_{\theta_1\theta_1} = \frac1{\Delta^2} \left[ g(t)^2 \left(
a^2 + f(s)^2\right)^2 + f(s)^2 \left(ac + b \right)^2 \right]
\nonumber\\
&& G^\prime_{\theta_1\theta_2} = \frac1{\Delta^2} \left[ b\left(
a^2g(t)^2 - c^2 f(s)^2\right) - ac \left( a^2 g(t)^2 + c^2 f(s)^2 +
2 f(s)^2 g(t)^2\right) \right]
\nonumber\\
&& G^\prime_{\theta_2\theta_2} = \frac1{\Delta^2} \left[ f(s)^2 \left(
c^2 + g(t)^2\right)^2 + g(t)^2 \left(ac - b\right)^2 \right]
\nonumber\\
&& {\cal A}_{\theta_1} = \frac1{2\Delta} \left[ ab - cf(s)^2\right]
\qquad \qquad \  {\cal A}_{\theta_2} = \frac1{2\Delta}
\left[ -bc - a g(t)^2\right] \nonumber\\
&& B^\prime_{\theta_1\theta_2} = -\frac{b}{\Delta} \qquad\qquad
\qquad\qquad\qquad
\phi^\prime = \phi_0 + \log \left[ \bar{f}(s)^2 \bar{g}(t)^2
\Delta\right]\nonumber\\
&& \det\left[ - G^\prime\right] = \frac{g(t)^2 f(s)^2}{\Delta^2}\, ,
\label{GBA}
\ea
where
\be
\Delta = b^2 + a^2 g(t)^2 + c^2 f(s)^2 + g(t)^2 f(s)^2 \ .
\ee
Obviously, for $a=c=0$, one gets back
the space-time solutions in (\ref{NW}) and (\ref{NWphi}).\\

It is also possible to make a Weyl rescaling in order to bring the effective
action  (\ref{Seffprime}) to the Einstein form
\be
S_{eff}^E =\int d^4 X \sqrt{-G^E} \left[
R^{E\, (4)} - \frac12 (\nabla \phi)^2 -
\frac1{12} e^{2\phi}
H^2 - \frac1{16} e^{\phi} {\rm Tr} F^2 \right]\, ,
\label{SeffE}
\ee
where the Einstein
metric $G^E_{\mu\nu} = e^{\phi} G_{\mu\nu}$ is used to contract the indices.

Notice that when $f$ and $g$ are given by eq. (\ref{fgtan}) one has
\ba
\phi^\prime &=& \phi_0 + \log \left[ b^2 \sin^2(t) \cos^2(s) + a^2
\cos^2(t) \cos^2(s)\right. \\
&&\qquad\qquad + \left. c^2 \sin^2(t)\sin^2(s) + \cos^2(t)\sin^2(s)
\right]\ . \nonumber
\ea
When the parameters $a,b$ and $c$ are non-zero, the dilaton is finite for
any value of the space-time coordinates $s,t$
(and therefore, the Weyl rescaling to the Einstein form does not
change the properties of the universe described above). When some of the
parameters $a,b,c$ are zero, the dilaton diverges in particular points in
space-time. Explicitly,
if $a=0$ the dilaton diverges at $t=0,\, s=0$; if $c=0$ it diverges
at $t=\pi/2 , \, s=\pi/2$; and if $b=0$ it diverges at
$t=\pi/2,\,s=0$ ($t , s \in [0,\pi/2]$).
The divergences of the dilaton appear when a curvature singularity is formed.

Let us discuss  briefly the target space
properties of some of the models  presented above. From eq. (\ref{GBA}),
we see that the volume of the universe, described by that
class of solutions, approaches zero when
$t\to 0$ (this is because $g(t=0)=0$ or $\infty$ for all four possible
choices of $g$). Therefore, the universe starts from a ``big bang.''
Later, the universe can either be an expanding and
contracting closed universe, in some cases collapsing at the end, or
an open universe. The different types of universes depend on the
different choices of the functions $f, g$ and the
parameters $a,b,c$. (We do not consider the models with
singularities that are built in the initial conditions and remain for any
time $t$, as it happens for instance
when $a=b=0$, $c \neq 0$, and for functions $f$ and $g$ given by
eq.~(\ref{fgtan}), or in the example of~\cite{KL}.)

Consider again the case when the functions $f$ and $g$
are given by eq. (\ref{fgtan}) (with the parameters $b$ and $c$ not both zero).
For this example, the backgrounds in (\ref{GBA})
describe an  anisotropic expanding and contracting universe, in the
presence of an antisymmetric tensor background, a Maxwell  field, and a
dilaton. The
topology of a spatial slice is $S^3$, the three sphere, and the universe
recollapses at the time $t=\frac{\pi}{2}$.

Next we discuss the formation of singularities in the universe described
above. There are different cases depending on the parameters $a,b$,
and $c$.
For $a=0$ the universe is singular at the ``big bang,''
namely, when $t=0$ a
singularity exists at $s=0$. For $b=0$ ($c=0$),
a naked singularity is about to form at $s=0$ ($s=\pi/2$) at the time
$t=\pi/2$, when the universe recollapses. For non-vanishing parameters $a,b$
and $c$ there are no singularities. Therefore, a naked singularity may
form only at the times when the universe collapses.

Because there is no singularity for non-vanishing parameters $a,b$ and $c$,
there are indications that one can {\em remove target space
singularities in string theory}
by continuous $O(d,d,\R)$ rotations \cite{GPa,Vremsin}.

Finally, we remark briefly on some consequences of the
duality group symmetries.
In the examples we have discussed, the matrix  $g_b$ in eq. (\ref{gb})
($g_{a,b,c}$ in eq. (\ref{gabc})) is an element of  the duality group
$O(2,2,\Z)$ ($O(3,3,\Z)$) if $b\in \Z$ ($a,b,c\in \Z$).
In particular, all the solutions (\ref{NW}) with
$b\in \Z$ are equivalent as CFTs to the $b=0$ case, for
which the background is a direct product of (a suitable
analytic continuation) of
two $D=2$ black hole  solutions discussed in section 4.5.
Moreover, all the solutions (\ref{GBA}) with $a,b,c\in \Z$ are also
equivalent CFTs. Therefore, these closed expanding universes, in the
presence of electromagnetic currents, are also equivalent as CFTs to the
direct product of (a suitable analytic continuation) of
two $D=2$ black hole  solutions, and an extra coordinate
compactified on a circle.

\end{subsection}

\begin{subsection}{Axial-Vector Duality as a Gauge Symmetry}
\setcounter{equation}{0}

The interpretation
of discrete symmetries in target space as gauge symmetries was
already discussed in the flat case in section 2.6.
Can we also interpret
the target space dualities in {\em curved backgrounds} as spontaneously
broken gauge symmetries (of an underlying gauge algebra)
in string theory? In this section we answer in the affirmative, concerning
particular elements of the $O(d,d,\Z)$ duality group, following~\cite{GKi}.
Namely, we show that any element of $O(d,d,\Z)$,
relating an axially gauged Abelian quotient
of a WZNW model to its vector
gauging \cite{Kir}, is an element of a spontaneously broken gauge group,
in the sense discussed in section 2.6.
In particular, this proves that the duality, relating the $\sigma$-model
background of an axially gauged Abelian
coset to a vectorially gauged coset, is an exact symmetry in string theory.
This is true for
compact groups, as well as for non-compact groups, and therefore, it has
important implications for the black-hole duality presented in section
4.5, and for the study of
singularities  in string theory discussed in section 4.7.
The extension to the full $O(d,d,\Z)$ might
be done along the lines of section 2.6 \cite{GMR1}.

This section is organized as follows:  In subsection 4.8.1 we start with
the simplest non-trivial case,  the $SU(2)$ (or $SL(2,\R)$)
model and its one-parameter sub-class of
marginal deformations. In subsection 4.8.2
we extend the discussion to a general group. In both cases,
we discuss the action of duality on the line of marginal
deformations, and its relation to a broken gauge transformation.
It will turn out that  the two boundaries of the deformation
line (which are related by this duality) correspond to the axial coset and
the vector coset (times a non-compact scalar field).

\begin{subsubsection}{$J\Jb$ Deformation of $SU(2)$ or $SL(2,\R)$ WZNW Models
and Duality as a Gauge Symmetry}

In this subsection we consider duality as a spontaneously broken
gauge symmetry for
the simplest nontrivial case, namely, duality acting on the deformation
line of $SU(2)$ or $SL(2,\R)$ WZNW models\footnote{We define
the conformal field theory corresponding to $SL(2,\R)$ to be the one
regularized by its Euclidean continuation. This regularization has been
shown to provide a consistent path integral prescription for the
$SL(2,\R)$ model \cite{Gaw}.}.

The worldsheet action of the  $SU(2)_k$ WZNW model is given
(in Euler angle parametrization as in (\ref{sl2par})) by
\ba
S[x,\tha,\thb]&=&S_1+S_a+S[x], \nonumber\\
S_1&=&\frac{k}{2\pi}\int d^2 z (\d\tha \db\tha + \d\thb \db\thb
+2\S (x) \d\thb \db\tha), \nonumber\\
S_a&=&\frac{k}{2\pi}\int d^2 z(\d\thb\db\tha - \d\tha\db\thb),\nonumber\\
S[x]&=&\frac{k}{2\pi}\int d^2 z\; \d x \db x -
\frac{1}{8\pi}\int d^2 z\; \p_0 R^{(2)},
\label{su2}
\ea
where $\S(x)=\cos 2x$, and $\p_0$ is a constant dilaton.
The action of the $SL(2,\R)_k$ WZNW model in (\ref{sl2wzw})
is obtained from (\ref{su2}) by taking
$x\rightarrow ix$ and $k\rightarrow -k$.

As discussed in subsection 4.2.1,
the antisymmetric term $S_a$ in (\ref{su2}) is (locally) a total
derivative, and therefore, may give only topological contributions,
depending on the periodicity of the coordinates $\th $.
To specify the periodicity, we follow subsection 4.2.1 and define
\be
\th=\thb-\tha\; , \qquad \tht=\tha+\thb\; ,
\label{th2}
\ee
such that
\be
\th\equiv \th+2\pi\; , \qquad  \tht\equiv\tht+2\pi\; .
\label{th1}
\ee
In these coordinates the action becomes
\be
S[x,\th,\tht]=\frac{1}{2\pi}\int d^2 z
(\d\th, \d\tht, \d x)\left( \begin{array}{clcr}
                                          E & 0\\
                                          0 & k
                                         \end{array}\right)
\left(\begin{array}{clcr} \db \th\\ \db\tht \\ \db x\end{array} \right)
-\frac{1}{8\pi}\int d^2 z \p_0 R^{(2)}.
\label{Ssu2}
\ee
Here we have organized the background into a $3\times 3$ block diagonal
matrix, where the  $2\times 2$ matrix block $E$ is
\be
E=\frac{k}{2}\left(\begin{array}{clcr}
        1-\S & 1+\S\\
      -(1+\S) & 1+\S \end{array}\right).
\label{Esu2}
\ee

The action $S$ in eq. (\ref{Ssu2}) is manifestly invariant under the
$U(1)_L\times U(1)_R$ affine symmetry generated by the chiral current $J$
and anti-chiral current $\Jb$ given by
\ba
J=\frac{k}{2}[-(1-\S)\d\th +(1+\S)\d\tht]\; , \nonumber\\
\Jb=\frac{k}{2}[(1-\S)\db\th +(1+\S)\db\tht]\; .
\label{JJ}
\ea
In addition, there are two extra chiral currents, and two extra
anti-chiral currents, completing the affine $SU(2)_L\times SU(2)_R$
(or $SL(2,\R)_L\times SL(2,\R)_R$) symmetry of  the WZNW model.

We can deform the action $S$ to new conformal backgrounds by
adding to it any truly marginal  deformation. We will focus on marginal
deformations that are obtained as a linear combination of chiral currents
times a linear combination of anti-chiral currents.
It is important to recall that all the
deformations that are equivalent under the action of the symmetry group
give rise to equivalent CFTs (although not necessarily to backgrounds that
are related by coordinate transformations). In the following we deform the
WZNW action with the $J\Jb$ marginal operator, as was done in refs.
\cite{HS,GKi}.

Deforming with $J\Jb$, the affine symmetry is broken to $U(1)_L\times
U(1)_R$. The $U(1)$ chiral and anti-chiral currents of the deformed theory
can be found and will be presented below.
Once choosing a $J\Jb$ deformation of the WZNW point (\ref{su2})
for specific $J$ and $\Jb$, the deformed theories are described in terms of
a one parameter family of $O(2,2,\R)$ rotations.

Let us describe these rotations.
The backgrounds corresponding to the $J\Jb$ deformation of the WZNW point,
with $J$ and $\Jb$ given in (\ref{JJ}), are constructed by an  $O(2,2,\R)$
transformation (\ref{tE}) of the WZNW background (\ref{Esu2}) by the
element \cite{HS,GKi}
\be
g_{\a}=
\left(\begin{array}{cccc} I&\cos^2\a(k-\tan\a)\e\\0&I\end{array}\right)
\left(\begin{array}{cccc} A(\a)&0\\0&(A(\a)^t)^{-1}\end{array}\right)
\left(\begin{array}{cccc} C(\a)&S(\a)\\S(\a)&C(\a)\end{array}\right)
\left(\begin{array}{cccc} I&-k\e\\0&I\end{array}\right),
\label{ga}
\ee
where
\ba
I=\left(\begin{array}{cccc} 1&0\\0&1\end{array}\right),\qquad
\e=\left(\begin{array}{cccc} 0&1\\-1&0\end{array}\right),&{}&\qquad
A(\a)=\left(\begin{array}{cccc} \cos\a&0\\0&\cos\a(1+k\tan\a)
\end{array}\right)\nonumber\\
C(\a)=\cos\a\; I, &{}&\qquad S(\a)=\sin\a\;\e.
\label{e}
\ea
Namely,
\be
\frac{1}{k}g_{\a}(E)\equiv E_{R(\a)}=\frac{1}{1+R^2\frac{1-\S}{1+\S}}
\left(\begin{array}{cccc} \frac{1-\S}{1+\S}&1\\-1&R^2\end{array}\right),
\label{ER}
\ee
where
\be
R(\a)^2=(1+k\tan\a)^2.
\label{R}
\ee
The parameter $R$ can be
interpreted, geometrically, as the radius of the Cartan torus.

The dilaton field $\p$ also transforms under $O(2,2,\R)$ rotations, as
discussed in sections 4.2, 4.3. At the WZNW
point the dilaton is the constant $\p=\p_0$ appearing in (\ref{su2}).
At the point $R$ the dilaton is (\ref{dtrans})
\be
\p (R)=\p_0+\frac{1}{2}\log\left(\frac{\det G(1)}{\det G(R)}\right),
\label{pR}
\ee
where $G(R)$ is the symmetric part of $E_R$.

Although we have generated the $R$-dependent backgrounds by $O(2,2,\R)$
rotations, which are correct only to leading order in $\alpha'$, it can be
shown \cite{GKi} that there is a scheme in which these
backgrounds are exact to all orders in $\alpha'$, for any $R$.

Two special backgrounds occur at the boundaries of the $R$-modulus space.
At $R=0$ ($\a=\tan^{-1}(-\frac{1}{k}$)) the background matrix is
\be
E_{R=0}=
\left(\begin{array}{cccc} \frac{1-\S}{1+\S}&1\\-1&0\end{array}\right)
=\left(\begin{array}{cccc} \frac{1-\S}{1+\S}&0\\0&0\end{array}\right)
\;\;\; {\rm mod} \;\; \Theta - {\rm shift}.
\label{pipa}
\ee
This background
corresponds to the direct product of the axially gauged coset
$SU(2)/U(1)_a$ (or $SL(2,\R)/U(1)_a$) and a free
scalar field in the singular compactification-radius limit
$r\to 0$ (which is equivalent to a non-compact free scalar field
via the $r\rightarrow 1/r$  circle duality).
The constant antisymmetric tensor in (\ref{pipa}) can be safely dropped
since the free scalar is non-compact.

At $R=\infty$ ($\a=\pi/2$) the background matrix is
\be
E_{R=\infty}=
\left(\begin{array}{cccc} 0&0\\ 0&\frac{1+\S}{1-\S}\end{array}\right).
\ee
This background corresponds to the direct product of the
vectorially gauged coset
$SU(2)/U(1)_v$ or $SL(2,\R)/U(1)_v$ and a free scalar field
at a compactification radius $r\to 0$~\footnote{The fact that at both
boundaries $R=0,\infty$ the decoupled scalar is at zero radius is because
in one limit the scalar is $\th$ and in the other limit it is the dual
field $\du\th$.}.

The $R=1$ ($\a=0$) point
corresponds to the original WZNW model. Around this point, and for an
infinitesimal deformation parameter $\delta\a$, the perturbed action is given
by
\be
S_{R=1+\delta R}=S_{R=1}+{\delta R^2 \over 4\pi k}\int d^2z J\Jb\; ,
\ee
where $J$ and $\Jb$ are given in (\ref{JJ}).
This deformation can be extended along the full $R$-line.
The $U(1)_{L}$ affine symmetry is generated by
$\th\rightarrow \th -\e$, $\tht\rightarrow \tht+\e /R^{2}$ with
the ($R$-dependent) chiral current
\be
J(R)=k\; {-(1-\S )\d\th +(1+\S )\d\tht\over 1+\S +R^{2}(1-\S)}\; ,
\ee
and the $U(1)_{R}$ affine symmetry is generated by
$\th\rightarrow \th +\epsilon$, $\tht\rightarrow\tht+\epsilon /R^{2}$ with
the anti-chiral current
\be
\Jb(R)=k\; {(1-\S )\db\th +(1+\S )\db\tht\over 1+\S +R^{2}(1-\S)}\; .
\ee
Therefore, the worldsheet action at the point $R+\delta R$ is given by
\be
S_{R+\delta R}=S_{R}+{\delta R^{2}\over 4\pi k}\int d^2z J(R)\Jb (R)\;,
\label{pp}
\ee
and writing the integration measure for the perturbed $\s$-model as
$\sqrt{G(1)}=\sqrt{G(R)}e^{\phi(R)}$ provides the variation of the dilaton
(\ref{pR}).

We now arrive to the important point of this section. The Weyl
transformation $J\rightarrow -J$ is given by a group rotation at the WZNW
point, and therefore, it is a symmetry of the WZNW model. Consequently,
the conformal perturbation
of the WZNW model by $\delta\a J\Jb$ is equivalent, infinitesimally, to the
conformal perturbation
by $-\delta\a J\Jb$. Therefore, the  points $\delta\a$ and $-\delta\a$
along the $\a$-modulus correspond to the same CFT. In string theory we say
that they are related by a spontaneously broken
$\Z_2$ gauge transformation in the
gauge  group of the enhanced symmetry point.

The spontaneously broken discrete symmetry is a target
space duality. This symmetry can be extended to finite $\a$, giving
rise to a $\Z_2$ element in the full $O(2,2,\Z)$ duality group, represented
by the matrix $g_D$
\ba
g_D&=&
\left(\begin{array}{cccc} 0&I\\ I&0\end{array}\right)
\left(\begin{array}{cccc} I&-\e\\0&I\end{array}\right)
\left(\begin{array}{cccc} e_2&e_1\\ e_1&e_2\end{array}\right)
\left(\begin{array}{cccc} I&\e\\0&I\end{array}\right)
\left(\begin{array}{cccc} 0&I\\ I&0\end{array}\right)
\nonumber\\
&=&\left(\begin{array}{cccc}
{\left(\begin{array}{cccc} 0&1\\ 0&1\end{array}\right)} &
{\left(\begin{array}{cccc} 1&0\\ 0&0\end{array}\right)} \\
{\left(\begin{array}{cccc} 1&-1\\ -1&1\end{array}\right)} &
{\left(\begin{array}{cccc} 0&0\\ 1&1\end{array}\right)}
\end{array}\right),
\label{gD}
\ea
where
\be
e_1=\left(\begin{array}{cccc} 1&0\\ 0&0\end{array}\right), \qquad
e_2=\left(\begin{array}{cccc} 0&0\\ 0&1\end{array}\right).
\ee
Here
$I$ is the 2-dimensional identity matrix, and $\e$ is given in (\ref{e}).
Up to a similarity transformation, $g_D$ is a factorized duality
(\ref{ERfd}). The element $g_D$ acts on $E_R$ by (\ref{tE}) and gives
\be
g_D(E_R)=E_{1/R}.
\ee
Therefore, duality takes the modulus $R$ to its inverse $1/R$.

A particular consequence is that the $R=0$ and $R=\infty$ boundary
points correspond to the same
CFT (to all orders in string perturbation theory).
Therefore,  the axial-vector duality of $SU(2)/U(1)$ and $SL(2,\R)/U(1)$ is
{\em exact}, and corresponds in string theory to a
spontaneously broken gauge symmetry of some
underlying string gauge algebra\footnote{The $\s$-models along the
$R<1$  half line
can be obtained as the backgrounds of the axially gauged $SU(2)\times
U(1)/U(1)_{diag}$ cosets, while the $R>1$ backgrounds can be obtained by the
vectorially gauged $SU(2)\times U(1)/U(1)_{diag}$ cosets~\cite{GKi}.
The $R\to 1/R$  duality relates these axial cosets to their vector
partners, and therefore, it is an axial-vector duality along the full
$R$-modulus, and in particular at its boundaries.}.
\end{subsubsection}

\begin{subsubsection}{$J\Jb$ Deformations and Duality
as a Gauge Symmetry for General Groups}

Here we describe $J\Jb$ deformations  of general WZNW models. These
conformal perturbations are related
to $O(d,d,\R)$ transformations \cite{HN,K2,GKi}. We  start
with a particular parametrization of a WZNW model on a group $G$. The group
$G$ can be semisimple. We  explicitly  indicate one of the
levels $k$, while the others are hidden in the action. The relevant level
is the one corresponding to the simple component of the group whose Cartan
we are deforming. We parametrize $g\in G$ as follows: we chose an element
in the Cartan sub-algebra, say $T^1$, and write it as
\be
g=e^{i\tha T^1}h(x)e^{i\thb T^1},
\ee
where $h\in G$ is independent of $\tha,\thb$.
Now, by  using the Polyakov-Wiegman formula \cite{PW} for the WZNW action
one finds (see for example \cite{Kir,K2})
\be
S[x^a,\tha,\thb]=S_1+S_a+S[x],
\label{wzwbeg}
\ee
\be
S_1=\frac{k}{2\pi}\int d^2 z \left(\d\tha \db\tha + \d\thb \db\thb
+2\S (x) \d\thb \db\tha+\G_a^1(x)\d x^a\db\tha+\G_a^2(x)\d\thb\db
x^a\right),
\ee
\be
S_a=\frac{k}{2\pi}\int d^2 z(\d\thb\db\tha - \d\tha\db\thb),
\ee
\be
S[x]=\frac{k}{2\pi}\int d^2 z\;\G_{ab}(x)\d x^a \db x^b -
\frac{1}{8\pi}\int d^2 z\;\p_0 R^{(2)},
\label{wzwend}
\ee
where $a=1,...,D-2$, $D={\rm dim}\,G$, and $\p_0$ is a constant dilaton. The
backgrounds $\S(x), \G^1(x), \G^2(x)$ and $\G(x)$ are
independent of the coordinates $\tha,\thb$, and therefore,
(\ref{wzwbeg})-(\ref{wzwend}) is of the form (\ref{Dd}) with $i,j=1$
(up to an overall $k$ factor).
With the coordinates $\th$ and $\tht$ defined in (\ref{th2}, \ref{th1}), the
action becomes
\be
S[x^a,\th,\tht]=\frac{1}{2\pi}\int d^2 z
(\d\th, \d\tht, \d x^a)\left( \begin{array}{cccc}
                                          E & F_b^2\\
                                          F_a^1 & F_{ab}
                                         \end{array}\right)
\left(\begin{array}{clcr} \db \th \\ \db \tht \\ \db x^b\end{array}
\right)
-\frac{1}{8\pi}\int d^2 z\;\p_0 R^{(2)},
\label{Swzw}
\ee
Here we have organized the background fields into a $D\times D$
background matrix with four blocks $E, F^1, F^2$ and $F$.
The $2\times 2$ background block $E$, the
$(D-2)\times 2$ block $F_b^2$, the $2\times (D-2)$ block
$F_a^1$, and the $(D-2)\times (D-2)$ block $F_{ab}$ are given by
\be
\left(\begin{array}{clcr}
                             E & F_b^2\\
                             F_a^1 & F_{ab}
                             \end{array}\right)=
\frac{k}{2}
\left(\begin{array}{clcr}
{\left(\begin{array}{clcr} 1-\S & 1+\S\\
      -(1+\S) & 1+\S \end{array}\right)} & {\left(\begin{array}{clcr}
                                           \G_b^2 \\ \G_b^2
                                            \end{array}\right)}\\
{\left(-\G_a^1 \;\;\;\;\;\;\; \G_a^1 \right)} & 2\G_{ab}
\end{array}\right).
\label{EFFF}
\ee

The action (\ref{Swzw}) is manifestly invariant under the $U(1)_L\times
U(1)_R$ affine symmetry generated by the chiral current $J$ and the
anti-chiral current $\Jb$ given by
\ba
J=\frac{k}{2}\left[-(1-\S)\d\th +(1+\S)\d\tht+\G_a^1\d x^a\right],
\nonumber\\
\Jb=\frac{k}{2}\left[(1-\S)\db\th +(1+\S)\db\tht+\G_a^2\db x^a\right].
\label{JJb}
\ea
In addition, there are extra $D-1$ chiral currents and
$D-1$ anti-chiral currents, completing the affine $G_L\times G_R$
symmetry of the WZNW model\footnote{
Actually, one can bring the background in (\ref{wzwbeg})-(\ref{wzwend})
into the form of eq. (\ref{Dd})
\cite{GR} with $i,j=1,...,r={\rm rank}\,G$ (up to an overall $k$ factor).
This can be done by explicitly
parametrizing the Cartan torus dependence of the WZNW model as
$$g=e^{i\sum_{i=1}^r \tha^i T^i}h(x^a)e^{i\sum_{i=1}^r \thb^i T^i},$$
where $\{T^i | i=1,...,r\}$ is a basis in the Cartan
sub-algebra, ${\rm Tr}(T^iT^j)=\eta^{ij}$, and $h\in G$ is independent of
$\tha^i,\thb^i$. In this form
the $r$ chiral currents and $r$ anti-chiral currents
corresponding to the left-handed and right-handed Cartan tori are
manifest.}.

We now deform the action (\ref{Swzw}) with the conformal perturbation
$J\Jb$, where $J$ and $\Jb$ are
given in (\ref{JJb}).
The $J\Jb$ deformation is equivalent to the action of
$g\in O(2,2,\R)$ transformations on the background matrices $E, F^1, F^2, F$.
The action of $g$ in (\ref{g})
on $E$ is given by (\ref{tE}). Here we need also the
action of $g$ on the background blocks $F^1,F^2, F$ given in eq.
(\ref{ghat}) \cite{GR}:
\ba
g(F^1)=F^1(cE+d)^{-1}, \qquad g(F^2)=(a-E'c)F^2,\nonumber\\
g(F)=F-F^1(cE+d)^{-1}cF^2,
\label{gF}
\ea
where $a,c,d$ are defined in (\ref{g}), and $E'$ is given in (\ref{tE}).

The modulus line of
$J\Jb$ conformal perturbations is generated by
acting on the background (\ref{EFFF}) with the one-parameter family
of  $O(2,2,\R)$, $g_{\a}$, given in (\ref{ga}).
Using eqs. (\ref{tE}, \ref{gF}) one finds
\be
\frac{1}{k}g_{\a}(E)\equiv E_{R(\a)},
\ee
\be
\frac{1}{k}g_{\a}(F^1)\equiv F^1_{R(\a)}
=\frac{(1+\S)^{-1}\G^1}{1+R^2\frac{1-\S}{1+\S}}
\left(-1,R^2\right),
\label{F1R}
\ee
\be
\frac{1}{k}g_{\a}(F^2)\equiv F^2_{R(\a)}
=\frac{(1+\S)^{-1}\G^2}{1+R^2\frac{1-\S}{1+\S}}
\left(\begin{array}{clcr} 1 \\ R^2 \end{array}\right),
\label{F2R}
\ee
\be
\frac{1}{k}g_{\a}(F)\equiv F_{R(\a)}
=\G+\frac{(R^2-1)(1+\S)^{-1}\G^1\G^2}{2(1+R^2\frac{1-\S}{1+\S})},
\label{FR}
\ee
where  $E_R$ and $R(\a)$ are given in (\ref{ER}, \ref{R}).

The constant dilaton $\p_0$ in the WZNW background (\ref{wzwend}) transforms
under $g_{\a}$ by eq. (\ref{pR}). For more details see sections 4.2, 4.3.

{}From eqs. (\ref{ER}, \ref{F1R}, \ref{F2R}, \ref{FR}) we see that,
up to an overall $k$ factor, the backgrounds are parametrized by a positive
real number $R^2 =(1+k\tan \a)^2$. As for the $SU(2)$ case, the parameter $R$
can be interpreted, geometrically, as the radius of the corresponding
circle in the deformed Cartan torus.

As for the $SU(2)$ or $SL(2,\R)$ cases,
the whole Cartan subalgebra survives along the deformation.
One can find that the explicit form of the ($R$-dependent) chiral
and anti-chiral currents we perturb with is
\be
J(R)=k\;{{-(1-\S)\d\th+(1+\S)\d\tht+\Gamma_{a}^{1}\d x^{a}}\over
 1+\S+R^{2}(1-\S)}\; ,
\ee
\be
{\bar J}(R)=k\;{{(1-\S)\db\th+(1+\S)\db\tht+\Gamma_{a}^{2}\db x^{a}}\over
 1+\S+R^{2}(1-\S)}\; ,
\ee
and it can be verified that eq. (\ref{pp}) is still valid\footnote{
The deformation described here is true for any background with a chiral
current and
an anti-chiral current; it is only for the purpose of discussing gauge
symmetries that we assume that a special point on the modulus line
(in our notation $R=1$) is a WZNW model.}.

We are now ready to
discuss target space duality in the $R$-line of conformal deformations.
Remarkably,  the action of the $O(2,2,\Z)$ element  $g_D$
(given in eq. (\ref{gD})) on the backgrounds $E,F^1,F^2,F$ in
eq. (\ref{EFFF}), gives a  simple transformation of
the modulus parameter $R$. By straightforward calculations, using
the transformations given in (\ref{tE}, \ref{gF}), one finds that
\be
g_D\left(
\left(\begin{array}{cccc}
                             E_R & F_R^2\\
                             F_R^1 & F_R
                             \end{array}\right)\right)=
\left(\begin{array}{cccc}
                             E_{1/R} & F_{1/R}^2\\
                             F_{1/R}^1 & F_{1/R}
                             \end{array}\right).
\ee
Therefore, duality relates the $\sigma$-model background at the modulus
point $R$ with the background at the modulus parameter $1/R$.

It is now easy to generalize the consequences described
for the $SU(2)$ and $SL(2,\R)$ cases in subsection 4.8.1.
At the fixed point $R=1$ the CFT has an
enhanced affine symmetry $G_L\times G_R$: it is
the original WZNW model.
Infinitesimally around the enhanced symmetry point, duality corresponds to
the transformation $\a\rightarrow -\a$. This transformation can be achieved
by a Weyl rotation  in $G_L$  (or $G_R$) that reflects $J\rightarrow
-J$ (or $\Jb\rightarrow -\Jb$). Duality is related to a Weyl reflection and
is, therefore, a spontaneously broken gauge symmetry
in the  gauge algebra of the associated string theory.

The duality transformation interchanges $R$ with $1/R$
and, in particular, identifies the two boundaries at $R=0$ and
$R=\infty$.
Let us show that these boundaries correspond,
respectively, to the direct product of the cosets $G/U(1)_a$ and $G/U(1)_v$
with a free, non-compact scalar field.

At $R=0$ ($\a=\tan^{-1}(-\frac{1}{k}$)) the background matrix is
\ba
E_{R=0}=
\left(\begin{array}{cccc} \frac{1-\S}{1+\S}&1\\-1&0\end{array}\right)
&=&
\left(\begin{array}{cccc} \frac{1-\S}{1+\S}&0\\0&0\end{array}\right)
\;\;\; {\rm mod} \;\; \Theta - {\rm shift},
\nonumber\\
F_{R=0}^1=(1+\S)^{-1}\G^1(-1,0),&{}& \qquad
F_{R=0}^2=(1+\S)^{-1}\G^2\left(\begin{array}{clcr} 1\\0\end{array}\right),
\nonumber \\
F_{R=0}&=&\G-\frac{1}{2}(1+\S)^{-1}\G^1\G^2.
\label{EFR0}
\ea
This background corresponds, as shown in section 4.2~\cite{RV,GR},
to the direct product of the axially gauged coset
$G/U(1)_a$ and a free scalar field in the compactification-radius
limit $r\to0$.

At $R=\infty$ ($\a=\pi/2$) the background matrix is
\ba
E_{R=\infty}=
\left(\begin{array}{clcr} 0&0\\ 0&\frac{1+\S}{1-\S}\end{array}\right) ,
&{}& F_{R=\infty}=\G+\frac{1}{2}(1-\S)^{-1}\G^1\G^2 ,
\nonumber\\
F_{R=\infty}^1=(1-\S)^{-1}\G^1(0,1), \qquad
&{}& F_{R=\infty}^2=(1-\S)^{-1}\G^2
\left(\begin{array}{clcr} 0\\1\end{array}\right) .
\label{EFRinf}
\ea
This background corresponds
to the direct product of the vectorially gauged coset
$G/U(1)_v$ and a free scalar field at $r\to 0$ (see section 4.2).

Therefore, we have obtained that
axial-vector duality is exact in general, since the
end-points are equivalent theories, analogously to
the $SU(2)$ and $SL(2,\R)$ cases.

\end{subsubsection}
\end{subsection}

\begin{subsection}{Topology Change and Duality in String Theory}
\setcounter{equation}{0}

In this section we discuss  topology change in
string theory and its relation with target space duality.
This issue will be addressed in particularly simple examples.
Here we present models in which connected paths in moduli space exist along
which the topology of the background metric changes.
It would be interesting to construct a CFT corresponding to a time
dependent background, where topology change occurs as a dynamical process.

The simplest example of topology change in string theory is observed in
the moduli space of $c=1$ backgrounds in figure 3.A. At the point where the
circle compactification line meets the orbicircle line -- the orbicircle
self-dual point $R_o=1$ -- the topology may
change from a circle to a segment. Moreover, at the $SU(2)$ enhanced
symmetry point -- the circle self-dual point $R_c=1$ -- the $\s$-model
description of the CFT can be given either in terms of a circle or,
alternatively, in terms of the three-dimensional sphere background of the
$SU(2)_1$ WZNW model.
In these examples one may connect semiclassical (i.e. large) backgrounds of
different topology. However, the actual
topology change occurs due to ``stringy
schizophrenia'' of backgrounds at the Planck scale: in some instances the
string cannot perceive in which world it lives.

Let us  now discuss  examples where topology change
occurs through a different mechanism. Namely, a single background is
associated to
each point of the connected path in the moduli space: no doubling of
the kind described above is invoked, rather, the topology changes when
classical curvature singularities arise.
We will describe the geometry along the $R$-line of section 4.8
and a  topology change in the extended moduli
space of $SU(2)$ or $SL(2,\R)$ \cite{GKi}.
Moreover, we will show that there is a
topology change in the moduli space of the cosmological string solutions
discussed in section 4.8.

We present at first  the description of the
target space geometry along the $R$-line of marginal deformations of the
$SU(2)$ WZNW model. The $\sigma$-model metric of the deformed
$SU(2)$ (\ref{ER}) (in the coordinates $\th$, $\tht$, $x$)
is given by
\be
G(R)=k\left(
\matrix{{{\tan}^{2}x\over  1+R^{2}{\tan}^{2}x}& 0&0\cr
0&{R^{2} \over 1+R^{2}{\tan}^{2}x}&0\cr
0&0&1\cr}\right).
\ee
The scalar curvature ${\cal R}$ is
\be
{\cal R}=-{2\over k}{2-5R^{2}+2(R^{4}-1){\sin}^{2}x \over
 (1+(R^{2}-1){\sin}^{2}x)^2} \; .
\label{rr}
\ee
The manifold is regular except at the end-points, where
\be
{\cal R}(R=0)=-{4\over k{\cos}^{2}x}\;\;,\;\;{\cal R}(R=\infty)=-{4\over
 k{\sin}^{2}x}\; .
\ee
At $R=1$ we get the constant curvature of $S^{3}$, ${\cal R}=6/k$.
The metric $G(R)$ has no conical singularities along the $R$-line.

Let us note that the geometrical data (metric, curvature, etc.) are
invariant under the $R\rightarrow 1/R$ duality, together with the
coordinate transformation $x\rightarrow \pi /2-x$.
In particular, the volume of the manifold as a function of $R$,
\be
V(R)\sim {R\;{\log}R\over R^{2}-1},
\ee
is duality invariant: $V(R)=V(1/R)$. The volume vanishes
only at the boundaries of moduli space, namely, at $R=0,\infty$.

Along the $0<R<\infty$ deformations line of the $SU(2)$ WZNW $\s$-model,
the topology of the background space is of the three sphere.
However, at the boundaries ($R=0,\infty$),
the topology is changed to that of
a product of a two-disc (corresponding to the
$SU(2)/U(1)$ conformal background) with a circle whose
radius shrinks to 0.

For $SL(2,\R)$, the trigonometric functions in (\ref{rr}) are replaced by the
corresponding hyperbolic functions. Here the manifold has a curvature
singularity for $0\leq R <1$; similar remarks apply to its Euclidean
version, the 3-$d$ hyperboloid.
The background at the boundary
is the direct product of
a degenerated circle and a semi-infinite ``cigar'' (at
$R\to\infty$) with the topology of the disk,
or the infinite ``trumpet'' (at $R\to 0$) with the topology of a cylinder;
these correspond to
the dual pair of the 2-$d$ Euclidean black-hole backgrounds (see section
4.5 and figure 1.F).

A topology change at the boundary of moduli space is not surprising.
However, the $R$-line of deformations is not the full story in the moduli
space of $G$ WZNW $\s$-models.
Indeed, any conformal $\s$-model with $d$ Abelian symmetries can be
transformed  to  new conformal backgrounds by
$O(d,d,\R)$ rotations (see section 4.3)\footnote{Recall that in
the bosonic string, the
$O(d,d,\R)$ rotations give only the leading order in $\a'$ of the conformal
backgrounds, but there exist higher order corrections that make them exact
\cite{GR,GPa,K2,GKi}.}.
In this larger moduli space of $\s$-models, there
are other interesting deformation lines, along which the topology might
change.

For example, in the moduli space of the
$SU(2)$ WZNW $\s$-model, there is a
one-parameter family of $O(2,2,\R)$ rotations
\be
\tilde{g}_{\a}=
\left(\begin{array}{cccc} I&k\e\\0&I\end{array}\right)
\left(\begin{array}{cccc} C(\a)&S(\a)\\S(\a)&C(\a)\end{array}\right)
\left(\begin{array}{cccc} I&-k\e\\0&I\end{array}\right)
\ee
($\e, C(\a), S(\a)$ are given in (\ref{e})),
that generate the backgrounds
\be
\tilde{g}_{\a}(E)=\frac{k}{\Delta}
\left(\begin{array}{cccc} 1-\S & B \\-B & 1+\S \end{array}\right),
\label{Eline}
\ee
where
\ba
\S=\cos 2x\; , &{}&\qquad
\Delta=\cos^2 \a (1+\S )+(\cos\a+k\sin\a)^2(1-\S )\; ,\nonumber\\
B&=&\frac{1}{k}\sin(2\a)-\left[\cos(2\a)+\frac{k}{2}\sin(2\a)\right]
(1-\S)+\Delta \; .
\ea
The dilaton transforms by eq. (\ref{pR}).

The ($\a$-dependent) metric is given by the line element
\bea
ds^2(\a)&=&\frac{2k}{\Delta(\a)}[\sin^2x d\th^2+\cos^2x d\tht^2]+kdx^2,
\nonumber\\
\Delta(\a)&=&2\cos^2 \a \cos^2x+2(\cos\a+k\sin\a)^2\sin^2x .
\label{line}
\eea
At the point $\a=0$ the background includes a metric of
the $SU(2)_k$ group manifold $S^3$ (as well as an antisymmetric
background).
Along the line $0<\a <\pi/2$ the background includes the
metric (\ref{line}) with the topology
of $S^3$ (as well as an antisymmetric background and a dilaton field).
At the point $\a=\pi/2$ the background metric is
\be
ds^2(\a =\pi/2)=\frac{1}{k}d\th^2+\frac{1}{k}\cot^2xd\tht^2+kdx^2.
\label{aline}
\ee
At this point the manifold has a topology of $D_2\times S^1_{1/k}$,
where $D_2$
is a two-disc and $S^1_{1/k}$ is a circle with radius $r^2=1/k$.
One may continue to deform this theory by, for example, changing the
compactification radius $r$ of the free scalar field $\th$.

It is remarkable that (for integer $k$) the neighborhood of the point
$\a=\pi/2$ is mapped to the neighborhood of
the point $\a=0$ by an element of $O(2,2,\Z)$ \cite{RV,GR,Kumar},
namely, by a target space generalized duality.
Therefore, {\em a region in the moduli space, where a topology change occurs,
is mapped to a region where there is no topology change at all}. A similar
phenomenon happens for more complicated examples in the moduli space of
Calabi-Yau compactifications \cite{Getal,Wcy}.

At this point, let us make two comments:

\vskip .1in
\noindent
(a) The $\s$-models along the $\a$--line (\ref{line})
have  conical singularities.
Therefore, to make sense of the CFTs along the $\a$--line one should
understand CFTs corresponding to backgrounds with (non-orbifold)
conical singularities.

\vskip .1in
\noindent
(b) After the topology is changed at $\a=\pi/2$,
one cannot get rid of the curvature singularity encountered at that point
simply by deforming the compactification radius of the free scalar field
$\th$.

\noindent
To cure both problems, one should look at the moduli space of WZNW
$\s$-models in higher dimensions.

Indeed, an even more interesting example of
topology change occurs in the moduli
space of the $SU(2)\times SL(2,\R)$ WZNW model. Here one can follow
deformation lines along which there is no scalar
curvature singularity, except at
the point where the topology is changed. Explicitly, one first deforms
$SU(2)\times SL(2,\R)$ to the background of the coset
$\left[{SU(2)\over U(1)}\times{SL(2,\R)\over U(1)}\times
U(1)\right]\times U(1)'$, where the background of the coset in the
brackets is given in (\ref{GBA}) together with (\ref{fgtan}),
and with $a=b=c=0$. Now, one can rotate the
background by $O(3,3,\R)$ transformations
into the backgrounds (\ref{GBA}) with $a,b,c$
simultaneously non-zero.  The resulting backgrounds correspond to the
product of cosmological solutions
in the presence of a gauge field (\ref{GBA}) times the circle $U(1)'$.
As shown in section 4.7 \cite{GPa}, these backgrounds have no scalar
curvature singularities.

Finally, and as before, the point $a=b=c=0$ is related by $O(3,3,\Z)$
target space dualities
to  points $a,b,c\in \Z$ (see the end of section 4.7)
where no topology change occurs.

\end{subsection}
\end{section}

\newpage

\begin{section}{Worldsheet/Target-Space Duality}

A classical action describing the propagation of an extended object of
dimension $d$ in a space-time of dimension $D$ is described classically by a
$d$-dimensional world-manifold field theory, with $D$ target space
coordinates. For strings, the $d=2$ worldsheet theory can actually be
quantized.
In this section we present some mathematical identities. They are
intriguing as they may suggest that the target space and the worldsheet are
actually part of a larger structure.

In the following it is shown \cite{GMR} that there exist
relations between pairs $(\Sigma, T)$, where $\Sigma$ denotes the
worldsheet and $T$ denotes the target space.
It is shown that in some cases the ``classical''\footnotemark
\footnotetext{In this section ``classical'' refers to the
partition function corresponding to
the zero-mode Hamiltonian in eq.~(\ref{ER13}).}
part of the partition function
describing pairs $(\Sigma, T)$ is related to the classical part of
the pair $(T, \Sigma)$, where the worldsheet and target space are
interchanged.
Most of the discussion in this section deals with the bosonic string.
The classical part of the pairs $(\Sigma, T)$ and $(T, \Sigma)$ is
manifestly identical if the following conditions are satisfied.
Either the target space is ``almost complex'' or the period
matrix $\tau$ describing the Riemann surface is ``real.''
The terms ``almost complex'' and ``real'' in this context will
be defined below.
This symmetry relates the dimension of the target space of a
string theory to the order of interaction, namely, the genus of
the worldsheet Riemann surface.
Duality relations between spaces of different dimensions were
discussed and utilized in the study of statistical mechanics
systems \cite{AERS}.

Here we present the mathematical identities in an
attempt to learn about the symmetries of string theories.
We first consider toroidal bosonic compactifications.
As an immediate consequence of the identities,
we will prove that $Sp(2d,\Z)\subset O(2d,2d,\Z)$
is a symmetry of the genus-$g$ partition functions
for almost complex toroidal backgrounds in $2d$ dimensions.
Moreover, we will prove that $O(d,d,\Z)\subset Sp(4d,\Z)$ is a symmetry
of the genus-$g$ partition functions
with real period matrices and any $d$-dimensional flat target space.
A proof that $O(d,d,\Z)$ is a
symmetry to all orders and for any Riemann surface is needed in order to
show that $O(d,d,\Z)$ is a symmetry valid to all orders in string
perturbation theory. This proof can be obtained by combining the methods
described in section 2 with the results of this section.

\begin{subsection}{Worldsheet/Target-Space Duality for Toroidal
Bosonic Compactifications}
\setcounter{equation}{0}

Let $E=G+B$ be a $2d\times 2d$ background matrix (see section 2)
\cite{GMR} of the form
\begin{equation}
E=\left(
\begin{array}{cc}
\sigma_2 & \sigma_1\\
-\sigma_1 & \sigma_2
\end{array}
\right),
\label{eq-ER1}
\end{equation}
where $\sigma_1$ and $\sigma_2$ are real, symmetric $d\times d$ matrices,
and $\sigma_2$ is positive definite.
The metric $G$ and antisymmetric background $B$ are, therefore,
\begin{equation}
G=\left(
\begin{array}{cc}
\sigma_2 & 0\\
0 & \sigma_2
\end{array}
\right),
\qquad
B=\left(
\begin{array}{cc}
0 & \sigma_1\\
-\sigma_1 & 0
\end{array}
\right).
\label{eq-ER2}
\end{equation}
Denote by $I_n$ the $n$-dimensional identity matrix, and by
$J_n$ the standard $2n$-dimensional antisymmetric matrix,
\begin{equation}
J_n=\left(
\begin{array}{cc}
0 & I_n\\
-I_n & 0
\end{array}
\right),
\qquad J_n^2=-I_{2n},
\label{eq-ER3}
\end{equation}
and define the complex $d\times d$ matrix $\sigma$ :
\begin{equation}
\sigma=\sigma_1+i\sigma_2 .
\label{eq-ER4}
\end{equation}
The specific form (\ref{eq-ER1}) of $E$ means that the target
space is ``almost complex'' in the sense that $E\equiv
-i\sigma$, i.e.\ :
\begin{equation}
G=({\Re}E)\otimes I_2 , \qquad B=(-{\Im}E)\otimes J .
\label{eq-ER5}
\end{equation}

In the loop expansion of string theory one sums over all
mappings from a genus-$g$ Riemann surface $\sum_g$ to the target
space $T^{2d}$.
In a canonical basis of cycles, $a_i$, $b_i$, where $i=1, \ldots, g$
(see section 2.3 and figure 2.A),
the Riemann surface is described by the period matrix $\tau=\tau_1+i\tau_2$.
The  $g\times g$ matrices $\tau_1$ and $\tau_2$ are real and symmetric,
and $\tau_2$ is positive  definite.
The classical piece of the multi-loop path integral will be
denoted by $Z_{class} (\sigma, \tau)$.
In what follows it is observed \cite{GMR} that
\begin{equation}
Z_{class}(\sigma, \tau)=Z_{class}(\tau, \sigma),
\label{eq-ER6}
\end{equation}
namely, the classical piece of the mapping from a Riemann surface
$\Sigma_g(\tau)$ to a target space $T^{2d}(\sigma)$ is equivalent
to the mapping from a Riemann surface $\Sigma_d(\sigma)$ to a
target space $T^{2g}(\tau)$.\footnote{
This result was observed first in the $d=1$, $g=1$ case \cite{DVV1}
(\ref{ER54}). In this case
it can be generalized to a triality while reintroducing the target space
complex structure.}

The result (\ref{eq-ER6}) can be useful in proving  the invariance of
genus-$g$ partition functions under generalized target space duality.
Notice that there is an isomorphism between $Sp(2g, \Z)$ (the
group of modular transformations of the Riemann surface)
generated by \cite{Mumford}
\begin{eqnarray}
T&\colon&\tau\to\tau+b,  \qquad b\ \mbox{ symmetric, integral,}
\nonumber \\
S&\colon&\tau\to -1/\tau ,  \nonumber \\
A&\colon&\tau\to m^t\tau m, \qquad m\in GL(g,\Z),
\label{eq-ER7}
\end{eqnarray}
and the symmetry transformations on the target space moduli
generated by
\begin{eqnarray}
T'&\colon&\sigma\to \sigma+b, \qquad b\ \mbox{ symmetric, integral,}
\nonumber \\
S'&\colon&\sigma\to -1/\sigma ,  \nonumber \\
A'&\colon&\sigma\to m^t\sigma m, \qquad  m\in GL(d,\Z) .
\label{eq-ER8}
\end{eqnarray}
As discussed in section 2,
invariance under $T'$ reflects the fact  that the $B$ terms
serve as two dimensional theta parameters.
Invariance under $S'$ is the inversion duality symmetry
of the closed string with $2d$ compactified dimensions.
The action $A'$ on the background $\sigma$ is equivalent to a
change of the basis of the lattice defining the target space torus.
There is, therefore, {\em an intriguing connection between target
space duality and worldsheet modular invariance}.
The duality symmetry $E\to 1/E$ (\ref{ER25})
$(\sigma\to -1/\sigma)$ is a
simple consequence of the symmetry $\tau\leftrightarrow\sigma$.
In fact, this is a way to prove that inversion duality is an exact
symmetry to all orders in loop expansion.
Moreover, it also follows
that the subgroup $Sp(2d,\Z)\subset O(2d,2d,\Z)$ is a symmetry to all orders in
loop expansion, for almost complex toroidal backgrounds. A proof that
$O(d,d,\Z)$ is a symmetry to all orders, and for any toroidal backgrounds,
is straightforward.

To prove eq. (\ref{eq-ER6}) we work in the notation of section 2.
The closed bosonic string action
for the compactified dimensions (setting the dilaton to zero, for the time
being) is
\begin{equation}
S=\frac{1}{4\pi} \int d^2\sigma(\sqrt{h} h^{\alpha\beta}
G_{\mu\nu}\partial_\alpha X^\mu \partial_\beta X^\nu
+\epsilon^{\alpha\beta} B_{\mu\nu}\partial_\alpha
X^\mu\partial_\beta X^\nu) ,
\label{eq-ER9}
\end{equation}
where $\alpha,\beta=0,1$; $\mu,\nu=1,\ldots, 2d$; $\sigma_0,
\sigma_1$ are the world-sheet parameters; $h^{\alpha\beta}$ is
the worldsheet metric;
$h=\det h_{\alpha\beta}$; $\epsilon^{\alpha\beta}$ is the
standard $2$-dimensional antisymmetric tensor.
Here, the compact space is a $2d$-dimensional torus
$T^{2d}=\R^{2d}/\pi\Lambda^{2d}$;
the lattice $\Lambda^{2d}$ is spanned by the basis
$\{ e^\mu_i \; | \; i=1, \ldots, 2d \}
$.
Thus $X^\mu$ is identified with $X^\mu+\pi n^i e^\mu_i$, where
$n^i$ are integers. Let us define the constant metric $G_{ij}$ and
the constant antisymmetric tensor $B_{ij}$ by
\begin{eqnarray}
G_{\mu\nu} & = & 2G_{ij}(e^{i*})_\mu(e^{j*})_\nu , \nonumber \\
B_{\mu\nu} & = & 2B_{ij}(e^{i*})_\mu(e^{j*})_\nu ,
\label{eq-ER10}
\end{eqnarray}
where $\{ e^{i*}\; | \; i=1, \ldots, 2d \}$ is the basis dual to $\{ e_i\}$.
In the orthonormal gauge,
$h_{\alpha\beta}=e^{\phi(z,\bar{z})}\delta_{\alpha\beta}$,
the action may be written as
\begin{equation}
S=\frac{1}{2\pi}\int d^2z E_{ij} \partial X^i\bar\partial X^j ,
\label{eq-ER11}
\end{equation}
where $E=G+B$; $X^i\equiv X^{\mu}(e^{i*})_{\mu}$.
This action depends on the period matrix $\tau$ through the complex
structure, used to define $z$ and $\bar{z}$.

The classical piece of the genus-$g$ partition function gives a
contribution \cite{A-GMV,ABMNV,gv,GMR}
\begin{eqnarray}
Z_{class}(G,B,\tau) & = & (\det
G)^{\frac{g}{2}}\sum_{n,m}\exp\{-\pi [n^{ia}(\tau_2^{-1})_{ab}
G_{ij} n^{jb} \nonumber \\
&& +m^{ia}(\tau_2+\tau_1\tau_2^{-1}\tau_1)_{ab} G_{ij}
m^{jb}\nonumber \\
& & -2m^{ia}(\tau_1\tau_2^{-1})_{ab}G_{ij} n^{jb}+2in^{ia}B_{ij}
m^{ja}]\} ,
\label{eq-ER12}
\end{eqnarray}
where $a,b=1, \ldots, g$; $n$ and $m$ are $2dg$-dimensional vectors
with integer components. (This is a  generalization of the result
presented for circle compactifications in section 2.3).
The normalization is such that $Z_{class} ={\rm Tr}(\exp(-\tau_2
H+i\tau_1 P))$ in the one-loop case $(g=1)$ ($H$ is the
Hamiltonian and $P$ is the worldsheet momentum).
In the path integral formulation at one loop, one gets $(\det G)^{1/2}$ by
integrating the zero modes $X_0$: it is the volume of the target space.

For $G$ and $B$ of the form (\ref{eq-ER2}) one gets
\begin{eqnarray}
Z_{class}(\sigma,\tau)& = & (\det\sigma_2)^g\sum_{m_1, m_2, n_1,
n_2}\exp\{-\pi[ n_1\tau_1^{-1}\sigma_2n_1+n_2\tau_2^{-1}\sigma_2n_2
\nonumber \\
&& + m_1(\tau_2+\tau_1\tau_2^{-1}\tau_2)\sigma_2m_1
+m_2(\tau_2+\tau_1\tau_2^{-1}\tau_1)\sigma_2m_2 \nonumber \\
&& - 2m_1\tau_1\tau_2^{-1} \sigma_2n_1-2m_2\tau_1\tau_2^{-1}\sigma_2
n_2 \nonumber \\
&& + 2in_1\sigma_1m_2-2in_2\sigma_1m_1]\} ,
\label{eq-ER13}
\end{eqnarray}
where $n_1$, $n_2$, $m_1$, $m_2$ are $dg$-dimensional vectors of
integer components.
A Poisson resummation on $m$ leads to
\begin{eqnarray}
Z_{class}(\sigma,\tau) & = & (\det\tau_2)^d\sum_{m_1, m_2, k_1,
k_2} \exp\{
-\pi[ k_1\sigma_2^{-1}\tau_2k_1+k_2\sigma_2^{-1}\tau_2k_2
\nonumber \\
&& + m_1(\sigma_2+\sigma_1\sigma_2^{-1}\sigma_1)\tau_2m_1
+m_2(\sigma_2+\sigma_1\sigma_2^{-1}\sigma_1)\tau_2m_2 \nonumber
\\
&& +2m_1\sigma_1 \sigma_2^{-1}\tau_2k_2-2m_2\sigma_1\sigma_2^{-1}\tau_2
k_1 \nonumber \\
&& -2ik_1\tau_1 m_1 -2ik_2\tau_1m_2]\} .
\label{eq-ER14}
\end{eqnarray}
After the resummation the partition function can be written in a
more compact form:
\begin{equation}
Z_{class} (\sigma,\tau)=(\det\tau_2)^d\sum_{m,k}\exp\{ i\pi[P_R^a
\tau_{ab} P_R^b - P_L^a\bar\tau_{ab} P_L^b]\},
\label{eq-ER15}
\end{equation}
where, as in (\ref{ERplpr}),
\begin{eqnarray}
P_R^a = (k^a-m^a E^t)e^*, \qquad P_L^a = (k^a+m^a E)e^*.
\label{eq-ER16}
\end{eqnarray}
Identifying $(k_1, k_2, m_1, m_2)$ in the sum (\ref{eq-ER14})
with $(n_1, n_2, -m_2, m_1)$ in the sum (\ref{eq-ER13}) we get the proof of
(\ref{eq-ER6}):
$$Z_{class}(\sigma, \tau)=Z_{class}(\tau, \sigma).$$

We recall that eq. (\ref{eq-ER6}) is proved for any element
$\tau$ of the Siegel upper-half-space (the space of symmetric, complex
$g\times g$ matrices with positive definite imaginary part).
Not all of them correspond to period matrices of Riemann surfaces.
The fact that eq. (\ref{eq-ER6}) is correct for any $\tau$
suggests that there might be a physical meaning to the partition
function of a flat target space for any $\tau$.

The target space structure in eq. (\ref{eq-ER6}) is restricted
to a background matrix of the form $\sigma$ (eq. (\ref{eq-ER2})
and eq. (\ref{eq-ER5})).
However, one can work with a general $d\times d$ background
$E=G+B$ and a subset of the $2g$-dimensional Siegel
upper-half-space of the form
\begin{equation}
\tau_2=\left(
\begin{array}{cc}
T_2 & 0\\
0 & T_2
\end{array}
\right), \qquad
\tau_1=\left(
\begin{array}{cc}
0 & T_1 \\
-T_1 & 0
\end{array}
\right),
\label{eq-ER18}
\end{equation}
where $T_2$ is a $g\times g$ positive definite symmetric matrix
and $T_1$ is a $g\times g$ antisymmetric matrix.
This subset generates a space isomorphic to $O(g,g,\R)/(O(g,\R)\times
O(g,\R))$.
In this case, $\tau_1+i\tau_2$ is equal to $J\otimes (T_2+T_1)$,
where $J$ is the complex structure ($J^2=1$).
The matrix $(T_2+T_1)$ is real and in this sense $\tau$ is ``real.''
Defining $T=T_2+T_1$, one finds
\begin{equation}
Z_{class} (E,T)=Z_{class}(T,E),
\label{eq-ER19}
\end{equation}
namely, the classical partition function of a flat background described
by $E$, and a period matrix $\tau$ described by $T$, is the same
as that of a background described by $T$ on a Riemann surface
described by $E$. The subgroup of the genus-$2g$ modular group that
keeps $\tau$ real is $O(g,g,\Z)\subset Sp(4g,\Z)$.
Therefore, as an immediate  consequence of (\ref{eq-ER19}), we get that
the target space duality group, $O(d,d,\Z)$,
is a symmetry to all orders, when the period matrices are real.

If the target space is ``almost-complex'' and the period matrix
is ``real'' then there is a triality relation
\begin{equation}
Z_{class}(E_g, \tau_{4d})=Z_{class}(E_{2d},
\tau_{2g})=Z_{class}(E_{4g}, \tau_d),
\label{eq-ER20}
\end{equation}
where $\tau_a$ and $E_b$ stand for the relevant $a\times a$
period matrix and the $b\times b$ background matrix, respectively.
It is generalized to a multiality if the period matrix
and the background matrix have a higher ``complex-real''
structure. For example, let
\begin{eqnarray}
E_{2^n d} & =&\gamma\otimes I_{2^n} , \nonumber \\
\tau_{2^m g} &=& it\otimes I_{2^m} ,
\label{eq-ER21}
\end{eqnarray}
where $\gamma$ ( $t$ ) are $d\times d$ ( $g\times g$ )
positive-definite symmetric matrices.
Then there is a $(m+n+1)$-ality symmetry
\begin{eqnarray}
\ldots &= & Z(t\otimes I_{2^{m+1}},\gamma\otimes I_{2^{n-1}}) =
Z(\gamma\otimes I_{2^n}, t\otimes I_{2^m}) \nonumber \\
& = & Z(t\otimes I_{2^{m-1}}, \gamma\otimes I_{2^{n+1}})
= Z(\gamma\otimes I_{2^{n+2}}, t\otimes I_{2^{m-2}})=\ldots
\label{eq-ER22}
\end{eqnarray}
The reduction in the number of handles on the worldsheet gives
birth to copies of the target space $\gamma$, and vice versa.
In case $m=1$, $n=0$ or $m=0$, $n=1$ we get the two
possibilities of duality symmetry mentioned before (eqs.
(\ref{eq-ER6}) and (\ref{eq-ER19})).
If $m=n=1$ then we get the triality symmetry of eq.
(\ref{eq-ER20}).

Additional observations on the target space - worldsheet coupling
can be obtained by the following considerations.
The exponent in equation (\ref{eq-ER12}) can be written as
\begin{equation}
(n-\bar\tau m)^t \tau_2^{-1} G(n-\tau m)+2in^t Bm=
N^t(H_1 G+iH_2 B)N,
\label{eq-ER23}
\end{equation}
where we group the $n$ and $m$ into a column vector
\begin{equation}
N=\left(
\begin{array}{cc}
n \\
m
\end{array}
\right),
\label{eq-ER24}
\end{equation}
and the matrices $H_1$ and $H_2$ are defined to be
\begin{equation}
H_1=\left(
\begin{array}{cc}
\tau_2^{-1} & -\tau_2^{-1}\tau_1 \\
-\tau_1\tau_2^{-1} & \tau_2+\tau_1\tau_2^{-1}\tau_1
\end{array}
\right),
\qquad
H_2=\left(
\begin{array}{cc}
0 & I \\
-I & 0
\end{array}
\right) .
\label{eq-ER25}
\end{equation}
The complex matrix $H=H_1+iH_2$ is the natural Hermitian form of
the Jacobian torus \cite{Mumford}.
A Riemann surface with $g$ handles is described by a
$g$-dimensional complex torus (called the ``Jacobian torus'')
\begin{equation}
J^g=\C^g/\Lambda, \qquad \Lambda=\Z^g\otimes\tau\Z^g.
\label{eq-ER26}
\end{equation}
Here $\Lambda$ is the lattice with vectors $\pi_R=n-\tau m$, where
$n,m$ are $g$-vectors with integer components.
The Hermitian form $\bar\pi_R\tau_2^{-1}\pi_R$
$(\bar\pi_R=n-\bar\tau m)$ is related to $H$ as
\begin{equation}
\bar\pi _R\tau_2^{-1}\pi_R= N^t HN.
\label{eq-ER27}
\end{equation}
The vectors $\pi_R$ $(\pi_L=n+\tau m)$ are ``equivalent'' to the
``momenta'' $P_R$ $(P_L)$ of eq. (\ref{eq-ER16}) ($\tau$ is
replaced by $E$ and $J$ stands for $i$).
Changing $\tau$ to $\sigma$ in eq. (\ref{eq-ER25}), $H_1$ is
nothing but the metric on the phase space $Z=(X', P)$ of the
compactified string theory \cite{GRV} (\ref{ER18}).
Therefore, a close relation between a genus-$g$ Riemann surface and a
$2g$-dimensional toroidal compactification is suggested.

It is interesting to notice that the metric of the target space, $G$,
``feels'' the symmetric piece of the Jacobian torus
Hermitian form, $H_1$, while
the antisymmetric background, $B$, ``feels'' only the
antisymmetric piece, $H_2$.
The constant dilaton background, $\phi$, is coupled on the worldsheet to the
two-dimensional   scalar curvature $R^{(2)}$.
The 2-$d$ integration of $R^{(2)}$ leads to $g-1$, where $g$ is
the genus.
There exists a functional of $\phi$, leading to $d-2$ (or
$d-26$ with the ghost sector); that is the beta function of the
dilaton. The full set of constant parameters we are dealing with is $(G,
B, \phi, \tau_1, \tau_2)$.
The parameter $\phi$ is ``neutral'' under the exchange of $E$ with $\tau$.
However, $\phi$ might transform to some $\phi'$ in order to
keep the overall factor correct (this is also true for the $O(d,d,\Z)$
duality group symmetry on target space in higher genus).
Indeed, it is the transformation of the dilaton under duality in
eq. (\ref{tdil}) which
provides the correct factors. (For a detailed study of the transformation of
the dilaton, we refer the reader to subsection 4.2.4).

Having pointed out similarities between the worldsheet and target
space moduli, one can go a step further and attempt to view both as
part of a larger space \cite{Malkin}. This can be approached by considering
\cite{Malkin,Osorio} the symmetries of the classical partition function
of a string moving in a general background $E$ on a worldsheet manifold
described by $\tau$.
The classical partition function can be written as \cite{Malkin}:
\bea
& & Z_{class}(E,\tau)= \nonumber \\
& & \sum_{n,m \in \Z^{dg}} \exp \left\{ -\pi \left [
(n^t,m^t) \left( \begin{array}{cc} G\otimes (\tau_2 +\tau_1
\tau_2^{-1}\tau_1) & G\otimes (\tau_1\tau_2^{-1}) +iB\otimes I \\
G\otimes (\tau_2^{-1}\tau_1) -i B\otimes I & G\otimes \tau_2^{-1}
\end{array} \right) \left(\begin{array}{c} n \\ m \end{array}\right)
\right] \right\}.\nonumber \\ & &
\label{Malk1}
\eea
After performing a Poisson resummation it assumes the form
\bea
& & Z_{class}(E,\tau) =  \nonumber \\
& & \sum_{n,k \in \Z^{dg}} \exp \left\{ -\pi \left [
(n^t,k^t) \left( \begin{array}{cc} \tau_2 \otimes (G-BG^{-1} B) &
-\tau_2\otimes (BG^{-1}) -i\tau_1\otimes I \\
\tau_2\otimes (G^{-1}B) -i \tau_1\otimes I & \tau_2\otimes G^{-1}
\end{array} \right) \left(\begin{array}{c} n \\ k \end{array}\right)
\right] \right\}.\nonumber \\ & &
\label{Malk2}
\eea
Both matrices in the exponents of (\ref{Malk1}) and (\ref{Malk2}) are
symmetric, and their imaginary part is positive definite.
As such, they are elements of the Siegel upper-half space of
dimension $2dg$.  One can define a map
\be
E_d\otimes H_g \to H_{2dg},
\label{GPR-map}
\ee
where $E_d$ is a $d$-dimensional background and $H_g$ is the
one defined in eq. (\ref{eq-ER25}) for genus $g$.
The map takes $E$ and $\tau$ into the matrix
appearing in the exponent of (\ref{Malk1}). One already knows that the
group $O(d,d,\Z)$ acts on $E$ and the group $Sp(2g,\Z)$ acts on $\tau$, and
both groups are symmetries of the partition function. Under the embedding
(\ref{GPR-map}), both
groups are mapped into some subgroups of $Sp(4dg,\Z)$, which is the
modular group of $H_{2dg}$.  This is the symmetry of the partition
function. A generic element in $Sp(4dg,\Z)$ will mix $E$ and $\tau$.
However, only for the cases discussed above (such as when $E$ represented an
almost complex flat target space) one was able to identify an element
of $Sp(4dg,\Z)$ which would cause the change
$Z_{class}(E,\tau)=Z_{class}(\tau,E)$. The
importance of the role played by the $Sp(4dg,\Z)$ is still unclear.

We have studied mathematical identities of the classical piece of the bosonic
string partition functions. The formal interchange of the worldsheet
with the target space does not apply in an obvious way to the quantum part
of the partition function. If it so happens that a string theory contains
only a finite number of target space particles, no
worldsheet oscillators exist, and the mathematical identities
may acquire some physical reality.
In the next section we discuss one such case.

\end{subsection}

\begin{subsection}{Worldsheet/Target-Space Duality of $N=2$ Strings}
\setcounter{equation}{0}

Symplectic structures emerged in considering worldsheet/target-space
duality in the case when the target space was flat.  A similar structure
was considered also in a case of non-flat backgrounds \cite{OV}.
The setting is a theory with {\em local} $N=2$ worldsheet
supersymmetry  \cite{Aetal}.  Gauge fixing in
such theories leads to the following ghost systems: one associated
with general coordinate invariance which contributes $-26$ units to the
central charge $c$, two  associated with the two local
supersymmetries, contributing $c=22$, and one associated with a $U(1)$
gauge symmetry, contributing $c=-2$.  All in all the matter system
needs to contribute $c=6$ to arrange for the cancellation of the total
central charge. A system of four bosonic coordinates and four fermions has
the required central charge.  The reason that this four
dimensional system has not been the prime candidate for a string
background is that $N=2$ supersymmetry requires that the space
has either a (4,0) or a (2,2) signature.

In the (2,2) signature the model is still rather interesting. We will
just state a few of its features.
The target space spectrum of this model consists
of only a single massless scalar particle\footnote{A similar claim is
made for $c=1$ systems coupled to 2-$d$ gravity.}. This particle is
associated with a deformation of the K\"{a}hler structure of the
target space. Furthermore, this model does not contain
any oscillators: its partition function consists purely of
what was called the classical part in the previous section. It is thus an
interesting system for checking the physical realization of
mathematical identities of the type discussed above.

It is suggested \cite{OV} to
consider as the target space of this model a Ricci-flat K\"{a}hler manifold
constructed by associating to a given
Riemann surface a 4-dimensional symplectic space, called ``the
cotangent bundle of the Riemann surface.'' A cotangent space to a given
manifold of dimension $d$ is a $2d$-dimensional space. The extra
$d$-dimensional variables can be considered as the conjugate momenta to
the original coordinates, with all the extra structure a phase space
entails.  The target space itself
has also moduli; it has an equal number of complex-deformation
moduli and K\"{a}hler-class moduli. Thus, structurewise, there appears a
threefold similarity in the case the ``target space genus'' equals the
genus $g$  of the worldsheet. The
similarity is between the given worldsheet moduli, the complex-structure
moduli of the target space, and the K\"{a}hler-class
moduli of the target space.
For a genus-one case the partition function is known and is triality
symmetric under interchanges among all three moduli.  It is
conjectured \cite{OV} that the triality is a general property of the theory.

\end{subsection}

\end{section}

\vskip .3in \noindent
{\bf Acknowledgements} \vskip .2in \noindent
A large part of the material contained in this report is based on work done
in collaboration with S. Ferrara, L. Girardello, E. Kiritsis,
N. Malkin, A. Pasquinucci, M. Ro\v{c}ek and
G. Veneziano. We also thank S. Elitzur and D. Kazhdan
for many useful discussions.
The work of AG is supported in part by the BSF - American-Israeli Bi-National
Science Foundation, and by an Alon fellowship.
The work of ER is supported in part by
the BSF - American-Israeli Bi-National Science Foundation and the Israeli
Academy of Sciences.
This work is supported in part by the Center for Microphysics and Cosmology.
MP would like to thank the Racah Institute of Physics,
the Hebrew University, for its kind hospitality.

\newpage

\begin{table}[h]


\caption{Summary of Target Space Duality in Different Settings}

\vskip 0.3in

{\scriptsize

\begin{tabular}{|c|c|c|c|c|}
\hline
Background & The moduli-generating group ${\cal G}$ & The moduli
space ${\cal M}$ & The duality group ${\cal G}_d$ & Remarks \\ \hline \hline
$d$-dim & & & & $\zZ_2$ is the \\
toroidal & $O(d,d,\rR)$ & ${O(d,d,\rR)\over
O(d,\rR) \times O(d,\rR)}$ & $O(d,d,\zZ)\otimes_S \zZ_2 $ &
worldsheet \\ backgrounds & & & & parity\\ \hline
$D=4$  & & & & $N=4$ \\
toroidal  &  $O(6,22,\rR)$ & ${O(6,22,\rR)\over
O(6,\rR) \times O(22,\rR)}$ & $O(6,22,\zZ)\otimes_S \zZ_2 $ & $D=4$
\\ heterotic & & & & SUSY \\ \hline
$D=4$  & & ${\cal M}={\cal K}\times {\cal Q}$ & & untwisted \\
$\zZ_N \subset SU(2)$ & & ${\cal K}$=K\"ahler manifold & &
sector \\
orbifolds & & ${\cal Q}$= quaternionic manifold & & \\
\cline{1-1}\cline{3-3}\cline{5-5} & & & &  $N=2$\\
$\zZ_2$ & ${O(2,2,\rR) \times O(4,4,\rR)}$ &
${O(2,2,\rR)\over O(2,\rR) \times O(2,\rR) } \times {O(4,4,\rR)\over
O(4,\rR)\times O(4,\rR)}$ & $O(2,2,\zZ) \times O(4,4,\zZ)$ &
$D=4$ \\ & & & & SUSY.\\
$\zZ_3,\zZ_4,\zZ_6$ & ${O(2,2,\rR) \times O(2,4,\rR)}$ &
${O(2,2,\rR)\over O(2,\rR) \times O(2,\rR) } \times {O(2,4,\rR)\over
O(2,\rR)\times O(4,\rR)}$ & $O(2,2,\zZ) \times O(2,4,\zZ)$ &
Submoduli \\ & & & & of $T^2\times {\cal K}_3$ \\
\hline & & & & \\
$T^2\times {\cal K}_3$ & $O(2,2,\rR)\times O(4,20,\rR)$ &
${O(2,2,\rR)\over O(2,\rR)\times O(2,\rR)} \times {O(4,20,\rR)\over
O(4,\rR)\times O(20,\rR)}$ & $O(2,2,\zZ) \times O(4,20,\zZ)$ & conjectured
\\ & & & & \\ \hline
$\zZ_N\subset SU(3)$  & & & & untwisted \\
orbifolds & & & & sector \\
\cline{1-1} \cline{5-5}
$\zZ_3$ & $SU(3,3,\rR)$ & ${SU(3,3,\rR)\over U(1)\times
SU(3,\rR)\times SU(3,\rR)}$ &
$SU(3,3,\zZ)$ & $N=1$\\ & & & & $D=4$ \\
$\zZ_6$ &  $SU(2,2,\rR)\times SU(1,1,\rR)$ & ${SU(2,2,\rR)\over U(1)\times
SU(2,\rR)\times SU(2,\rR)} \times {SU(1,1,\rR)\over U(1)}$ &
$SU(2,2,\zZ)\times SU(1,1,\zZ)$ & SUSY. \\ & & & & \\
$\zZ_7,\zZ_8,\zZ_{12}$ & $SU(1,1,\rR)^3$ &
$\left({SU(1,1,\rR)\over U(1)}\right)^3$ & $SU(1,1,\zZ)^3$ & Submoduli \\
& & & &  of CY  \\
$\zZ_4$ & $SU(2,2,\rR)\times O(2,2,\rR)$ & ${SU(2,2,\rR)\over U(1)\times
SU(2,\rR) \times SU(2,\rR)} \times { O(2,2,\rR) \over O(2,\rR)
\times O(2,\rR)}$
& $SU(2,2,\zZ) \times O(2,2,\zZ)$ & spaces \\ & & & & \\
$\zZ_{6'},\zZ_{8'},\zZ_{12'}$ & $SU(1,1,\rR)^2 \times O(2,2,\rR) $ &
$ \left( {SU(1,1,\rR)\over U(1)}\right)^2 \times {O(2,2,\rR)\over
O(2,\rR)\times O(2,\rR)}$ & $SU(1,1,\zZ)\times O(2,2,\zZ)$ & \\ & &
& & \\
\hline
Calabi-Yau & it does not exist in general & special K\"ahler
manifold, & mirror symmetry and & \\
spaces & & ${\cal M}={\cal M}_{(1,1)}\times {\cal M}_{(2,1)}$,
& monodromy group,& \\  & & unknown in general  & unknown in general & \\
\hline Curved & & & & axial-vector \\
backgrounds  & $O(d,d,\rR)$ & ${O(d,d,\rR)\over {\cal H}}$ &
$O(d,d,\zZ)$ & duality \\
with $U(1)^d$ & & & &  is in ${\cal G}_d$\\
symmetries & & $O(d,\rR)_{diag} \subseteq {\cal H} \subseteq
O(d,\rR)\times O(d,\rR)$ & \\ & & & & \\ \hline
\end{tabular}}

\end{table}

\newpage

\font\twelvebf=cmbx12
\renewcommand{\Large}{\normalsize}
\renewcommand{\bf}{\sc}
\newcommand{\np}{Nucl.\ Phys.\ }
\newcommand{\pr}{Phys.\ Rev.\ }
\newcommand{\cmp}{Commun.\ Math.\ Phys.\ }
\newcommand{\pl}{Phys.\ Lett.\ }

\newpage
\noindent
{\bf FIGURE CAPTIONS}

\begin{description}

\item[1.A:]
A string cannot tell if it is moving on circle (a) or circle (b).

\item[1.B:] Every point in the shaded region corresponds to an allowed
string background. Points (a),(b),(c) correspond to isolated single
backgrounds.

\item[1.C:]
The line describes the space of compactifications on a circle
of radius $R_c$. Arrows connect physically equivalent points related by
duality.

\item[1.D:]
An orbicircle obtained from a circle by identifying $X$ with $-X$.

\item[1.E:] From the point of view of a string, a circle (at a special
radius) is equivalent to an orbifold (at another special radius).

\item[1.F:] String propagation on (a), the semi-infinite cigar, is
equivalent to propagation on (b), the infinite trumpet.

\item[2.A:] Genus-$g$ Riemann surface with cycles $a, b$.

\item[2.B:] The $\rho=\tau$ slice of the fundamental domain of 2-$d$
toroidal backgrounds. The points (a) and (b) are the enhanced symmetry
points $SU(2)^2$ and $SU(3)$, respectively. The shaded region takes into
account also the $B\to -B$ symmetry.

\item[3.A:] The $c=1$ moduli space. Arrows connect physically
equivalent points related by duality. The self-dual point, $R_c=1$,
possesses an
enhanced $SU(2)_L\times SU(2)_R$ affine symmetry. The self-dual point,
$R_o=1$,
possesses an enhanced $U(1)_L\times U(1)_R$ affine symmetry; it is
equivalent to $R_c=2\sqrt{2}$.

\item[4.A:]
The maximal extension of the black hole solution in Kruskal coordinates.

\end{description}

\newpage

\end{document}